\documentclass[journal, twoside, onecolumn, final]{IEEEtran}
\usepackage{pdfpages}
\usepackage[utf8x]{inputenc}
\usepackage[T1]{fontenc}
\usepackage{setspace} 
\usepackage[cmex10]{amsmath}
\usepackage{amssymb}
\usepackage{mathtools}
\usepackage[caption=false]{subfig}
\usepackage{graphicx}
\usepackage{multirow}

\title{Angular Resolution of Closely-Spaced Targets with Antenna Arrays}

\DeclareMathOperator{\lin}{lin}
\DeclareMathOperator{\diag}{diag}
\DeclareMathOperator{\E}{E}
\DeclareMathOperator{\var}{var}
\DeclareMathOperator{\trace}{trace}
\DeclareMathOperator{\grad}{grad}
\DeclareMathOperator{\sign}{sign}

\newenvironment{bottompar}{\par\vspace*{\fill}}{\clearpage}

\begin{document}
\pagenumbering{roman}

\maketitle
\vspace{3cm}
\begin{center}
Approved dissertation from the Department of Electrical Engineering\\
of the Rheinisch-Westfälische Technische Hochschule Aachen\\
to obtain the academic degree of Doctor of Natural Sciences
\end{center}
\vspace{5cm}
\begin{center}
Presented by Diplom-Mathematician Ulrich Nickel\\
from Burg auf Fehmarn\\
~\\
Authorized translation from German by Dr. David F. Crouse\\
U. S. Naval Research Laboratory\\
April 2019
\end{center}

\begin{bottompar}
\noindent Advisor: Professor Dr. ret. nat. H. Lueg\\
Thesis Committee Member: Professor Dr.-Ing. H. D. Lüke\\
Day of the Oral Examination: 17. December 1982
\end{bottompar}
\newpage

I would like to heartily thank Professor Dr. ret. nat. H. Lueg for his nurturing questions and friendly support of this work. I would also like to thank Professor Dr. -Ing. H. D. Lüke for his acceptance of a role on my thesis committee.
\vspace{3cm}

My thanks also go out to Dr. ret. nat. H. Springer and Dr. -Ing. W. D. Wirth, who made the completion of this work at the FFM possible. I would like to thank Dr. Wirth and my colleagues for the numerous suggestions and discussions.
\newpage

\tableofcontents

\section*{Mathematical Symbols and Terminology}
\addcontentsline{toc}{section}{Mathematical Symbols and Terminology}
\begin{center}
\begin{tabular}{cl}
$\bigwedge\limits_{x\in A}:H(x)$						& ``For all $x\in A$ the statement $H(x)$ is true''\\
$\bigvee\limits_{x\in A}:H(x)$							& ``There exists an $x\in A$ such that $H(x)$ is\\
												&true''\\
$\wedge$, $\vee$,									& ``and'', ``or'' (non-exclusive or)\\
$\mathbb{N}$, $\mathbb{Z}$, $\mathbb{R}$, $\mathbb{C}$	&Set of natural, integer, real and complex\\
												&numbers\\
$\mathbb{R}_{+}$									&The set of positive, real numbers\\
$\mathbb{C}^{N\times M}$ $\mathbb{R}^{N\times M}$		&The set of $M\times N$ matrices\\
												&over $\mathbb{C}$, $\mathbb{R}$\\
$A\backslash B:=\{x\in A|x\notin B\}$	\!\!\!\!\!\!\!\!					&The difference set\\
$\bar{A}$											&The closure of the set $A$\\
$A\oplus B$										&The direct sum of $A$ and $B$\\
$\mathbf{z}^*$										&A complex conjugate, transposed\\
												&vector.\\
$\lVert\ldots\rVert$									&The quadratic sum norm in $\mathbb{C}^N$\\
												&$\left(\mathbb{R}^N\right)$ ($2$-norm)\\
$\mathbf{z}\sim\mathcal{N}_{\mathbb{C}^N}\left(\mathbf{s},\mathbf{R}\right)$&``$\mathbf{z}$ is a normally distributed \\
												&random variable in $\mathbb{C}^N$ with \\
												&expected value $\mathbf{s}$ and covariance\\
												&matrix $\mathbf{R}$''\\
$z\sim\mathcal{R}(M)$								&``$z$ is uniformly distributed on the\\
												&set $M$'' ($M\subset\mathbb{R}$ or $\mathbb{R}^N$)\\
$I_M(x)$											&Indicator function,\\
												&$I_M(x)=\left\{\begin{IEEEeqnarraybox}[\relax][c]{c.s}
1&if $x\in M$\\
0&otherwise
\end{IEEEeqnarraybox}\right.$\\
$\delta_{ik}$										&Kroneker Symbol,\\
												&$\delta_{ik}=\left\{\begin{IEEEeqnarraybox}[\relax][c]{c.s}
1&if $i= k$\\
0&if $i\neq k$
\end{IEEEeqnarraybox}\right.$\\
$:=$, $:\Longleftrightarrow$							&Defining equality, equivalence,\\
												& $x:=y$ ``$x$ is defined by $y$''\\
$f:$$A\rightarrow B\atop x\mapsto f(x)$					&$f$ is the maping from set $A$ to\\
												&set $B$ and every $x\in A$ is assigned\\
												&to an $f(x)\in B$, also $f\atop A\rightarrow B$\\
$f_x:=\frac{\partial f}{\partial x}$							&Partial derivative with respect to $x$\\
$f_{\mathbf{x}}:=\left(\begin{IEEEeqnarraybox}[\relax][c]{c}
\IEEEstrut
\frac{\partial f}{\partial x_1}\\
\vdots\\
\frac{\partial f}{\partial x_N}
\IEEEstrut
\end{IEEEeqnarraybox}\right)$							&The gradient from $f$\\
$P(x)$											&Probability distribution of $x$\\
$p(x)$											&Probability density of $x$\\
$P(M)$											&Probability of the event $M$\\
												&$P(M)=\int\limits_MdP(x)$\\
$N$												&Number of antennas\\
$M$												&Number of targets\\
$\lambda$										&Wavelength\\
$(x_i,y_i),i=1,\ldots,N$								&Coordinates of the antenna elements\\
												&either in centimeters or in units of $\frac{2\pi}{\lambda}$\\
$\mathbf{z}=\mathbf{s}+\mathbf{n}$						&Measurement vector consisting of\\
												&signal and noise.
\end{tabular}
\begin{tabular}{cl}
$\omega$											&A direction in the form of a direction cosine. \\
												& $\omega=(u,v)$ for planar arrays.\\
$\mathbf{a}(\omega)$								&The antenna transfer function for a direction $\omega$\\
$\mathbf{A}(\boldsymbol{\omega})$						&Antenna transfer matrix for $M$ directions $\boldsymbol{\omega}$\\
$\mathbf{b}$										&Vector of $M$ complex amplitudes\\
$\mathbf{B}$										&$:=\E\left\{\mathbf{b}\mathbf{b}^*\right\}$\\
$\mathbf{R}$										&$:=\E\left\{\mathbf{z}\mathbf{z}^*\right\}$\\
$BW$											&Half beamwidth, $BW=0.887\frac{\lambda}{D}$ radians or $51\frac{\lambda}{D}$\\
												&degrees, where $D$ is the aperture in centimeters.\\
$V^M$											&$:=\{\boldsymbol{\omega}|\bigwedge\limits_{i\in1,\ldots,M}u_i^2+v_i^2\leq1\}$ \\
												&The viewing area of $M$ directions\\
$V_{\wp}(\boldsymbol{\omega}_g)$						&$:=\{\boldsymbol{\omega}\in V^M|\!\!\!\bigwedge\limits_{i\in1,\ldots,M}\!\!\!\!\!(u_i\!-\!u_{ig})^2\!+\!(v_i\!-\!v_{ig})^2<\wp^2\}$\\
												&$\wp$ is the area around $\boldsymbol{\omega}_g$\\
$M(\boldsymbol{\omega}_g)$							&$:=\{\boldsymbol{\omega}\in V_{BW/2}(\boldsymbol{\omega}_g)|u_1<u_2<\ldots u_M\}$\\
$\substack{N\\ \diag(\alpha_i)\\ i=1}$						&$:=\left(\begin{IEEEeqnarraybox}[\relax][c]{c.c.c}
\alpha_1&&0\\
&\ddots&\\
0&&\alpha_N
\end{IEEEeqnarraybox}\right)$\\
$\mathbf{I}$										&$:=\diag(1)$, the identity matrix\\
$\boldsymbol{\Gamma}$								&$:=\mathbf{I}-\mathbf{A}(\mathbf{A}^*\mathbf{A})^{-1}\mathbf{A}^*$ A projection matrix that\\
												&projects onto $\lin H (A)^{\bot}$.\\
$\lin H (\mathbf{x}_1,\ldots,\mathbf{x}_k)$					&$:=\left\{\mathbf{y}\left|\bigvee\limits_{\alpha_1\ldots\alpha_k}:\mathbf{y}=\sum_{i=1}^k\alpha_1\mathbf{x}_i\right.\right\}$ is the linear hull of $\mathbf{x}_1,\ldots,\mathbf{x}_k$.
\end{tabular}\\

\end{center}

\pagenumbering{arabic}
\setcounter{page}{1}

\section{Introduction}\label{Sec1}

Electronically steerable antenna arrays, whereby the signal received by each individual element is accessible, provide more information about the received signal than reflector antennas. This work concerns itself with the utilization of this additional information to angularly resolve closely-spaced point targets.

The detection and location of a target with conventional signal-processing techniques consists, in principle, of scanning a beam over the target and determining the maximum of the radar scan pattern. For localization, a full scan does not have to be explicitly performed. Rather, an estimated value of the slope of the radar scan pattern can be measured (monopulse processing). By such processing, the angular resolution, that is the realization, for example, that two and not one nor three targets are present, is limited by the beamwidth of the antenna response such that even by an arbitrarily good signal-to-noise ratio (SNR), two closely-spaced targets yield only a single maximum in the radar scan pattern. By a finite SNR the resolution limit, which is statistically defined by confidence intervals, is naturally larger.

The aperture size and frequency, which play a decisive roles in determining the antenna response of the system, are generally limited by design considerations.  Thus, for a fixed aperture size and frequency, attempts have been made to reduce the beamwidth. With reflector antennas, this has been done by changing the shape of the surface, for antenna arrays, simply through the use of complex weights (super gain antennas). However, such designs stand out for their sensitivity to receiver noise, which is not taken into account in such techniques.

The stochastic formulation of the resolution problem by a given type of noise and the development of a corresponding optimal estimation procedure appears to be needed. The more precise the signal and interference models that are chosen, the more accurate the resolution procedure will be, as long as the data matches the model. However, the increased precision leads to a worsening of the estimates when a model mismatch exists. This dilemma of precision versus robustness is common to all superresolution algorithms. Consequently, the determination of the correct parameterization  of the model greatly influences the performance of the algorithm because without \emph{a priori} information, no superresolution beyond conventional limits is possible.

In this work, the signals are assumed to be plane waves originating from infinitely distant point targets. This assumptions appears to be reliable for radar applications. The signals are spatially sampled by an antenna array, whereby the receiver noise is additive Gaussian. These are the fundamental assumptions in the model used here.

There exists no satisfactory optimality criterion for the assumed model. Therefore,  a solution using the classical maximum-likelihood (ML) principles is investigated in order to develop a procedure that can be realized in a real radar. The realizability of the procedure must be judged according to the desired application. Here, the suppression of multipath error by low-altitude targets and the detection and tracking of groups of aircraft or jammers is the intended application. The procedure being studied is not significantly worse than the corresponding estimation and tests using the ML principle, which is used as an "asymptotically optimal" algorithm.

The resolution algorithm studied here is formulated for electronically steerable array antennas. However, it is also suitable for the resolution of planar waves using spatially separated antennas as well as the spectral-line resolution of signals sampled in time.

Increasing the resolution capability of a system always necessitates a higher SNR than needed for the pure detection or localization of a target. These high values  are attained by a target flying close to the radar, as every halving of the distance increases the SNR by $12\,\text{dB}$ according to the $r^4$ law. Viewing the $3$-$\text{dB}$ distance of a target as the range of the radar and demanding $18$ to $23\,\text{dB}$ for the resolution of two targets separated by a half beamwidth (See Chapters 8 and 10), this SNR is reached by $0.4$ to $0.3$ of the $3$-$\text{dB}$ distance. One can always assume that a higher SNR is given for the resolution of jammers.

\section{Model Assumptions and Terminology}\label{Sec2}
\begin{center}
\textbf{Summary of Chapter 2}
\end{center}
First, the underlying point-target model is discussed. This can be written in terms of the measured antenna-array outputs $\mathbf{z}$ as $\mathbf{z}=\mathbf{A}\mathbf{b}+\mathbf{n}$, where $\mathbf{b}$ is a vector of unknown, complex amplitudes, and $\mathbf{A}$ is the  transfer matrix of the antenna array with the form given in (2.1-2). The additive noise on all channels of the array is represented by $\mathbf{n}$.

In this chapter, Signal Models $1$--$4$, differing in the distributions of the complex amplitudes, and various antenna-array configurations are given for use in the simulations. Covariance matrices of the noise $\E\{\mathbf{n}\mathbf{n}^*\}$ and the corresponding covariance matrices of the antenna outputs $\E\{\mathbf{z}\mathbf{z}^*\}$ are calculated for various types of noise. These covariance matrices will be needed in chapters $4$, $5$, and $7$ to calculate some expected values. They are primarily needed for the computational generation of normally distributed random variables for noise in the simulations.
~\\
\hrule

At a particular sensing point (antenna element), an additive combination of signals and noise is measured. When dealing with radar, the signal generally originates from an active transmitter, which is not important for the overall signal model.

\subsection{Signal Model}

The targets that are to be resolved should be in the far field of the radar, and their width should be significantly less than half the beamwidth; that is, they are point targets. Theoretically, the danger exists that under this signal model extended targets are resolved into individual scatterers. However, with the parameters of radars that are currently in use, the ability to resolve individual scatterers is far from their obtainable resolution: with $2^\circ$ beamwidth a target of $40\,\text{m}$ width takes up approximately $5.7\%$ of the beamwidth at a distance of $20\,\text{km}$.

Considering an isotropic element at position $(x,y)$ on the antenna aperture, neglecting noise, at the output of a narrowband receiver one gets the complex signal (the down-converted quadrature components) from $M$ targets:
\begin{equation}
s=\sum_{i=1}^M\beta_ie^{j\varphi_i}e^{-j\frac{2\pi}{\lambda}\left(xu_i+yv_i\right)}\label{eq2-1-1}
\end{equation}
where $\beta_i$, $\varphi_i$ are the amplitude and phase of the $i$th target, $\lambda$ is the wavelength, and $u_i$ is the direction cosine of the azimuth $\vartheta_i$ and  elevation $\psi_i$ of the $i$th target ($u_i=\sin \vartheta_i \cos \psi_i, v_i=\sin \psi_i,i=1\ldots M$).

Combining $\beta_ie^{j\varphi_i}$ together into a ``complex amplitude'' $\mathbf{b}_i$, the signal vector representing the output of an antenna array with $N$ elements on positions $(x_k,y_k),k=1\ldots N$ is
\begin{equation}
\mathbf{s}=\mathbf{A}\mathbf{b}\in\mathbb{C}^N\label{eq2-1-2}
\end{equation}
with
\begin{align*}
\mathbf{b}\in&\mathbb{C}^M\\
a_{ki}=&e^{-j\frac{2\pi}{\lambda}(x_ku_i+y_kv_i)}\quad(i=1\ldots M,k=1\ldots N).
\end{align*}
To simplify the notation, the antenna elements shall be expressed in units of $2\pi/\lambda$ so that
\begin{equation*}
a_{ki}=e^{-j(x_ku_i+y_kv_i)}
\end{equation*}
as long as a fixed frequency is considered.

\begin{figure}
\centering
\subfloat[ELAN $25$]{\includegraphics[width=0.4\textwidth]{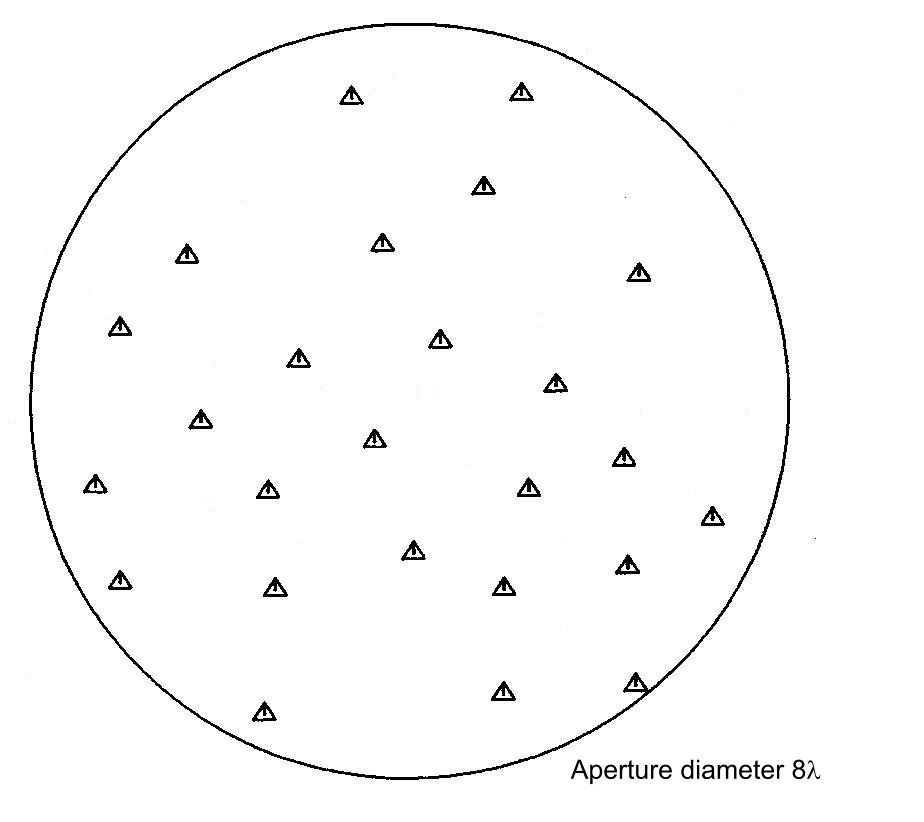}}\quad\quad
\subfloat[ELAN $39$]{\includegraphics[width=0.4\textwidth]{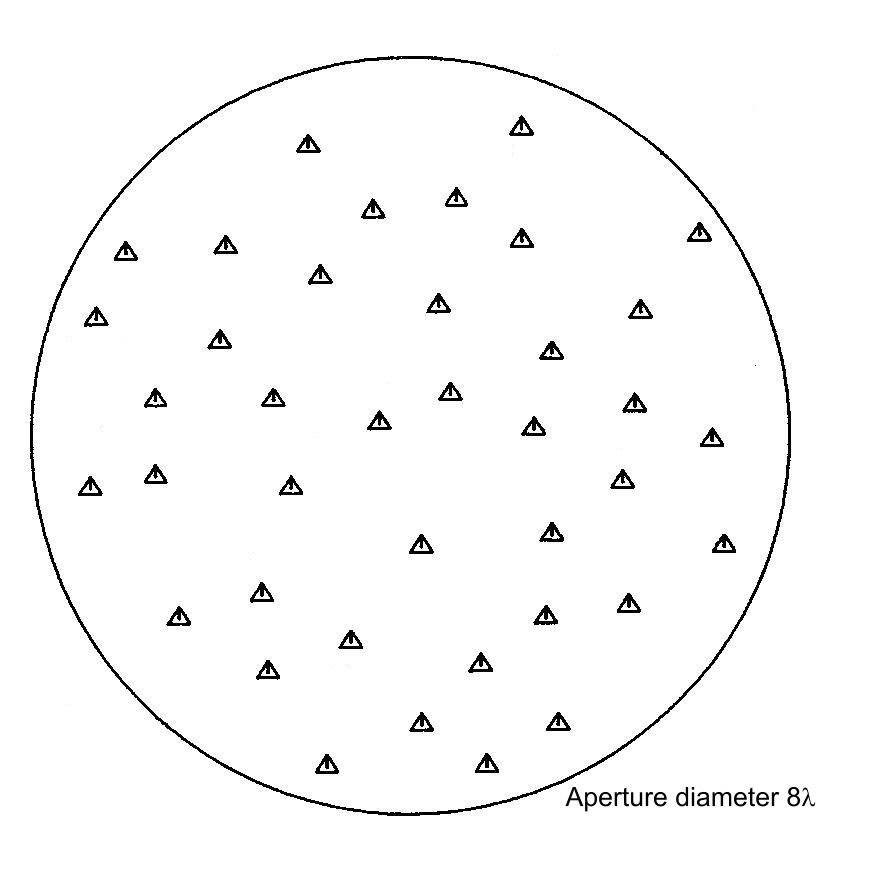}}
\caption{Layout of the receiver antenna arrays. The diameter of the antennas is 8 times the wavelength. \label{Fig2-1}}
\end{figure}

\begin{figure}
\centering
\includegraphics[width=0.4\textwidth]{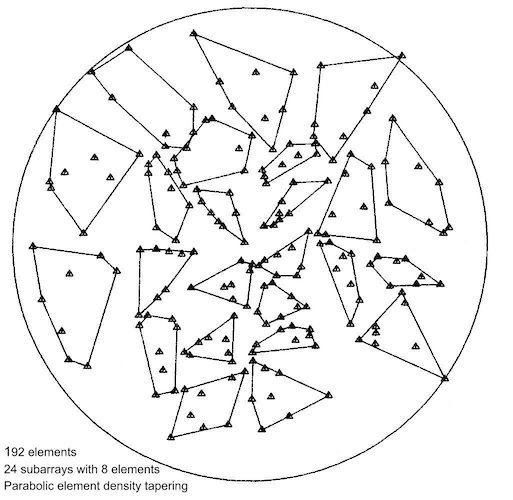}
\caption{Layout of the receiver antenna array ELAN 192.  ELAN 192 consists of $192$ antenna elements combined into $24$ subarrays, each with $8$ elements, diameter $37\lambda$, and parabolic element density tapering.\label{Fig2-2}}
\end{figure}

The following different examples of antennas shall be considered
\begin{itemize}
\item ELAN $k$ L: A linear antenna with $k$ elements regularly spaced $\lambda/2$ apart (Nyquist sampling).
\item ELAN 6: A planar array with diameter $4\lambda$. 
\includegraphics[width=0.25\textwidth]{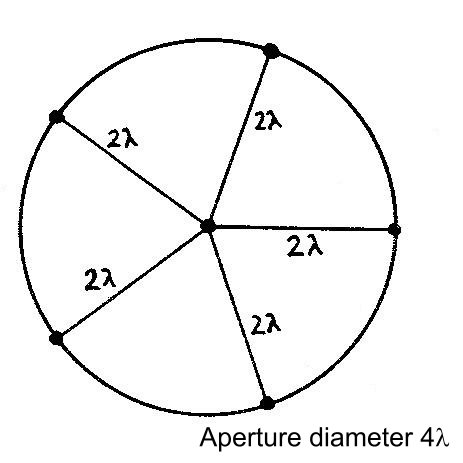}
\\
\item ELAN 25, 29 and ELAN 192 as shown in Figs. \ref{Fig2-1} and \ref{Fig2-2}.
\end{itemize}
The antenna ELAN 192 is designed combining the elements into $24$ subarrays. The arrangement of the subarrays matches that used in the FFM's electronic radar system ELRA.\footnote{The dissertation references an image of the ELRA system, which is was omitted from the dissertation. Additionally, ELRA=elektronisches Radar (electronic radar), and was an important project at the FFM for a long time.}

The $3$-$\text{dB}$ beamwidth $BW$ of these antennae is approximated using the formulae $BW=0.887\lambda/D$ radians or respectively $BW=51\lambda/D$, with $D$ degrees ($D$ is the aperture diameter in centimeters). This formula is exact for linear, continuously filled antennas with a $\sin x/x$ response. The ELAN antennas all have a considerably larger beamwidth. In the literature, the $0-0$-width of the  $\sin x/x$ characteristic is often used. This must be taken into account when considering the results.

For the direction cosines, the relation $u^2+v^2\leq 1$ holds, so that for $M$ directions, given as
\begin{equation*}
\boldsymbol{\omega}:=\left(\begin{IEEEeqnarraybox}[\relax][c]{c}
\IEEEstrut
\mathbf{u}\\
\mathbf{v}
\IEEEstrut
\end{IEEEeqnarraybox}\right)=(u_1\ldots u_M,v_1 \ldots v_M)^T,
\end{equation*}
the following relation is true
\begin{equation*}
\boldsymbol{\omega}\in V^M:=\left\{\left.\left(\begin{IEEEeqnarraybox}[\relax][c]{c}
\IEEEstrut
\mathbf{u}\\
\mathbf{v}
\IEEEstrut
\end{IEEEeqnarraybox}\right)\in\mathbb{R}^{2M}\right|\bigwedge\limits_{i\in \{1 \ldots M\}} : u_i^2+v_i^2\leq1\right\}.
\end{equation*}
For the $M$ given directions $\boldsymbol{\omega}_g$ denotes a $\wp$ neighborhood with
\begin{equation}
V_\wp(\boldsymbol{\omega}_g):=\left\{\boldsymbol{\omega}\left|\bigwedge\limits_{i\in1,\ldots,M}(u_i-u_{ig})^2+(v_i-v_{ig})^2<\wp^2\right.\right\}\label{eq2-1-3}
\end{equation}
where $\wp\in\mathbb{R}_{+}$.

The signal structure assumed in (2.1-2) is actually distorted by the whole bandwidth of the receiver, the spread of the target, the response of the individual array elements, and their coupling. Therefore, the algorithm derived from this model should be as robust as possible with regard to such distortions (See Chapters 5.4, 7.3). 

When considering a sequence of subsequent measurements, the amplitudes $b_i$ will experience fluctuations. These fluctuations depend on the aspect-angle-dependent reflective properties of the aerial targets originating on the one hand from their ever-present roll and pitch motions (slow fluctuations) and on the other hand from reflections off of propellers (if present), vibrations of the fuselage, or reflections from the ionized gas jets of jet engines (fast fluctuations) \cite{ref1}.

The fluctuations strongly depend on the type of aircraft, and they depend even more so on the wavelength and revisit rate. The revisit rate of an electronic radar can be very irregular, for example, depending on the load condition of the system. Thus, the superresolution algorithm should not make any assumptions regarding the fluctuations.

The following four target models shall be used as examples during the analysis:

\begin{itemize}
\item \underline{Model 1: Deterministic Target}\hfill (2.1-4)\\
$\beta_i$ is constant; $\varphi_i$ is constant or changes with Doppler angle, ($i=1\ldots M$)

\item \underline{Model 2: Normally-Distributed Phase Fluctuations}\\
$\beta_i$ is constant, $\varphi_i\sim\mathcal{N}_{\mathbb{R}}(\bar{\varphi}_i,\sigma_i^2)$. This results in:
\begin{align*}
\E\left\{b_i\right\}=&\beta_ie^{-\sigma_i^2/2}e^{j\bar{\varphi}_i}\\
\E\left\{b_ib_k^*\right\}=&\left\{\begin{IEEEeqnarraybox}[\relax][c]{c's}
\beta_i\beta_ke^{-(\sigma_i^2+\sigma_k^2)/2}e^{j(\bar{\varphi}_i-\bar{\varphi}_k)}&for $i\neq k$\\
\beta_i^2&for $i=k$
\end{IEEEeqnarraybox}\right. .
\end{align*}
Since $\mathbf{B}:=\E\left\{\mathbf{b}\mathbf{b}^*\right\}$, having the form ``diagonal matrix plus dyad,'' $\mathbf{B}$ is a regular matrix.

\item \underline{Model 3: Uncorrelated Targets with Fixed Amplitudes}\\
$\beta_i$ is constant; $\varphi_i$ is distributed uniformly on $(0,2\pi)$. Thus holds
\begin{align*}
\E\left\{\mathbf{b}\right\}=&\mathbf{0}\\
\E\left\{\mathbf{b}\mathbf{b}^*\right\}=&{M \atop{\diag(\beta_i^2)\atop i=1}}.
\end{align*}
\item \underline{Model 4: Uncorrelated Rayleigh Targets} (Swerling II-Model) \cite{ref1}\\
$\beta_i$ are independently Rayleight distributed and $\varphi_i$ are independently uniformly distributed on $(0,2\pi)$, meaning that
\begin{equation*}
b_i\sim\mathcal{N}_{\mathbb{C}}(0,\sigma_i^2),\quad i=1\ldots M
\end{equation*}
which is equivalent to
\begin{equation*}
\mathbf{b}\sim\mathcal{N}_{\mathbb{C}^M}(\mathbf{0},\mathbf{B})
\end{equation*}
with
\begin{equation*}
\mathbf{B}={M \atop{\diag(\sigma_i^2)\atop i=1}}.
\end{equation*}
\end{itemize}

Models $1$ and $4$ represent two extremes. Model $1$ with a constant amplitude is a limiting case of Model $2$. Theorems for stochastic signals are thus also valid for deterministic signals in the limit.

It is hard to specify conditions for the validity of these models, because little measured data is known. With regard to the RADICORD experiments of the FFM \cite{ref2}, Model $2$ was determined to hold for a jet engine aircraft with $\sigma(\varphi)\approx 15^\circ$ from pulse-to pulse (a $2\,\text{ms}$ pulse repetition period). 

\subsection{Noise/ Interference}\setcounter{equation}{0}

The receive array should consists of $N$ independent, isotropic elements so that the sampled values of the noise are uncorrelated, normally distributed random variables when white Gaussian noise in the receiver is assumed. At the output of the antenna array, one has the data vector
\begin{equation*}
\mathbf{z}=\mathbf{s}+\mathbf{n}\in\mathbf{C}^N
\end{equation*}
with
\begin{equation*}
\mathbf{n}\sim\mathcal{N}_{\mathbb{C}^N}(\mathbf{0},\sigma^2\mathbf{I}).
\end{equation*}

In most cases, the normalization $\sigma^2=1$ can be performed (An exception is just Chapter 7). Thus one has
\begin{itemize}
\item For Signal Model 1:\\
\begin{equation*}
\mathbf{z}\sim\mathcal{N}_{\mathbb{C}^N}(\mathbf{s},\mathbf{I})
\end{equation*}
\item For Model 2:
\begin{align*}
\E\left\{\mathbf{z}\right\}=&\mathbf{A}\E\left\{\mathbf{b}\right\}\\
\E\left\{\mathbf{z}\mathbf{z}^*\right\}=&\mathbf{I}+\mathbf{A}\mathbf{B}\mathbf{A}^*
\end{align*}
with
\begin{equation*}
B_{ik}=\left\{\begin{IEEEeqnarraybox}[\relax][c]{c's}
\beta_i\beta_ke^{j(\bar{\varphi}_i-\bar{\varphi}_k)}e^{-\frac{\sigma_i^2+\sigma_k^2}{2}}&For $i\neq k$\\
\beta_i^2&For $i=k$
\end{IEEEeqnarraybox}\right.
\end{equation*}

\item For Model 3:
\begin{align*}
\E\left\{\mathbf{z}\right\}=&\mathbf{0}\\
\E\left\{\mathbf{z}\mathbf{z}^*\right\}=&\mathbf{I}+\mathbf{A}\mathbf{B}\mathbf{A}^*
\end{align*}
with
\begin{equation*}
B_{ij}=\beta_i^2\delta_{ij}
\end{equation*}
\item For Model 4:
\begin{equation*}
\mathbf{z}\sim\mathcal{N}_{\mathbb{C}^N}(\mathbf{0},\mathbf{I}+\mathbf{A}\mathbf{B}\mathbf{A}^*)
\end{equation*}
with
\begin{equation*}
B_{ij}=\sigma_i^2\delta_{ij}.
\end{equation*}
\end{itemize}

These assumptions about the noise are generally not satisfied. The array elements are not all the same, so that the covariance matrix of $\mathbf{n}$ is diagonal. One also has external interference so correlations arise such that $\mathbf{n}\sim\mathcal{N}(\mathbf{0},\mathbf{R})$ must be assumed.

During the formulation of the resolution problem in Chapter 3, it is initially assumed that $\mathbf{n}\sim\mathcal{N}(\mathbf{0},\mathbf{R})$. However, since $\mathbf{R}$ is generally unknown and also not easy to estimate, the derivation of the superresolution method assumed that $\mathbf{n}\sim\mathcal{N}(\mathbf{0},\mathbf{I})$ and the robustness of this assumption is demanded. However, when performing superresolution, that is, when investigating the fine structure of the signal, the noise background must be better known than when simply detecting a target.

For the simulations, external interference is always modeled as white-noise jamming; that is, they are distributed according to Model 4. To avoid the case that a jammer happens to fall on a null of the antenna response, such jammers are modeled as a collection of closely-spaced point jammers so that in the limiting case one has ``angularly limited'' white noise. At the array element at position $(x,y)$, one receives the jamming signal
\begin{equation*}
n_S(x,y)=\int_{V}e^{-j(xu+yv)}\,db(u,v)
\end{equation*}
whereby $b(\omega)$ is a complex-valued stochastic process with
\begin{align*}
\E\{b(\omega)\}=&0\\
\E\{db(\omega)db^*(\omega')\}=&B(\omega)\delta(\omega-\omega')d\omega\,d\omega'
\end{align*}
so that
\begin{align}
R(x,y,x',y'):=&\E\left\{n_S(x,y)n_S^*(x',y')\right\}\notag\\
=&\int_Ve^{-j((x-x')u+(y-y')v)}B(u,v)du\,dv\label{eq2-2-1}
\end{align}

Only constant power densities are considered: With regard to a linear array on the interval $[\bar{u}-s,\bar{u}+s]$ with
\begin{equation*}
B(u)=\frac{\wp^2}{2s}I_{[-s,s]}(u-\bar{u}),\quad0<s<1
\end{equation*}
so that
\begin{equation}
R(x,x')=\wp^2\frac{\sin (x-x')s}{(x-x')s}e^{-j(x-x')\bar{u}}.\label{eq2-2-2}
\end{equation}
or with regard to a planar array on a disk of radius $r$:
\begin{equation*}
B(u,v)=\frac{\wp^2}{\pi r^2}I_{\{(u,v)|u^2+v^2\leq r^2\}}(u-\bar{u},v-\bar{v}),\quad0<r<1
\end{equation*}
so that
\begin{equation}
R(x,y,x',y')=\wp^2\Lambda_1\left(r\sqrt{(x-x')^2+(y-y')^2}\right)e^{-j((x-x')\bar{u}(y-y')\bar{v})}\label{eq2-2-3}
\end{equation}
where
\begin{equation*}
\Lambda_1(x)=2\frac{J_1(x)}{x}
\end{equation*}
and where $J_1(x)$ is a first order Bessel function of the first kind. For a summation of such jammers with the receiver noise, one can thus write $n\sim\mathcal{N}(\mathbf{0},\mathbf{I}+\mathbf{R})$.

\section{Detection and Estimation Problems}\label{Sec3}
\begin{center}
\textbf{Summary of Chapter 3}
\end{center}
In order to formulate the resolution problem as a multihypothesis test, the signal model of Chapter 2 is formulated in terms of hypotheses. That is, as sets of probability distributions to which the measurement vector must be assigned. Subsequently, a solution of this multihypothesis test is given using a sequence of $2$-hypothesis tests. In such tests, one assumes that the number of targets $M$ is equal to $1, 2, 3,\ldots$; one estimates the associated directions and then tests whether the assumed number of targets is sufficient. Well-known tests, such as the likelihood ratio test, can be used for the $2$-hypothesis tests. The likelihood ratio test has the advantage that it asymptotically maintains a fixed probability that one chooses the null hypothesis in the event that a signal is actually present (a Type $1$ error). The ability to maintain a fixed Type-$1$ error probability can only be extended to multihypothesis tests under certain conditions. It is shown that the multihypothesis test is asymptotically bounded below a particular probability of overestimating the number of targets. With the hypothesis complexity one cannot expect more than an asymptotic optimality.

The technique developed here for improved resolution will be subsequently compared with suggestions from the literature.
~\\
\hrule
~\\
The assumed Signal Model $\mathbf{z}=\mathbf{A}\mathbf{b}+\mathbf{n}$ is the basis upon which the resolution shall be achieved. In this model, the distribution of the random variable $\mathbf{n}$ is assumed to be known, whereas the parameters $\boldsymbol{\omega}$, $\mathbf{b}$, and the number of signals $M$ are unknown.

\subsection{Resolution as a Decision Problem}\label{Sec3-1}
\setcounter{equation}{0}

Resolution, that is the decision that precisely $M$ targets, neither more nor fewer, are present in specific directions according to the signal model means the assignment of the measured data, $\mathbf{z}_1,\ldots,\mathbf{z}_K$ to a class of distributions. The distribution class should be independent of the amplitude fluctuations, be they according to Models $2$, $3$, or $4$. This can be achieved by regarding the complex amplitudes $\mathbf{b}$ not as random but rather as an unknown, deterministic sequence $\mathbf{b}_1,\ldots,\mathbf{b}_K$. Thus, the set of the samples of cardinality $K$ is 
\begin{equation}
\mathcal{Z}^K:=\left\{\left[\begin{IEEEeqnarraybox}[\relax][c]{c}
\mathbf{z}_1\\
\vdots\\
\mathbf{z}_K
\end{IEEEeqnarraybox}\right]\in\mathbb{C}^N\times\ldots\mathbb{C}^N|\mathbf{z}_i\sim\mathcal{N}(\mathbf{s}_i,\mathbf{R}),i=1,\ldots,K\right\}.\label{eq3-1-1}
\end{equation}
Resolution then means the assignment of the samples $\mathbf{z}_1,\ldots,\mathbf{z}_K$ to the hypotheses (the sets of probability distributions)
\begin{align}
W^K_m(\boldsymbol{\omega}):=&\left\{P_{\mathbf{z}_1,\ldots,\mathbf{z}_K}\left|\bigvee\limits_{\mathbf{b}_i\in\mathbb{C}^M}:\mathbf{z}_i\sim\mathcal{N}\left(\mathbf{A}(\boldsymbol{\omega})\mathbf{b}_i,\mathbf{R}\right),i=1,\ldots,K\right.\right\}\notag\\
=&\left\{P_{\mathbf{z}_1,\ldots,\mathbf{z}_K}\left|\bigwedge\limits_{i\in\{1,\ldots,K\}}:\mathbf{z}_i\sim\mathcal{N}(\mathbf{s}_i,\mathbf{R})\wedge\mathbf{s}_i\in H_M(\boldsymbol{\omega})\right.\right\}\label{3-1-2}
\end{align}
whereby
\begin{align*}
H_M(\boldsymbol{\omega}):=&\left\{\mathbf{s}\in\mathbb{C}^N\left|\bigvee\limits_{\mathbf{b}\in\mathbb{C}^M}:\mathbf{s}=\mathbf{A}(\boldsymbol{\omega})\mathbf{b}\right.\right\}\\
=&\lin H(\mathbf{a}(\boldsymbol{\omega}_1),\ldots,\mathbf{a}(\boldsymbol{\omega}_M)).
\end{align*}

Though a hypothesis is actually a set of distributions, the parameterized set $H_M(\boldsymbol{\omega})$ shall subsequently be denoted as a ``hypothesis,'' because it uniquely characterizes $W^K_M(\boldsymbol{\omega})$. The assignment to the hypothesis  $W^K_M(\boldsymbol{\omega})$, or respectively $H_M(\boldsymbol{\omega})$, is represented by the assignment of the parameter $(\boldsymbol{\omega},M):=\boldsymbol{\vartheta}$.

A general starting point for the decision $\boldsymbol{\vartheta}$ is the assignment of costs $L(\boldsymbol{\vartheta},\boldsymbol{\vartheta}_0)$ to every given target configuration $\boldsymbol{\vartheta}_0$ and the minimization of the expected cost $R$. The decision criterion $\hat{\boldsymbol{\vartheta}}(\mathbf{z}_1,\ldots,\mathbf{z}_K)$ is subsequently chosen so that
\begin{align}
R(\hat{\boldsymbol{\vartheta}},\boldsymbol{\vartheta}_0):=&\E_{\mathbf{z}}\left\{L(\hat{\boldsymbol{\vartheta}},\boldsymbol{\vartheta}_0)\right\}=\min ! \quad\quad\text{(Bayes' Decision)}.\label{eq3-1-3}
\end{align}

However, this procedure, though often described in the literature, is not always good. It generally requires the use of arbitrary prior distributions and the treatment of all parameters as random variables, which is not justified for direction estimation. Moreover, the assignment of costs for use in radars does not make sense. Rather, controlling the error probability is important.

Therefore, though just as arbitrary, the ML principle shall be considered, as it is easier to use. Estimation and tests according to the principle of likelihood maximization possess special asymptotic optimality criteria, which is why they are often referred to as "optimal algorithms'' in the signal-processing community.

To apply the ML principle, we shall first consider the multihypothesis test between the hypotheses
\begin{align}
H_M:=&\bigcup\limits_{\boldsymbol{\omega}\in \Omega}H_M(\boldsymbol{\omega})\label{eq3-1-4}\\
=&\left\{\mathbf{s}\in\mathbb{C}^N\left|\bigvee_{\mathbf{b}\in\mathbb{C}^M}\bigvee_{\boldsymbol{\omega}\in\Omega}:\mathbf{s}=\mathbf{A}(\boldsymbol{\omega})\mathbf{b}\right.\right\}\notag
\end{align}
for $M\in \mathbb{N}$, whereby $\Omega\in V^M$ is a yet-to-be specified region of allowable directions of arrival, which could be the entire field of view $V^M$.

These hypotheses have the following properties
\begin{enumerate}
\item \begin{equation}
H_N=\mathbb{C}^N\label{eq3-1-5}
\end{equation}
meaning that at most $N$ targets can be resolved.
\item
\begin{equation}
H_0\subset H_1 \subset\ldots\subset H_{N-1}\subset H_N\label{eq3-1-6}
\end{equation}
because for every signal present, a signal from an arbitrary direction with amplitude $0$ can be added. The hypotheses thus overlap.
\end{enumerate}

A multihypothesis test is consequently the mapping of the sample space $\mathcal{Z}^K$ into the decision space, which by a non-randomized $2$-hypothesis test consists of $\{0,1\}$, and here consists of $\{1,2,\ldots, N\}$. Such a multihypothesis test can be decomposed into a series of $2$-hypothesis tests, $H_i$ versus $H_k$ for $i\neq k$ from which the complete decision can be made \cite{ref18}. The multihypothesis test can be created from the combination of the tests $\varphi$ and $\psi$ as follows:
\begin{equation*}
\mathcal{Z}^K \xrightarrow{\varphi}\{0,1\}^{N(N-1)/2}\xrightarrow{\psi}\{1,\ldots,N\}.
\end{equation*}
Due to the structure \eqref{eq3-1-6} of the hypotheses, the test $\varphi$ can be simplified. One must execute tests $H_i$ against $K_i:=\mathbb{C}^N-H_i$ for $i=1,\ldots,N$. This means that
\begin{equation}
\mathcal{Z}^K\xrightarrow{\varphi}\{0,1\}^N\xrightarrow{\psi}\{1,\ldots,N\}\label{eq3-1-7}
\end{equation}
so that $\varphi(\mathbf{z}_1,\ldots,\mathbf{z}_k)=(\varphi_1(\mathbf{z}_1,\ldots,\mathbf{z}_k),\ldots,\varphi_N(\mathbf{z}_1,\ldots,\mathbf{z}_k))$.

How the test $\varphi$ is to be chosen depends strongly on the choice of $\varphi$. In order to be able to control the error probability, for every test $\varphi_i$ for $H_i$ against $K_i$, a ``uniformly best test of level $\alpha$'' should be the goal. That means, a test that does not exceed a specified probability of a Type-$1$ error and that, among all tests of level $\alpha$, minimizes the probability of a Type-$2$ error (optimal in a Neyman-Pearson sense).

Because composite hypotheses $H_i$ against alternatives $K_i$ are present and these are generally nonlinear subsets of $\mathbb{C}^N$, such uniformly best tests generally do not exists. However, one can find asymptotically (for large numbers of samples) best tests.

The likelihood ratio tests for $H_i$ against $K_i$ has the form
\begin{equation*}
\varphi_i(\mathbf{z}_i,\ldots,\mathbf{z}_k)=\left\{\begin{IEEEeqnarraybox}[\relax][c]{c's}
0&if $2\ln T_i(\mathbf{z}_1,\ldots,\mathbf{z}_k)\leq\eta$\\
1&if $2\ln T_i(\mathbf{z}_1,\ldots,\mathbf{z}_k)>\eta$
\end{IEEEeqnarraybox}\right.
\end{equation*}
whereby $0$ denotes the acceptance of hypothesis $H_i$, $1$ the acceptance of the alternative $K_i$, and $T_i$ is the likelihood ratio
\begin{equation}
T_i(\mathbf{z}_1,\ldots,\mathbf{z}_k):=\frac{\sup\limits_{\mathbf{s}\in\mathbb{C}^N}p(\mathbf{z}_1,\ldots,\mathbf{z}_k;\mathbf{s})}{\sup\limits_{\mathbf{s}\in H_i}p(\mathbf{z}_1,\ldots,\mathbf{z}_k;\mathbf{s})}.
\end{equation}

If $H_i$ is parameterized by $\boldsymbol{\vartheta}$, then $\sup\limits_{\boldsymbol{\vartheta}\in H_i}p(\mathbf{z}_1,\ldots,\mathbf{z}_K|\mathbf{s}(\boldsymbol{\vartheta}))$ is the ML estimate for the parameter  $\boldsymbol{\vartheta}$ (see \cite[pp. 92]{ref3}). The decision problem thus decomposes into a sequence of (continuous) parameter-estimation steps for every hypothetical number of targets $i$, from which the test value (test statistic) $T_i$ is calculated and a sequence of 2-hypothesis tests is executed. From those results, using $\psi$, the number of targets $M$ is determined. The associated target directions $\boldsymbol{\omega}$ are consequently the ML estimates associated with $M$ targets.

In order to apply the test and estimation theory, a number of terms shall be defined.

For estimation functions $\hat{\boldsymbol{\vartheta}}=\hat{\boldsymbol{\vartheta}}(\mathbf{z}_1,\ldots,\mathbf{z}_k))$ for a parameter $\boldsymbol{\vartheta}_0$, the following asymptotic optimality criteria are defined:
\begin{enumerate}
\item
\begin{align}
\left\{\hat{\boldsymbol{\vartheta}}_k\right\}\text{ \textbf{is consistent}}:\Longleftrightarrow&\hat{\boldsymbol{\vartheta}}_k\rightarrow\boldsymbol{\vartheta}_0\text{ converges stochastically for }K\rightarrow\infty\notag\\
(\Longleftrightarrow&\bigwedge\limits_{\epsilon\in\mathbb{R}_{+}}\lim_{k\rightarrow\infty}P\left\{\lVert\hat{\boldsymbol{\vartheta}}_k-\boldsymbol{\vartheta}_0\rVert\geq\epsilon\right\}=0)\label{eq3-1-9}
\end{align}
\item If $\left\{\hat{\boldsymbol{\vartheta}}_k\right\}$ is consistent, then the covariance matrix, if it exists, can be estimated by the \textbf{Cramér-Rao lower bound} (CRLB)
\begin{equation*}
\E\left[\left(\hat{\boldsymbol{\vartheta}}_k-\boldsymbol{\vartheta}_0\right)\left(\hat{\boldsymbol{\vartheta}}_k-\boldsymbol{\vartheta}_0\right)^T\right]\geq\frac{1}{k}\mathbf{F}^{-1}(\boldsymbol{\vartheta}_0)
\end{equation*}
whereby
\begin{equation}
F_{i,j}(\boldsymbol{\vartheta}_0)=\left.\int\frac{\partial \ln p(\mathbf{z};\boldsymbol{\vartheta})}{\partial\vartheta_i}\frac{\partial \ln p(\mathbf{z};\boldsymbol{\vartheta})}{\partial\vartheta_j}p(\mathbf{z};\boldsymbol{\vartheta})d\mathbf{z}\right|_{\boldsymbol{\vartheta}_0}\label{Eq3-1-10}
\end{equation}
is the Fisher Information Matrix (FIM). (Precise requirements are given in \cite[Sec. 2.27]{ref3}. In this instance, ``$\geq$'' means positive semidefiniteness of the difference matrix).
\item The CRLB is the motivation for the definition
\begin{align}
\left\{\hat{\boldsymbol{\vartheta}}_k\right\}\text{ is \textbf{asymptotically efficient}}:\Longleftrightarrow&\text{The PDF of }\sqrt{k}(\hat{\boldsymbol{\vartheta}}_k-\boldsymbol{\vartheta}_0)\notag\\
&\text{converges to }\mathcal{N}\left(\mathbf{0},\mathbf{F}^{-1}(\boldsymbol{\vartheta}_0)\right).\label{eq3-1-11}
\end{align}
Every consistent ML estimate $\hat{\boldsymbol{\vartheta}}_k$ is asymptotically efficient (\cite[Sec. 2.32]{ref3}).

In test problems, there a reduction in the measured data $\mathbf{z}$ is usually performed. That is, a statistic $T(\mathbf{z})$ is formed.

\item  A data reduction without ``information loss over a parameter $\boldsymbol{\vartheta}$'' is considered sufficient:
\begin{align}
T(\mathbf{z})\text{ is \textbf{sufficient} for }\boldsymbol{\vartheta}:\Longleftrightarrow&P_{\boldsymbol{\vartheta}}\left\{\mathbf{z}\in B|T=t\right\}\notag\\
\scriptsize(\text{meaning in terms of }P_{\boldsymbol{\vartheta}}(\mathbf{z}))&\text{  is independent of }\boldsymbol{\vartheta}\text{ for all results }B.\label{eq3-1-12}
\end{align}

The sufficiency of an $r$-dimensional parameter $\boldsymbol{\vartheta}$ necessitates an $r$-dimensional statistic (assuming that a true parameterization is present, meaning that $P_{\boldsymbol{\vartheta}}\neq P_{\boldsymbol{\vartheta}'}$ for $\boldsymbol{\vartheta}\neq\boldsymbol{\vartheta}'$).

\item In the event that no sufficient statistic can be found, one can often require an invariance in terms of the measurement system:
Letting $G$ be a group of transformations onto $\mathcal{Z}$, 
\begin{align*}
T(\mathbf{z})\textbf{ is invariant in terms of }G:\Longleftrightarrow&\bigwedge\limits_{g\in G}:T(g(\mathbf{z}))=T(\mathbf{z})
\end{align*}
\begin{align*}
T(\mathbf{z})\text{ is maximally invariant with respect to }G:&\Longleftrightarrow T\text{ is invariant with}\atop\text{respect to $G$ and}\notag\\
&\bigwedge\limits_{\mathbf{x},\mathbf{z}\in\mathcal{Z}}:\left(T(\mathbf{x})=T(\mathbf{z})\Rightarrow\bigvee\limits_{g\in G}:g(\mathbf{x})=\mathbf{z}\right).
\end{align*}
This is due to \cite[pg. 30ff]{ref3}.
\end{enumerate}

\subsection{The Estimation Problem}\label{Sec3-2}
\setcounter{equation}{0}

For $\mathbf{z}\sim\mathcal{N}(\mathbf{s},\mathbf{R})$, $\mathbf{s}=\mathbf{A}\mathbf{b}$ corresponds to the likelihood function, \cite{ref30}:
\begin{equation*}
L(\boldsymbol{\vartheta})=p(\mathbf{z},\mathbf{s}(\boldsymbol{\vartheta}))=\frac{1}{\pi^N|\mathbf{R}|}e^{-(\mathbf{z}-\mathbf{A}\mathbf{b})^*\mathbf{R}^{-1}(\mathbf{z}-\mathbf{A}.\mathbf{b})}
\end{equation*}
Here the parameters to be estimated are $\boldsymbol{\vartheta}=(\mathbf{u},\mathbf{v},\mathbf{b})$. One can also equivalently minimize $-\ln L(\boldsymbol{\vartheta})$ as the function
\begin{equation*}
\tilde{Q}(\boldsymbol{\vartheta})=(\mathbf{z}-\mathbf{A}\mathbf{b})^*\mathbf{R}^{-1}(\mathbf{z}-\mathbf{A}\mathbf{b}).
\end{equation*}
The minimum with respect to $\mathbf{b}$ can be immediately determined. It is
\begin{equation*}
\mathbf{b}=\left(\mathbf{A}^*\mathbf{R}^{-1}\mathbf{A}\right)^{-1}\mathbf{A}^*\mathbf{R}^{-1}\mathbf{z}
\end{equation*}
This was determined using the method of least squares for  $\min\limits_{\mathbf{x}}(\mathbf{y}-\mathbf{C}\mathbf{x})^*(\mathbf{y}-\mathbf{C}\mathbf{x})$, giving the result $\mathbf{x}=(\mathbf{C}^*\mathbf{C})^{-1}\mathbf{C}^*\mathbf{y}$ and applied with $\mathbf{y}=\mathbf{L}^*\mathbf{z}$, $\mathbf{C}=\mathbf{L}^*\mathbf{A}$ for the decomposition $\mathbf{R}^{-1}=\mathbf{L}\mathbf{L}^*$.

Substituting $\hat{\mathbf{b}}$ into $\tilde{Q}$ and simplifying, one gets
\begin{align}
Q(\mathbf{u},\mathbf{v})=&\min\limits_{\mathbf{b}}\tilde{Q}(\mathbf{u},\mathbf{v},\mathbf{b})=\mathbf{z}^*\left(\mathbf{R}^{-1}-\mathbf{R}^{-1}\mathbf{A}\left(\mathbf{A}^*\mathbf{R}^{-1}\mathbf{A}\right)^{-1}\mathbf{A}^*\mathbf{R}^{-1}\right)\mathbf{z}.\label{eq3-2-1}
\end{align}

In general, the SNR is not sufficient for a reasonable estimation accuracy, so one must use multiple temporally sequential samples $\mathbf{z}_1,\ldots,\mathbf{z}_K$. Considering the measurement data structure in \eqref{eq3-1-1}, the likelihood function is
\begin{equation*}
L_K(\boldsymbol{\vartheta})=\prod_{i=1}^Kp(\mathbf{z}_i,\mathbf{s}_i(\boldsymbol{\vartheta}))
\end{equation*}
because the measurement noise is independent. The quantities here are $\boldsymbol{\vartheta}=(\mathbf{u},\mathbf{v},\mathbf{b}_1,\ldots,\mathbf{b}_K)$ and $\mathbf{s}_i(\boldsymbol{\vartheta})=\mathbf{A}(\boldsymbol{\omega})\mathbf{b}_i$. This is equivalent to a single-sample estimation problem with
\begin{equation*}
\tilde{\mathbf{s}}=\left(\begin{IEEEeqnarraybox}[\relax][c]{c}
\IEEEstrut
\mathbf{s}_1\\
\vdots\\
\mathbf{s}_K
\IEEEstrut
\end{IEEEeqnarraybox}\right)\in\mathbb{C}^{NK}\text{ with }\tilde{\mathbf{z}}\sim\mathcal{N}_{\mathbb{C}^{NK}}(\tilde{\mathbf{s}},\tilde{\mathbf{R}}),\tilde{\mathbf{R}}=\left(\begin{IEEEeqnarraybox}[\relax][c]{c.c.c}
\mathbf{R}&\ldots&\mathbf{0}\\
\vdots&\ddots\vdots\\
\mathbf{0}&\ldots&\mathbf{R}
\end{IEEEeqnarraybox}\right)
\end{equation*}
instead of $K$ samples of $\mathbf{s}\in\mathbb{C}^N$. Both formulations lead to
\begin{equation}
L_K(\boldsymbol{\vartheta})=\frac{1}{\pi^{NK}|\mathbf{R}|^K}e^{-\sum_{i=1}^K\left(\mathbf{z}_i-\mathbf{A}\mathbf{b}_i\right)^*\mathbf{R}^{-1}\left(\mathbf{z}_i-\mathbf{A}\mathbf{b}_1\right)}.\label{eq3-2-2}
\end{equation}
The maximization of $L_K(\boldsymbol{\vartheta})$ is now equivalent to the minimization of the exponent. Thus
\begin{align}
\min\limits_{\boldsymbol{\vartheta}}\tilde{Q}_K=&\min\limits_{\boldsymbol{\omega}}\sum_{i=1}^K\min_{\mathbf{b}_i}(\mathbf{z}_i-\mathbf{A}\mathbf{b}_i)^*\mathbf{R}^{-1}(\mathbf{z}_i-\mathbf{A}\mathbf{b}_i)\notag\\
=&\min\limits_{\boldsymbol{\omega}}\sum_{i=1}^K\mathbf{z}_i^*\!\!\left(\mathbf{R}^{-1}-\mathbf{R}^{-1}\mathbf{A}\left(\mathbf{A}^*\mathbf{R}^{-1}\mathbf{A}\right)^{-1}\mathbf{A}^*\mathbf{R}^{-1}\right)\mathbf{z}_i\notag\\
=&\min\limits_{\boldsymbol{\omega}}Q_K(\boldsymbol{\omega})\label{eq3-2-3}
\end{align}
 with
 \begin{equation*}
 Q_K(\boldsymbol{\omega}):=\sum_{i=1}^K\mathbf{z}_i^*\mathbf{\Gamma}\mathbf{z}_i
 \end{equation*}
 and
 \begin{equation*}
 \boldsymbol{\Gamma}:=\mathbf{R}^{-1}-\mathbf{R}^{-1}\mathbf{A}\left(\mathbf{A}^*\mathbf{R}^{-1}\mathbf{A}\right)\mathbf{A}^*\mathbf{R}^{-1}.
 \end{equation*}

In practice, the minimization of $Q_K(\boldsymbol{\omega})$ is undesirable, as one must simultaneously save and process a very large amount of data $\mathbf{z}_1,\ldots,\mathbf{z}_K$. For large sample sizes, an equivalent estimation procedure can be found:
\begin{align*}
\frac{1}{K}Q_K(\boldsymbol{\omega})\rightarrow&<Q>(\boldsymbol{\omega})\text{ for }K\rightarrow\infty
\end{align*}
where
\begin{equation*}
<Q>(\boldsymbol{\omega}):=\lim\limits_{T\rightarrow\infty}\frac{1}{T}\int_0^TQ_T(\boldsymbol{\omega})dt
\end{equation*}
is the time average. It is
\begin{align}
<Q>(\boldsymbol{\omega})=&<\mathbf{z}^*\boldsymbol{\Gamma}\mathbf{z}>\notag\\
=&<(\mathbf{A}\mathbf{b}+\mathbf{n})^*\boldsymbol{\Gamma}(\mathbf{A}\mathbf{b}+\mathbf{n})>\notag\\
=&<\mathbf{b}^*\mathbf{A}^*\mathbf{\Gamma}\mathbf{A}\mathbf{b}>+<\mathbf{n}^*\mathbf{\Gamma}\mathbf{A}\mathbf{b}>+<\mathbf{b}^*\mathbf{A}^*\mathbf{\Gamma}\mathbf{n}>+<\mathbf{n}^*\mathbf{\Gamma}\mathbf{n}>.\label{eq3-2-4}
\end{align}
The noise $\mathbf{n}$ can be assumed to be ergodic  so that if in addition the logical physical requirement
\begin{equation*}
\frac{1}{K}\sum_{i=1}^K\mathbf{b}_i\mathbf{b}_i^*\rightarrow\mathbf{B}
\end{equation*}
is fulfilled for $K\rightarrow\infty$, then
\begin{align*}
<\mathbf{n}^*\boldsymbol{\Gamma}\mathbf{n}>=&\E\left\{\mathbf{n}^*\boldsymbol{\Gamma}\mathbf{n}\right\}\\
=&\E\left\{\trace\boldsymbol{\Gamma}\mathbf{n}\mathbf{n}^*\right\}\\
=&\trace\boldsymbol{\Gamma}\mathbf{R}
\end{align*}
and it is almost certain that 
\begin{equation*}
<\mathbf{n}^*\boldsymbol{\Gamma}\mathbf{A}\mathbf{b}>=0 
\end{equation*}
(it can be proven using the Kolmogorov Inequality)
and
\begin{equation*}
<\mathbf{b}^*\mathbf{A}^*\boldsymbol{\Gamma}\mathbf{A}\mathbf{b}>=\trace\mathbf{A}^*\boldsymbol{\Gamma}\mathbf{A}\mathbf{B}.
\end{equation*}

If the $\mathbf{b}_i$ can be considered values in an ergodic stochastic process, then instead of $<Q>$ and $<\mathbf{b}\mathbf{b}^*>$, one can write $\E\left\{Q\right\}$ and $\E\left\{\mathbf{b}\mathbf{b}^*\right\}$.

The function $<Q>$, or respectively  $\E\left\{Q\right\}$, can be minimized using a stochastic approximation algorithm (the Robbins-Monroe Algorithm):
\begin{equation*}
\boldsymbol{\omega}_{k+1}=\boldsymbol{\omega_k}-\mathbf{a}_k\,\text{grad }Q(\boldsymbol{\omega}_k,\mathbf{z}_k)\quad k=1,2,\ldots
\end{equation*}
with
\begin{equation*}
Q(\boldsymbol{\omega}_k,\mathbf{z}_k)=\mathbf{z}_k^*\boldsymbol{\Gamma}(\boldsymbol{\omega}_k)\mathbf{z}_k.
\end{equation*}
If the algorithm converges, then it must converge to a local minimum of $<Q>$, (or, respectively, $E\left\{Q\right\}$) \cite{ref23}.

In the event that the global minimum is found, then this procedure produces the same estimation result as the ``optimal'' ML estimation (3.2-3). However, here, at every step only a single piece of data $\mathbf{z}_k$ is used.

\subsection{The Detection Problem}\label{Sec3-3}
\setcounter{equation}{0}

According to Chapter 3.1, the multihypothesis test has the form
\begin{equation*}
\mathcal{Z}^K\xrightarrow{\varphi}\{0,1\}\xrightarrow{\psi}\{1,\ldots,N\}
\end{equation*}
with $\varphi=(\varphi_1,\ldots,\varphi_N)$ and $\varphi_i$ tests $H_i$ against $K_i=\mathbb{C}^N\backslash H_i$ for $(i=1\ldots N)$. According to (3.1-4), $H_i$ corresponds to the statement that ``the number of targets is $\leq i$,'' and $K_i$ that ``the number of targets is $>i$.''

For every $\varphi_i$, there is the probability of a Type-$1$ error:
\begin{equation*}
\alpha_i=P_{\boldsymbol{\vartheta}}\left\{\varphi_i(\mathbf{z}_i,\ldots,\mathbf{z}_K)=1\right\}\text{ for }\boldsymbol{\vartheta}\in H_i
\end{equation*}
which is the probability that a decision that the ``number of targets is $>i$'' is made in the instance where actually $M\leq i$. Similarly, the probability of a type two error is
\begin{equation*}
\beta_i=P_{\boldsymbol{\vartheta}}\left\{\varphi_i(\mathbf{z}_1,\ldots,\mathbf{z}_K)=0\right\}\text{ for }\boldsymbol{\vartheta}\in K_i
\end{equation*}
which is the probability that a decision that the ``number of targets is $\leq i$'' is made in the case that $M>i$. The detection probability for $H_i$ is consequently
\begin{equation*}
P_{E,i}=1-\alpha_i.
\end{equation*}

As an expansion of this definition, the error probability for the entire decision $\psi(\varphi(\mathbf{z}_1,\ldots,\mathbf{z}_K))$ shall be defined
\begin{itemize}
\item Type-$1$ Error:
\begin{equation*}
\bar{\alpha}_i:=P_{\boldsymbol{\vartheta}}\left\{\psi\left(\varphi\left(\mathbf{z}_1,\ldots,\mathbf{z}_K\right)\right)>M\right\}
\end{equation*}
when $\boldsymbol{\vartheta}\in H_i \backslash \bigcup\limits^{i-1}_{j=1}H_j=H_i\backslash H_{i-1}$.
\item Type-$2$ Error:
\begin{equation*}
\bar{\beta}_i:=P_{\boldsymbol{\vartheta}}\left\{\psi\left(\varphi\left(\mathbf{z}_1,\ldots,\mathbf{z}_K\right)\right)<M\right\}
\end{equation*}
when $\boldsymbol{\vartheta}\in H_i\backslash H_{i-1}$
\end{itemize}
and the detection probability is
\begin{equation*}
\bar{P}_{E,i}:=P_{\boldsymbol{\vartheta}}\left\{\psi\left(\varphi\left(\mathbf{z}_1,\ldots,\mathbf{z}_K\right)\right)=M\right\}\quad\boldsymbol{\vartheta}\in H_i\backslash H_{i-1}.
\end{equation*}

The set $H_i\backslash H_{i-1}$ corresponds to the statement ``no fewer than $i$ targets are present.'' In $H_i\backslash H_{i-1}$, there are no signals with zero complex amplitude (otherwise, if $b_i=0$ for $\mathbf{s}=\sum_{k=1}^i\mathbf{a}_k\mathbf{b}_k$ then $\mathbf{s}\in H_{i-1}$ would be true). The family $\{H_i\backslash H_{i-1}\}_{i=1\ldots,N}$ is thus a disjoint decomposition of $\mathbb{C}^N$. In the event that all tests $\varphi_i$ have the same $\alpha$, then it is possible to construct a $\psi$ that does not depend upon the full test sequence $\varphi_1,\ldots,\varphi_N$.

In the event that the level $\alpha$ is held by all $\varphi_i$, then the decision ``The number of targets present is $M$'' is determined by the sequence $\varphi_1=\varphi_2=\ldots\varphi_{M-1}=1,\varphi_M=0$. The test $\varphi_{M+1}$ would put the decision in question if $\varphi_{M+1}=1$, but $P_{\boldsymbol{\varphi}}\left\{\varphi_{M+1}=1\right\}\leq\alpha$ in the event that $\boldsymbol{\varphi}\in H_M\backslash H_{M-1}$, so that $\boldsymbol{\varphi}\in H_{M+1}\varphi_M$. $\varphi_{M+1}$ yields no additional information, because all of the tests are ``equally good,'' meaning that they are all of level $\alpha$. Thus, the multihypothesis test can be defined as
\begin{equation}
\psi(\varphi_1,\ldots,\varphi_N)=\min\left\{\left. i\in\left\{1,\ldots,N\right\}\right|\varphi_i=0\right\}\label{eq3-3-1}
\end{equation}
so that the test can be performed sequentially. One begins with $\varphi_1$, $\varphi_2$, and so on and one ends the test as soon as hypothesis $H_i$ is accepted. Considering the requisite parameter estimation for every single $\varphi_i$, this is a considerable computational savings. The decision ``the number of targets is $M$'' is also determined by the sequence $\varphi_N=\varphi_{N-1}=\ldots\varphi_M=0,\varphi_{M-1}=1$ if for all $i$, $\beta_i\leq \beta$. In this case, the test $\varphi_{M-2}$ also provides no additional information. In this situation, the test can be performed descending sequentially from $\varphi_N$, $\varphi_{N-1}$ to the first rejected hypothesis. For the application considered here, the test with the specified level $\alpha$ is preferable, because $N$ is normally very large, while $M$ is very small. Subsequently, only a test of the form given in (3.3-1) shall be considered.

The detection probability for the test in (3.3-1) is
\begin{align}
\bar{P}_E=&P_{\boldsymbol{\vartheta}}\left\{\psi(\varphi_1,\ldots,\varphi_M)=M\right\}\notag\\
=&P_{\boldsymbol{\vartheta}}\left\{\varphi_M=0\wedge\varphi_{M-1}=\varphi_{M-2}=\ldots =\varphi_1=1\right\}\label{eq3-3-2}
\end{align}
for $\boldsymbol{\vartheta}\in H_M\backslash H_{M-1}$, and the total probability of a Type-$1$ error is
\begin{equation*}
\bar{\alpha}=P_{\boldsymbol{\vartheta}}\left\{\varphi_M=\ldots=\varphi_1=1\right\}
\end{equation*}
for ($\boldsymbol{\vartheta}\in H_M\backslash H_{M-1}$).

If all tests $\varphi_i$ are independent from the other decisions $\varphi_K$ ($i\neq k$), then 
\begin{align}
\bar{\alpha}_M=&P_{\boldsymbol{\vartheta}}\left\{\varphi_M=1\right\}\ldots P_{\boldsymbol{\vartheta}}\left\{\varphi_1=1\right\}\label{eq3-3-3}\\
=&\alpha_M(1-\beta_{M-1})\ldots(1-\beta_1)\notag\\
\leq& \alpha\notag
\end{align}
according to the assumptions. In this case, the entire test is subject to an error level of $\alpha$!

The independence of the tests $\varphi_i$ can only be obtained if for each test new data $\mathbf{z}_i$ is used. Otherwise, only selected data would be used for a test $\varphi_M$, namely, the realization $\mathbf{z}_1,\ldots,\mathbf{z}_K$ for $\varphi_1=\ldots\varphi_{M-1}=1$. As shall be shown in Chapter 7, the use of new data makes the estimation convenient to a stochastic approximation.

The test developed here has no optimality properties; the use of the full sequence $\varphi_1,\ldots,\varphi_N$ can provide a test with increased angular resolution. The test (3.3-1) is, however, the most computationally efficient with regard to the parameter-estimation part. 

\subsection{Previous Superresolution Techniques}\label{Sec3-4}

From the vast literature on resolution, those works that use the Signal Model $\mathbf{z}=\mathbf{A}\mathbf{b}+\mathbf{n}$, that is ``spectral-line models'' shall be considered. The superresolution techniques for continuous spectra generally begin by estimating the covariance matrix $\mathbf{R}=\E\left\{\mathbf{z}\mathbf{z}^*\right\}$ and create the spectrum $\hat{S}(\omega)$ therefrom. As spectral lines, one generally estimates the $M$ largest maxima from $\hat{S}(\omega)$ ($M$-fold $1$-dimensional maximization). The determination of the number $M$ in this problem has not yet been satisfactorily resolved. An overview of this process is available in \cite{ref4,ref5}.

The method described here for the Model $\mathbf{z}=\mathbf{A}\mathbf{b}+\mathbf{n}$ utilizing aspects of estimation and detection theory was first considered in $1967$ \cite{ref6,ref7,ref8}. The resolution problem was formulated as a Bayes' decision problem as in (3.1-3). In \cite{ref8}, we come to ML estimation via minimization of the $Q$ function, as in (3.2-3). However, only for a single sample $K=1$. In \cite{ref8}, the minimization of the $Q$ function was performed using a grid search, as in Chapter 4.4; for the test statistics, the estimated values of $\hat{\boldsymbol{\omega}}$ and $\hat{\boldsymbol{b}}$ were suggested.

The minimization of the $Q$ function was also performed in \cite{ref9,ref10,ref11,ref12}, whereby in \cite{ref12} a grid search was also used; in \cite{ref9} an additional random search and gradient method, in \cite{ref10,ref11} through the use of unspecified conditional loops for the solution from $\text{grad} Q=0$ for deterministic targets. In \cite{ref13}, one tried to analytically find the minimum of the $Q$ function for special antenna layouts with few elements given a single sample. The problem of regularity, given in (4.2-1) and (5.1-1) is discussed there.

The use of the single-sample ML estimation the time time/frequency domain is given in \cite{ref14} and \cite{ref15}. In \cite{ref14}, an approximation of the $Q$ function for widely-spaced targets was used and minimized with a grid search. In \cite{ref15}, the detection problem was also mentioned. A sequence of $2$-hypothesis tests was suggested, whereby the minimum of the $Q$ function was used as the test statistic. The bound $\eta$ was specified solely by the total noise power.

An application of ML estimation for a uniform data structure from radio astronomy is provided in  \cite{ref16}. The minimization of the (single sample) $Q$ function was performed using a gradient method. As a detection test, a sequence of randomized $2$-hypothesis tests were considered, whereby the hypothesis with the highest probability was chosen. The probability of making a false decision was not considered.

The most extensive study of the $Q$ function is found in \cite{ref17}. Here, for the first time, two different signal models were considered, namely Models $1$ and $4$. The statement in \cite{ref17} that the ML estimate in Signal Model $4$ converges to that of Signal Model $1$ in high-SNR scenarios is, however, false. As multihypothesis test, \cite{ref17} also suggests a sequential test that essentially uses the minimum of the $Q$ function, but always works with the same data set. A consistency condition along the lines of a common level $\alpha$ is not mentioned.

The general structure of an $M$-hypothesis test as a combination of $N(N-1)/2$ 2-hypothesis tests, as in Chapter \ref{Sec3-1} is studied in \cite{ref18}. There the entire sequence of 2-hypothesis tests is required.

In the previous literature, arbitrary two-dimensional antenna-array layouts, sequential parameter-estimation procedures using multiple samples for different signal models, and a sequential multihypothesis  for for various signal models that maintain a constant error bound have not yet been considered.

\section{Resolution Using Information Only from Spatial Samples}\label{Sec4}
\setcounter{figure}{0}
\begin{center}
\textbf{Summary of Chapter 4}
\end{center}

ML direction estimation must be performed to calculate the likelihood ratio used in the test procedure of Chapter 3. This estimation leads to the minimization of the $Q$ function in (4.1-1). The chapter focusses on the properties of the $Q$ function. The minimization of this function can be interpreted as the maximization of simultaneously formed, decoupled sum beams. For the case of a single target, this reduces to the problem of the maximization of the conventional sum beam.

A sufficient condition of the existence of a unique global minimum depending on the layout of the antenna elements, the ``strongly $M$-regular elements layout'' (4.2-1) is given. Thus, one can analyze where a particular array layout allows for the resolution of $M$ targets with this procedure.

The accuracy of the direction estimates of the targets is analyzed based upon the form of the minimum point of the $Q$ function. The accuracy is stochastically determined by the Cramér-Rao lower bound. Using this bound, curves for an analysis of the resolution capability as a function of the target-separation distance and of the SNR are presented. The relative phase between the targets significantly influences the accuracy of the estimates.

The minimization of the $Q$ function using a grid search, which is often discussed in the literature, is subsequently considered. It is found to be of little use in radar applications.
~\\
\hrule
~\\

For a sufficiently large $N$, as described in Chapter \ref{Sec3}, the resolution with a single sample $\mathbf{z}$ is possible (monopulse estimation). However, a test of the form (3.3-1) does not possess a probability of error that is simple to estimate. In the following, $\mathbf{z}$ is a given vector in $\mathbb{C}^N$.

\subsection{The $Q$-Function and its Basic Properties}\label{Sec4-1}
\setcounter{equation}{0}

In order to estimate the directions $\mathbf{u}$ and $\mathbf{v}$, one must minimize the function $Q(\mathbf{u},\mathbf{v})$ of (3.2-1). In the following, only receiver noise without jamming is considered so that $\mathbf{R}=\mathbf{I}$ and according to (3.2-1) $Q$ has the form
\begin{align}
Q(\mathbf{u},\mathbf{v})=&\min\limits_{\mathbf{b}}(\mathbf{z}-\mathbf{A}\mathbf{b})^*(\mathbf{z}-\mathbf{A}\mathbf{b})\label{eq4-1-1}\\
=&\left\lVert\mathbf{z}-\mathbf{A}\left(\mathbf{A}^*\mathbf{A}\right)^{-1}\mathbf{A}^*\mathbf{z} \right\rVert^2\label{eq4-1-2}\\
=&\lVert\boldsymbol{\Gamma}\mathbf{z}\rVert^2\notag\\
=&\mathbf{z}^*\boldsymbol{\Gamma}\mathbf{z}\label{eq4-1-3}
\end{align}
with $\boldsymbol{\Gamma}=\mathbf{I}-\mathbf{A}\left(\mathbf{A}^*\mathbf{A}\right)^{-1}\mathbf{A}^*$.

The matrix $\boldsymbol{\Gamma}$ is a projection matrix, because $\boldsymbol{\Gamma}^2=\boldsymbol{\Gamma}$ and $\boldsymbol{\Gamma}=\boldsymbol{\Gamma}^*$, and it projects on the space $S^\bot$ when $S=\lin H(\mathbf{a}_1,\ldots,\mathbf{a}_M)$. $\boldsymbol{\Gamma}$ decomposes $\mathbb{C}^N$ into $\mathbb{C}^N=S\oplus S^\bot$.

Concrete interpretations of the minimization of $Q$:
\begin{itemize}
\item With $Q$, the residual energy after the signal extraction is minimized (because $\boldsymbol{\Gamma}\mathbf{z}\in S^\bot$).
\item $S$ is chosen such that $\mathbf{z}$ and $\boldsymbol{\Gamma}\mathbf{z}$ are maximally orthogonal.
\item The minimization of $\mathbf{z}^*\boldsymbol{\Gamma}\mathbf{z}$ is equivalent to the maximization of $\mathbf{z}^*\mathbf{A}\left(\mathbf{A}^*\mathbf{A}\right)^{-1}\mathbf{A}^*\mathbf{z}$. The term $\mathbf{a}_i^*\mathbf{z}$ represents the formation of a conventional sum beam in the direction $(u_i,v_i)$. Thus, $\mathbf{A}^*\mathbf{z}$ is the formation of $M$ simultaneous sum beams. The minimization of $Q$ is thus equivalent to the maximization of $M$ simultaneous sum beams that have been decoupled using the $M\times M$ matrix $\left(\mathbf{A}^*\mathbf{A}\right)^{-1}$.

For $M=1$ the term
\begin{equation*}
\mathbf{z}^*\mathbf{a}\left(\mathbf{a}^*\mathbf{a}\right)^{-1}\mathbf{a}^*\mathbf{z}=\frac{|\mathbf{z}^*\mathbf{a}|^2}{N}
\end{equation*}
is the conventional signal-processing technique.
\end{itemize}

Properties of the $Q$ function:
\begin{enumerate}
\item \underline{Periodicity}\hfill (4.1-4)

Let the $x$-coordinates of the antenna elements lie on a grid a distance $d$ apart, such that $x_i=k_id$, $k_i\in\mathbb{Z}$, $i=1\ldots N$. Then
\begin{equation*}
Q(u_1,\ldots,u_j,\ldots,u_M,v_1,\ldots v_M)=Q(u_1,\ldots,u_j+n\frac{\lambda}{d},\ldots,u_M,v_1,\ldots, v_M)
\end{equation*}
for all $j=1\ldots M$ and $n\in\mathcal{Z}$. A similar identity applies to the $y$ coordinates and $\mathbf{v}$.

Proof:

By definition:
\begin{align*}
a_{ik}(u_k+n\frac{\lambda}{s},v_k)=&e^{-j\frac{2\pi}{\lambda}\left(x_i\left(u_k+n\frac{\lambda}{d}\right)+y_iv_k\right)}\\
=&e^{-j\frac{2\pi}{\lambda}\left(x_iu_k+y_iv_k\right)-j\frac{2\pi}{\lambda}k_idn\frac{\lambda}{d}}\\
=&e^{-j\frac{2\pi}{\lambda}\left(x_iu_k+y_iv_k\right)}\\
=&a_{ik}(u_k,v_k).
\end{align*}

If $d=\frac{\lambda}{2}$ (Nyquist sampling), then one has exactly a period of $2$. That is, uniqueness in the visible region $V^1$.

\item\underline{Symmetry}\hfill (4.1-5)

Let $\rho$ be an arbitrary permutation from $(1,\ldots,M)$. Then,
\begin{equation*}
Q(\omega_1,\ldots,\omega_M)=Q(\omega_{\rho(1)},\ldots,\omega_{\rho(M)})
\end{equation*}
for $\omega_i=(u_i,v_i)$. This means that the numbering of the targets is arbitrary.

Proof:

Let $\rho$ be an arbitrary permutation from $(1,\ldots,M)$. Then,
\begin{equation*}
\mathbf{A}(\omega_{\rho(1)},\ldots,\omega_{\rho(M)})=\mathbf{A}(\boldsymbol{\omega})\mathbf{P}
\end{equation*}
where $\mathbf{P}$ is the corresponding permutation matrix ($\mathbf{P}^{-1}=\mathbf{P}$, $\mathbf{P}^T=\mathbf{P}$).

Thus, it is sufficient to only consider $Q$ in a particular region such as $\left\{(u,v)|u_1\leq u_2\leq\ldots \leq u_M\right\}$. In the event that the function is also periodic, the region to consider must be additionally constrained.

\item\underline{Singularities}\hfill (4.1-6)

In the formation in (4.1-2) and (4.1-3), $Q$ is not defined for $\omega_i=\omega_j$ for $i\neq j$ ($\omega_k=(u_k,v_k)$), because $\mathbf{A}^*\mathbf{A}$ becomes singular. In the formation of (4.1-1), the value of $Q$ is well defined and 
\begin{equation*}
Q_M(\omega_q,\ldots,\omega_j,\ldots,\omega_M)=Q_{M+k}(\omega_1,\ldots,\!\!\!\!\!\!\underbrace{\omega_j,\ldots, \omega_j}_{k+1\text{ equal arguments}} \!\!\!\!\!\!\ldots, \omega_M)
\end{equation*}
holds. The proof is in Appendix 1, (A.1-1).

\item\underline{Differentiability}\hfill (4.1-7)

Define
\begin{equation*}
\gamma:=\left\{\boldsymbol{\omega}\in V^M\left|\bigvee\limits_{i,k\in \{1\ldots M\} \atop i\neq k}:(u_i,v_i)=(u_k,v_k)\right.\right\}
\end{equation*}
Then $Q$ is infinitely differentiable on $V^M\backslash\gamma$. The first and second derivatives are in Appendix 1 in (A.1-4) and (A.1-5) (The proof is in Appendix 1).

\item\underline{The Influence of the Directional Response of the Individual Elements}\hfill (4.1-8)

If all of the antenna elements have the same directional response $f(u,v)$, then the $Q$ function is the same as that of an array of omnidirectional array elements. The estimation of the complex amplitudes (and thus the condition according to (4.3-3)) is, however, different.

\textbf{Proof:}
The elements of the transfer matrix for a directional response of $f(u,v)$ of the individual elements $a_{ki}$ are $f_i$ ($f_i=f(u_i,v_i)$). This means that instead of the matrix $\mathbf{A}$, one has $\mathbf{A}\mathbf{F}$ where $\mathbf{F}=\diag\limits_{i=1}^M(f_i)$.

However, $\mathbf{A}\mathbf{F}\left(\left(\mathbf{A}\mathbf{F}\right)^*\left(\mathbf{A}\mathbf{F}\right)\right)^{-1}\left(\mathbf{A}\mathbf{F}\right)^*=\mathbf{A}\left(\mathbf{A}^*\mathbf{A}\right)^{-1}\mathbf{A}^*$ such that $Q^F=Q$. Of course, $\left(\left(\mathbf{A}\mathbf{F}\right)^*\left(\mathbf{A}\mathbf{F}\right)\right)^{-1}\left(\mathbf{A}\mathbf{F}\right)^{-1}=\mathbf{F}^{-1}\left(\mathbf{A}^*\mathbf{A}\right)^{-1}\mathbf{A}^*$ so that $\mathbf{b}^F=\mathbf{F}^{-1}\hat{\mathbf{b}}$ (whereby the index $F$ always indicates the estimation with the transfer matrix $\mathbf{A}\mathbf{F}$).

\begin{figure}
\centering
\includegraphics[width=0.6\textwidth]{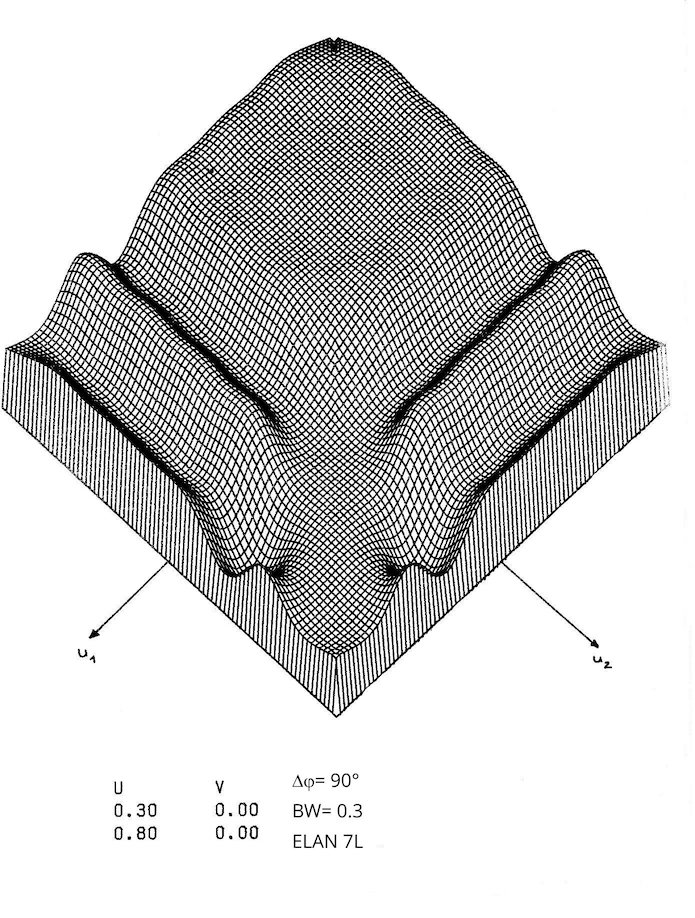}
\caption{$Q$-surface for two targets using ELAN 7 L.\label{Fig4-1}}
\end{figure}
\begin{figure}
\centering
\includegraphics[width=0.6\textwidth]{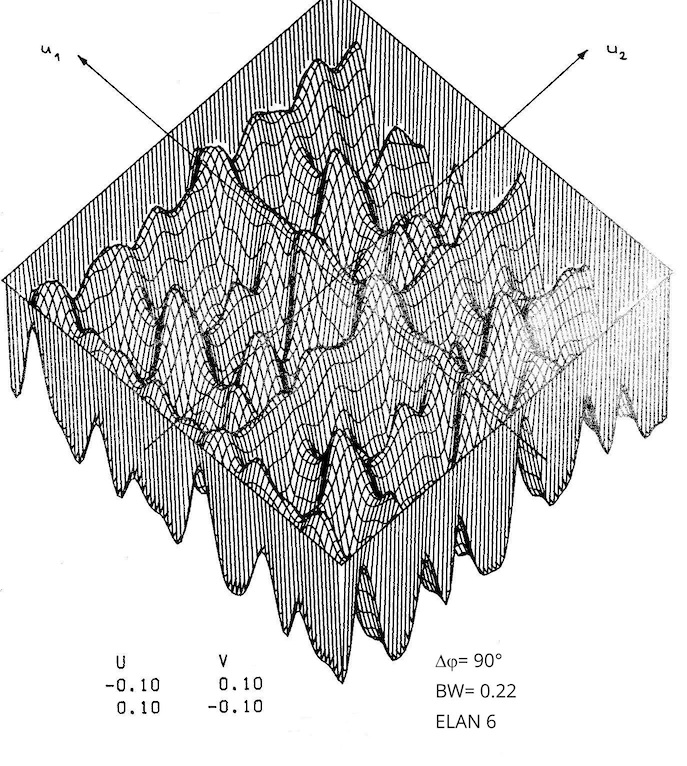}
\caption{$Q$-surface for two targets using ELAN 6.\label{Fig4-2}}
\end{figure}
\begin{figure}
\centering
\includegraphics[width=0.6\textwidth]{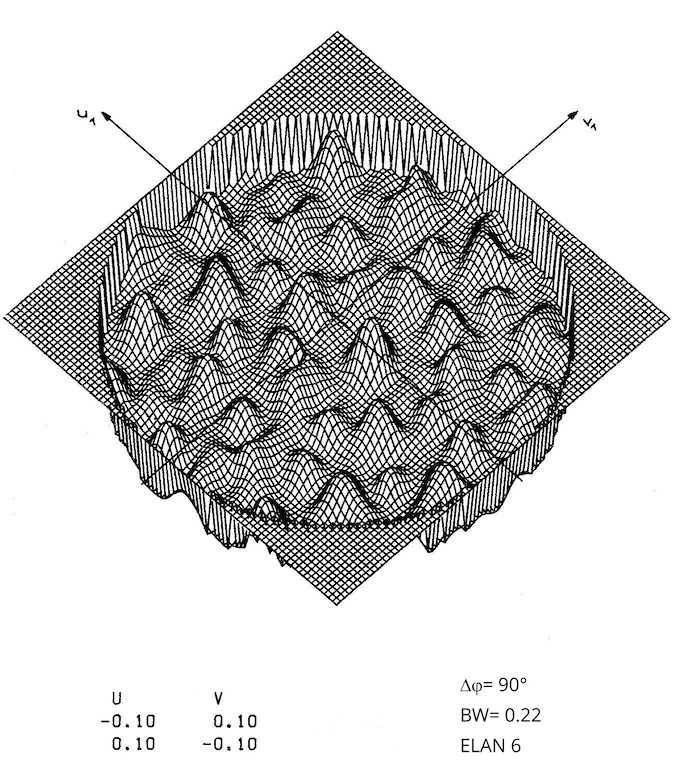}
\caption{$Q$-surface for two targets using ELAN 6.\label{Fig4-3}}
\end{figure}
\begin{figure}
\centering
\includegraphics[width=0.6\textwidth]{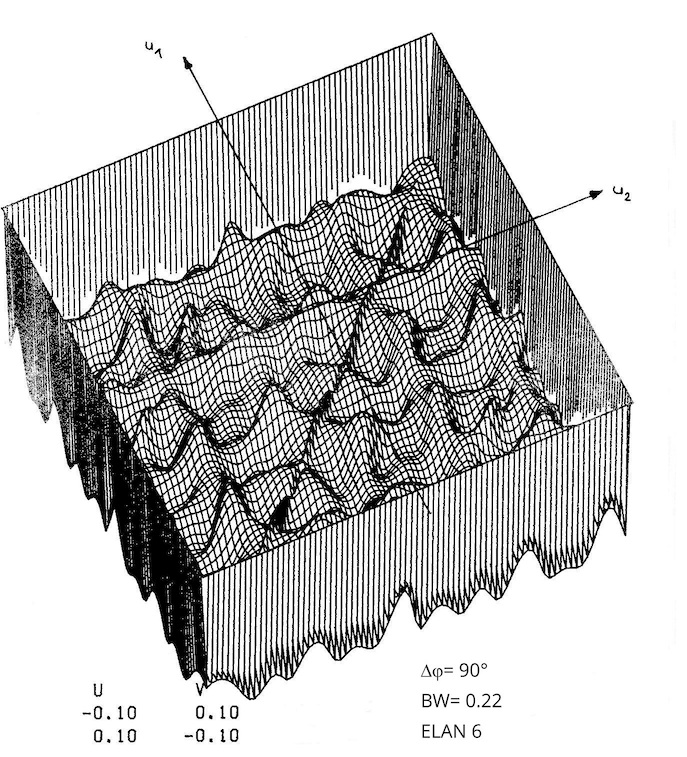}
\caption{$Q$-surface for two targets using ELAN 6.\label{Fig4-4}}
\end{figure}

The $Q$-function for the direction estimation of two targets when two targets are present without noise is shown in Figs. \ref{Fig4-1} through \ref{Fig4-4}.

The $Q$ function for the linear array ELAN 7L is shown in Fig. \ref{Fig4-1}. One can recognize the symmetry with respect to $\gamma$ about the line $u_1=u_2$. For planar arrays, $Q(u_1,u_2,v_1,v_2)$ is a hypersurface in $\mathbb{R}^5$ and only cuts of this surface can be displayed. These are shown in Figs. \ref{Fig4-2}, \ref{Fig4-3}, and \ref{Fig4-4} for ELAN 6. Because one is primarily interested in the minimum of $Q$, the $Q$ function has been flipped and is shown from below so that the minimum appears as the maximum. At the edge of the visible region, the $Q$ function is set to zero ($=$ the minimum of $Q$ as a reference). The surfaces can thus be interpreted as those of the function $\mathbf{z}^*\mathbf{A}\left(\mathbf{A}^*\mathbf{A}\right)^{-1}\mathbf{A}^*\mathbf{z}$. Figure \ref{Fig4-2} shows $F(u_1,u_2)=Q(u_1,u_2,v_1^{\text{ex}},v_2^{\text{ex}})$, and Fig. \ref{Fig4-3} shows $F(u_1,v_1)=Q(u_1,u_2^{\text{ex}},v_1,v_2^{\text{ex}})$, whereby the given target directions are $\mathbf{u}^{\text{ex}}=(-0.1,0.1)$, $\mathbf{v}^{\text{ex}}=(0.1,-0.1)$. The region $\gamma$ does not appear in Figure \ref{Fig4-2} at all because for all $u_1,u_2$, $\omega_1\neq\omega_2$. In Fig. \ref{Fig4-3}, it only appears as a peak with $(u_1,v_1)=(0.1,-0.1)$. The maximum (minimum of $Q$) is always exactly $\boldsymbol{\omega}_1=\boldsymbol{\omega}^{\text{ex}}$. The case is shown in Fig. \ref{Fig4-4} as in Fig. \ref{Fig4-2}, but the cut goes through the point $(0,0)$; thus $F(u_1,v_1)=Q(u_1,0,v_1,0)$. One can recognize the symmetry and the line $\gamma$ as in Fig. \ref{Fig4-1}. The plot in Fig. \ref{Fig4-3} resembles the response of the antenna array ELAN 6; it shows the directional response of the decoupled sum beams when the direction $\omega_2$ is constant.
\end{enumerate}

\subsection{The Existence and Uniqueness of the Solution}\label{Sec4-2}

The question arises, whether at least in a definition region that is restricted according to (4.1-4) and (4.1-5), the function $Q$ possesses at least a unique global minimum. For that, it is required that the number of measurements $2N$ be larger than the number of parameters $\mathbf{u}$, $\mathbf{v}$, $\mathbf{b}$ that are to be determined. With respect to a linear array, this is $2N\geq 3M$ and with respect to a planar array, this is $2N\geq4M$. Additionally,  there are certain demands on the ordering of the elements. An ordering, for which the $Q$-function possesses a clear global minimum in a particular region allowing the resolution of $M$ targets, is denoted as a ``strongly $M$-regular element arrangement.''

\textbf{Definition:}\hfill (4.2-1)

Let $\Omega\subset V^1$. An arrangement $x_1,\ldots,x_N,y_1,\ldots,y_N$ of $N$ antenna-array elements is strongly regular $\Longleftrightarrow$ For all $\omega_i$ pairwise different, $\boldsymbol{a}(\omega_1),\ldots,\mathbf{a}(\omega_{2M})$ are linearly independent.

For a strongly $M$-regular element arrangement, it is necessary that $N\geq2M$. For a strongly 1-regular arrangement, every two direction vectors $\mathbf{a}(\omega_1)$, $\mathbf{a}(\omega_2)$ must be linearly independent. Thus,
\begin{equation*}
\left|\begin{IEEEeqnarraybox}[\relax][c]{c'c}
\mathbf{a}_1^*\mathbf{a}_1&\mathbf{a}_1^*\mathbf{a}_2\\
\mathbf{a}_2^*\mathbf{a}_1&\mathbf{a}_2^*\mathbf{a}_2
\end{IEEEeqnarraybox}\right|\neq0.
\end{equation*}
This means that because $\mathbf{a}^*_1\mathbf{a}_1=\mathbf{a}_2^*\mathbf{a}_2=N$ and $\left|\mathbf{a}_1^*\mathbf{a}_2\right|^2=N^2\left|f(\omega_2-\omega_1)\right|^2$, if $f(\omega)$ is the directional response of the element arrangement, then $|f(\omega_2-\omega_1)|^2\neq1$, which means that the antenna array cannot have a secondary main beam.

\underline{Theorem (4.2-2)}:
\begin{itemize}
\item \textbf{Preconditions}

Let $x_i,y_i\quad(i=1,\ldots, N)$ be a strongly $M$-regular array-element arrangement in the region $\Omega\subset V^1$.

\item \textbf{Assertions}

\begin{itemize}
\item (i) The representation $\mathbf{s}=\mathbf{A}\mathbf{b}$ is unique except for the ordering of the results. More specifically
\begin{equation*}
\left(\bigvee\limits_{\substack{\boldsymbol{\omega},\tilde{\boldsymbol{\omega}}\in\Omega^M\\ \omega_k\text{ pairwise different}\\\tilde{\omega}_k\text{pairwise different}}}\bigvee\limits_{\mathbf{b},\tilde{\mathbf{b}}\in\mathbb{C}^M\backslash\{\mathbf{0}\}}:\mathbf{A}(\boldsymbol{\omega})\mathbf{b}=\mathbf{A}(\tilde{\boldsymbol{\omega}})\tilde{\mathbf{b}}\right)\Longrightarrow
\left(\bigvee\limits_{\mathbf{P}\text{ permutation}}:\mathbf{P}\boldsymbol{\omega}=\tilde{\boldsymbol{\omega}}\right)
\end{equation*} 

\item (ii) In each symmetry region of $\Omega^M$ according to (4.1-5), the $Q$ function has a unique minimum for $M$ targets as the noise goes to zero. This means that
\begin{equation*}
\left(Q(\boldsymbol{\omega})=\min!\wedge Q(\tilde{\boldsymbol{\omega}})=\min!\right)\Longrightarrow\bigvee\limits_{\mathbf{P}\text{ permutation}}:\mathbf{P}\boldsymbol{\omega}=\tilde{\boldsymbol{\omega}}.
\end{equation*}
\end{itemize}

\item \textbf{Proof}

\begin{itemize}
\item \underline{For (i)}: The proof in indirect by showing that $\neg (i) \Longrightarrow \neg\text{Preconditions}$.

Let $\omega_1,\ldots,\omega_M$ be pairwise different and $\tilde{\omega}_1,\ldots,\tilde{\omega}_M$ also be pairwise different, and $\omega_1,\ldots,\omega_M,\tilde{\omega}_1,\ldots,\tilde{\omega}_M$ are also pairwise different for $(1\leq r\leq M)$ and $\mathbf{A}(\boldsymbol{\omega})\mathbf{b}=\mathbf{A}(\tilde{\boldsymbol{\omega}})\tilde{\mathbf{b}}$. Without a loss of generality, let $b_1\neq0$, specifically $b_1=1$. Then
\begin{equation*}
\mathbf{a}(\omega_1)=\sum_{i=1}^r\mathbf{a}(\omega_i)b_i-\sum_{i=2}^M\mathbf{a}(\omega_i)c_i
\end{equation*}
for an appropriate $c_i\in\mathbb{C}$. Consequently, $\mathbf{a}(\tilde{\omega}_1),\ldots,\mathbf{a}(\tilde{\omega}_r),\mathbf{a}(\omega_1),\ldots,\mathbf{a}(\omega_r)$ are linearly independent, which is a contradiction!

\item \underline{For (ii)}:

One can say that $Q=\mathbf{s}\boldsymbol{\Gamma}\mathbf{s}=\min!\Longleftrightarrow\mathbf{s}^*\boldsymbol{\Gamma}\mathbf{s}=\mathbf{0}$ but
\begin{align*}
\mathbf{s}^*\boldsymbol{\Gamma}\mathbf{s}=\mathbf{s}^*\tilde{\Gamma}\mathbf{s}=0&\Longleftrightarrow\boldsymbol{\Gamma}\mathbf{s}=\tilde{\boldsymbol{\Gamma}}\mathbf{s}=\mathbf{0}\\
&\Longleftrightarrow\bigvee\limits_{\mathbf{b},\tilde{\mathbf{b}}\in\mathbb{C}^M\backslash\{\mathbf{0}\}}:\mathbf{s}=\mathbf{A}\mathbf{b}=\tilde{\mathbf{A}}\tilde{\mathbf{b}}
\end{align*}
from which the  assertion (i) follows.
\end{itemize}
\end{itemize}

Given a very good SNR and a strongly $M$-regular antenna array, the directions and amplitudes of $M$ targets can be determined nearly perfectly. With large numbers of elements, this holds asymptotically because the influence of the noise becomes smaller. This reduction in the noise influence is because for an arbitrary $\boldsymbol{\omega}$, $\lim\limits_{N\rightarrow\infty}\frac{1}{N-M}\mathbf{n}^*\boldsymbol{\Gamma}\mathbf{n}=1$, almost certainly. Thus, $Q\approx\min!\Longleftrightarrow\mathbf{s}^*\boldsymbol{\Gamma}\mathbf{s}=0$.

There remains the important question of which arrangements of elements are strongly $M$-regular. This question can not be answered in a general context here; only examples of strongly $M$-regular  arrays can be given.
\begin{enumerate}
\item Every linear array with $N$ elements $(N\geq 2M)$ and $2M$ elements with a constant spacing $d$ (for example $\lambda/2$) is strongly $M$-regular in an appropriate region $\Omega\subset V^M$.

This is because every $2M$ vectors are linearly independent if they are already independent in the $2M$ components of a uniform grid. For these $2M$ components, the matrix
\begin{equation*}
\left(\begin{IEEEeqnarraybox}[\relax][c]{c'c'c}
\IEEEstrut
1				&\ldots	&1\\
e^{-jdu_1}			&\ldots	&e^{-jdu_{2M}}\\
e^{-j2du_1}		&\ldots	&e^{-j2du_{2M}}\\
\vdots			&\ddots	&\vdots\\
e^{-j(2M-1)du_1}	&\ldots	&e^{-j(2M-1)du_{2M}}
\IEEEstrut
\end{IEEEeqnarraybox}\right)
\end{equation*}
must be regular. This is a van der Monde matrix of the form
\begin{equation*}
\left(\begin{IEEEeqnarraybox}[\relax][c]{c'c'c}
\IEEEstrut
1				&\ldots	&1\\
\gamma_1		&\ldots	&\gamma_{2M}\\
\vdots			&\ddots	&\vdots\\
\gamma_1^{2M-1}	&\ldots	&\gamma_{2M}^{2M-1}
\IEEEstrut
\end{IEEEeqnarraybox}\right)
\end{equation*}
for $\gamma_i=e^{-jdu_i}$, which must be regular if $\gamma_i\neq \gamma_k$ for all $i\neq k$. This means that within a period, according to (4.1-4), the vectors are linearly independent.

\item Every planar-element arrangement with $N\geq(2M)^2$ that has $2N\times 2M$ elements on a rectangular grid is strongly regular in a suitable region of $\Omega\subset V^M$.

To show this, it suffices to show the linear independence of the $\mathbf{a}$-vectors with respect to the grid points. Then one can write
\begin{align*}
\mathbf{a}(u,v)=&\left(\begin{IEEEeqnarraybox}[\relax][c]{c}
\IEEEstrut
e^{-jx_1u}\left(\begin{IEEEeqnarraybox}[\relax][c]{c}
			\IEEEstrut
			e^{-jy_1v}\\
			\vdots\\
			e^{-jy_{2M}v}
			\IEEEstrut
			\end{IEEEeqnarraybox}\right)\\
\hline\\
\vdots\\
\hline\\
e^{-jx_{2M}u}\left(\begin{IEEEeqnarraybox}[\relax][c]{c}
			\IEEEstrut
			e^{-jy_1v}\\
			\vdots\\
			e^{-jy_{2M}v}
			\IEEEstrut
			\end{IEEEeqnarraybox}\right)
\IEEEstrut
\end{IEEEeqnarraybox}\right)\\
=&\mathbf{a}_{(x)}(u)\otimes\mathbf{a}_{(y)}(v)
\end{align*}
if
\begin{equation*}
\mathbf{a}_{(x)}(u)=\left(\begin{IEEEeqnarraybox}[\relax][c]{c}
e^{-jx_1u}\\
\vdots\\
e^{-jx_{2M}u}
\end{IEEEeqnarraybox}\right)
\end{equation*}
and $\otimes$ is the Kronecker product (tensor product) operator.

Let $\omega_1\ldots\omega_{2M}$ be pairwise different, which means that $u_1,\ldots, u_r$ are pairwise different and $ v_r\ldots v_{2M}$ are pairwise different for $1\leq r\leq 2M$. Then, according to the first example of strongly $M$-regular arrays, $\mathbf{a}_{(x)}(u_1)\ldots\mathbf{a}_{(x)}(u_r)$ are linearly independent and $\mathbf{a}_{(y)}(v_r),\ldots,\mathbf{a}_{(y)}(v_{2M})$ are linearly independent. Furthermore, the $r\times(2M-r)$ direction vectors $(\mathbf{a}_{(x)}(u_i)\otimes\mathbf{a}_{(y)}(v_k))$ for $i=1,\ldots, r$ and $k=r,\ldots, 2M$ are linearly independent, because
\begin{align*}
\det\left(\left(\mathbf{A}_{(x)}\otimes\mathbf{A}_{(y)}\right)^*\left(\mathbf{A}_{(x)}\otimes\mathbf{A}_{(y)}\right)\right)=&\det\left(\left(\mathbf{A}^*_{(x)}\mathbf{A}_{(x)}\right)\otimes\left(\mathbf{A}^*_{(y)}\mathbf{A}_{(y)}\right)\right)\\
=&\det\left(\mathbf{A}^*_{(x)}\mathbf{A}_{(x)}\right)^r\det\left(\mathbf{A}^*_{(y)}\mathbf{A}_{(y)}\right)^{2M-r}\\
\neq&0
\end{align*}
as proven, for example, in \cite[pg. 348]{ref20}, using the abbreviations
\begin{align*}
\mathbf{A}_{(x)}=&\left(\mathbf{a}_{(x)}(u_1)\ldots\mathbf{a}_{(x)}(u_r)\right)\\
\mathbf{A}_{(y)}=&\left(\mathbf{a}_{(y)}(v_r)\ldots\mathbf{a}_{(x)}(v_{2M})\right)
\end{align*}
and thus, $\mathbf{A}(\omega_1),\ldots,\mathbf{A}(\omega_{2M})$ are linearly independent.
\end{enumerate}

The condition of strong $M$-regularity is sufficient for the resolvability of $M$ targets, but it is not always necessary. The number of antenna elements $N\geq2M$ is, however, necessary, because even though a 3-element linear array possesses as many measurements as unknowns of interest, for every direction pair $u_1, u_2$, there are always two directions $\tilde{u}_1,\tilde{u}_2$ and amplitudes $\mathbf{b},\tilde{\mathbf{b}}$ to specify so that $\mathbf{A}\mathbf{b}=\tilde{\mathbf{A}}\tilde{\mathbf{b}}$. When considering the projection of the element positions of a planar array onto all possible lines in the antenna-array plane, every $2$-regular planar array  must always possess more than three projected element positions. Thus, the following element positions are not strongly $2$-regular.
\begin{center}
\includegraphics[width=0.6\textwidth]{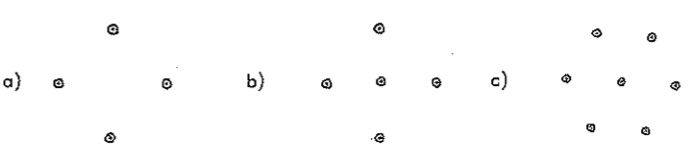}
\end{center}
With a sufficiently large number of randomly distributed positions, the probability of a strong-$M$ regularity approaches $1$ if $N\gg M$.

\subsection{The Behavior of $Q$ at the Minimum}\label{Sec4-3}
\setcounter{equation}{0}

The question arises how the conditioning of the given minimization problem depends on the target and antenna parameters. This can be studied using techniques from differential geometry by studying the curvature of the surface $Q(\boldsymbol{\omega})$ or respectively the surface $\E\left\{Q(\boldsymbol{\omega}\right)\}$ at the minimum with good SNR. Additionally, one can also consider the FIM (3.1-10), whereby one obtains an estimate of the variance of the desired parameter. For deterministic signals, the two approaches are closely related.

\underline{Definition}: (the second fundamental tensor)\hfill (4.3-1)

Let $\phi:\Omega\rightarrow\mathbb{R}^{M+1}$, $\boldsymbol{\omega}\mapsto(Q(\boldsymbol{\omega}),\boldsymbol{\omega})^T$, $(\Omega\in\mathbb{R}^M)$ be a twice-differentiable surface in $\mathbb{R}^{M+1}$.
\begin{itemize}
\item (i) The matrix
\begin{equation*}
\mathbf{S}(\boldsymbol{\omega}_0)=\left.Q_{\boldsymbol{\omega}\boldsymbol{\omega}}\left(\mathbf{I}+Q_{\boldsymbol{\omega}}Q_{\boldsymbol{\omega}}^T\right)^{-1}\right|_{\boldsymbol{\omega}_0}
\end{equation*}
is called the ``\underline{matrix of the second fundamental tensor}'' of $\phi$ (F-tensor).

For $M=2$, this can be found in the usual books on differential geometry, for example \cite[Theorem 3.5.5]{ref21}; for $M>2$ the definition still holds, but the derivation comes with a considerable theoretical effort, as in \cite[Pgs. 104, 109]{ref22}.

\item (ii) The eigenvalues/ eigenvectors of $\mathbf{S}$ are called the principal curvatures/ the principal curvature directions.
\item (iii) $\frac{1}{M}\trace\mathbf{S}$ is called the average curvature.
\end{itemize}

Interpretations of the fundamental tensor:

For every cut direction $\mathbf{w}$ ($\mathbf{w}\in\mathbb{R}^M$, $\lVert\mathbf{w}\rVert=1$) through the surface $\phi$, the term $\mathbf{w}^T\mathbf{S}(\boldsymbol{\omega}_0)\mathbf{w}$ is the geodetic curvature of the path on the surface $\phi$ caused by the cut passing through point $\boldsymbol{\omega}_0$. (Thus, the geodetic curvature is by definition the area of the parallelogram formed by the velocity and acceleration vectors of the path.)

At the minimum $\boldsymbol{\omega}_0$ with $Q_{\boldsymbol{\omega}}(\boldsymbol{\omega}_0)=\mathbf{0}$, the relation $S(\boldsymbol{\omega}_0)=Q_{\boldsymbol{\omega}\boldsymbol{\omega}}(\boldsymbol{\omega}_0)$ holds and the eigenvalues and eigenvectors determine the shape of the minimum. At the minimum, the principal curvature directions are the directions of the steepest and flattest slopes.

\underline{Theorem (4.3-2)}:
\begin{itemize}
\item \textbf{Preconditions}

Under Signal Model $1$ (deterministic signal), $\mathbf{n}\sim\mathcal{N}(\mathbf{0},\mathbf{I})$. Let $\mathbf{S}(\boldsymbol{\omega}_0)$ be the matrix of the second fundamental tensor of $Q$ without noise, or of $\E\left\{Q\right\}$ ($\mathbf{S}$ is the same in both cases), at the position of the global minimum $\boldsymbol{\omega}_0$.

Let $\mathbf{F}$ be the FIM of the estimate $(\boldsymbol{\omega},\mathbf{b})$ from $p(\mathbf{z};\boldsymbol{\omega},\mathbf{b})$ according to (3.1-10). Thus,
\begin{equation*}
\mathbf{F}=\left(\begin{IEEEeqnarraybox}[\relax][c]{c'c}
\IEEEstrut
\E\left\{(\ln p)_{\boldsymbol{\omega}}(\ln p)_{\boldsymbol{\omega}}^T\right\}&\E\left\{(\ln p)_{\boldsymbol{\omega}}(\ln p)_{\mathbf{b}}^T\right\}\\
\E\left\{(\ln p)_{\mathbf{b}}(\ln p)_{\boldsymbol{\omega}}^T\right\}&\E\left\{(\ln p)_{\mathbf{b}}(\ln p)_{\mathbf{b}}^T\right\}
\IEEEstrut
\end{IEEEeqnarraybox}\right)
\end{equation*}
(The subscripts $\boldsymbol{\omega},\mathbf{b}$ mean ``derivative with respect to $\mathbf{\omega}$ or respectively $\mathbf{b}$'')

\item \textbf{Assertions}

\begin{equation*}
\mathbf{S}^{-1}(\boldsymbol{\omega}_0)=\left[\mathbf{F}^{-1}\right]_{|\boldsymbol{\omega}\times\boldsymbol{\omega}}(\boldsymbol{\omega}_0)
\end{equation*}
where $\left[~\right]_{|\boldsymbol{\omega}\times\boldsymbol{\omega}}$ selects the submatrix of $\mathbf{F}^{-1}$ corresponding to the second derivative with respect to $\boldsymbol{\omega}$ in $\mathbf{F}$.

\item \textbf{Proof}
See Appendix 2 for the proof.
\end{itemize}

Thus, the variance of a direction estimate $u_i$ is inversely proportional to the curvature in the cutting direction of the $u_i$-axis.

For a linear antenna and $M=2$, the shape of the minimum through the primary curvature direction can be determined and the dependence of the given parameters can be discussed. For this case, the second derivative according to
(A.1-6) in Appendix 1 is
\begin{equation*}
Q_{\mathbf{u}\mathbf{u}}=2\text{Re }\left(\begin{IEEEeqnarraybox}[\relax][c]{c'c}
\IEEEstrut
|b_1|^2c_{11}			&|b_1||b_2|e^{j(\varphi_2-\varphi_1)}c_{12}\\
|b_1||b_2|e^{j(\varphi_1-\varphi_2)}c_{21}	&|b_2|^2c_{22}
\IEEEstrut
\end{IEEEeqnarraybox}\right)
\end{equation*}
where $c_{ik}=\mathbf{a}_{iu_i}^*\boldsymbol{\Gamma}\mathbf{a}_{ku_k}$ and $b_i=|b_i|e^{j\varphi_i}$ for $i,k=1,2$. The quantity $c_{ik}$ only depends on the target directions and the element positions. For an antenna that is symmetric about the origin, $-x_{-k}=x_k$ ($k=-s,\ldots,s$; $N=2s+1$; $x_0=0$) one gets
\begin{align*}
c_{11}=c_{22}=&\sum_{k=-s}^sx_k^2-\frac{N|d|^2}{N^2-|f|^2}\\
c_{12}=c_{21}=&h-\frac{f|d|^2}{N^2-|f|}
\end{align*}
with sum pattern
\begin{equation*}
f(u_2-u_1)=\mathbf{a}_1^*\mathbf{a}_2=\sum_{k=-s}^s\cos x_k(u_2-u_1)
\end{equation*}
and difference pattern
\begin{equation*}
d(u_2-u_1)=\mathbf{a}_1^*\mathbf{D}\mathbf{a}_2=j\sum_{k=-s}^sx_k\sin x_k(u_2-u_1)=-jf'(u_2-u_1)
\end{equation*}
and
\begin{equation*}
h(u_2-u_1)=\mathbf{a}_1^*\mathbf{D}^2\mathbf{a}_2=\sum_{k=-s}^sx_k^2\cos x_k(u_2-u_1)=-f''(u_2-u_1)
\end{equation*}
where 
\begin{equation*}
\mathbf{D}=\diag\limits_{i=1}^N (x_i)
\end{equation*}
For the above, the relations
\begin{equation*}
\left(\mathbf{A}^*\mathbf{A}\right)^{-1}=\frac{1}{N^2-|f|^2}\left(\begin{IEEEeqnarraybox}[\relax][c]{c}
N-f\\
-f^*N
\end{IEEEeqnarraybox}\right)
\end{equation*}
and $\mathbf{a}_i^*\mathbf{D}\mathbf{a}_i=0$ ($i=1,2$) were used

The eigenvalues of the matrix $\left(\begin{IEEEeqnarraybox}[\relax][c]{c'c}
\alpha&\gamma\\
\gamma&\beta
\end{IEEEeqnarraybox}\right)$ are
\begin{equation*}
\lambda_{1,2}=\frac{\alpha+\beta}{2}\pm\sqrt{\left(\frac{\alpha-\beta}{2}\right)^2+\gamma^2}
\end{equation*}
and the non-normalized eigenvectors are
\begin{equation*}
\Psi_{1,2}=\left(\begin{IEEEeqnarraybox}[\relax][c]{c}
\frac{-\gamma}{\frac{\alpha-\beta}{2}\mp\sqrt{\left(\frac{\alpha-\beta}{2}\right)^2+\gamma^2}}\\
1
\end{IEEEeqnarraybox}\right)
\end{equation*}

\underline{Special Cases}\hfill (4.3-3)
\begin{enumerate}
\item \underline{Targets of Equal Intensity $|b_1|=|b_2|$}:

In this instance, $\alpha=\beta$ and the primary curvature directions are $\left(\begin{IEEEeqnarraybox}[\relax][c]{c}
1\\
1
\end{IEEEeqnarraybox}\right)$ and $\left(\begin{IEEEeqnarraybox}[\relax][c]{c}
1\\
-1
\end{IEEEeqnarraybox}\right)$ with the primary curvatures $\lambda_{1,2}=\alpha\pm|\gamma|$.

Contour lines of $Q$ near the minimum thus look like ellipses with the principal axis directions $\left(\begin{IEEEeqnarraybox}[\relax][c]{c}
1\\
1
\end{IEEEeqnarraybox}\right)$ and $\left(\begin{IEEEeqnarraybox}[\relax][c]{c}
1\\
-1
\end{IEEEeqnarraybox}\right)$. The quantity $\gamma$ thus describes the eccentricity of the ellipses. If one has orthogonal signals $\varphi_1-\varphi_2=\frac{\pi}{2}$, then $\gamma=0$. That means that the minimum is approximately circular in shape. With $\varphi_1-\varphi_2=0,\pi$, the eccentricity is maximized.
The position of the ellipses is qualitatively shown in Fig. \ref{Fig4-5}. For targets that have equal phases ($\varphi_1=\varphi_2$), the distance between the targets can be determined quite well; however, the central point between them cannot be determined as well. For targets having opposite phases $\varphi_1-\varphi_2=\pi$, the central point between the targets can be determined well, but the distance between the targets cannot be determined to as accurate a degree.

\begin{figure}
\centering
\includegraphics[width=0.4\textwidth]{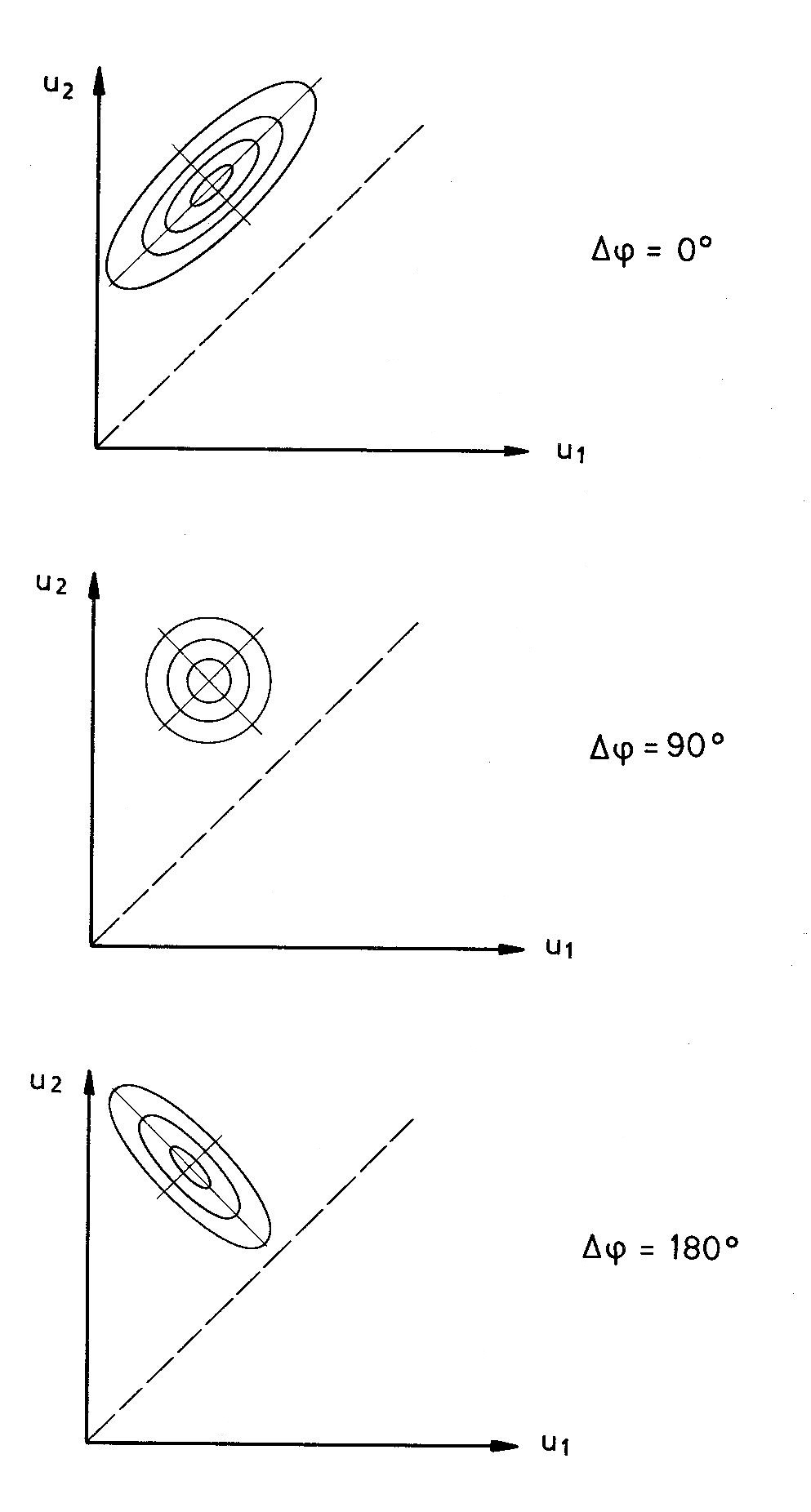}
\caption{\label{Fig4-5}}
\end{figure}

Figures \ref{Fig4-6} and \ref{Fig4-7} show a cut through the $Q$-surface of a linear array in the direction $\left(\begin{IEEEeqnarraybox}[\relax][c]{c}
1\\
1
\end{IEEEeqnarraybox}\right)$ ($45^\circ$ cutting angle) and $\left(\begin{IEEEeqnarraybox}[\relax][c]{c}
-1\\
1
\end{IEEEeqnarraybox}\right)$. ($135^\circ$ cutting angle) for phase differences $\Delta\varphi=0,\frac{\pi}{4},\frac{\pi}{2},\frac{3\pi}{4}$, (Curves $1$, $2$, $3$, $4$, $5$). The large difference in slope with the various phase differences is striking.

\begin{figure}
\centering
\includegraphics[width=0.6\textwidth]{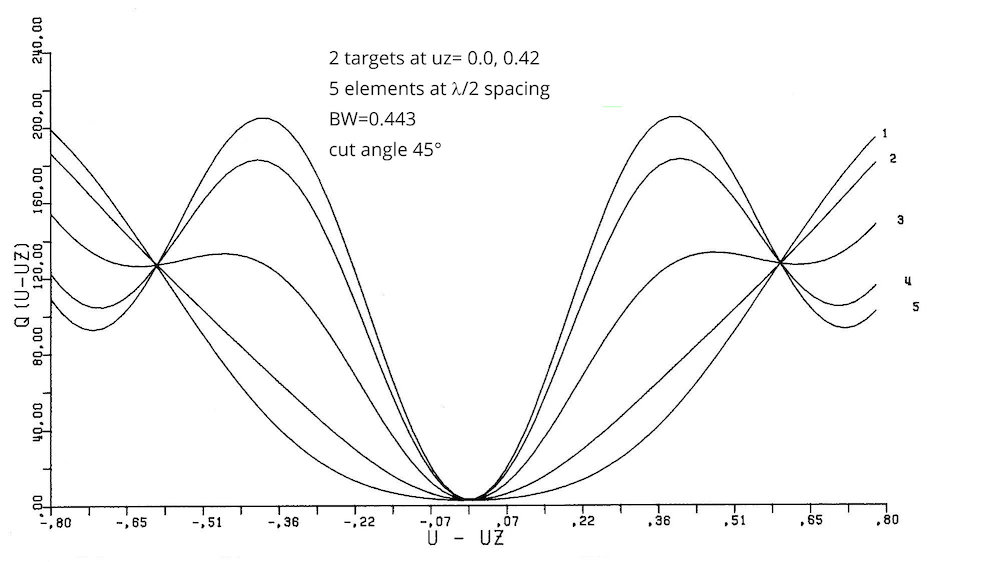}
\caption{\label{Fig4-6}}
\end{figure}

\begin{figure}
\centering
\includegraphics[width=0.6\textwidth]{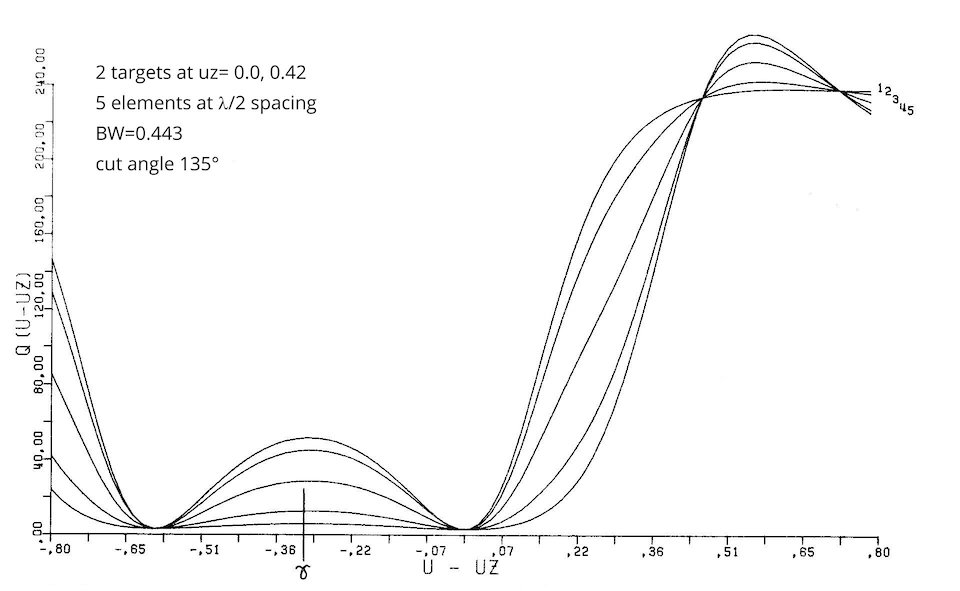}
\caption{\label{Fig4-7}}
\end{figure}

\item \underline{$|b_1|$ or $|b_2|\Longrightarrow0$}:

In this instance, the matrix $\mathbf{S}$ is singular and $\left(\begin{IEEEeqnarraybox}[\relax][c]{c}
1\\
0
\end{IEEEeqnarraybox}\right)$ and $\left(\begin{IEEEeqnarraybox}[\relax][c]{c}
0\\
1
\end{IEEEeqnarraybox}\right)$ are the principal curvature directions. Differing signal amplitudes rotate the principal curvature directions in a direction that is more parallel to the coordinate system.
\end{enumerate}

In order to find an ordering of array elements that is good for target resolution, one can maximize the mean curvature independently of the random phase relation of the targets. According to (A.1-6) in Appendix 1, the mean curvature is:
\begin{equation*}
\frac{1}{M}\trace \mathbf{S}=\frac{2}{M}\sum_{k=1}^M|b_k|^2\left(\mathbf{a}_k^*\mathbf{D}_x\boldsymbol{\Gamma}\mathbf{D}_x\mathbf{a}_k+\mathbf{a}^*_k\mathbf{D}_y\boldsymbol{\Gamma}\mathbf{D}_y\mathbf{a}_k\right)
\end{equation*}
with
\begin{equation*}
\mathbf{D}_x=\diag\limits_{k=1}^N(x_k)
\end{equation*}
and $\mathbf{D}_y$ defined analogously.
 
 If this expression is to be maximized independently of the $|b_k|^2$, then $\mathbf{a}_k^*\mathbf{D}\boldsymbol{\Gamma}\mathbf{D}\mathbf{a}_k$ must be maximized. The occurs under the two conditions
 \begin{enumerate}
 \item $\boldsymbol{\Gamma}\mathbf{D}\mathbf{a}_k=\mathbf{D}\mathbf{a}_k$, meaning that $\mathbf{d}\mathbf{a}_k \bot \lin H(\mathbf{a}_1\ldots\mathbf{a}_M$) for all $k=1\ldots M$, and $\mathbf{a}_i^*\mathbf{D}\mathbf{a}_k=0$ for all $i,k=1\ldots M$.
 \item $\lVert\mathbf{D}\mathbf{a}_k\rVert=\text{maximal} !$
 \end{enumerate}
 
 Because $\mathbf{a}_k^*\mathbf{D}_x^2\mathbf{a}_k=\sum_{i=1}^Nx_i^2$ (and analogously for $y$), Condition 2 means that the elements should be placed as closely to the edge of the aperture as possible. Condition $1$ means that the targets should lie as close to the null points of the difference beam as possible. This means placed on the maxima of the main beam and sidelobes of the sum beam pattern in the $M$ target directions.
 
 Of course, this also holds if outside interference is present. Consequently, low sidelobes are often more important than a high resolution capability for a radar.  These considerations show that the demand for a good localization, specifically Condition 2 for $k=1$, can also be generalized to $M$ targets.
 
 Using the Cramér-Rao lower bound, one can now specify what is the best estimation performance one can expect. From the CRLB inequality, one obtains for a high SNR
 \begin{equation*}
 \E\left\{\left(\hat{\omega}_i-\omega_i^{\text{ex}}\right)^2\right\}\geq\left(Q^{-1}_{\boldsymbol{\omega}\boldsymbol{\omega}}\right)_{ii}
 \end{equation*}
 for Signal Model $1$. For Signal Model$ 4$, using (3.1-10), 
 \begin{align*}
 p(\mathbf{z};\boldsymbol{\omega},\boldsymbol{\wp})=&\frac{1}{\pi^N|\mathbf{R}|}e^{-\mathbf{z}^*\mathbf{R}^{-1}\mathbf{z}}\\
 \mathbf{R}=&\mathbf{I}+\mathbf{A}\mathbf{B}\mathbf{A}^*\\
 \mathbf{B}=&\diag_{i=1}^M(\wp_i^2)
 \end{align*}
 and one can compute the FIM. This is
 \begin{align*}
 \frac{\partial \ln p}{\partial \vartheta_i}=&-\frac{\partial}{\partial \vartheta_i}\ln|\mathbf{R}|-\mathbf{z}^*\frac{\partial \mathbf{R}^{-1}}{\partial\vartheta_i}\mathbf{z}\\
 =&-\frac{\frac{\partial|\mathbf{R}|}{\partial \vartheta_i}}{|\mathbf{R}|}+\mathbf{z}^*\mathbf{R}^{-1}\frac{\partial \mathbf{R}}{\partial \vartheta_i}\mathbf{R}^{-1}\mathbf{z}
 \end{align*}
 where Lemma 1 of Appendix 1 was used. Using the rule that
 \begin{equation*}
 \frac{\partial|\mathbf{R}(\boldsymbol{\vartheta})|}{\partial \vartheta_i}=\left.|\mathbf{R}|\trace\left(\mathbf{R}^{-1}\frac{\partial \mathbf{R}}{\partial \vartheta_i}\right)\right|_{\boldsymbol{\vartheta}},
\end{equation*}
 one can write
 \begin{equation*}
\frac{\ln p}{\partial \vartheta_i}=-\trace\left(\mathbf{R}^{-1}\frac{\partial\mathbf{R}}{\partial\vartheta_i}\right)+\mathbf{z}^*\mathbf{R}^{-1}\frac{\partial\mathbf{R}}{\partial\vartheta_i}\mathbf{R}^{-1}\mathbf{z}
 \end{equation*}
 where as in Appendix \ref{Appen2}, the FIM can be computed. Figure \ref{Fig4-8} shows this for two targets of equal amplitude with ELAN 11 L.
 
 \begin{figure}
\centering
\includegraphics[width=0.5\textwidth]{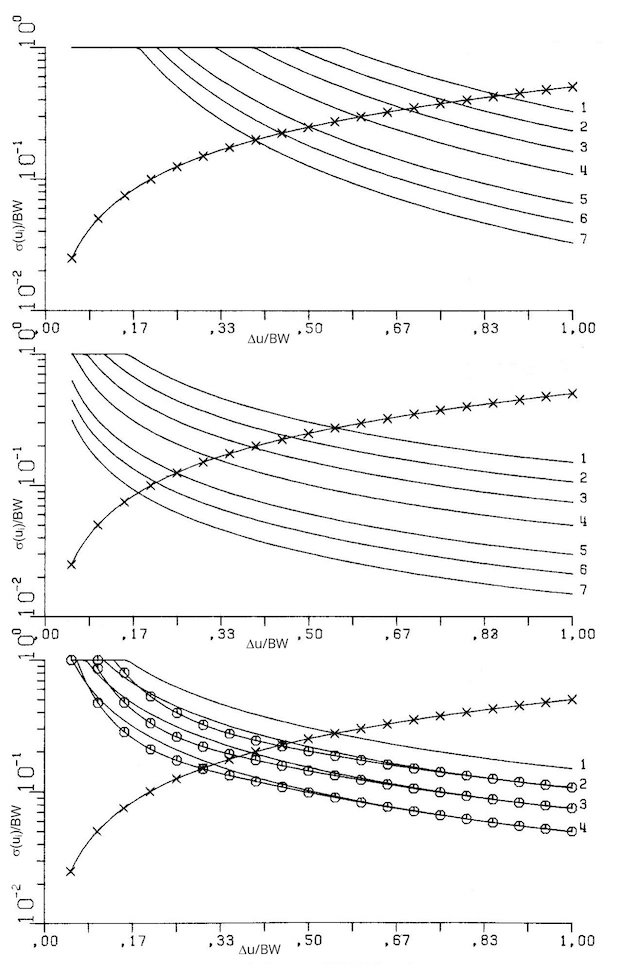}
\caption{\label{Fig4-8}}
\end{figure}
 
What is shown is $\sqrt{\var(\hat{u}_i)}/BW$ with respect to $|u_1-u_2|/BW$. Subplots (a) and (b) show curves for Signal model 1, whereby curves 1,2,3,4,5,6,7 are shown for an average SNR on the individual elements of $3,6,9,12.5,17,20,23\,\text{dB}$. Here, SNR is given as $\frac{|b_1|^2+|b_2|^2}{\E\left\{|n_1|^2\right\}}$. If one demands that $\sqrt{\var(\hat{u}_i)}\leq\frac{|u_1-u_2|}{2}$, then the standard deviation must lie below the curved marked with $X$ in the figure. Using the curve with $X$'s, one can read off the necessary SNR or the obtainable separation.
 
The curves in (c) show the CRLB for Signal Model $4$ and Signal Model $1$ for $\Delta\varphi=90^\circ$. These curves are almost the same. Because both of these models represent extremes, one can assume that the curves can be used as approximations for the other models.

The curves also show the integration gain (consider (3.1-10)) using multiple samples for ML estimation. If one takes the SNR of Curve $1$ for a single sample, then Curve $2$ corresponds to using $K=2$ samples, Curve $3$ $K=4$ samples, and Curve $4$ $K=9$ samples.

For large SNR, one achieves near equality to the CRLB, because $Q\approx\E\left\{Q\right\}-\sigma^2(N-M)$ under Signal Model $1$.

The curves hardly change if instead of ELAN 11 L, linear arrays having a large number of elements with $\lambda/2$ spacing are used and the total SNR is held constant (the mean SNR on the individual elements decreases by $10\log N$). For the curves for ELAN 11 L, one obtains a relative deviation $\delta$,
\begin{equation*}
\delta:=\frac{\sigma_{\text{ELAN NL}}(\Delta u)-\sigma_{\text{ELAN 11 L}}(\Delta u)}{\sigma_{\text{ELAN 11 L}}(\Delta u)}
\end{equation*}
 of
 \begin{center}
 \begin{tabular}{|c|cccc|}
 \hline
$\delta/BW$	&$N=51$	&$N=101$	&$N=501$	&$N=1001$\\
\hline
 $u/BW=0.3$	&+0.15	&+0.17		&+0.19		&+0.19\\
 $u/BW=0.5$	&+0.15	&+0.17		&+0.19		&+0.19\\
 $u/BW=0.7$	&+0.15	&+0.17		&+0.19		&+0.19\\
 \hline
 \end{tabular}
 \end{center}
where the values are given for $\Delta\varphi=90^\circ$.

Thus, the curves give a rough SNR estimate for large antennas. The necessary total SNR of circa $16\,\text{db}$ according to (b) for $\Delta\varphi=90^\circ$ for the separation of two targets a half beamwidth apart (Curve 2) also holds for ELAN 1001 L.
 
\subsection{Minimization Techniques}\label{Sec4-4}
\setcounter{equation}{0}
From Chapter \ref{Sec4-1} and \ref{Sec4-2}, one can see that the $Q$ function can be used to estimate the target directions when using an antenna whose elements are ordered so that they possess a strong-$M$ regularity. However, the Cramér-Rao inequality (Fig. \ref{Fig4-8}) shows that a very good SNR is necessary for such a resolution (For example, $17\,\text{dB}$ for two targets a distance of $BW/2$ apart with $\Delta\varphi=0$ according to Fig. \ref{Fig4-8}!).

To start, one needs a good initial estimate for the minimum of $Q$. This can be obtained using conventional directional estimation with sum beams. For all of the examples considered, within a halfbeamwidth radius of the true target positions $(\mathbf{u}^{\text{ex}},\mathbf{v}^{\text{ex}})$,
\begin{equation}
M\left(\mathbf{u}^{\text{ex}},\mathbf{v}^{\text{ex}}\right):=\left\{\bigwedge\limits_{i\in \{1\ldots M\}}:\left(u_i-u_i^{\text{ex}}\right)^2+\left(v_i-v_i^{\text{ex}}\right)^2<\left(\frac{BW}{2}\right)^2\wedge u_1<u_2<\ldots u_M\right\}\quad(\text{$BW=3\,\text{dB}$ beamwidth})\label{Eq4-4-1}
\end{equation}
the global minimum was always clear, as in Figs. \ref{Fig4-6} and Fig. \ref{Fig4-7}.

With closely-spaced targets, one first considers a rough direction estimate $(u_g,v_g)$ and minimizes $Q$ on $M(u_g,v_g)$.

\begin{figure}
\centering
\includegraphics{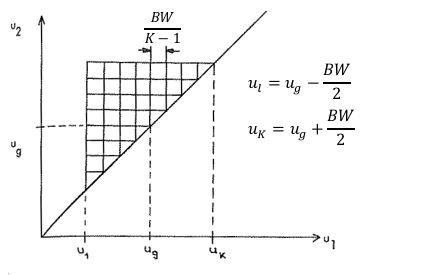}
\caption{\label{Fig4-9}}
\end{figure}

The simplest minimization method is the computation of the $Q$ function on various grid points $(u_i,v_i)$, $i=1\ldots K$. For example as with a linear array in Fig. \ref{Fig4-9}.
Such grid minimization can be performed particularly easily: One computed a bundle of $K$ beams in directions $\omega_1,\ldots, \omega_K$ in parallel and then evaluates the $Q$ function with them by decoupling sets of $M$ beams with $\left(\mathbf{A}^*\mathbf{A}\right)^{-1}$. The values of the matrix $\left(\mathbf{A}^*\mathbf{A}\right)^{-1}$ can be computed ahead of time and read from a table. The number of grid points ($=$ the number of $Q$ function evaluations) thus represents $\binom{K}{M}$ and increases rapidly with increasingly fine grids.

\begin{figure}
\centering
\includegraphics[width=0.6\textwidth]{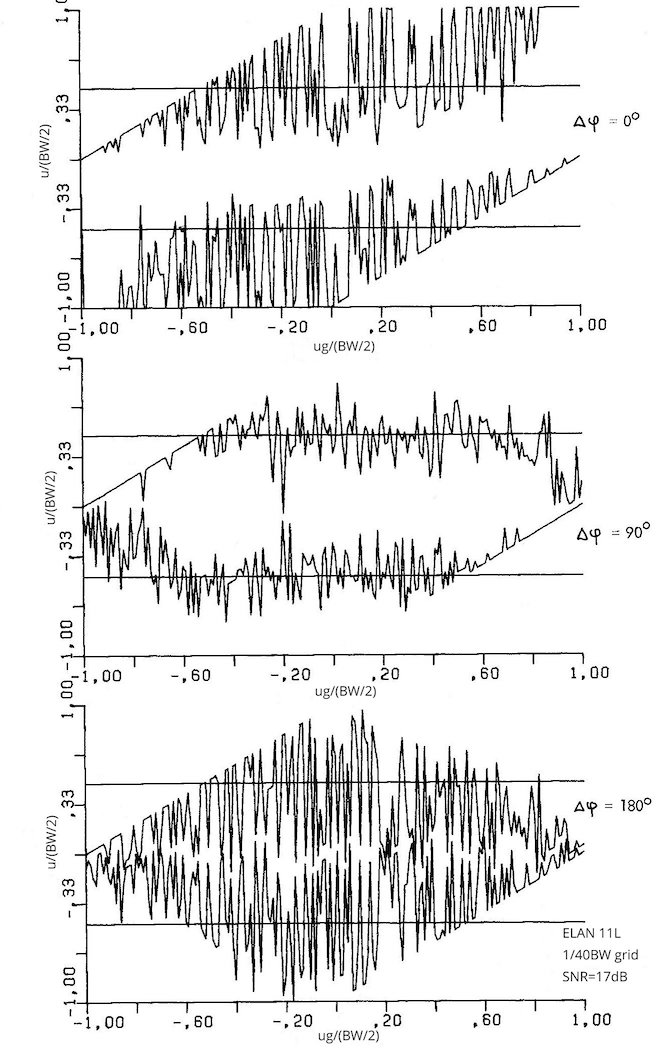}
\caption{\label{Fig4-10}}
\end{figure}

\begin{figure}
\centering
\includegraphics[width=0.6\textwidth]{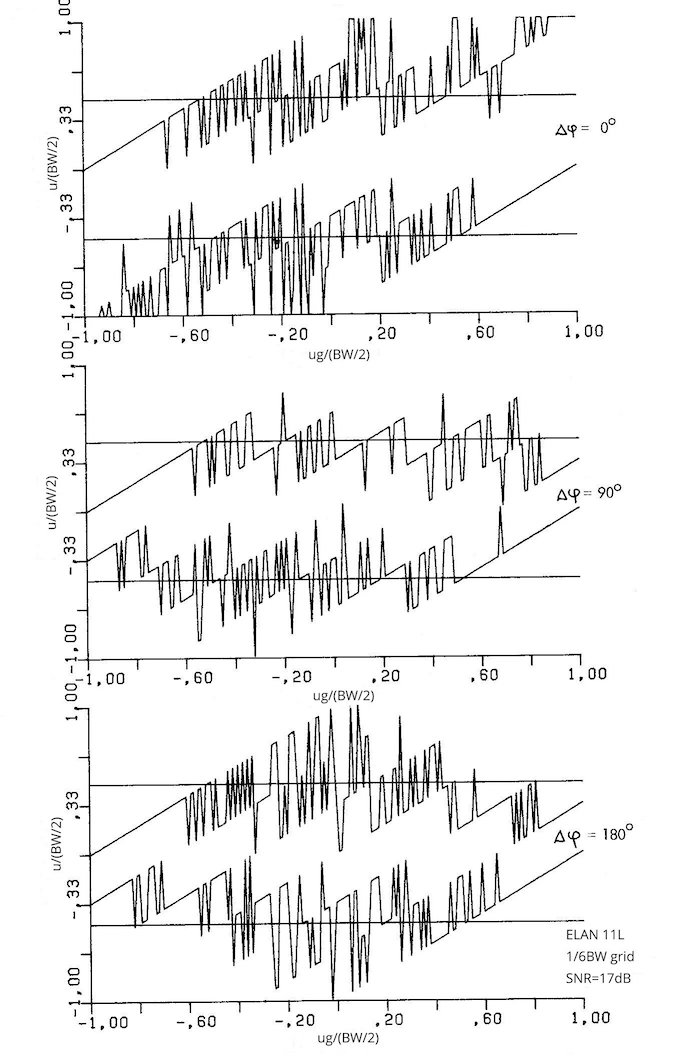}
\caption{\label{Fig4-11}}
\end{figure}

\begin{figure}
\centering
\includegraphics[width=0.6\textwidth]{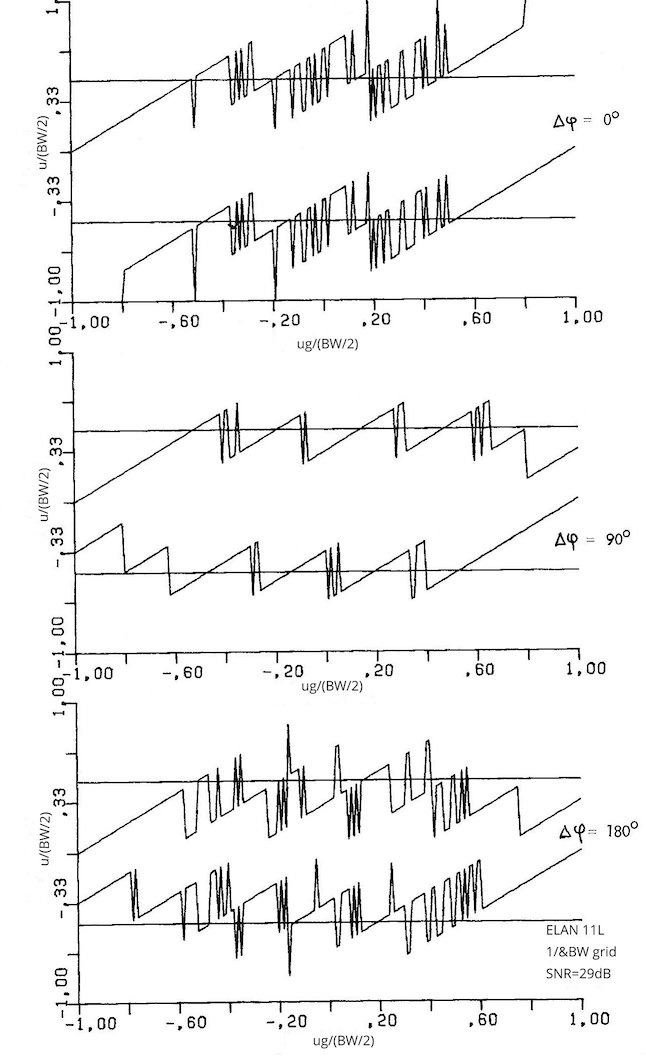}
\caption{\label{Fig4-12}}
\end{figure}

The accuracy of such estimation for a $1/40\,BW$ grid with ELAN 11 L is shown in Fig. \ref{Fig4-10}¯. The estimated direction per half $3$-$\text{dB}$ beamwidth is shown with respect to the rough estimate $u_g$. The value $u_g$ is determined by the raster position with respect to the target direction. because the targets lie $0.5BW$ apart, the targets are only both in the search region with $u_b/BW=0.5$. The same situation is shown in Fig. \ref{Fig4-11} with a $1/6BW$ grid. The mean SNR at the individual elements is $17\,\text{dB}$ in both instances. The estimates are shown on a $1/6 BW$ grid with an SNR of $29\,\text{dB}$ in Fig. \ref{Fig4-12}. The high SNR from the CRLB (here for two targets) being above $17\,\text{dB}$ for $\Delta\varphi=0^\circ,180^\circ$ is quite necessary. The feasibility of such a superresolution with a single pulse is thus greatly limited.

The assessment of the quality of an estimation method as in Figs. \ref{Fig4-10} and \ref{Fig4-11} is somewhat problematic. The dispersion of the minimum $\hat{\boldsymbol{\omega}}$ is in this instance on the order of the entire search area. By quantizing such a strongly clipped random variable, the usual definition of a mean-squared error $\E\left\{\left(\omega_i-\omega_i^{\text{ex}}\right)\right\}$ is not particularly informative. Indeed, the simulations show that in the case of Figs. \ref{Fig4-10} and \ref{Fig4-11} with $\Delta\varphi=0^\circ,180^\circ$, the mean-squared error decreases as the grid becomes increasingly coarse. Thus, the average error indications in the literature \cite{ref8,ref12,ref14} have little validity fior the actual usability of the method. In such instances, it would be better to provide probabilities for the position of the minimum on the grid. With low dispersion of $\hat{\boldsymbol{\omega}}$ (as in Fig. \ref{Fig4-12}) the mean-squared error is, however, appropriate.

With planar arrays, the computational complexity of this procedure increases dramatically: If one uses a $1/6BW$ grid in azimuth and elevation, then one has $15$ look directions. That means that $\binom{15}{2}=171$ is the number of $Q$-function evaluations in contrast to $\binom{7}{2}=21$ $Q$-function evaluations for a linear array. The minimization of $Q$ using a gradient method is theoretically possible. However, the entire data vector $\mathbf{z}$ must be saved and used in every iteration. A raster search or a stochastic approximation would thus be preferable. As long as the SNR is not very good (for $M=2$, for example, over $30\,\text{dB}$), elaborate, precise methods are not worth the effort. The resolution of multiple targets with a single pulse appears to be unrealistic due to the high-SNR demands.

\section{Stochastic Approximation}\label{Sec5}
\setcounter{figure}{0}
\begin{center}
\textbf{Summary of Chapter 5}
\end{center}

The question of whether a computationally feasible direction-estimation procedure exists is the decisive question for the application of a superresolution procedure. In this chapter, a stochastic approximation is suggested, which can be interpreted as an iterative recursion of simultaneous, decoupled difference beams. In the case of a single target, this procedure reduces itself to the iteration over the usual difference beam. The measurement vectors do not need to all be saved in the procedure, as only the current measured data goes into the computation via the simultaneous beams.

Initially, the conditions for the convergence of this procedure are given. A ``weak $M$-regularity'' (5.1-1) proves sufficient for the antenna layout. Subsequently, one can specify the loss compared to the optimal, ML procedure via the computation of the asymptotic distribution of the estimated directions. Using the optimal choice of the iteration convergence factor $\mu$, this loss is very small.

Thereafter, the procedure is tested through simulation on various deviations from the signal and noise models. It turns out to be relatively insensitive to model mismatches. 
~\\
\hrule
~\\

The SNR of a single pulse is generally not sufficient to resolve multiple targets, as in Chapter \ref{Sec4}. If multiple samples (pulses) are used $\mathbf{z}_1,\mathbf{z}_2,\ldots$, then it is desirable if the data is only used serially and if processors already installed in the radar for building sum-and-difference beams can be used. Thus, instead of processing a big vector $\mathbf{z}_k$, only vectors of length $M$ are used. This can be achieved by extending the gradient procedure of Chapter \ref{Sec4} to a stochastic approximation algorithm:
\begin{equation*}
\boldsymbol{\omega}_{n+1}=\boldsymbol{\omega}_n-a_n\grad Q(\boldsymbol{\omega}_n,\mathbf{z}_n)\quad n=1,2,3,\ldots
\end{equation*}
for an appropriate series of coefficients $a_n$ and
\begin{equation*}
Q(\boldsymbol{\omega}_n,\mathbf{z}_n)=\mathbf{z}_n^*\boldsymbol{\Gamma}(\boldsymbol{\omega}_n)\mathbf{z}_n
\end{equation*}

According to \eqref{eqA-1-4} in Appendix \ref{Appen1},
\begin{equation*}
Q_{u_i}=2\text{Re }j\hat{b}_i^*\left(\mathbf{a}_i^*\mathbf{D}_x\mathbf{z}-\mathbf{a}_I^*\mathbf{D}_x\mathbf{A}\hat{\mathbf{b}}\right)\tag{*}
\end{equation*}
where $\hat{\mathbf{b}}:=\left(\mathbf{A}^*\mathbf{A}\right)^{-1}\mathbf{A}^*\mathbf{z}$. An analogous solution applies to $Q_{v_i}$.

\begin{figure}
\centering
\includegraphics{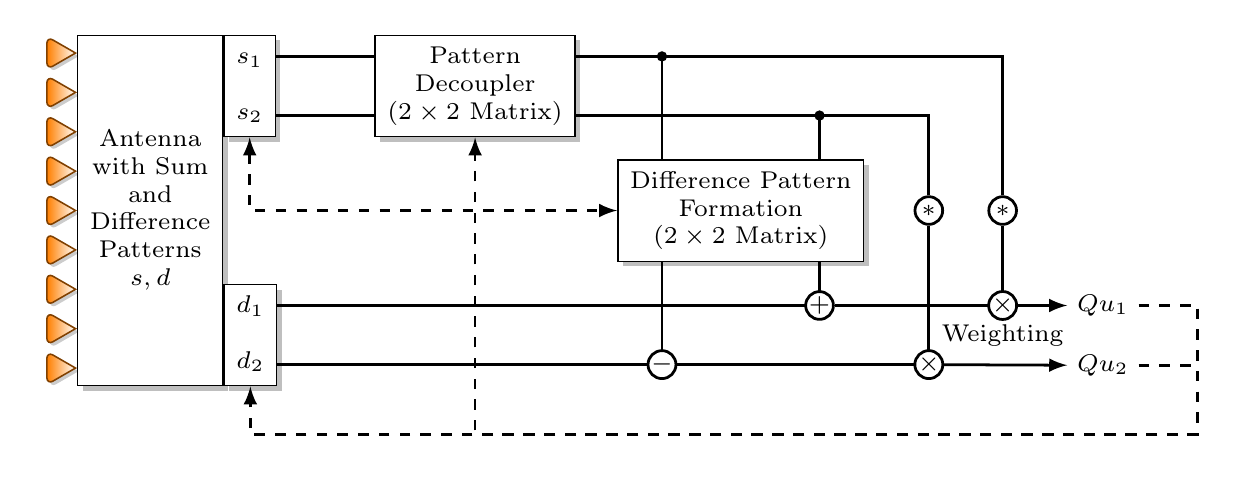}
\caption{Generalized difference beams for two targets.\label{Fig5-1}}
\end{figure}

Because $\mathbf{a}_i\mathbf{D}_x\mathbf{z}$ represents the construction of a difference beam response with respect to the $x$-axis in the direction $\omega_i$, the gradient of $Q$ is just computed from simultaneous sum-and-difference beamformer outputs
\begin{equation*}
d_{xi}:=\mathbf{a}_i^*\mathbf{D}_x\mathbf{z},d_{y_i}:=\mathbf{a}_i\mathbf{D}_y\mathbf{z},s_i=\mathbf{a}_i^*\mathbf{z}\quad(i=1\ldots M)
\end{equation*}
that are then multiplied by the previously known matrix functions $\mathbf{A}^*\mathbf{D}_x\mathbf{A}$, $\mathbf{A}^*\mathbf{D}_y\mathbf{A}$, $\left(\mathbf{A}^*\mathbf{A}\right)^{-1}$ (all $M\times M$ matrices). The necessary computing processes for the computation of (*) for two targets with a linear antenna are shown in Fig. \ref{Fig5-1}.

\subsection{Convergence and Asymptotic Behavior}\label{Sec5-1}
\setcounter{equation}{0}

The convergence and asymptotic behaviour of stochastic approximation algorithms depends decisively on the function  $\E\{Q(\boldsymbol{\omega})\}$, respectively $<Q(\boldsymbol{\omega})>$. Therefore, before this statement can be proven, these functions shall be investigated. According to (A.4-1) and Chapter 2.2, if stochastic signals are present,
\begin{align*}
\E\{Q\}=&\trace\mathbf{A}^*\boldsymbol{\Gamma}\mathbf{A}\mathbf{B}+\trace\boldsymbol{\Gamma}\quad\text{with $\mathbf{B}=\E\{\mathbf{b}\mathbf{b}^*\}$ Hermitian and regular}\\
=&\trace\mathbf{A}^*\boldsymbol{\Gamma}\mathbf{A}\mathbf{B}+N-M
\end{align*}
If deterministic signals are present, then $\mathbf{B}=<\mathbf{b}\mathbf{b}^*>$ can be singular. Because this is a boundary case of Model 2, in the following, stochastic signals are assume, so that $\mathbf{B}$ is regular.

The function $\E\{Q\}$ also possesses the properties (4.1-4), (4.1-5), (4.1-6), (4.1-7), (4.1-8), because these are properties of the projection matrix $\boldsymbol{\Gamma}$. If $\mathbf{B}$ is singular, then the statement on the unique solution of (4.2-2) also holds. If $\mathbf{B}$ is regular, then the requirements can be weakened, whereby the uniqueness of the representation $\mathbf{s}=\mathbf{A}\mathbf{b}$ is no longer required. Arrays that have a clear minimum of $\E\{Q\}$ should thus be weakly $M$-regular.

\underline{Definition (5.1-1)}:

Let $\Omega\subset V^1$. An ordering $(x_1\ldots x_N),(y_1\ldots y_N)$ of $N$ antenna elements is
\begin{equation*}
\text{Weakly $M$-regular on $\Omega$}\Longleftrightarrow\bigwedge\limits_{\omega_1\ldots\omega_{M+1}\in\Omega \atop \text{pairwise different}}:(\mathbf{a}(\omega_1),\ldots,\mathbf{a}(\omega_{M+1})) \text{ are linearly independent.}
\end{equation*}

Thus the following theorem holds:\\
\underline{Theorem (5.1-2)}:
\begin{itemize}
\item \textbf{Preconditions}

Let $(x_i,y_i)$, $i=1,\ldots, N$ be a weakly $M$-regular antenna array on the region $\Omega\subset V^1$. Let $\E\{\mathbf{s}\mathbf{s}^*|\}=\mathbf{A}\mathbf{B}\mathbf{A}^*$ with $\mathbf{B}$ Hermitian and regular.

\item \textbf{Assertions}

In every symmetric region of $\Omega^M$, according to (4.1-5), $\E\{Q\}$ possesses a clear minimum. That means
\begin{equation*}
\left(\E\{Q(\boldsymbol{\omega})\}=\min! \wedge \E\{Q(\tilde{\boldsymbol{\omega}})\}=\min!\right)\Longrightarrow \bigvee\limits_{\mathbf{P}\text{ permutation}}:\mathbf{P}\boldsymbol{\omega}=\tilde{\boldsymbol{\omega}}.
\end{equation*}

Because the ML-estimation asymptotically minimizes $E\{Q\}$, this theorem gives the conditions for consistency of the ML estimation (3.2-3). A comparison of this theorem with Theorem (4.2-2) also shows how far temporal sampling can replace spatial sampling.

\item \textbf{Proof}

Because $\mathbf{B}$ is positive definite and Hermitian, there exists a lower-triangular Cholesky decomposition matrix $\mathbf{L}$ with $\mathbf{B}=\mathbf{L}\mathbf{L}^*$ as follows:
\begin{align*}
\E\{Q(\boldsymbol{\omega})\}=\min!\Longrightarrow0=&\trace\mathbf{A}^*\boldsymbol{\Gamma}\mathbf{A}\mathbf{L}\mathbf{L}^*\\
=&\trace\mathbf{L}^*\mathbf{A}^*\boldsymbol{\Gamma}^*\boldsymbol{\Gamma}\mathbf{A}\mathbf{L}\\
=&\sum_{i=1}^M\lVert \left(\mathbf{\Gamma}\mathbf{A}\mathbf{L}\right)_{\text{$i$th row}}\rVert^2\quad(\boldsymbol{\Gamma}=\boldsymbol{\Gamma}^*=\boldsymbol{\Gamma}^2)\\
=&\sum_{i=1}^M\lVert\boldsymbol{\Gamma}\mathbf{A}\mathbf{l}_i\rVert^2\quad \text{with $\mathbf{l}_i=(\mathbf{L})_{\text{$i$th row}}$}.
\end{align*}

Because $\mathbf{L}$ is triangular, it follows for $i=M,\ldots,1$ backwards that
\begin{equation*}
\bigwedge\limits_{i\in\{1\ldots\}}:\lVert\boldsymbol{\Gamma}\mathbf{a}_i\rVert^2=0
\end{equation*}
but $\boldsymbol{\Gamma}\mathbf{A}=\tilde{\boldsymbol{\Gamma}}\mathbf{A}=\mathbf{0}\Longrightarrow\lin H(\mathbf{A}(\boldsymbol{\omega}))=\lin H(\mathbf{A}(\tilde{\boldsymbol{\omega}}))$. If there were to be a $\omega_i$ with $\omega_i\neq\tilde{\omega}_k$ for all $k=1\ldots M$, then there would be $\mathbf{b}\in\mathbb{C}^M$ with $\mathbf{a}(\omega_i)=\sum_{k=1}^M\mathbf{a}(\tilde{\boldsymbol{\omega}}_k)b_k$. That means, $(\mathbf{a}(\tilde{\omega}_1),\ldots,\mathbf{a}(\tilde{\omega}_M))$ are linearly independent, which is a contradiction!
\end{itemize}

According to the comments on Theorem (4.2-2), it is also true that every linear antenna with $M+1$ elements on a regular grid is $M$-regular. Moreover, every planar array with $(M+1)\times(M+1)$ elements on a regular grid is $M$-regular in the neighborhood of the main beam. Most antenna arrays are also weakly $2$-regular:

\underline{Theorem (5.1-3)}:
\begin{itemize}
\item (i) Every linear array with $N\geq3$ is weakly $2$-regular in a $\wp$-neighborhood of the main beam $\Omega_\wp=\left\{u\left||u|\leq \wp BW\right.\right\}$, whereby $\wp$ depends on the element spacing.
\item (ii) A planar array is precisely weakly $2$-regular in the neighborhood of the main beam $\Omega_\wp=\{u^2+v^2\leq \wp^2 BW^2\}$ if the projection of the element positions on every line passing through the planar array forms a weakly $2$-regular order on $\Omega_\wp=\left\{u\left||u|\leq \wp BW\right.\right\}$.
\item (iii) If an ordering for $\boldsymbol{\omega}\in \Omega^{M+1}$ is weakly $M$-regular, then it is also weakly $M$-regular for $\boldsymbol{\omega}\in(\omega_o+\Omega) \times\ldots \times(\omega_0+\Omega)$.
\end{itemize}
See Appendix 3 for the proof.

Thus, the linear $3$-element array with $\lambda/2$ spacing between elements, for example, is weakly $2$-regular, but not strongly $2$-regular. The element arrangements (b) and (c) at the end of Chapter 4.2 are likewise weakly $2$-regular. Thus, one has a simple criterion by which to construct antenna arrays with a minimal number of elements such that two targets can be resolved.

It must be shown that the stochastic approximation algorithm actually converges to the minimum of $\E\{Q\}$.

Since the function $\E\{Q\}$ does not possess a clear minimum, the approximation algorithm must, as in (4.4-1), be defined within a $\wp$-neighborhood of the global minimum. This means that the sequence $\boldsymbol{\omega}_n$ is subject to restrictions. For such demands, the conditions of Ljung \cite{ref23} are given. Theorem 4 of \cite{ref23} shall be formulated for the case and notation considered here:

\setcounter{equation}{3}

Let there be measurement data of the form $\mathbf{z}_n=\mathbf{C}\mathbf{e}_n$, $\mathbf{z}\in\mathbb{C}^N$, $\mathbf{C}\in\mathbb{C}^{N\times K}$, and $\mathbf{e}_n$ a sequence of random or deterministic vectors in $\mathbb{C}^K$. Let $\Omega\subset\mathbb{R}^M$, $\Omega$ be open and connected, $\Omega_2\subset\Omega_1\subset\Omega$ with $\Omega_1$ open and bounded and with $\Omega_2$ compact. Considering the iteration specification
\begin{equation}
\boldsymbol{\omega}_{n+1}=\left[\boldsymbol{\omega}_n+\frac{1}{n}\mathbf{G}(\boldsymbol{\omega}_n\mathbf{z}_n)\right]_{\Omega_1,\Omega_2}\quad n=1,2,\ldots
\label{Eq5-1-4}
\end{equation}
where
\begin{equation*}
\left[\boldsymbol{\omega}\right]_{\Omega_1,\Omega_2}:=\left\{\begin{IEEEeqnarraybox}[\relax][c]{c's}
\boldsymbol{\omega}&for $\boldsymbol{\omega}\in\Omega_1$\\
\text{a value in $\Omega_2$}&if $\boldsymbol{\omega}\notin\Omega_1$
\end{IEEEeqnarraybox}\right.
\end{equation*}
and $\mathbf{G}:\mathbb{R}^M\times\mathbb{C}^N\rightarrow\mathbb{R}^M$ is a deterministic function. The following preconditions apply to $\Omega$:
\begin{itemize}
\item \underline{A1}: 

$\mathbf{G}(\boldsymbol{\omega},\mathbf{z})$ is locally Lipschitz-continuous on $\Omega$. That means
\begin{equation*}
\bigwedge\limits_{\boldsymbol{\omega}\in\Omega}\bigvee\limits_{\epsilon\in\mathbb{R}_{+}}\bigwedge\limits_{\boldsymbol{\omega}_1,\boldsymbol{\omega}_2\in V_\epsilon(\boldsymbol{\omega})}:\left\lVert\mathbf{G}(\boldsymbol{\omega}_1,\mathbf{z})-\mathbf{G}(\boldsymbol{\omega_2},\mathbf{z})\right\rVert^2<K_\epsilon(\boldsymbol{\omega},\mathbf{z})\lVert\boldsymbol{\omega}_1-\boldsymbol{\omega}_2\rVert^2
\end{equation*}
where $V_\epsilon(\boldsymbol{\omega}):=\text{an open sphere around $\boldsymbol{\omega}$ with radius $\epsilon$}$.

\item \underline{A2}:

\begin{equation*}
\frac{1}{n}\sum_{k=1}^n\mathbf{G}(\boldsymbol{\omega},\mathbf{z}_k)\rightarrow \mathbf{f}(\boldsymbol{\omega})\quad\text{for $n\rightarrow\infty$ for all $\boldsymbol{\omega}\in\Omega$}
\end{equation*}

\item \underline{A3}:

\begin{equation*}
\frac{1}{n}\sum_{k=1}^nK_\epsilon(\boldsymbol{\omega},\mathbf{z}_k)\quad\text{converges for $n\rightarrow\infty$ for all $\boldsymbol{\omega}\in\Omega$}.
\end{equation*}
\end{itemize}

Consider the differential equation
\begin{equation}
\dot{\boldsymbol{\omega}}=\mathbf{f}(\boldsymbol{\omega})\quad(\boldsymbol{\omega}\in\Omega).
\label{Eq5-1-5}
\end{equation}
Let $\bar{\boldsymbol{\omega}}$ be the only stationary point in $\Omega$ (that means $\mathbf{f}^{-1}\{\mathbf{0}\}\cap\Omega=\{\bar{\boldsymbol{\omega}}\}$) with attraction region $\Omega_A\supset\bar{\Omega}$ (equals the set of all points for which the trajectories $\bar{\boldsymbol{\omega}}$ converge). Then, the following holds (as per Ljung, 1977)

\underline{Theorem (5.1-6)}: (Ljung, 1977)

If A1, A2, and A3 are fulfilled, and all trajectories of the differential equation (5.1-5) that begin in $\Omega_2$ never leave $\Omega_1$ (meaning that for every solution from $\{\dot{\boldsymbol{\omega}}=\mathbf{f}(\boldsymbol{\omega}), \boldsymbol{\omega}(0)=\boldsymbol{\omega}_0\in\Omega_2\}$ that $w(t)\in\Omega_1$ for $t\in(0\infty)$), then $\boldsymbol{\omega}_n\rightarrow\bar{\omega}$ as $n\rightarrow\infty$. In the event that $\mathbf{e}_n$ is stochastic and A2, A3 hold with probability $1$, then the convergence also occurs with probability $1$.

\begin{figure}
\centering
\includegraphics[width=0.5\textwidth]{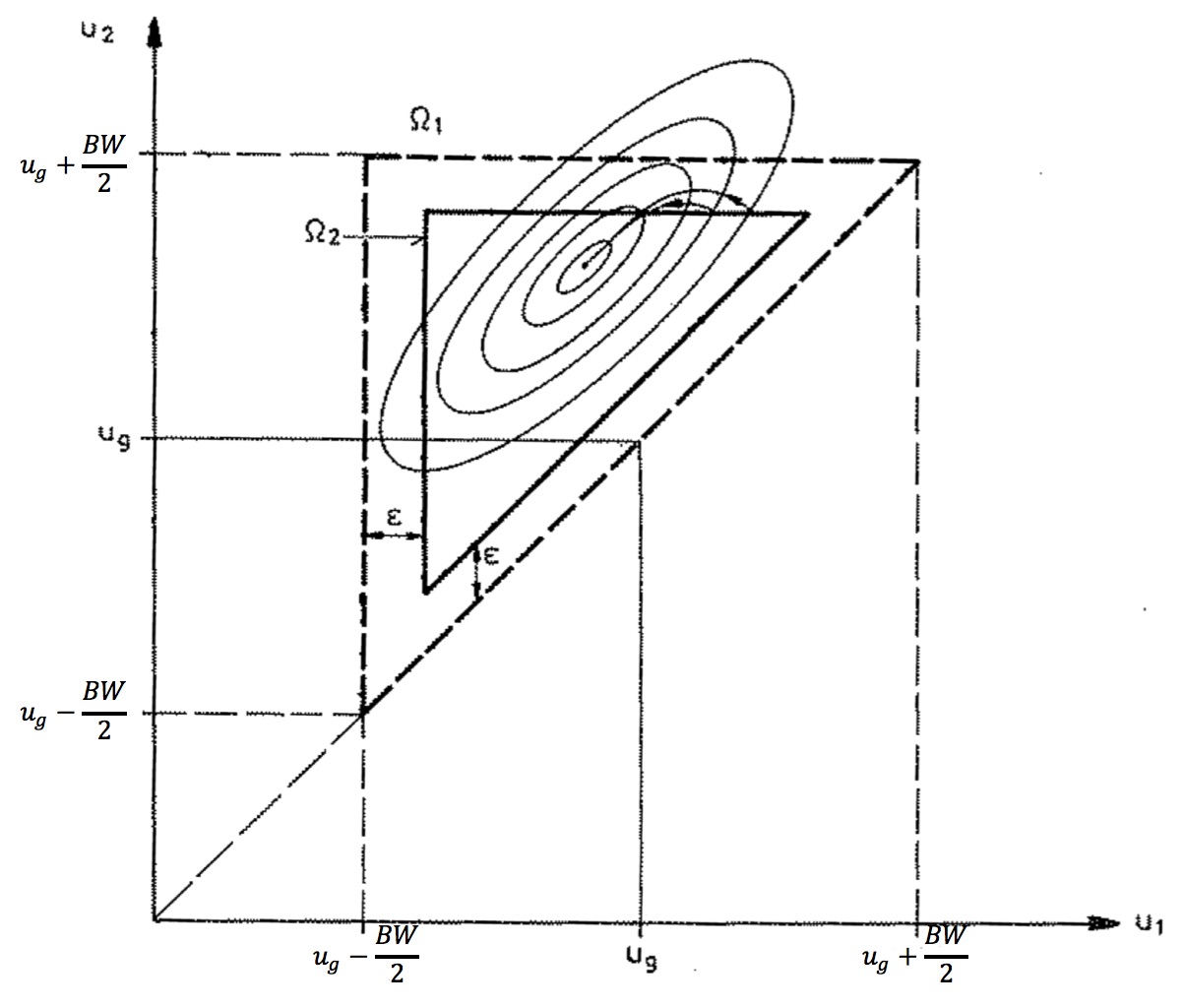}
\caption{\label{Fig5-2}}
\end{figure}

\underline{Application to the $Q$-Function}:
\begin{enumerate}
\item The given measurement data $\mathbf{z}_i=\mathbf{A}\mathbf{b}_i+\mathbf{n}_i$ corresponds to the desired structure with
\begin{align*}
\mathbf{C}=&(\mathbf{A},\mathbf{I})\in\mathbb{C}^{N\times(M+N)}\\
\mathbf{e}_i=&\left(\begin{IEEEeqnarraybox}[\relax][c]{c}
\IEEEstrut
\mathbf{b}_i\\
\mathbf{n}_i
\IEEEstrut
\end{IEEEeqnarraybox}\right)\in\mathbb{C}^{M+N}.
\end{align*}

\item If $\mathbf{G}(\boldsymbol{\omega},\mathbf{z})=-\grad Q(\boldsymbol{\omega},\mathbf{z})$, then A1 holds, because $Q$ is twice differentiable on $V^M\backslash \gamma$ and 
\begin{equation*}
\left\lVert Q_{\boldsymbol{\omega}}(\boldsymbol{\omega}_1)-Q_{\boldsymbol{\omega}}(\boldsymbol{\omega}_2)\right\rVert^2 \leq \left\lVert Q_{\boldsymbol{\omega}\boldsymbol{\omega}}(\tilde{\boldsymbol{\omega}})\right\rVert^2\left\lVert\boldsymbol{\omega}_1-\boldsymbol{\omega}_2\right\rVert^2
\end{equation*}
for a matrix norm that matches a vector 2-norm and $\tilde{\boldsymbol{\omega}}$ is between $\boldsymbol{\omega}_1$ and $\boldsymbol{\omega}_2$.

The Lipschitz-constant $K$ can be found by estimating the elements of $Q_{\boldsymbol{\omega}\boldsymbol{\omega}}$ in a neighborhood $V_\epsilon(\boldsymbol{\omega})$ with $\boldsymbol{\omega}_1,\boldsymbol{\omega}_2,\tilde{\boldsymbol{\omega}}\in V_\epsilon(\boldsymbol{\omega})$. For example,
\begin{align*}
\left\lVert Q_{\boldsymbol{\omega}\boldsymbol{\omega}}\right\rVert^2=&\sum_{i,k=1}^MQ_{\omega_i\omega_k}^2\\
=&\sum_{i,k=1}^M\left(\mathbf{z}^*\boldsymbol{\Gamma}_{\omega_i\omega_k}\mathbf{z}\right)^2\\
\leq&\sum_{i,k=1}^M\left(\sum_{r,l=1}^M\Gamma_{r,l\,\omega_i\omega_k}^2\right)^2\left\lVert\mathbf{z}\right\rVert^4\\
=&\left(\sum_{i,k=1}^M\trace \boldsymbol{\Gamma}^2_{\omega_i\omega_k}\right)^2\lVert\mathbf{z}\rVert^4\\
\leq&\max\limits_{V_\epsilon(\boldsymbol{\omega})}\left(\sum \trace \boldsymbol{\Gamma}^2_{\omega_i\omega_k}\right)^2\lVert\mathbf{z}\rVert^2\\
=&K_\epsilon(\boldsymbol{\omega})\lVert\mathbf{z}\rVert^4
\end{align*}
whereby $\epsilon<\text{dist}(\boldsymbol{\omega},\gamma)$, and $\gamma$ is according to (4.1-7).

\item
\begin{align*}
\frac{1}{n}\sum_{k=1}^n\text{grad}Q(\boldsymbol{\omega},\mathbf{z}_k)=&\frac{1}{n}\sum_{k=1}^n\mathbf{z}_k^*\boldsymbol{\Gamma}_{\boldsymbol{\omega}}\mathbf{z}_k\\
=&\trace \boldsymbol{\Gamma}_{\boldsymbol{\omega}}\frac{1}{n}\sum_{k=1}^n\mathbf{z}_k\mathbf{z}_k^*
\end{align*}
This limit value exists according to (3.2-4) and
\begin{equation*}
\frac{1}{n}\sum_{k=1}^n\grad Q(\boldsymbol{\omega},\mathbf{z}_k)\rightarrow\grad \E\{Q(\boldsymbol{\omega})\}\quad (n\rightarrow\infty)
\end{equation*}
Thus, A2 holds.

\item The Lipschitz-constant always has the form $K_\epsilon\lVert\mathbf{z}\rVert^4$. Thus, for A3, the existence of the fourth moment of $\mathbf{z}$ is required. This is a limitation on the allowed amplitude distributions of the signals. The density from $|b|$ must be stronger than $1/x^5$ for $x\rightarrow\infty$ to drop out. In practice, this is not a problem, because the input signals are bounded and thus all moments exist.

\item Decisive for the convergence behavior is the differential equation $\dot{\boldsymbol{\omega}}=-\E\left\{\grad Q(\boldsymbol{\omega})\right\}$. that is, the function $\E\{Q(\boldsymbol{\omega})\}$. It determines the choice of regions $\Omega,\Omega_1,\Omega_2$. One gets an approximation for the minimum by using conventional localization of the targets for the coarse directions $\boldsymbol{\omega}_g$. If all targets are within a beamwidth, then using (5.1-2), one can predict that the minimum of $\E\{Q\}$ on $M(\boldsymbol{\omega}_g)$ (see (4.4-1)) is unique. All considered cuts through $Q$ and $\E\{Q\}$ corresponding to Figs. \ref{Fig4-6} and \ref{Fig4-7} with arbitrary cut angles, show that $M(\boldsymbol{\omega}_g)$ is also a part of the attraction region $\Omega_A$. One can thus choose $\Omega=\Omega_1=M(\boldsymbol{\Omega}_g)$. $\Omega_2$ is now correctly chosen, in case a ``potential surface'' $F$ of $\E\{Q\}$ lies exactly ``between $\Omega_1$ and $\Omega_2$,'' that is $F\subset\Omega_q$, $F\cap\Omega_2=\emptyset$ and $f\subset\Omega_1\backslash\Omega_2$. However, this is generally not possible, because the position of the minimum in $\Omega_1$ is unknown. If one instead uses
\setcounter{equation}{6}
\begin{equation}
\Omega_2:=\left\{\boldsymbol{\omega}\left|\bigwedge\limits_{i\in\{1\ldots M\}}:\lVert\omega_i-\omega_{ig}\rVert^2\leq\left(\frac{BW}{2}(1-\epsilon)\right)^2\wedge\bigwedge\limits_{i\neq k}\left\lVert\omega_i-\omega_k\right\rVert^2\geq\left(\frac{BW}{2}\epsilon\right)^2\right.\right\}
\end{equation}
then the condition is fulfilled if $\boldsymbol{\omega}_{\text{ex}}$ is an element of $\Omega_2$ and $\epsilon$ is sufficiently large. With a linear antenna and $M=2$, the following relations in Figure \ref{Fig5-2} are demanded.
\end{enumerate}

With this choice of $\Omega_2$, if one passes through an ellipsoid to approach the minimum of $\E\{Q\}$, it is necessary that this ellipsoid not be too eccentric with respect to the chosen $\epsilon$. With deterministic signals, the estimation is already poorly conditioned (consider Chapter 4.3) so that the stochastic approximation converges too slowly for the most practical cases, even with the correct $\Omega_2$. With stochastic signals, the curvature at the minimum of $\E\{Q\}$ is
\begin{equation*}
\E\left\{Q_{\boldsymbol{\omega}\boldsymbol{\omega}}\right\}=\trace \Gamma_{\boldsymbol{\omega}\boldsymbol{\omega}}\E\left\{\mathbf{z}\mathbf{z}^*\right\}.
\end{equation*}
For Signal Models $3$ and $4$ (uncorrelated signals), the matrix $\mathbf{B}$ in the equation
\begin{equation*}
\E\left\{\mathbf{z}\mathbf{z}^*\right\}=\mathbf{I}+\mathbf{A}\mathbf{B}\mathbf{A}^*
\end{equation*}
is diagonal so that
\begin{align*}
\E\left\{Q_{\omega_i\omega_k}\right\}=&\trace \mathbf{A}^*\boldsymbol{\Gamma}_{\omega_i\omega_k}\mathbf{A}\mathbf{B}\quad\text{from \eqref{eqA-1-6} and \eqref{eqA-1-7}}\\
=&\sum_{l=1}^M\wp_l^2\mathbf{a}_l^*\boldsymbol{\Gamma}_{\omega_i\omega_k}\mathbf{a}_l\quad\text{when $\mathbf{B}=\diag(\wp_l^2)$}\\
=&\wp_i^22\text{Re }\mathbf{a}_{i\omega_i}^*\boldsymbol{\Gamma}\mathbf{a}_{i,\omega_i}\delta_{ik}\quad\text{from \eqref{eqA-1-7}}.
\end{align*}
That means $\E\{Q_{\boldsymbol{\omega}\boldsymbol{\omega}}\}$ is diagonal, the principle directions of curvature are the coordinate axes and the eccentricity is essentially determined by the power of the targets $\wp_l^2$. These target powers are going to be nearly equal for large targets. If not, the estimate is poorly conditioned. If the minimum is rotationally symmetric, then all trajectories are approximately straight lines and all trajectories from $\Omega_1$ remain in $\Omega_1$, because $\Omega_1$ is convex.

By a sufficiently good conditioning of the minimization problem, the following holds:\\
\underline{Corollary (5.1-8)} (Convergence of the stochastic approximation)
\begin{itemize}
\item \textbf{Preconditions}
$\E\left\{\mathbf{z}\mathbf{z}^*\mathbf{z}\mathbf{z}^*\right\}$ exists, $\mu\in\mathbb{R}_{+}$. All trajectories of $\dot{\boldsymbol{\omega}}=-\grad \E\{Q\}$ that begin in $\Omega_2$ end up in in $\Omega_1$.

\item \textbf{Assertions}

For
\begin{equation*}
\boldsymbol{\omega}_{n+1}=\left[\boldsymbol{\omega}_n-\frac{\mu}{n}\grad Q(\boldsymbol{\omega}_n,\mathbf{z}_n)\right]_{\Omega_1\Omega_2}
\end{equation*}
it is true that $\boldsymbol{\omega}_n\rightarrow\bar{\boldsymbol{\omega}}$, almost certainly (for $\grad \E\{Q\}(\bar{\boldsymbol{\omega}})=\mathbf{0}$).
\end{itemize}

For convergence, correction vectors $\mathbf{G}$, other than $\mathbf{G}=\grad Q$, are possible as long as the conditions of Theorem (5-1-6) are fulfilled. One can, for example, replace the derivatives $\mathbf{a}_\omega=-j\mathbf{D}\mathbf{a}$ ($\mathbf{D}=\diag(x_i)$ for $\omega=u$, $D=\diag (y_i)$ if $\omega=v$) by the difference beam $\tilde{\mathbf{D}}\mathbf{a}$ with $\tilde{\mathbf{D}}=\diag(\sign x_i)$ (The difference beam in $v$ with $y_i$ is analogous). and then use these in the computation of the gradient. If the antenna elements are symmetric about $(x_0,y_0)=(0,0)$, then this difference beam passes through $0$ at the maximum point of the sum beam and the following holds for a correction-vector $\tilde{\mathbf{G}}$ constructed in such a manner
\begin{equation*}
\tilde{\mathbf{G}}(\boldsymbol{\omega})=\mathbf{0}\Longleftrightarrow Q(\boldsymbol{\omega})=\min!
\end{equation*}
This equation further shows that beyond the CRLB, an analysis of the surface $\E\{Q\}$ according to (4.3-3) also provides information. Attempting to make statements about the distance from the exact minimum after an infinite number of iterations is quite futile. One can, however, draw conclusions about the asymptotic distribution of the estimate and the asymptotic convergence rate, because stochastic approximations are often asymptotically normally distributed. One can then compare the covariance matrix of the asymptotic distribution with the FIM and check the loss with respect to ``optimal'' ML estimation. An asymptotic convergence rate $\beta$ means that for a sequence $x_n\rightarrow x$, $\beta$ is the largest number for which $n^\beta(x_n-x)$ converges to a fixed distribution.

Initially, a theorem from \cite{ref24} regarding the asymptotic normal distribution shall be cited in a simplified manner:\\
\underline{Theorem (5.1-9)} (Fabian, 1967)
\begin{itemize}
\item \textbf{Preconditions}
\begin{equation*}
\boldsymbol{\epsilon}_{n+1}=\boldsymbol{\epsilon}_n-\frac{1}{n}\mathbf{G}_n\boldsymbol{\epsilon}_n+\frac{1}{n}\mathbf{r}_n\label{FabEq1}
\end{equation*}
where $\boldsymbol{\epsilon}_n,\mathbf{r}_n$ are sequences of random vectors in $\mathbb{R}^M$, and $\mathbf{G}_n$ is a sequence of matrices in $\mathbb{R}^{M\times M}$. Also,
\begin{equation*}
\boldsymbol{\epsilon}_n\rightarrow \mathbf{0}, \mathbf{G}_n\rightarrow\mathbf{G}\quad\text{almost certainly}\label{FabEq2}
\end{equation*}
and $\mathbf{G}$ is positive definite, $\mathbf{U}^T\mathbf{G}\mathbf{U}=\boldsymbol{\Lambda}$ is diagonal and
\begin{equation*}
\min\limits_i\Lambda_{ii}>\frac{1}{2}\label{FabEq3}
\end{equation*}

\begin{equation*}
\E\left\{\mathbf{r}_n|\boldsymbol{\epsilon}_n|\right\}=\mathbf{0}\text{ and }c_1>\left\lVert\E\left\{\left.\mathbf{r}_n\mathbf{r}_n^T\right|\boldsymbol{\epsilon}_n\right\}-\boldsymbol{\Sigma}\right\rVert\rightarrow0\text{ for a }\boldsymbol{\Sigma}\in\mathbb{R}^{M\times M}\label{FabEq4}
\end{equation*}

\begin{equation*}
\E\left\{\left\lVert\mathbf{r}_n\right\rVert^2\right\}\leq c_2.\label{FabEq5}
\end{equation*}

\item \textbf{Assertions}

\begin{equation*}
\sqrt{n}\boldsymbol{\epsilon}_n\sim\mathcal{N}_{\mathbb{R}^M}\left(\mathbf{0},\mathbf{U}\mathbf{M}\mathbf{U}^T\right)
\end{equation*}
with
\begin{equation*}
M_{ij}=\frac{\left(\mathbf{U}\boldsymbol{\Sigma}\mathbf{U}\right)_{ij}}{\Lambda_{ii}+\Lambda_{jj}-1}\quad(i,j=1\ldots M)
\end{equation*}
\end{itemize}

This theorem can be applied as follows:
\begin{enumerate}
\item It suffices to consider the algorithm (5.1-4) without projection onto the region $\Omega_2$, because only the asymptotic portion of the sequence $\boldsymbol{\Omega}_n$ is of interest and thus $\boldsymbol{\omega}_n\in\Omega_2$.
\item If $\boldsymbol{\omega}_n\rightarrow\bar{\boldsymbol{\omega}}$ almost certainly holds (5.1-8), then considering the sequence $\boldsymbol{\epsilon}_n:=\boldsymbol{\omega}_n-\bar{\boldsymbol{\omega}}$, one can say that $\boldsymbol{\epsilon}_n\rightarrow\mathbf{0}$ and
\begin{align*}
\boldsymbol{\epsilon}_{n+1}=&\boldsymbol{\epsilon}_n-\frac{\mu}{n}\grad Q(\bar{\boldsymbol{\omega}}+\boldsymbol{\epsilon}_n,\mathbf{z}_n)\\
=&\boldsymbol{\epsilon}_n-\frac{\mu}{n}\E\left\{\left.Q_{\boldsymbol{\omega}}(\bar{\boldsymbol{\omega}}+\boldsymbol{\epsilon}_n,\mathbf{z}_n)\right|\boldsymbol{\epsilon}_n\right\}+\frac{\mu}{n}\left[\E\left\{\left.Q_{\boldsymbol{\omega}}(\bar{\boldsymbol{\omega}}+\boldsymbol{\epsilon}_n,\mathbf{z})n\right|\boldsymbol{\epsilon}_n\right\}-Q_{\boldsymbol{\omega}}(\bar{\boldsymbol{\omega}}+\boldsymbol{\epsilon}_n,\mathbf{z}_n)\right].
\end{align*}

The required linearization of the iteration about $\bar{\boldsymbol{\omega}}$ in the precondition with error $\mathbf{r}_n$ can be achieved through a Taylor expansion.

\item $\E\left\{\left.Q_{\boldsymbol{\omega}}(\bar{\boldsymbol{\omega}}+\boldsymbol{\epsilon}_n,\mathbf{z}_n)\right|\boldsymbol{\epsilon}_n\right\}$ can be brought into the form $\mathbf{G}_n\boldsymbol{\epsilon}_n$ via a Taylor expansion:
\begin{equation*}
Q_{\boldsymbol{\omega}}(\bar{\boldsymbol{\omega}}+\boldsymbol{\epsilon}_n)=Q_{\boldsymbol{\omega}}(\bar{\boldsymbol{\omega}})+Q_{\boldsymbol{\omega}\boldsymbol{\omega}}(\bar{\boldsymbol{\omega}})\boldsymbol{\epsilon}_n+\frac{1}{2}\boldsymbol{\epsilon}_nQ_{\boldsymbol{\omega}\boldsymbol{\omega}\boldsymbol{\omega}}(\boldsymbol{\xi}_n)\boldsymbol{\epsilon}_n
\end{equation*}
where
\begin{equation*}
\xi_{n,i}=\bar{\omega}_i+\vartheta_i\epsilon_{n,i}\quad \vartheta_i\in(0,1).
\end{equation*}
Consequently,
\begin{equation*}
\E\left\{\left.Q_{\boldsymbol{\omega}}(\bar{\boldsymbol{\omega}}+\boldsymbol{\epsilon}_n,\mathbf{z}_n)\right|\boldsymbol{\epsilon}_n\right\}=\mathbf{0}+\E\left\{Q_{\boldsymbol{\omega}\boldsymbol{\omega}}(\bar{\boldsymbol{\omega}})\right\}\boldsymbol{\epsilon}_n+\frac{1}{2}\boldsymbol{\epsilon}_n^T\mathbf{M}_n\boldsymbol{\epsilon}_n
\end{equation*}
because $\E\left\{Q_{\boldsymbol{\omega}}(\bar{\boldsymbol{\omega}})\right\}=\mathbf{0}$ and $\mathbf{M}_n:=\E\left\{\left.Q_{\boldsymbol{\omega}\boldsymbol{\omega}\boldsymbol{\omega}}(\boldsymbol{\xi}_n)\right|\boldsymbol{\epsilon}_n\right\}$.

Let $\mu\left.\left[\E\left\{\left.Q_{\boldsymbol{\omega}}\right|\boldsymbol{\epsilon}_n\right\}-Q_{\boldsymbol{\omega}}\right]\right|_{\boldsymbol{\omega}_n}=:r_n$. then the precondition holds with
\begin{equation*}
\mathbf{G}_n:=\mu\left(\E\left\{Q_{\boldsymbol{\omega}\boldsymbol{\omega}}(\bar{\boldsymbol{\omega}})\right\}+\frac{1}{2}\boldsymbol{\epsilon}_n\mathbf{M}_n\right).
\end{equation*}

\item Because $\boldsymbol{\epsilon}\rightarrow\mathbf{0}$ and $Q_{\boldsymbol{\omega}\boldsymbol{\omega}\boldsymbol{\omega}}$ is continuous and bounded on $\Omega_2$, then $\mathbf{G}_n\rightarrow\mu\E\left\{Q_{\boldsymbol{\omega}\boldsymbol{\omega}}(\bar{\boldsymbol{\omega}})\right\}$ almost certainly holds and thus the corresponding precondition is satisfied.

\item Let $\lambda$ be the smallest eigenvalue of $\E\left\{Q_{\boldsymbol{\omega}\boldsymbol{\omega}}(\bar{\boldsymbol{\omega}})\right\}$ so the required condition on $\mu$ is that $\mu\lambda>\frac{1}{2}$. Thus, $\mu$ can not be chosen to be arbitrarily small.

\item By the structure of the problem, $\E\left\{\mathbf{r}_n|\boldsymbol{\epsilon}_n\right\}=\mathbf{0}$. Furthermore, for $\mathbf{r}_n=\mu\left(\E\left\{\left.Q_{\boldsymbol{\omega}}\right|\boldsymbol{\epsilon}_n\right\}-Q_{\boldsymbol{\omega}}\right)$, one can write
\begin{equation*}
\E\left\{\left.\mathbf{r}_n\mathbf{r}_n^T\right|\boldsymbol{\epsilon}_n\right\}=\mu^2\left(\E\left\{\left.Q_{\boldsymbol{\omega}}Q_{\boldsymbol{\omega}}^T\right|\boldsymbol{\epsilon}_n\right\}-\E\left\{\left.Q_{\boldsymbol{\omega}}\right|\boldsymbol{\epsilon}_n\right\}\E\left\{\left.Q_{\boldsymbol{\omega}}\right|\boldsymbol{\epsilon}_n\right\}^T\right)
\end{equation*}
but $\E\left\{\left.Q_{\boldsymbol{\omega}}\left(\bar{\boldsymbol{\omega}}+\boldsymbol{\epsilon}_n\right)\right|\boldsymbol{\epsilon}_n\right\}\rightarrow 0$ almost certainly, thus
\begin{equation*}
\E\left\{\left.\mathbf{r}_n\mathbf{r}_n^T\right|\boldsymbol{\epsilon}_n\right\}\rightarrow\mu^2\left.\E\left\{Q_{\boldsymbol{\omega}}Q_{\boldsymbol{\omega}}^T\right\}\right|_{\bar{\boldsymbol{\omega}}}=:\boldsymbol{\Sigma}
\end{equation*}
and thus the precondition is true.

\item Because $\boldsymbol{\epsilon}_n\in\Omega_2$ is bounded, $\E\left\{\left.\lVert\mathbf{r}_n\rVert^2\right|\boldsymbol{\epsilon}_n\right\}$ is also bounded and thus $\E\left\{\lVert\mathbf{r}_n\rVert^2\right\}$ is bounded and thus the precondition holds.
\end{enumerate}

Consequently the theorem is usable. The convergence rate of the stochastic approximation algorithm (5.1-8) is thus at least $\sqrt{n}$.

The magnitudes of the quantities $\mathbf{G}_n,\mathbf{r}_n$ in the theorem, that is, the matrices $\mathbf{G}=\mu\E\left\{Q_{\boldsymbol{\omega}\boldsymbol{\omega}}(\bar{\boldsymbol{\omega}})\right\}$ and $\boldsymbol{\Sigma}=\mu^2\left.\E\left\{Q_{\boldsymbol{\omega}\boldsymbol{\omega}}^T\right\}\right|_{\bar{\boldsymbol{\omega}}}$ determine the asymptotic variance.

For deterministic targets $\mathbf{z}=\mathbf{s}+\mathbf{n}$, where $\mathbf{s}$ is constant, the asymptotic variance can be well compared to the CRLB. In this instance,
\begin{equation*}
\E\left\{Q_{\boldsymbol{\omega}\boldsymbol{\omega}}(\bar{\boldsymbol{\omega}})\right\}=\mathbf{s}^*\boldsymbol{\Gamma}_{\boldsymbol{\omega}\boldsymbol{\omega}}\mathbf{s}
\end{equation*}
and according to Appendix 2 (A.2-1),
\begin{equation*}
\E\left\{Q_{\boldsymbol{\omega}}Q_{\boldsymbol{\omega}}^T(\bar{\boldsymbol{\omega}})\right\}=\mathbf{s}^*\boldsymbol{\Gamma}_{\boldsymbol{\omega}\boldsymbol{\omega}}\mathbf{s}.
\end{equation*}
If $\lambda_i$ are the eigenvalues of $\mathbf{s}^*\boldsymbol{\Gamma}_{\boldsymbol{\omega}\boldsymbol{\omega}}\mathbf{s}$, then from Theorem (5.1-9)
\begin{equation*}
M_{ik}=\delta_{ik}\frac{\mu^2\lambda_i}{2\mu\lambda_i-1}=\text{diagonal $(i,k=1\ldots M)$}
\end{equation*}
and the eigenvectors of the asymptotic covariance matrix are equal to the primary curvature directions. The maximum eigenvalue of $\mathbf{M}$ is minimized for the choice $\bar{\mu}=\frac{1}{\lambda_{\text{min}}}$ as illustrated:
\begin{center}
\includegraphics[width=0.5\textwidth]{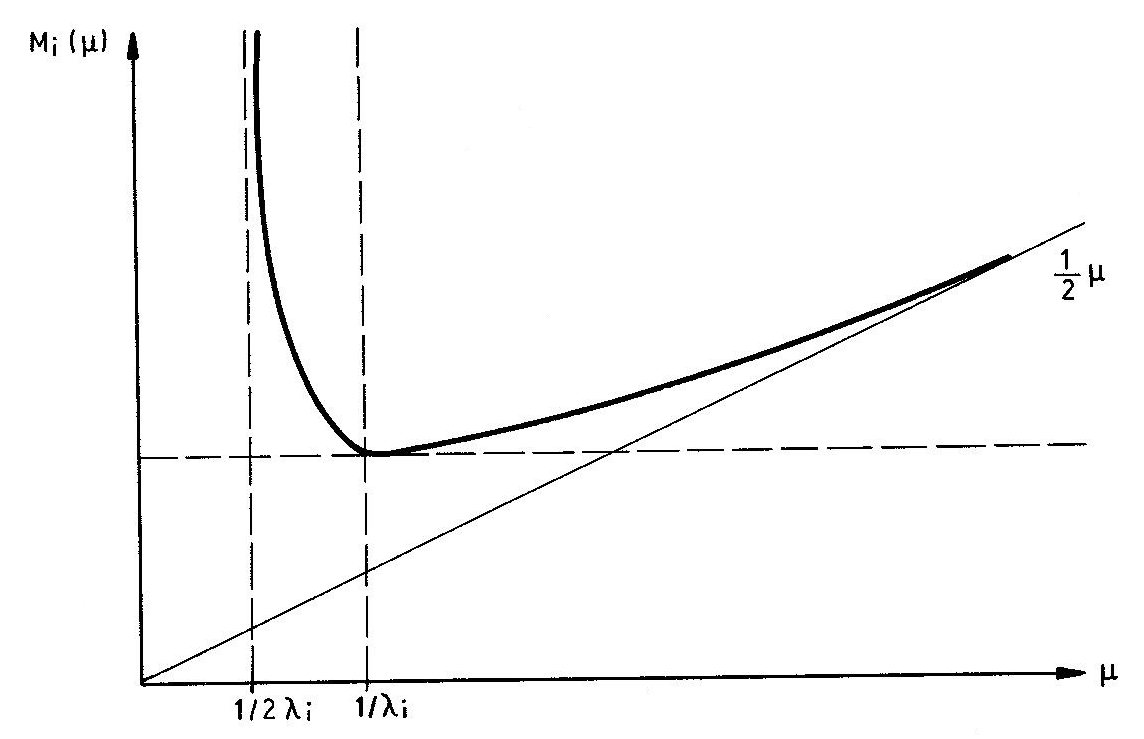}
\end{center}

If $\lambda_1<\lambda_2<\lambda_3\ldots$, then with $\mu=\frac{1}{\lambda_1}$, the $M_1>M_2>\ldots$. If the average dispersion $\trace\mathbf{M}$ is to be minimized, then $\mu<\frac{1}{\lambda_{\text{min}}}$ must be chosen.

\begin{figure}
\centering
\includegraphics[width=0.5\textwidth]{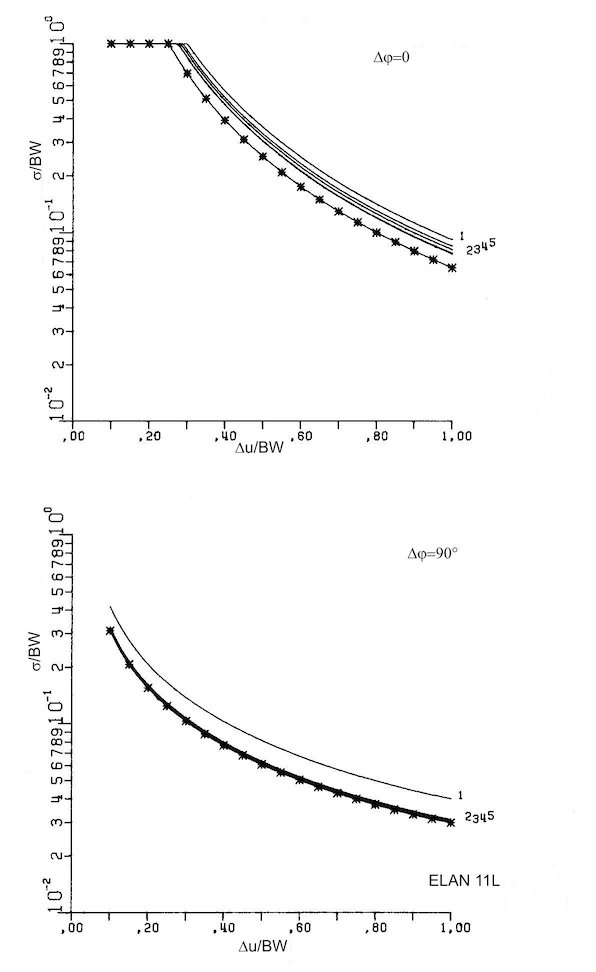}
\caption{Asymptotic dispersion of ML and the stochastic approximation \label{Fig5-3}}
\end{figure}

Because $M_i(\mu)\leq\frac{1}{\lambda_{\text{min}}}$, the stochastic approximation is generally not an asymptotically efficient estimator unlike the ML estimation (for Signal Model 1). However, the loss is not very large as can be seen in Fig. \ref{Fig5-3} for ELAN 11 L. The line with asterisks is the average dispersion according to the CRLB, that is $\left(\frac{1}{\lambda_1}+\frac{1}{\lambda_2}\right)/2$ and curves $1,2,3,4,5$ are the average dispersion of the stochastic approximation $(M_1+M_2)/2$ over ($u_1-u_2)/BW$ for $\mu=0.6\bar{\mu},0.8\bar{\mu},\bar{\mu},1.2\bar{\mu},1.4\bar{\mu}$ ($\bar{\mu}=\frac{1}{\lambda_{\text{min}}}$). One recognizes that with $\Delta\varphi=0^\circ$, $\bar{\mu}$ does not minimize the average dispersion, rather a somewhat smaller value does.

For stochastic, uncorrelated signals under Signal Models 3 and 4, the asymptotic covariance matrix can also be easily computed. In these instances $\E\left\{Q_{\boldsymbol{\omega}\boldsymbol{\omega}}\right\}$ is diagonal, so $\mathbf{U}=\mathbf{I}$, and $\boldsymbol{\Sigma}$ is then non-diagonal, so $\mathbf{M}$ is non-diagonal.

\begin{figure}
\centering
\includegraphics[width=0.6\textwidth]{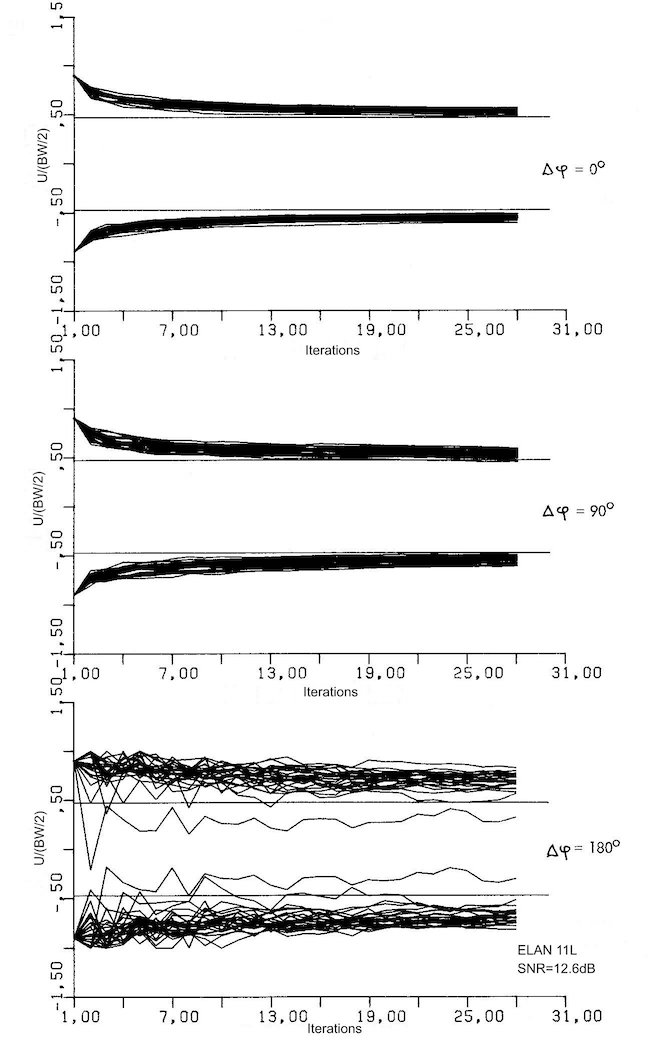}
\caption{\label{Fig5-4}}
\end{figure}

The convergence of the algorithm from (5.1-8) is shown in Fig. \ref{Fig5-4} for ELAN 11L, Signal Model 1 for\\ $\left(|b_1|^2+|b_2|^2\right)/\E\left\{|n|^2\right\}=12\text{dB}$.

\begin{figure}
\centering
\includegraphics[width=0.6\textwidth]{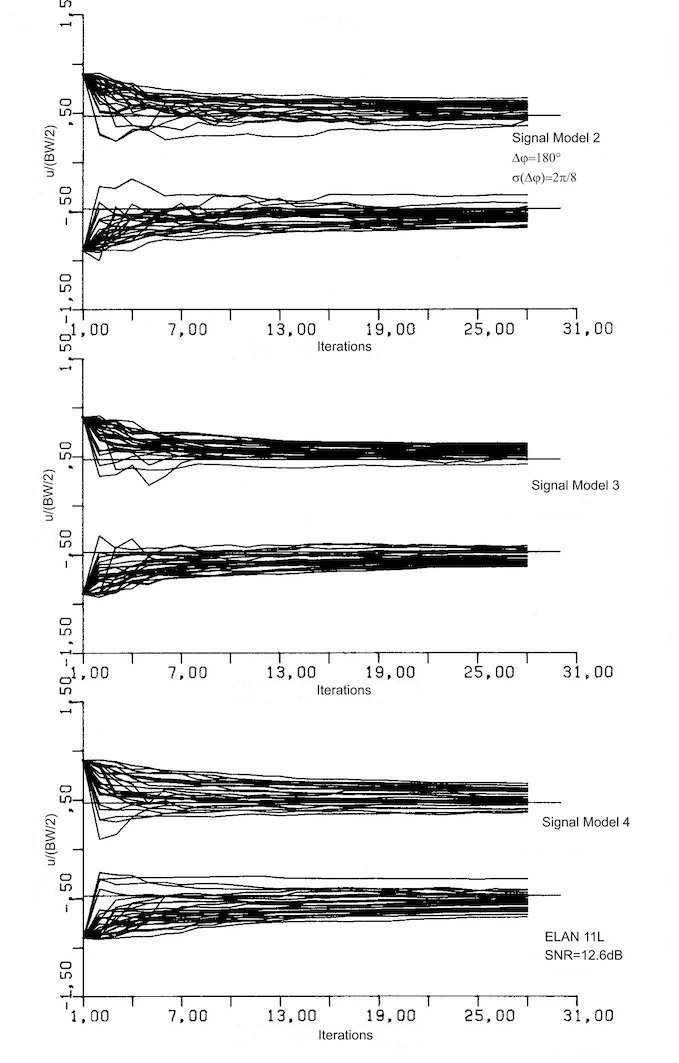}
\caption{\label{Fig5-5}}
\end{figure}

As a starting value, a widely separated configuration is always chosen as $\mathbf{u}_0=\left(u_g-0.9BW/2,u_g+0.9BW/2\right)$ where $u_g$ is a coarse estimate of the target directions, so that the matrix $\mathbf{A}^*\mathbf{A}$ has a good condition number. With this choice of starting values, if $\Delta\varphi=\pi$, then there is practically no convergence, because the starting value already provides a strong extraction of the signal portion. That is, $\lVert\boldsymbol{\Gamma}\mathbf{s}\rVert$ is small, consider (4.3-3,1) or respectively Fig. \ref{Fig4-7}. The algorithm actually converges in this instance, but only with $\sqrt{n}$ while with $\Delta\phi=0,\pi/2$, the rough position of the minimum is found faster. Fig. \ref{Fig5-5} shows the same configuration for Signal Model 2 with $\sigma(\Delta\varphi)=2\pi/8$, and Signal Models 3 and $4$. The plots show that it is good to demand phase fluctuations. That is, a wider bandwidth or a longer sampling period, so that the worst-case phase relation $\Delta\varphi=180^\circ$ is averaged out.

The determination of a specific $\mu$ is problematic with this algorithm, because the optimal $\mu$ is sensitive to the underlying signal model and target configuration. Thus, $\mu$ must be estimated using a small number of measurements before the iteration can be estimated. From the first three measurements, the term
\begin{equation*}
\text{\ae}=\frac{1}{3}\sum_{i=1}^3\lVert Q(\boldsymbol{\omega}_0,\mathbf{z}_i)\rVert\quad(\boldsymbol{\omega}_0=\text{starting value of the iteration})
\end{equation*}
is computed and
\setcounter{equation}{9}
\begin{equation}
\mu=\frac{\delta}{\text{ae}}\frac{BW}{2}
\label{Eq5-1-10}
\end{equation}
is chosen with $\delta=0.9$ for linear arrays and $\delta=1.8$ for planar arrays, and the iteration with $n=3$ is begun. This choice of $\mu$ proved to be good through simulation. However, in Fig. \ref{Fig5-4}, Signal Model 4 shows that such a choice of a fixed $\mu$ is not equally good in each iteration step. At the beginning of the iterations, the fluctuations of $\lVert\grad Q\rVert$ should possibly be more strongly dampened, and that later $\mu$ would be increased. Such an adaptation of $\mu$ on the achieved distance to the minimum provides a good, complicated procedure with deterministic quadratic functions (for example, Newton-Raphson or Fletcher-Powell algorithms). A simple, random choice of $\mu$ that depends on $Q(\boldsymbol{\omega}_N)\grad Q(\boldsymbol{\omega}_n)$ shall be investigated in the next subsection.

\subsection{Acceleration of the Initial Convergence}

In many applications, the number of iterations should be so small that the asymptotic behavior is barely seen. Although the asymptotic convergence rate of $\sqrt{n}$ itself cannot be improved, in such instances, the algorithm can be modified so that for $n\geq n_0$ the probability that $\boldsymbol{\omega}_n$ is in the vicinity of the exact value is large. In this sense, the initial convergence rate should be accelerated. Because theoretical statements to this effect are difficult to make, one is forced to rely on simulation.

The poor convergence of the algorithm occurs primarily with Signal Model $4$ and is explained by the fluctuation of the magnitude of the correction vector. It is thus logical that the correction vector be dampened or limited (``soft-'' or ``hard-limiter''). Assuming that the fluctuations of $\lVert\grad Q\rvert$ depend on $Q$, for example, that they grow with $f(Q)$, then the correction vector $\mathbf{G}=\frac{\grad Q}{f(Q)}(\boldsymbol{\omega})$ for a suitable $f\neq 0$ is reasonable to use. This assumption for Signal Model 4 is based on the following lemma:

\underline{Lemma (5.2-1)}
\begin{itemize}
\item \textbf{Assertion}:  If $\mathbf{z}\sim\mathcal{N}(\mathbf{0},\mathbf{R})$, then for all $\boldsymbol{\omega}\in V^M$ 
\begin{equation*}
\frac{1}{N}\E\left\{Q\right\}^2\leq\var Q\leq\E\left\{Q\right\}^2
\end{equation*}

\item \textbf{Proof}:
According to Appendix \ref{Appen4},
\begin{align*}
\E\left\{Q\right\}=&\trace \boldsymbol{\Gamma}\mathbf{R}\\
\E\left\{Q^2\right\}=&\trace\boldsymbol{\Gamma}\mathbf{R}\boldsymbol{\Gamma}\mathbf{R}-\left(\trace\boldsymbol{\Gamma}\mathbf{R}\right)^2
\end{align*}
therefore
\begin{equation*}
\var(Q)=\trace\boldsymbol{\Gamma}\mathbf{R}\boldsymbol{\Gamma}\mathbf{R}
\end{equation*}

Let $\mathbf{R}:=\mathbf{L}\mathbf{L}^*$ (using, for example, a Cholesky decomposition), $\mathbf{M}:=\boldsymbol{\Gamma}\mathbf{L}$, so that $\boldsymbol{\Gamma}=\boldsymbol{\Gamma}^2=\boldsymbol{\Gamma}^*$.
\begin{equation*}
\E\left\{Q\right\}=\left\lVert\mathbf{M}\right\rVert^2
\end{equation*}
\begin{align*}
\var(Q)=&\lVert\mathbf{M}\mathbf{M}^*\rVert^2\quad\text{with $\lVert\mathbf{M}\rVert^2=\sum_{i,k}|m_{ik}|^2$}\\
=&\sum_{i,j}\left|\sum_{k}m_{ik}m_{jk}^*\right|^2.
\end{align*}
The proof comes from the Schwartz inequality, where it can be shown that
\begin{equation*}
\frac{1}{N}\lVert\mathbf{M}\rVert^4\leq\lVert\mathbf{M}\mathbf{M}^*\rVert^2\leq\lVert\mathbf{M}\rVert^4.
\end{equation*}
\end{itemize}

Additionally one can introduce an upper bound on the magnitude of the correction vector $\mathbf{G}$. Because there are a great many ways of changing the procedure, in addition to the original algorithm, only four variants shall be studied here. All have the form $\boldsymbol{\omega}_{n+1}=\boldsymbol{\omega}_n-\frac{\mu}{n}\boldsymbol{G}(\boldsymbol{\omega}_n)$, where the different variants are
\setcounter{equation}{1}
\begin{enumerate}
\item \begin{equation}
\mathbf{G}=\frac{\grad Q}{Q}=\grad(\ln Q)\label{Eq5-2-2}
\end{equation}
\item \begin{equation*}
\mathbf{G}=\frac{\grad Q}{1+Q^2}=\grad(\arctan Q)
\end{equation*}
\item\begin{equation*}
\mathbf{G}=\left\{\begin{IEEEeqnarraybox}[\relax][c]{c's}
\grad Q&If $\lVert\grad Q\rVert<\eta$\\
\eta\frac{\grad Q}{\lVert\grad Q\rVert} &otherwise
\end{IEEEeqnarraybox}\right.
\end{equation*}
\item \begin{equation*}
\mathbf{G}=\left\{\begin{IEEEeqnarraybox}[\relax][c]{c's}
\frac{\grad Q}{\lVert\grad Q\rVert} &if $\grad Q \neq 0$\\
0& otherwise
\end{IEEEeqnarraybox}\right.
\end{equation*}
\end{enumerate}
and the original algorithm is $\mathbf{G}=\grad Q$. 

\underline{Comments}
\underline{To 1}:

This is a generalization of the usual monopulse ratio $w=\text{Re }\frac{D(\omega)}{S(\omega)}$ where $M=1$ ($D,S=$ difference beam and sum beam outputs). Considering the maximization of $T=\mathbf{z}^*\mathbf{A}\left(\mathbf{A}^*\mathbf{A}\right)^{-1}\mathbf{A}^*\mathbf{z}$ (generalized sum beam pattern, see Chapter \ref{Sec4-1}) instead of the minimization of $Q=\mathbf{z}^*\boldsymbol{\Gamma}\mathbf{z}$, then for $M=1$:
\begin{align*}
T=&\frac{\mathbf{z}^*\mathbf{a}\mathbf{a}^*\mathbf{z}}{N}=\frac{|S|^2}{N}(\omega)\\
T'=&\frac{\mathbf{z}^*\mathbf{a}_{\omega}\mathbf{a}\mathbf{z}+\mathbf{z}^*\mathbf{a}\mathbf{a}_{\omega}\mathbf{z}}{N}=2\text{Re }\frac{
\mathbf{a}_{\omega}^*\mathbf{a}\mathbf{a}^*\mathbf{a}}{N}\quad(\mathbf{a}_\omega:=\frac{\partial}{\partial \omega}\mathbf{a})\\
\frac{T'}{T}=&2\text{Re }\frac{\mathbf{a}_{\omega}^*\mathbf{z}}{\mathbf{a}^*\mathbf{z}}=2\text{Re }\frac{D(\omega)}{S(\omega)}.
\end{align*}
The monopulse ratio is used for locating a target, because it is approximately a linear function of $\omega$. Its variance is, however, very large (specifically $\infty$, \cite{ref26}). The direction of the correction remains by Versions 1 and 2 the same $\frac{Q_n}{\lVert Q_n\rVert}$!

\underline{To 3}:

Such a bounding corresponds to an implicit $Q(\omega)$-dependent bounding of the input data $\mathbf{z}_k$, because only when $\mathbf{z}_k$ is ``large'' can $\mathbf{z}^*\boldsymbol{\Gamma}_{\boldsymbol{\omega}}\mathbf{z}$ be ``large''. For $\eta\rightarrow\infty$, one obtains the original algorithm.

\underline{To 4}:

For $M=1$, this corresponds to a generalized sign control algorithm. For a given $\mathbf{z}$ the correction vector is not a continuous function at the point $\bar{\boldsymbol{\omega}}$ with $\grad Q(\bar{\boldsymbol{\omega}})=\mathbf{0}$

The convergence points of this algorithm are according to Theorem (5.1-6), the only stable stationary points of the associated differential equation (5.1-5), if convergence can be obtained at all. One must consequently investigate the expected value of the correction  magnitudes $1,\ldots ,4$.

\underline{To 1,2}:

It is questionable whether the expected value exists at all. For example, the expected value of the ratio of two normally distributed random variables does not exist \cite{ref25}. However the existence of the expected value of the monopulse ratio is known \cite{ref26}. It is to be feared that the normalization with $Q$ or respectively $1+Q^2$ increases the fluctuations.

\underline{To 3}:

For a large $\eta$, this algorithm is the same as the original algorithm, because
\begin{align*}
\E\left\{\mathbf{G}\right\}=&\int\limits_{\{\Vert Q_{\boldsymbol{\omega}}\rVert\leq\eta\}}Q_{\boldsymbol{\omega}}(\mathbf{z})dP(\mathbf{z})+\eta\int\limits_{\{\Vert Q_{\boldsymbol{\omega}}\rVert>\eta\}}\frac{Q_{\boldsymbol{\omega}}}{\lVert Q_{\boldsymbol{\omega}}\rVert}dP(\mathbf{z})\\
=&\E\left\{Q_{\boldsymbol{\omega}}\right\}+\int\limits_{\{\Vert Q_{\boldsymbol{\omega}}\rVert>\eta\}}\left(\eta\frac{Q_{\boldsymbol{\omega}}}{\lVert Q_{\boldsymbol{\omega}}\rVert}-Q_{\boldsymbol{\omega}}\right)dP(\mathbf{z})
\end{align*}
and therefore
\begin{align*}
\lVert\E\left\{Q_{\boldsymbol{\omega}}\right\}-\E\left\{\mathbf{G}\right\}\rVert^2=&\left\lVert\int\left(\eta\frac{Q_{\boldsymbol{\omega}}}{\lVert Q_{\boldsymbol{\omega}}\rVert}-Q_{\boldsymbol{\omega}}\right)I_{\{\lVert Q_{\boldsymbol{\omega}}\rVert>\eta\}}(\mathbf{z})dP(\mathbf{z})\right\rVert^2\\
\leq&\int\left\lVert\left(\eta\frac{Q_{\boldsymbol{\omega}}}{\lVert Q_{\boldsymbol{\omega}}\rVert}-Q_{\boldsymbol{\omega}}\right)I_{\{\lVert Q_{\boldsymbol{\omega}}\rVert>\eta\}}(\mathbf{z})\right\rVert^2dP(\mathbf{z})
\end{align*}
where the Schwartz inequality was used for each summand of $\lVert\rVert^2$. Continuing the simplification,
\begin{align*}
=&\int\left\lVert\frac{Q_{\boldsymbol{\omega}}}{\lVert Q_{\boldsymbol{\omega}}\rVert}\left(\eta-\lVert Q_{\boldsymbol{\omega}}\rVert\right)\right\rVert^2I_{\{\lVert Q_{\boldsymbol{\omega}}\rVert>\eta\}}(\mathbf{z})dP(\mathbf{z})\\
\leq&\int\left(\eta-\lVert Q_{\boldsymbol{\omega}}\rVert\right)^2dP(\mathbf{z})P\left\{\lVert Q_{\boldsymbol{\omega}}\rVert>\eta\right\}
\end{align*}
where the Schwartz inequality was used and whereby the first factor is minimized for $\eta=\E\left\{\lVert Q_{\boldsymbol{\omega}}\rVert\right\}$.

If $P\left\{\lVert Q_{\boldsymbol{\omega}}\rVert>\eta\right\}$ falls sufficiently fast for $\eta\rightarrow\infty$ (for example with $e^{-\eta}$), then using an appropriate $\eta$, an arbitrarily good agreement with the value $\bar{\boldsymbol{\omega}}$ with $\E\left\{Q_{\boldsymbol{\omega}}\right\}(\boldsymbol{\omega})=\mathbf{0}$ can be achieved. With the bounding through $\eta$, the expected value of the estimate is steered. If the correction vector is a linear function of $\boldsymbol{\omega}$, it is shown in \cite{ref27} that such a saturation for an appropriate $\eta$ provides a robust estimator.

\underline{To 4}:

As in 1, 2, it is questionable whether $\E\left\{\mathbf{G}\right\}$ exists. For a constant input, the differential equation (5.1-5) does not have a global solution, because the right side of the differential equation is not continuous. For an arbitrary starting point, a local solution that can be continued until a point $\bar{\boldsymbol{\omega}}$ with $\grad Q(\boldsymbol{\omega})=\mathbf{0}$ exists. A convergence with probability $1$ cannot necessarily be expected. However, there can be a convergence in distribution (weak convergence) of the $\boldsymbol{\omega}_n$.

\begin{figure}
\centering
\includegraphics[width=0.6\textwidth]{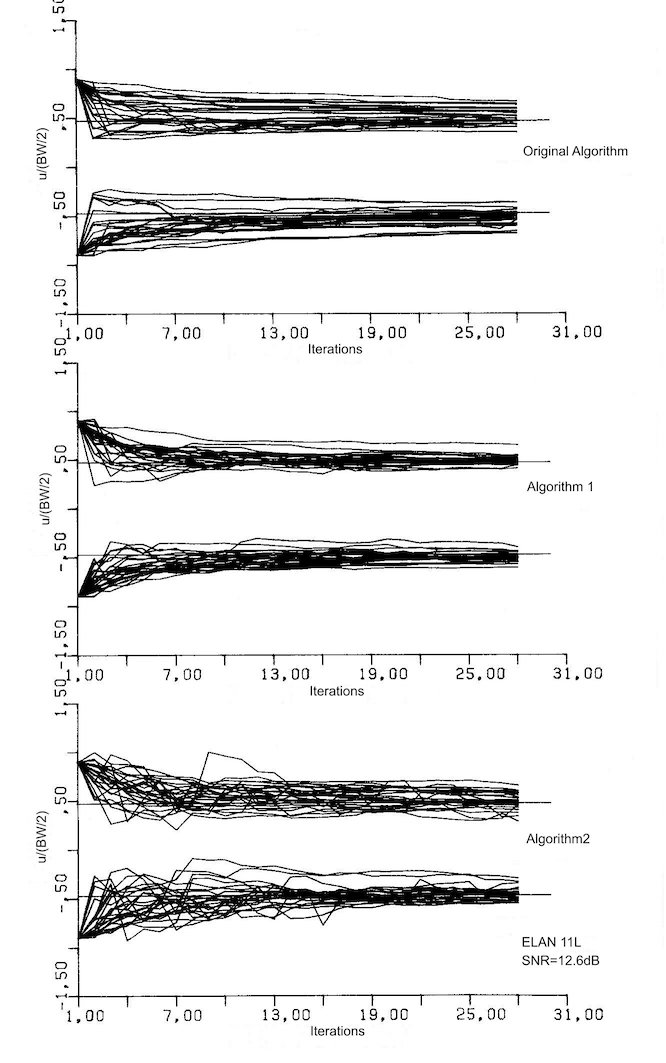}
\caption{\label{Fig5-6}}
\end{figure}

\begin{figure}
\centering
\includegraphics[width=0.6\textwidth]{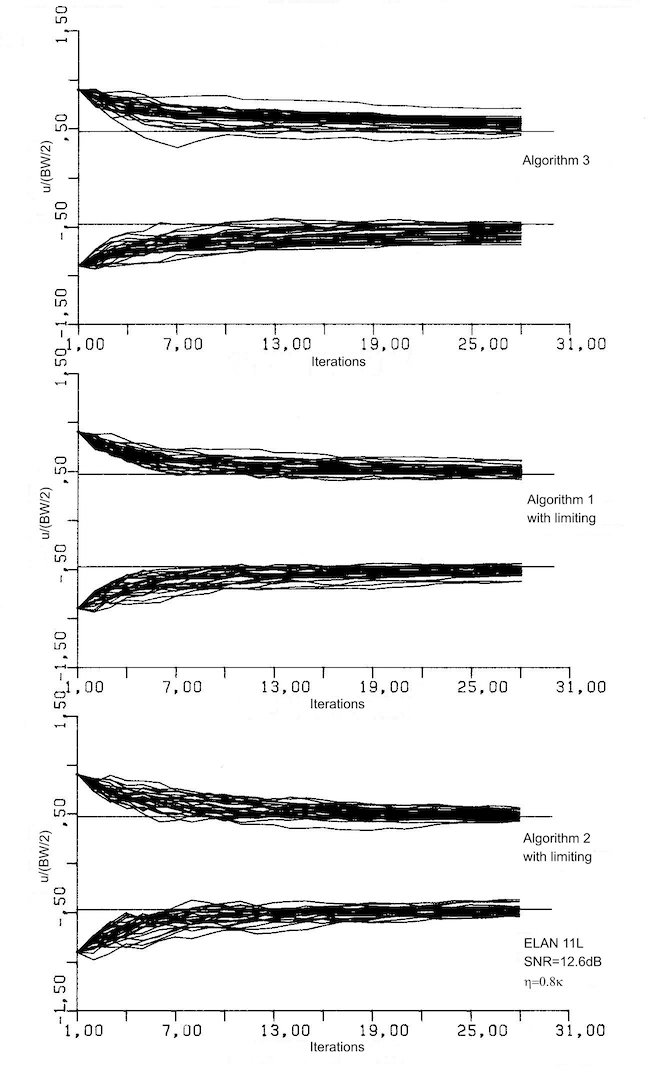}
\caption{\label{Fig5-7}}
\end{figure}

These four methods are now compared for the particularly critical Signal Model 4, whereby $\mu$ is chosen according to (5.1-10). The separation of the targets is $0.45BW$ with ELAN 11 L, the average SNR on an individual array element is $12\,\text{dB}$.  The convergence for the original algorithm and Versions $1$ and $2$ is shown in Fig. \ref{Fig5-6}. With Version 1, dampening of the fluctuations in the first two iteration steps leads to better convergence. With Version 2, the probability for a large correction near the minimum is still high. The dampening of the $Q$-function for deterministic inputs through the $\ln$ or $\arctan$ functions thus does not transfer to stochastic inputs. The same cases as Fig. \ref{Fig5-6} are shown in \ref{Fig5-7},  but all correction vectors have a saturation bound $\eta$ like Version 3 ($\eta=0.8\text{\ae}$, $\text{\ae}=\frac{1}{3}\sum_{i=1}^3\left\lVert\grad Q(\boldsymbol{\omega}_0,\mathbf{z}_i)\right\rVert$ as in (5.1-10)). The introduction of such a saturation is the decisive method of dampening the fluctuations.

\begin{figure}
\centering
\includegraphics[width=0.6\textwidth]{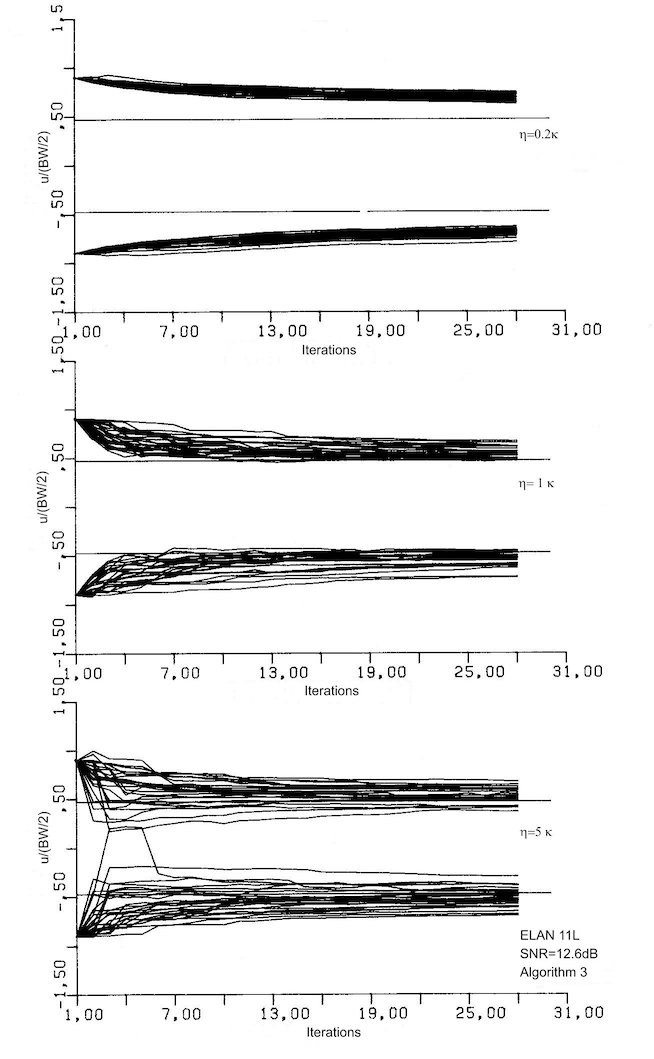}
\caption{\label{Fig5-8}}
\end{figure}

Using Version $3$ with the saturation bound $\eta$ in Fig. \ref{Fig5-8} shows a compromise exists between dampening fluctuations and estimator bias with a particular number of iterations exists (Array and target configurations are the same as in Figs. \ref{Fig5-6} and \ref{Fig5-7}).

\begin{figure}
\centering
\includegraphics[width=0.6\textwidth]{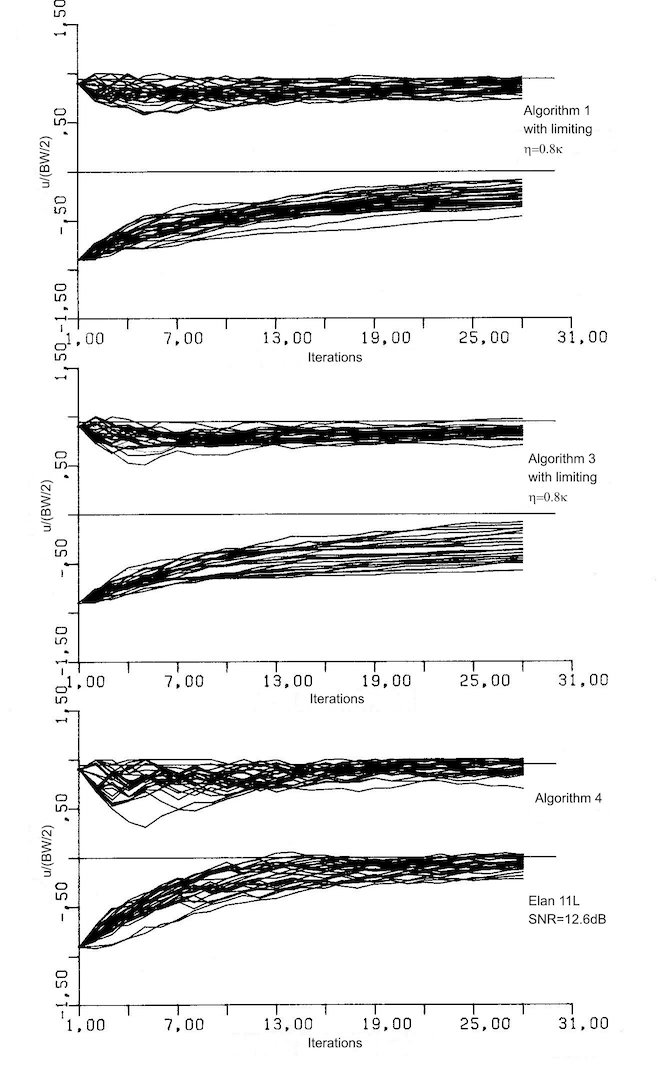}
\caption{\label{Fig5-9}}
\end{figure}

\begin{figure}
\centering
\includegraphics[width=0.6\textwidth]{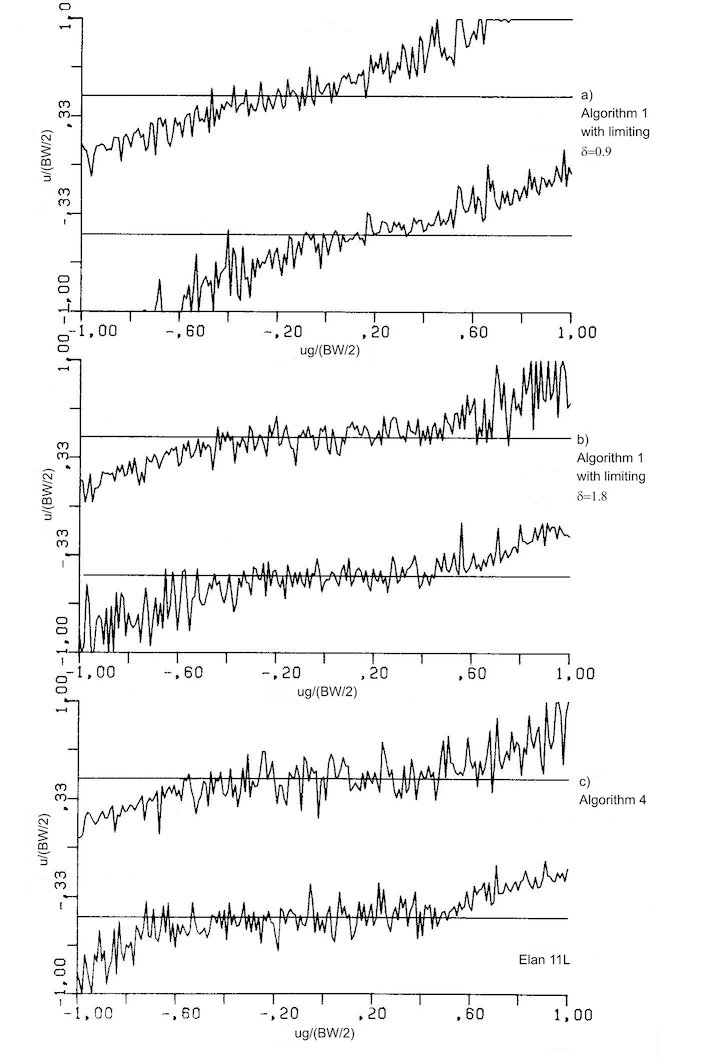}
\caption{\label{Fig5-10}}
\end{figure}

The behavior given a poor initial estimate when using Versions $3$ and $1$ with a poor initial estimate and bounding $(\eta=0.8\text{\ae})$ and also of Version $4$ is shown in Fig. \ref{Fig5-9}. With a good starting value, Version $4$ has a higher residual noise than the other versions, but with a bad initial estimate, it converges particularly fast to a bias-free process. This is also shown in Fig. \ref{Fig5-10}. There, the estimate after $17$ iterations (residual noise) is shown as a function of the initial direction estimate $\mathbf{u}_0=\left(\mathbf{u}_g-0.9BW,\mathbf{u}_g+0.9BW\right)$.

Given target positions $\mathbf{u}^{\text{ex}}=(-0.45BW,0.45BW)$, both targets are also in the search region with $u_g=-0.45BW$. The curves (a) and (b) show that for Algorithm 1 (with bounding $\eta=0.8\text{\ae}$) the convergence with a poor initialization can be improved using a larger value of $\delta$ in \eqref{Eq5-1-10}. The convergence to the minimum takes place very cautiously, in time increments $t_n=\sum_{k=1}^n\frac{1}{k}$, where $n$ is the number of iterations \cite{ref23}.

\subsection{Application to Planar Antenna Arrays}

Though in the mathematical formulation, no difference needs to be made between linear and planar antenna arrays, significant differences become evident through simulation.

\subsubsection{Limiting the Search Area}

The necessary restriction with planar arrays to limit the search areas to a symmetric region with $\epsilon$-spacing from surface $u_i=u_k$, proves to be superfluous in practice for planar arrays because the case $u_i\approx u_k$ \emph{and} $v_i\approx v_k$ $(i\neq k)$ so rarely occurs. (In $\mathbb{R}^M$ one has $\dim\left\{\left.\mathbf{u}\in\mathbb{R}^M\right|u_i=u_k,i\neq k\right\}\leq M-1$, but in $\mathbb{R}^{2M}$ it is $\dim\left\{\left.(\mathbf{u},\mathbf{v})\right|u_i=u_k\wedge v_i=v_k,i\neq k\right\}\leq2M-2!$). This simplifies the cumbersome determination of the defined region of iteration, because by simultaneous azimuth and elevation estimation, there is no natural numbering of the targets.

\subsubsection{The Choice of the Initial Estimate and Poor Convergence}

\begin{figure}
\centering
\includegraphics[width=0.6\textwidth]{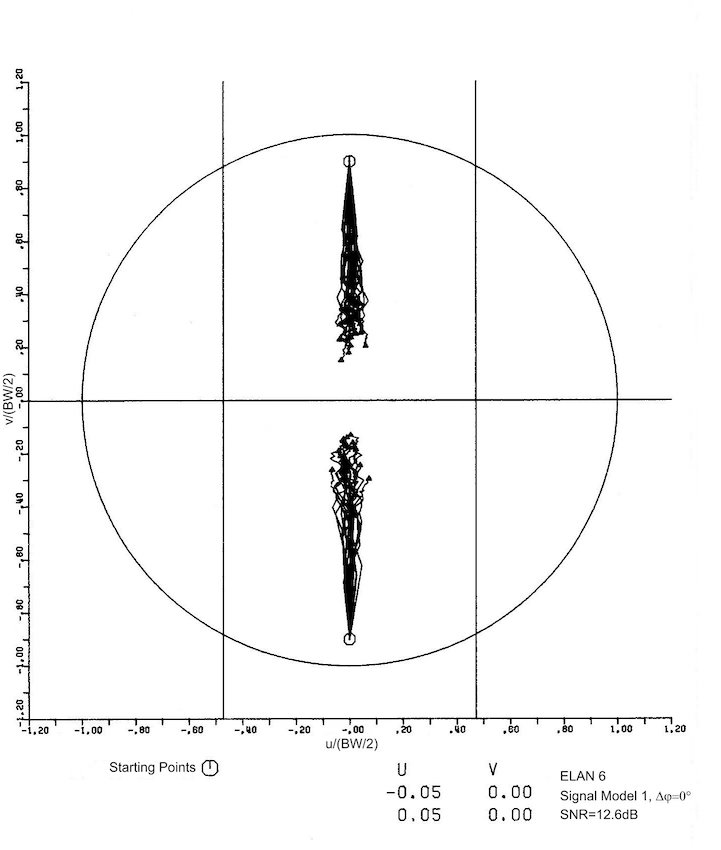}
\caption{Convergence of the stochastic approximation with two targets.\label{Fig5-11}}
\end{figure}

\begin{figure}
\centering
\includegraphics[width=0.6\textwidth]{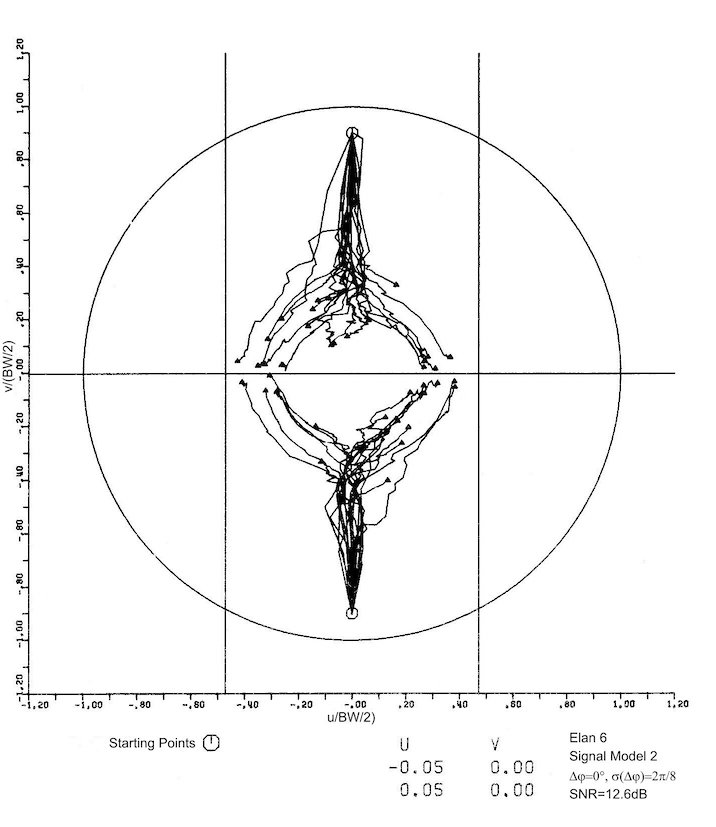}
\caption{Convergence of the stochastic approximation with two targets.\label{Fig5-12}}
\end{figure}

The starting value $\boldsymbol{\omega}_0$ should generally be chosen such that $\lVert\boldsymbol{\omega}_0-\boldsymbol{\omega}_{\text{ex}}\rVert$ is small and the starting directions are as far apart as possible. With two targets, prior knowledge about the orientation in the $u-v$-plane may accelerate the convergence. The poor convergence for Signal Model 1 in Fig. \ref{Fig5-4} with $\Delta\varphi=180^\circ$ arises with other phase relations with additional dependence on the directions. There are more instances in which the strong $M$-regularity of the ordering is poorly fulfilled. For example if $\Delta\varphi=0^\circ$, the order $\boldsymbol{\omega}_1=(-u,u,0,0)$ can be barely distinguished from the ordering $\boldsymbol{\omega}_1=(0,0,-v,v)$. That means that $\left\lVert\grad |\E\left\{Q\right\}\right\rVert(\boldsymbol{\omega}_2)$ is small for $\lVert\grad \E\left\{Q\right\}\rVert(\boldsymbol{\omega}_1)=0$. This is shown in Fig. \ref{Fig5-11} in the $(u-v)$ plane for ELAN 6. The starting value of the iteration is $\boldsymbol{\omega}_2=\left(0,0,-0.9\frac{BW}{2},0.9\frac{BW}{2}\right)$, the actual target position is $\boldsymbol{\omega}_1=\left(-\frac{BW}{4},\frac{BW}{4},0,0\right)$ with $\Delta\varphi=0^\circ$ and SNR=$12\,\text{dB}$. Although in the region of $\boldsymbol{\omega}_2$, the value $\left\lVert\grad\E Q\right\rVert\neq0$, the slope is so small that the iteration does not move.  The same configuration is shown in Fig. \ref{Fig5-12}, but for Signal Model 2 with $\Delta\varphi\sim\mathcal{N}\left(0,\left(\frac{2\pi}{8}\right)^2\right)$. This supports the demand for higher bandwidths and more frequent revisit periods.

\subsubsection{Attenuating the Fluctuations of $\grad Q$}
\begin{figure}
\centering
\includegraphics[width=0.6\textwidth]{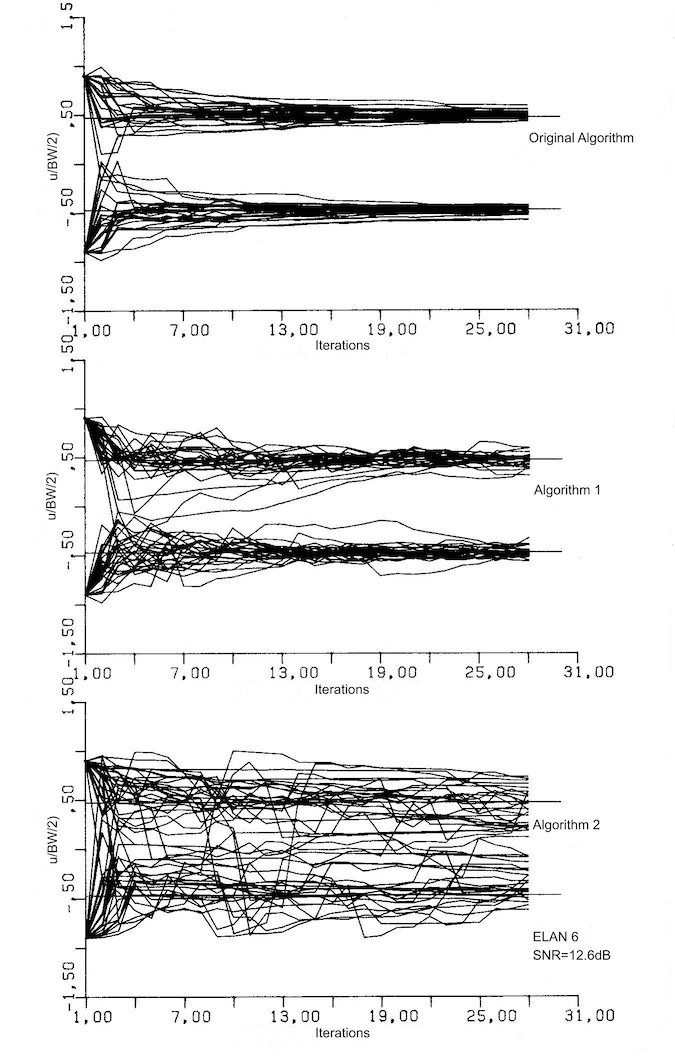}
\caption{\label{Fig5-13}}
\end{figure}

Versions $3$ and $4$ in Chapter 5.2 (limiting $\lVert\grad Q\rVert$) are also suitable iteration procedures with planar arrays. Versions $1$ and $2$ (normalization $\frac{\grad Q}{Q}$ and $\frac{\grad Q}{1+Q^2}$), however, behave differently. The convergence for the $u$-direction for ELAN 6 is shown in Fig. \ref{Fig5-13}. Version $1$ provides an estimate that is uniformly jumpy, as Version $2$ was in the linear case (See Fig. \ref{Fig5-6}).

\begin{figure}
\centering
\includegraphics[width=0.6\textwidth]{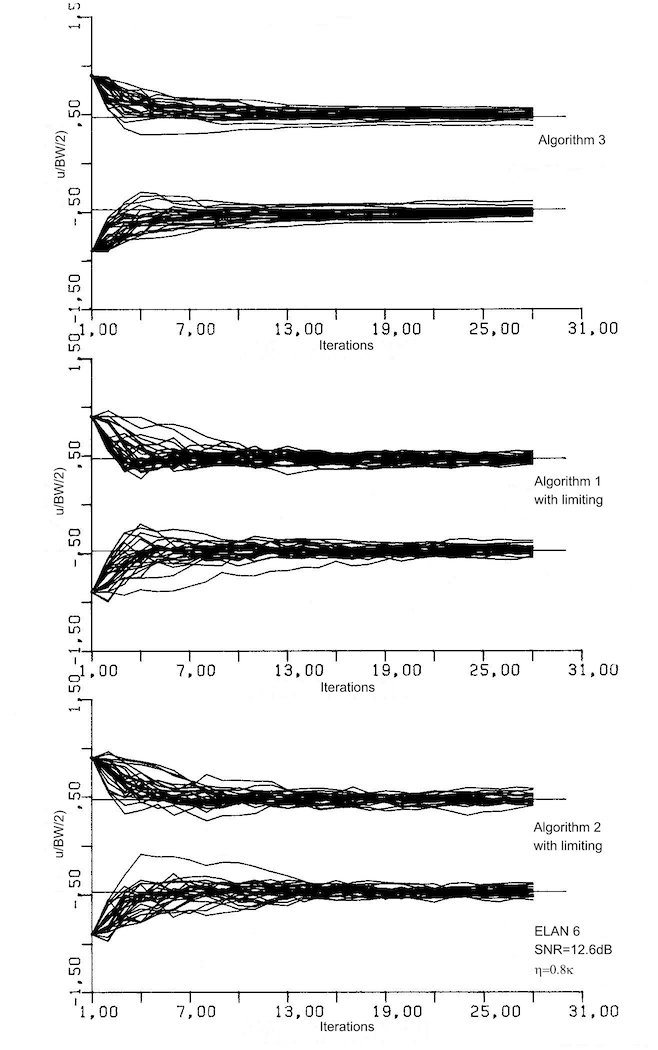}
\caption{\label{Fig5-14}}
\end{figure}

With regard to Version $2$, the choice of $\mu$ is so small that no convergence can be seen after $30$ iterations. By bounding the magnitude of the correction vector as in Version 3, one can again attain convergence. This is shown in Figure \ref{Fig5-14} ($\eta=0.8\text{\ae}$).

In any case, a limitation of the correction vector should be applied. Whether Version $1$, $2$, or $4$ should be used depends on the desired properties of the estimate. These three procedures provide an unbiased estimate after a small number of iterations.

\begin{figure}
\centering
\includegraphics[width=0.6\textwidth]{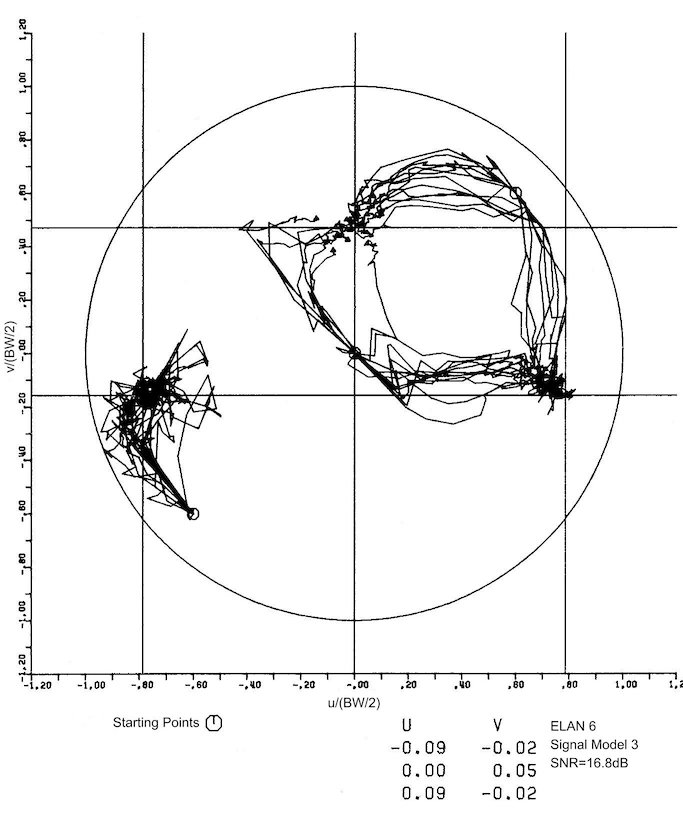}
\caption{Convergence of the stochastic approximation with three targets. \label{Fig5-15}}
\end{figure}

An example of the estimation of three targets is shown in Fig. \ref{Fig5-15} for Algorithm $3$ with $\eta=0.8\text{\ae}$. Initial values of the iteration are
\begin{align*}
\omega_1^{(0)}=&(-0.6 BW/2,-0.6BW/2)\\
\omega_2^{(0)}=&(0,0)\\
\omega_3^{(0)}=&(0.6BW/2,0.6BW/2)
\end{align*}

The average SNR at an individual element is $16.8\,\text{dB}$. One recognizes that the algorithm tries a different numbering of targets with different Monte Carlo runs.

\subsection{The Influence of Perturbations}

For simplicity, in the following subsection, only the original algorithm with bounding $(\eta=0.8\text{\ae})$ is analyzed.

\subsubsection{Coupling and Subarrays}

\begin{figure}
\centering
\includegraphics[width=0.6\textwidth]{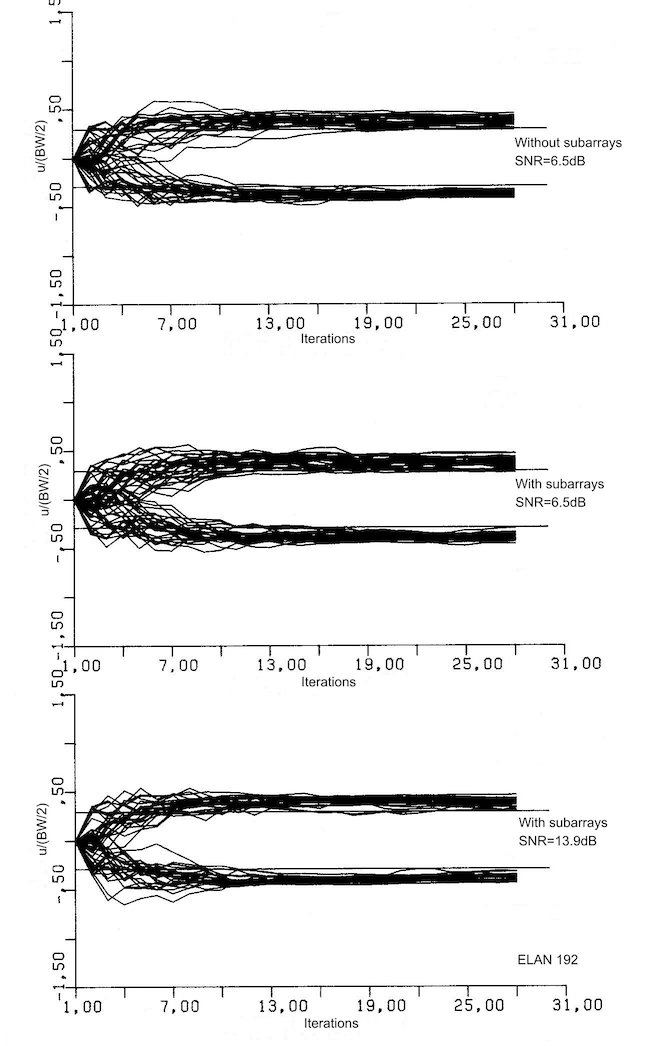}
\caption{\label{Fig5-16}}
\end{figure}

\begin{figure}
\centering
\includegraphics[width=0.6\textwidth]{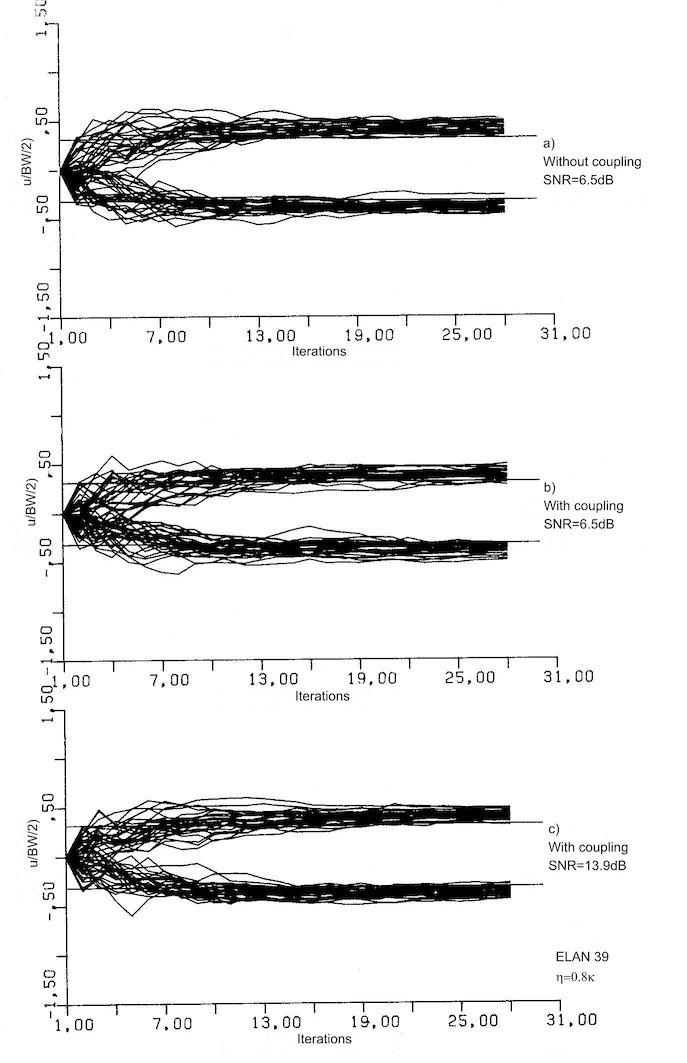}
\caption{\label{Fig5-17}}
\end{figure}

The influence of the coupling can be interpreted as a change in the characteristic responses of the individual elements. According to (4.1-8), if all of the individual elements have the same characteristic response, then this has no effect on the form of the $Q$ function. Thus, coupling has no influence as long as its effect on every element is the same, for example, by a large array on a regular grid. Additionally, preprocessing of the element outputs by summing them to form subarrays does not change the $Q$ function as long as the subarrays all have the same response (As new element coordinates $\mathbf{x},\mathbf{y}$, one must use the phase centers of the subarrays). With different element characteristic responses, if the number of elements is sufficiently large the distortion of the $Q$ function and thus the worsening of the estimate is not very large, because the antenna response differences are mitigated through averaging. This is shown in Fig. \ref{Fig5-16} for the antenna ELAN 192 with the subarray ordering that is used in the ELRA system of the FFM with $24$ subarrays of $8$ elements. The convergence for Signal Model 4 is shown. The position of the targets and $\mathbf{u}=\left(-0.3\frac{BW}{2},0.3\frac{BW}{2}\right)$, $\mathbf{v}=\left(-0.3\frac{BW}{2},0.3\frac{BW}{2}\right)$; the starting value for the iteration is $\mathbf{u}_0=(0,0)$, $\mathbf{v}_0=\left(-0.9\frac{BW}{2},0.9\frac{BW}{2}\right)$. The influence of coupling on ELAN 39 is shown in Fig. \ref{Fig5-17}. The coupling is simulated using the coupling matrix $\mathbf{C}$, (signal $\mathbf{C}\mathbf{s}$ instead of $\mathbf{s}$) that was generated using a program developed at the FFM for the V-shaped dipoles in the ELRA array (in \cite{ref28} this program is used). In such a model, each antenna element is represented using the equivalent circuit 
\begin{center}
\includegraphics{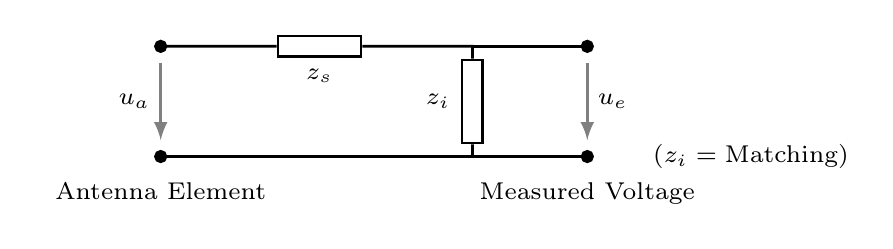}
\end{center}
so that $u_e=\frac{z_i}{z_s+z_i}u_a$ and $\mathbf{C}=z_i\left(\mathbf{Z}_s+z_i\mathbf{I}\right)^{-1}$.

In both instances, the procedure appears to not be sensitive to deviations from isotropic element responses as long as the number of elements used is sufficiently large.

\subsubsection{Extended Targets and Bandwidth}

According to Signal Model 4, extended targets can be approximately simulated using normal numbers with an appropriate covariance matrix according to (2.2-2) and (2.2-3). The bandwidth of the receiver can be similarly simulated. If one considers the target to be a pure noise source, then an antenna element receives a signal with frequency $\bar{f}$ and bandwidth $2B$ with the following frequency mixture that is delayed with respect to the element at $(0,0)$:
\begin{equation*}
z_{x,y}=\int_{\bar{f}-B}^{\bar{f}+B}b(f)e^{-j\frac{2\pi}{c}f(xu+yv)}df.
\end{equation*}
The frequency-component weightings $b(f)$ are independent for $f\neq f'$ and are zero on average (for example, with uniformly distributed initial phases). At element $(0,0)$, one has the power density $\wp^2$. Then $b$ can be formulated as a stochastic process with $\E\left\{b(f)\right\}=0$ and $\E\left\{b(f)b^*(f')\right\}=\frac{\wp^2}{2B}\delta(f-f')$ so that
\begin{align*}
\E\left\{E{z_{x,y}  z*_{x',y'}}\right\}=&\frac{\wp^2}{2B}\int_{\bar{f}-B}^{\bar{f}+B}e^{-j\frac{2\pi}{c}f\left((x-x')u+(y-y')v\right)}df\\
=&\wp^2\frac{\sin\frac{2\pi}{c}\left((x-x')Bu+(y-y')Bv\right)}{\frac{2\pi}{c}\left((x-x')Bu+(y-y')Bv\right)}e^{-j\frac{2\pi}{c}\bar{f}\left((x-x')u+(y-y')v\right)}.
\end{align*}
One obtains the same expressions for the correlation of a bandwidth-limited impulse $s(t)=\frac{\sin2\pi Bt}{2\pi Bt}e^{j2\pi ft}$.

\begin{figure}
\centering
\includegraphics[width=0.6\textwidth]{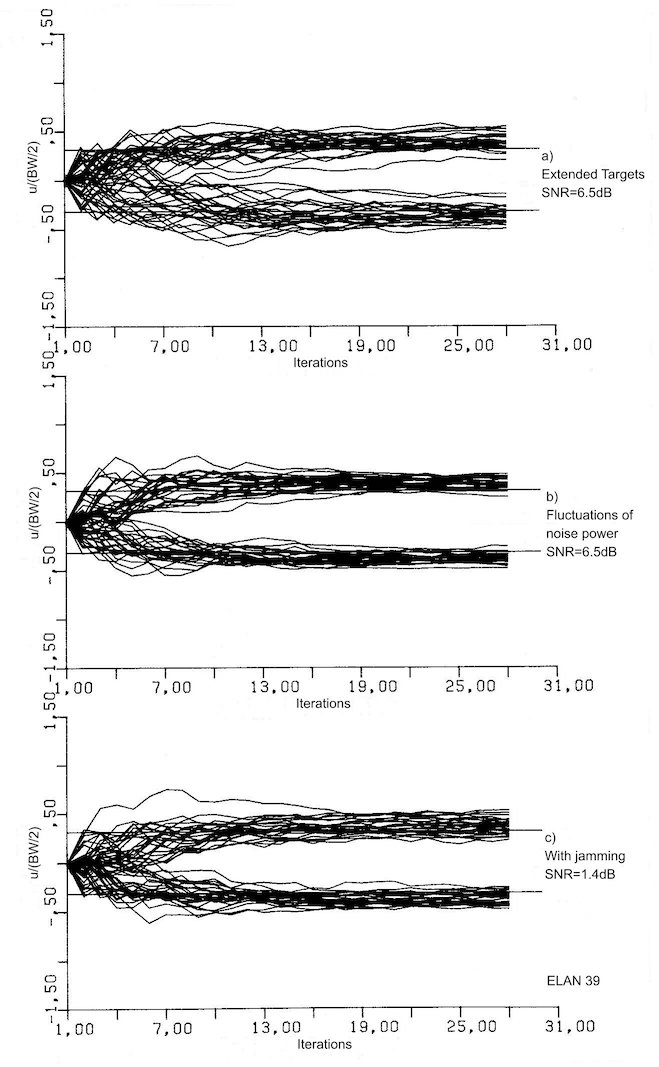}
\caption{\label{Fig5-18}}
\end{figure}

Thus, the bandwidth of the receiver with noise sources according to (2.2-2) functions like an extended quadratic target with the direction-dependent bandwidth $2s=\frac{2Bu}{\bar{f}}$. For small $s$, the covariance matrix of $\mathbf{z}$ only changes a little so that $\E\{Q\}=\trace \boldsymbol{\Gamma}\mathbf{R}$ is also only slightly perturbed. With ELAN 7 L, for example, with two extended targets of width $s=0.2BW$, there was no noticeable difference in the plot of $\E\{Q\}$ corresponding to Fig. \ref{Fig4-1}; the minimum of $\E\{Q\}$ was displaced by less than $0.05BW$. Simulations show a larger difference with planar arrays. In Fig. \ref{Fig5-18}, curve (a) shows this in comparison to Curve (a) in Fig. \ref{Fig5-17} for ELAN 39. In Fig. \ref{Fig5-18} (a), two extended targets with $r=0.2BW$ are assumed.

In general,  these effects can be neglected for targets extending only a few percent of the beamwidth and less than $1\%$ of the bandwidth.

\subsubsection{Correlated Noise (External Interference) and Amplifier Fluctuations}

If the noise vector $\mathbf{n}$ has the properties
\begin{align*}
\E\left\{\mathbf{n}\mathbf{n}^*\right\}=\mathbf{W}\neq \mathbf{I}\quad\text{but}\quad\E\left\{\mathbf{n}\right\}=\mathbf{0},\quad\E\left\{\mathbf{n}^*\mathbf{s}\right\}=0,
\end{align*}
then one gets a displacement of the expected value of the estimate so that
\begin{equation*}
\E\left\{Q\right\}=\trace\boldsymbol{\Gamma}\mathbf{A}\mathbf{B}\mathbf{A}^*+\trace\boldsymbol{\Gamma}\mathbf{W}\quad\text{and}\quad\trace\boldsymbol{\Gamma}\mathbf{W}\neq\text{constant}.
\end{equation*}
If one has external jammers that are sufficiently far from the directions to be estimated, then the jammers just go into the $Q$ function dampened by the sidelobes, because $\boldsymbol{\Gamma}$ represents the construction of simultaneous sum beams. If, for example, the jammer power is on the order of the target power, then the displacement is small. For example, with ELAN 11 L with two targets at $\mathbf{u}=\left(0,\frac{BW}{2}\right)$ and an extended target according to (2.2-2) with $\bar{u}=0.5$, $s=\frac{BW}{2}$, the displacement is less than $0.02BW$ ($\text{SNR}=3\,\text{dB}$). In this case, the form of the minimum remains almost unchanged. However, the values of $\E\left\{Q_{\text{min}}\right\}$ with jamming are much larger than $N-M=\E\left\{Q_{\text{min}}\right\}$ without jamming, which can become noticeable when a test uses $Q_{\text{min}}$ (see Chapter \ref{Sec7}). If one has fluctuations in the receiver noise power, then $\mathbf{W}=\diag\sigma_i^2$. For the simulation, noise-power fluctuations of $1\,\text{dB}$ are assumed and $\sigma_i^2\sim\mathcal{R}\left(\left[\sigma^2(1-1/4),\sigma^2(1+1/4)\right]\right)$ is chosen. The convergence in the $u$ direction for the same target configuration and antenna array as in Fig \ref{Fig5-17} (a) are shown in Fig. \ref{Fig5-18} for the the following scenarios:
\begin{itemize}
\item (a) Two extended targets according to (2.2-3) with $r=0.2BW$.
\item (b) Two point targets with $1$-$\text{dB}$ noise-power fluctuations at the receiver.
\item (c) Two point targets and also an extended jammer of the same power according to (2.2-3) with $(u,v)=(-0.5,0.1)$ and $r=0.9BW$.
\end{itemize}

If the external jammers become stronger, then additional measures for interference suppression must be taken, such as a pre-filtering of the simultaneous sum-and-difference beam outputs or through the use of an appropriate projection matrix as in (3.2-3).

\subsubsection{Quantization and Clipping}

If the input data $\mathbf{z}$ is quantized into real and imaginary parts having the same quantization levels $k\Delta$ $(k=\ldots-1,0,1,2,\ldots)$, then if $\Delta$ is small, this acts as an increased noise, because for $\mathbf{z}_q=\mathbf{z}+\boldsymbol{\epsilon}$,
\begin{equation*}
\E\left\{\mathbf{z}_q\mathbf{z}_q^*\right\}=\E\left\{\mathbf{z}\mathbf{z}^*\right\}+\E\left\{\boldsymbol{\epsilon}\boldsymbol{\epsilon}^*\right\}
\end{equation*}
and if $\text{Re }\epsilon_i\sim\mathcal{R}(\Delta)$, $\text{Im }\epsilon_i\sim\mathbb{R}(\Delta)$ can be assumed to be independent, then $\E\left\{\boldsymbol{\epsilon}\boldsymbol{\epsilon}^*\right\}=\diag\left(\frac{\Delta^3}{6}\right)$.

\begin{figure}
\centering
\includegraphics[width=0.6\textwidth]{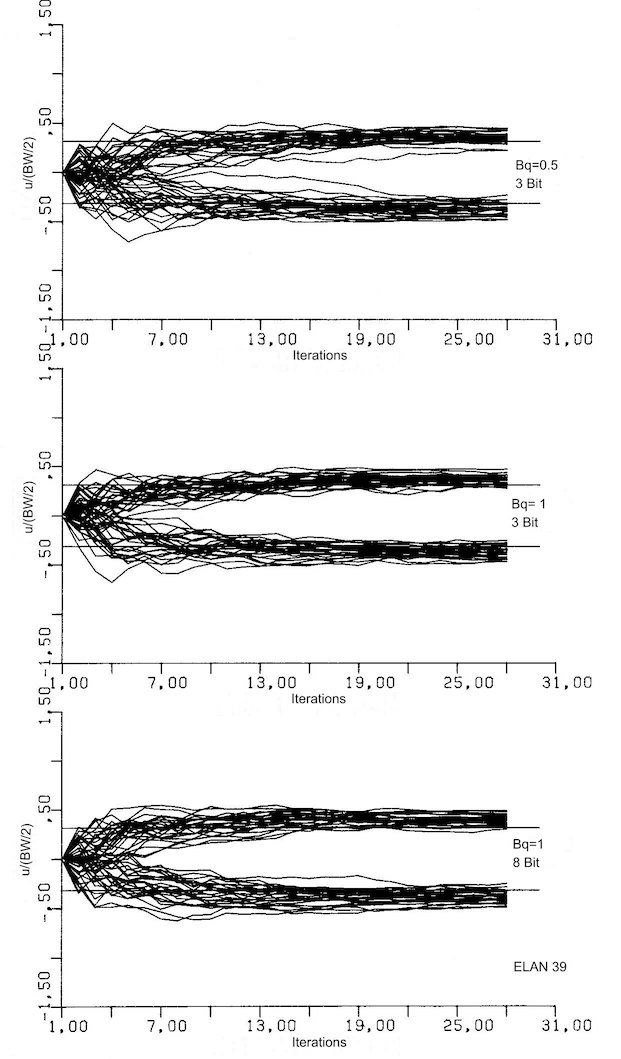}
\caption{\label{Fig5-19}}
\end{figure}

With rougher quantization, the assumption that the real and imaginary parts of the error are independent is no longer correct, and one gets correlations. That means that $\E\left\{\boldsymbol{\epsilon}\boldsymbol{\epsilon}^*\right\}$ is not diagonal and thus the function $\E\left\{Q\right\}$ is distorted. Such a distortion arises notably through clipping of the input data. The $Q$ function is, however, not particularly sensitive to such distortions. With ELAN 11 L, simulations of $\E\left\{\mathbf{z}_q\mathbf{z}_q^*\right\}$ showed that the quantization and clipping only caused a very small displacement of the minimum of $\E\left\{Q\right\}$. For example, with $3$-bit quantization and clipping to $\left|z_{q_i}\right|\leq\sqrt{\E\left\{z_iz_i^*\right\}}$, the displacement is less than $0.012BW$ (Separation of the targets $BW/2$, $\text{SNR}=12.5\,\text{dB}$). However, the minimum became notably flatter. This reduction in the curvature of the minimum is particularly strong with sharper clipping. This is shown in Fig. \ref{Fig5-19} for the same target and antenna-array situation as in Fig. \ref{Fig5-17} (a) for different quantizations and clippings $B_q$ when $\left|z_{q_i}\right|\leq B_q\sqrt{\E\left\{z_qz_q^*\right\}}$.

In all, the simulations show that the procedure is quite robust with low levels of interference and distortion.

\section{Averaged Grid Search}\label{Sec6}
\setcounter{figure}{0}
\begin{center}
\textbf{Summary of Chapter 6}
\end{center}

This chapter studies whether the suggested minimization procedure utilizing a grid search can be improved when the results are subsequently averaged. For linear antennas, the computational load of such an averaged grid search is higher than for the stochastic approximation when maintaining the same accuracy. For planar antennas,  the complexity is so much higher that stochastic approximation is generally preferred.
~\\
\hrule
~\\

The simplest realization of the grid search described in Chapter \ref{Sec4-4} implies that the method can also be used when multiple samples are taken. This can occur by performing a grid minimization for each sample $\mathbf{z}_k$, and then averaging the resulting estimation values $\hat{\boldsymbol{\omega}}_k$ ($k=1,\ldots, K$). One can perform such minimization on a rather coarse grid, because the dispersion of $\hat{\boldsymbol{\omega}}_k$ is quite large. Moreover, one can hope that with phase fluctuations, the worst case scenario (for example, $\Delta\varphi=0^\circ,180^\circ$ when $M=2$) averages out.
\begin{figure}
\centering
\includegraphics[width=0.6\textwidth]{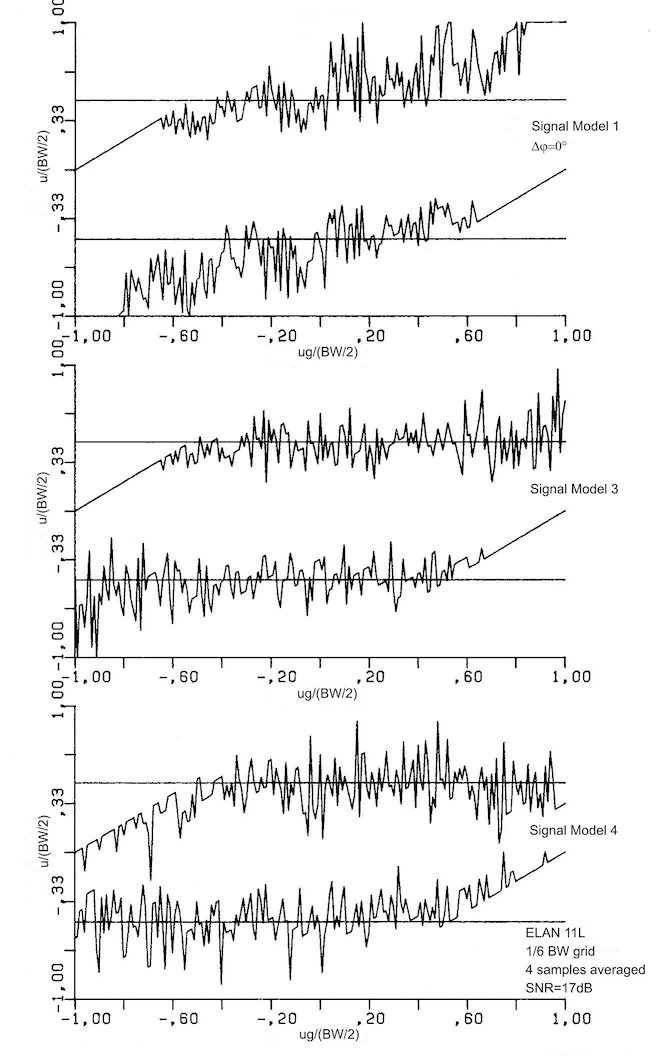}
\caption{\label{Fig6-1}}
\end{figure}

The same situations as in Fig. \ref{Fig4-11} (ELAN 11 L, $1/6 BW$ grid, distance between targets $0.5BW$, grid spans from $u_g-\frac{BW}{2}$ to $u_g+\frac{BW}{2}$) for Signal Models $1$, $3$, and $4$ averaging four samples are shown in Fig. \ref{Fig6-1}.

Whether the averaged grid search or the stochastic approximation is to be preferred depends on the specific application. The grid search requires a number of fixed sum beams that are jointly steerable. The stochastic approximation, on the other hand, requires independent sum-and-difference beams that are steerable during the iteration. Assuming that the hardware for each scenario has the same cost, one can compare both procedures by the number of their basic operations (that means, multiplications and division).

The number of basic operations (complex multiplications) are shown in the following table:
\begin{center}
\begin{tabular}{|l|c|c|}
\hline
						&stochastic approximation		& averaged grid search\\
\hline
\hline
						&$2i(M^2+M)$				&\\
\multirow{-2}{*}{linear array}	&$+i$ real roots			&\multirow{-2}{*}{$m\binom{k}{M}(M^2+M)$}\\
\hline
						&$3i\left(M^2+\frac{M}{2}\right)$&\\
\multirow{-2}{*}{planar array}	&$+i$ real roots 			&\multirow{-2}{*}{$m\binom{k}{M}(M^2+M)$}\\
\hline
\end{tabular}\quad\quad(6.1)
\end{center}
with $i$ being the number of iterations, $k$ the number of beams on the grid, and $m$ the number of samples. A real multiplication counts as $1/4$ of a complex multiplication. It is assumed that the saturation bound $\eta$ (See (5.2-2)) is reached after half of the specified number of iterations. The comparison to determine whether a saturation bound has been passed, as in the grid search, is assumed to have no cost. The number of beams in the grid is with regard to a linear antenna with a $\frac{1}{n}BW$ grid $k=n+1$, with regard to a planar antenna with a triangular grid with $\frac{1}{n}BW$ spacing as follows:
\begin{center}
\includegraphics[width=0.5\textwidth]{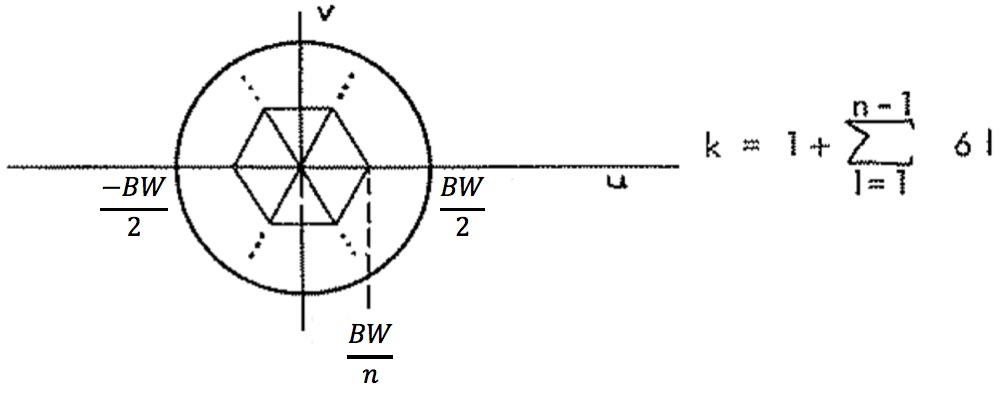}
\end{center}

For a linear antenna with, for example two largets and five beams, the average grid search with $60m$ operations is the same complexity as the stochastic approximation with $i12$ operations for $m=3, i=15$ and $m=4,i=20$, and so on. By this number of iterations, one can see from Fig. \ref{Fig5-10} that the stochastic approximation is better.

The averaged grid search will only be of interest in special cases. The computational complexity with linear arrays is with the same higher accuracy as that of the stochastic approximation. With planar arrays, the computational complexity is significantly higher.

\section{Construction of the Multihypothesis Test}\label{Sec7}
\setcounter{figure}{0}
\begin{center}
\textbf{Summary of Chapter 7}
\end{center}

With the suboptimal direction estimation algorithm of Chapter 5, which utilizes a stochastic approximation, a likelihood ratio test can no longer be used to determine the correct number of targets.  Thus, this chapter develops an appropriate suboptimal test. The minimal value of the $Q$ function averaged over a number of measurement vectors is suitable and can, moreover, be calculated without considerable computational effort. This test value measures the residual power after the subtraction of the estimated signal.

In order to have a multihypothesis test like that described in Chapter 3, the test must hold to an approximate probability of a Type-$1$ error. This is attained through the specification of a decision threshold by the approximation of the distribution of the test statistic as a $\chi^2$ random variable. Is is demonstrated through simulation that the total multihypothesis test approximately respects this error level, as demanded in Chapter 3.

For a target with amplitude fluctuations according to Signal Model $4$ (``Swerling-II Signal Model''), an approximation of the detection probability of the multihypothesis test is calculated and its validity proven through simulation. The curves for the detection probability can also be used with other signal models, as their detection probabilities proved higher in simulations.

Subsequently, the behavior of the test given deviations in the assumptions in the signal and noise models are tested through simulations, whereby it is observed that deviations from the noise model are the most critical.
~\\
\hrule
~\\

In Chapter 3.3 it was shown that the sequential multiple-hypothesis test has the following steps. For $M+1,2,\ldots, N$:
\begin{enumerate}
\item Test $H_M$ against $K_M=\mathbb{C}^N \backslash H_M$. This means compute $\varphi_M:\mathcal{Z}^K\rightarrow\{0,1\}, (\mathbf{z}_1,\ldots, \mathbf{z}_K)\mapsto\varphi_M(\mathbf{z}_1,\ldots,\mathbf{z}_K)$ for $H_M=\left\{\mathbf{s}\in\mathbb{C}^N\left|\bigvee\limits_{\mathbf{b}\in\mathbb{C}^M}\bigvee\limits_{\boldsymbol{\omega}\in\Omega^M}:\mathbf{s}=\mathbf{A}\mathbf{b}\right.\right\}$.
\item If $\phi_M=0$, then decide $\hat{M}:=M$ and stop.
\end{enumerate}

In order to satisfy the total error level $\alpha$ of such a multihypothesis test, according to Chapter 3.3, each $\phi_M$ must satisfy this error bound. The construction of such a test $\phi_M$ shall be considered here. It shall be shown that such a test can be used with the estimated parameters rather than focussing on particularly efficient tests.

\subsection{The $\bar{Q}$ Statistic and its Relationship with the Likelihood Ratio}\label{Sec7-1}
\setcounter{equation}{0}

The desired test should be as simple as possible. Using the estimated parameters $\hat{\boldsymbol{\omega}}(\mathbf{z}_1,\ldots,\mathbf{z}_k)$, a test statistic $T(\hat{\boldsymbol{\omega}}(\mathbf{z}_1,\ldots,\mathbf{z}_k))$ should be compared to a threshold $\eta\in\mathbb{R}$. Such a statistic cannot be sufficient for $H_M$, because $H_M$ is an $M$-dimensional partial set of $\mathbb{C}^N$. A sufficient statistic must be at least an $M$-component vector.

The formulation of the hypotheses
\begin{equation*}
H_M=\left\{\mathbf{s}\in\mathbb{C}^N\left|\bigvee\limits_{\mathbf{b}\in\mathbb{C}^M}\bigvee\limits_{\boldsymbol{\omega}\in\Omega}:\mathbf{s}=\mathbf{A}\mathbf{b}\right.\right\},\Omega\in V^M
\end{equation*}
clearly contains parameters on which the decision is dependent. Instead of a sufficient statistic, a statistic that is invariant to a group of transformations is sought, whereby the invariance with respect to the transformations arbitrarily eliminates parameters that depend on the measuring instrument. In this instance, it is demanded that the statistic be independent of the complex amplitudes $\mathbf{b}_1,\ldots,\mathbf{b}_K$. That is, not dependent on the specific point $(\mathbf{A}\mathbf{b}_1,\ldots,\mathbf{A}\mathbf{b}_k)$, rather only the space $S=\bigtimes_{k=1}^K\lin H(\mathbf{A})$. The magnitude of the test statistic should thus be invariant to all affine transformations on $S$; these are matrices
\begin{equation*}
\mathbf{C}=\left(\begin{IEEEeqnarraybox}[\relax][c]{c'c'c'c}
\IEEEstrut
\mathbf{C}_1&0&\ldots&0\\
0&\mathbf{C}_2&\ldots&0\\
\vdots&\vdots&\ddots&\vdots\\
0&0&\ldots&\mathbf{C}_K
\IEEEstrut
\end{IEEEeqnarraybox}\right)
\end{equation*}
with affine transformation $\mathbf{C}_i\in\mathbb{C}^{N\times N}$ that only transforms the components in $\lin H(\mathbf{A})$. That is $\mathbf{C}_i\mathbf{z}=\tilde{\mathbf{z}}_A+\mathbf{z}_A^\bot$ if $\mathbf{z}=\mathbf{z}_A+\mathbf{z}_A^\bot$ for $\mathbb{C}^N=\lin H(\mathbf{A})\oplus\lin H(\mathbf{A})^\bot$.

In the residual space $S^\bot\subset\mathbb{C}^{NK}$, the test statistic should be independent of the rotation of the coordinate system; that is independent of unitary transformations. Let $\mathbf{G}$ be the set of all such transformations. For a given $\mathbf{A}=\mathbf{A}(\boldsymbol{\omega}),\mathbb{C}^{NK}=S\oplus S^\bot$, then
\begin{equation*}
G:=\left\{\mathbf{H}:\mathbb{C}^{NK}\rightarrow\mathbb{C}^{NK}\left|
\bigvee\limits_{\mathbf{C}_k\in\mathbb{C}^{N\times N} \atop \mathbf{C}_k| \lin H(\mathbf{A})^\bot=\mathbf{I}}
\bigvee\limits_{\mathbf{U}:S^\bot\rightarrow S^\bot\atop \text{is unitary}}:\mathbf{H}\left(\begin{IEEEeqnarraybox}[\relax][c]{c}
\IEEEstrut
\mathbf{z}_1\\
\vdots\\
\mathbf{z}_K
\IEEEstrut
\end{IEEEeqnarraybox}\right)=\left(\begin{IEEEeqnarraybox}[\relax][c]{c}
\IEEEstrut
\mathbf{C}_1\mathbf{z}_{1A}\\
\vdots\\
\mathbf{C}_K\mathbf{z}_{KA}
\IEEEstrut
\end{IEEEeqnarraybox}\right)+\mathbf{U}\left(\begin{IEEEeqnarraybox}[\relax][c]{c}
\IEEEstrut
\mathbf{z}_{1A}^\bot\\
\vdots\\
\mathbf{z}_{KA}^\bot
\IEEEstrut
\end{IEEEeqnarraybox}\right)
\right.\right\}
\end{equation*}
Then $G$ is a group, because unitary and affine transformations form a group, and such transformations only decompose $\mathbb{C}^{NK}$ into orthogonal parts.

The $Q$-function is thus, with regard to $G$, an invariant reduction of the data $(\mathbf{z}_1,\ldots,\mathbf{z}_k)$.

\underline{Theorem (7.1-1):}
For all $\boldsymbol{\omega}$,
\begin{equation*}
\bar{Q}:=\frac{1}{K}\sum_{k=1}^K\left\lVert\boldsymbol{\Gamma}(\boldsymbol{\omega})\mathbf{z}_k\right\rVert^2
\end{equation*}
is a maximally invariant statistic with respect to the transformation group $G_{\boldsymbol{\omega}}$.

Proof:
\begin{itemize}
\item Part (i) Invariance:
Let $\mathbf{H}\in G$. Thus,
\begin{equation*}
\mathbf{H}=\left(\begin{IEEEeqnarraybox}[\relax][c]{c'c'c'c}
\IEEEstrut
\mathbf{C}_1&0&\ldots&0\\
0&\mathbf{C}_2&\ldots&0\\
\vdots&\vdots&\ddots&\vdots\\
0&0&\ldots&\mathbf{C}_K
\IEEEstrut
\end{IEEEeqnarraybox}\right)+\mathbf{U}
\end{equation*}
and $\mathbf{z}_k=\mathbf{z}_{kA}+\mathbf{z}_{kA}^\bot$ for all $k=1\ldots K$.

\begin{align*}
\bar{Q}\left(\mathbf{H}\left(\begin{IEEEeqnarraybox}[\relax][c]{c}
\IEEEstrut
\mathbf{z}_1\\
\vdots\\
\mathbf{z}_K
\IEEEstrut
\end{IEEEeqnarraybox}\right)\right)=&\frac{1}{K}\left\lVert\left(\begin{IEEEeqnarraybox}[\relax][c]{c'c'c'c}
\IEEEstrut
\boldsymbol{\Gamma}&0&\ldots&0\\
0&\boldsymbol{\Gamma}&\ldots&0\\
\vdots&\vdots&\ddots&\vdots\\
0&0&\ldots&\boldsymbol{\Gamma}
\IEEEstrut
\end{IEEEeqnarraybox}\right)\left(\left(\begin{IEEEeqnarraybox}[\relax][c]{c'c'c'c}
\IEEEstrut
\mathbf{C}_1&0&\ldots&0\\
0&\mathbf{C}_2&\ldots&0\\
\vdots&\vdots&\ddots&\vdots\\
0&0&\ldots&\mathbf{C}_K
\IEEEstrut
\end{IEEEeqnarraybox}\right)\left(\begin{IEEEeqnarraybox}[\relax][c]{c}
\IEEEstrut
\mathbf{z}_{1A}\\
\vdots\\
\mathbf{z}_{KA}
\IEEEstrut
\end{IEEEeqnarraybox}\right)+\mathbf{U}\left(\begin{IEEEeqnarraybox}[\relax][c]{c}
\IEEEstrut
\mathbf{z}_{1A}^\bot\\
\vdots\\
\mathbf{z}_{KA}^\bot
\IEEEstrut
\end{IEEEeqnarraybox}\right)\right)\right\rVert^2\\
=&\frac{1}{K}\left\lVert\left(\begin{IEEEeqnarraybox}[\relax][c]{c'c'c'c}
\IEEEstrut
\boldsymbol{\Gamma}&0&\ldots&0\\
0&\boldsymbol{\Gamma}&\ldots&0\\
\vdots&\vdots&\ddots&\vdots\\
0&0&\ldots&\boldsymbol{\Gamma}
\IEEEstrut
\end{IEEEeqnarraybox}\right)\mathbf{U}\left(\begin{IEEEeqnarraybox}[\relax][c]{c}
\IEEEstrut
\mathbf{z}_{1A}^\bot\\
\vdots\\
\mathbf{z}_{KA}^\bot
\IEEEstrut
\end{IEEEeqnarraybox}\right)\right\rVert^2
\end{align*}
where the simplification is because $\mathbf{C}_i\mathbf{z}_{iA}\in\lin H(\mathbf{A})$ and thus $\boldsymbol{\Gamma}\mathbf{C}_i\mathbf{z}_{iA}=\mathbf{0}$. A further simplification is
\begin{align*}
\bar{Q}\left(\mathbf{H}\left(\begin{IEEEeqnarraybox}[\relax][c]{c}
\IEEEstrut
\mathbf{z}_1\\
\vdots\\
\mathbf{z}_K
\IEEEstrut
\end{IEEEeqnarraybox}\right)\right)=\left\lVert
\mathbf{U}\left(\begin{IEEEeqnarraybox}[\relax][c]{c}
\IEEEstrut
\mathbf{z}_{1A}^\bot\\
\vdots\\
\mathbf{z}_{KA}^\bot
\IEEEstrut
\end{IEEEeqnarraybox}\right)
\right\rVert^2,
\end{align*}
which is because $\left(\begin{IEEEeqnarraybox}[\relax][c]{c'c'c}
\IEEEstrut
\boldsymbol{\Gamma}&&\\
&\ddots&\\
&&\boldsymbol{\Gamma}
\IEEEstrut
\end{IEEEeqnarraybox}\right)$ is a projection onto $S^\bot$ and $\mathbf{U}\left(\begin{IEEEeqnarraybox}[\relax][c]{c}
\IEEEstrut
\mathbf{z}_{1A}^\bot\\
\vdots\\
\mathbf{z}_{KA}^\bot
\IEEEstrut
\end{IEEEeqnarraybox}\right)\in S^\bot$. Continuing,
\begin{align*}
\bar{Q}\left(\mathbf{H}\left(\begin{IEEEeqnarraybox}[\relax][c]{c}
\IEEEstrut
\mathbf{z}_1\\
\vdots\\
\mathbf{z}_K
\IEEEstrut
\end{IEEEeqnarraybox}\right)\right)=&\frac{1}{K}\left\lVert\left(\begin{IEEEeqnarraybox}[\relax][c]{c}
\IEEEstrut
\mathbf{z}_{1A}^\bot\\
\vdots\\
\mathbf{z}_{KA}^\bot
\IEEEstrut
\end{IEEEeqnarraybox}\right)\right\rVert^2\quad \text{because $\mathbf{U}$ is unitary}\\
=&\frac{1}{K}\left\lVert\left(\begin{IEEEeqnarraybox}[\relax][c]{c'c'c'c}
\IEEEstrut
\boldsymbol{\Gamma}&0&\ldots&0\\
0&\boldsymbol{\Gamma}&\ldots&0\\
\vdots&\vdots&\ddots&\vdots\\
0&0&\ldots&\boldsymbol{\Gamma}
\IEEEstrut
\end{IEEEeqnarraybox}\right)\left(\begin{IEEEeqnarraybox}[\relax][c]{c}
\IEEEstrut
\mathbf{z}_1\\
\vdots\\
\mathbf{z}_K
\IEEEstrut
\end{IEEEeqnarraybox}\right)\right\rVert^2\\
=&\bar{Q}(\mathbf{z}_1\ldots\mathbf{z}_K).
\end{align*}

\item Part (ii), Maximality:
We have to show:
\begin{equation*}
\bigwedge\limits_{\mathbf{z},\mathbf{y}\in\mathbb{C}^{NK}}:\bar{Q}(\mathbf{z})=\bar{Q}(\mathbf{y})\Longrightarrow\bigvee\limits_{\mathbf{H}\in G}:\mathbf{z}=\mathbf{H}\mathbf{y}.
\end{equation*}

Let $\bar{Q}(\mathbf{z})=\bar{Q}(\mathbf{y})$; thus,
\begin{equation*}
  \begin{array}{r@{\,}l}
\left\lVert\left(\begin{IEEEeqnarraybox}[\relax][c]{c'c'c'c}
\IEEEstrut
\boldsymbol{\Gamma}&0&\ldots&0\\
0&\boldsymbol{\Gamma}&\ldots&0\\
\vdots&\vdots&\ddots&\vdots\\
0&0&\ldots&\boldsymbol{\Gamma}
\IEEEstrut
\end{IEEEeqnarraybox}\right)\left(\begin{IEEEeqnarraybox}[\relax][c]{c}
\IEEEstrut
\mathbf{z}_1\\
\vdots\\
\mathbf{z}_K
\IEEEstrut
\end{IEEEeqnarraybox}\right)
\right\rVert^2=\left\lVert\left(\begin{IEEEeqnarraybox}[\relax][c]{c'c'c'c}
\IEEEstrut
\boldsymbol{\Gamma}&0&\ldots&0\\
0&\boldsymbol{\Gamma}&\ldots&0\\
\vdots&\vdots&\ddots&\vdots\\
0&0&\ldots&\boldsymbol{\Gamma}
\IEEEstrut
\end{IEEEeqnarraybox}\right)\left(\begin{IEEEeqnarraybox}[\relax][c]{c}
\IEEEstrut
\mathbf{y}_1\\
\vdots\\
\mathbf{y}_K
\IEEEstrut
\end{IEEEeqnarraybox}\right)
\right\rVert^2
\\
    \begin{matrix}
      \mspace{0mu}\underbrace{\rule{3.8cm}{0pt}}_{\in S^\bot} &     
      \quad\quad\underbrace{\rule{3.8cm}{0pt}}_{\in S^\bot}\quad
    \end{matrix}
\end{array}
\end{equation*}
and there exists a matrix $\mathbf{U}$ unitary on $S^\bot$ such that
\begin{align*}
\left(\begin{IEEEeqnarraybox}[\relax][c]{c'c'c'c}
\IEEEstrut
\boldsymbol{\Gamma}&0&\ldots&0\\
0&\boldsymbol{\Gamma}&\ldots&0\\
\vdots&\vdots&\ddots&\vdots\\
0&0&\ldots&\boldsymbol{\Gamma}
\IEEEstrut
\end{IEEEeqnarraybox}\right)\left(\begin{IEEEeqnarraybox}[\relax][c]{c}
\IEEEstrut
\mathbf{z}_1\\
\vdots\\
\mathbf{z}_K
\IEEEstrut
\end{IEEEeqnarraybox}\right)=&\left(\begin{IEEEeqnarraybox}[\relax][c]{c'c'c'c}
\IEEEstrut
\boldsymbol{\Gamma}&0&\ldots&0\\
0&\boldsymbol{\Gamma}&\ldots&0\\
\vdots&\vdots&\ddots&\vdots\\
0&0&\ldots&\boldsymbol{\Gamma}
\IEEEstrut
\end{IEEEeqnarraybox}\right)\left(\begin{IEEEeqnarraybox}[\relax][c]{c}
\IEEEstrut
\mathbf{y}_1\\
\vdots\\
\mathbf{y}_K
\IEEEstrut
\end{IEEEeqnarraybox}\right)\\
=&\left(\begin{IEEEeqnarraybox}[\relax][c]{c'c'c'c}
\IEEEstrut
\boldsymbol{\Gamma}&0&\ldots&0\\
0&\boldsymbol{\Gamma}&\ldots&0\\
\vdots&\vdots&\ddots&\vdots\\
0&0&\ldots&\boldsymbol{\Gamma}
\IEEEstrut
\end{IEEEeqnarraybox}\right)\mathbf{U}\left(\begin{IEEEeqnarraybox}[\relax][c]{c}
\IEEEstrut
\mathbf{y}_1\\
\vdots\\
\mathbf{y}_K
\IEEEstrut
\end{IEEEeqnarraybox}\right)
\end{align*}
because $\boldsymbol{\Gamma}$ projects onto $S^\bot$ and $\mathbf{U}$ only functions on $S^\bot$. Therefore,
\begin{equation*}
\left(\begin{IEEEeqnarraybox}[\relax][c]{c}
\IEEEstrut
\mathbf{z}_{1A}^\bot\\
\vdots\\
\mathbf{z}_{KA}^\bot
\IEEEstrut
\end{IEEEeqnarraybox}\right)=\mathbf{U}\left(\begin{IEEEeqnarraybox}[\relax][c]{c}
\IEEEstrut
\mathbf{y}_{1A}^\bot\\
\vdots\\
\mathbf{y}_{KA}^\bot
\IEEEstrut
\end{IEEEeqnarraybox}\right).
\end{equation*}
For those components projected onto $\lin H(A)$, there exists  constantly affine transformations $\mathbf{C}_i$ so that $\mathbf{z}_{iA}=\mathbf{C}_i\mathbf{y}_{iA}$. Thus, there exists an $\mathbf{H}\in G$.
\end{itemize}

The demand that there be independence in the sequence of complex amplitudes $\mathbf{b}_1,\ldots,\mathbf{b}_K$ leads directly to the statistic $\bar{Q}$. All other statistics with this property are directly related to $\bar{Q}$. The statistic $\bar{Q}$ is maximally invariant. This means that all other $G$ invariant statistics $T(\mathbf{z}_1,\ldots,\mathbf{z}_K)$ are related to $\bar{Q}$ through a function $g$ such that $T=g(\bar{Q})$ \cite[pg. 33, Theorem 1.3]{ref3}. This simplifies the search for more efficient tests.

It is clear that $\bar{Q}$ is the average residual energy after signal extraction for a signal assumed to be in direction $\boldsymbol{\omega}$. Only this residual energy is independent of $\mathbf{b}_1,\ldots,\mathbf{b}_K$ if $\boldsymbol{\omega}$ is known. For the choice of $\eta$, this is more desirable than the use of the signal energy itself as in \cite{ref18}.

There remains the question whether the data reduction realized through $\bar{Q}$ suffices to make a decision. For this to be true, the distribution of $\bar{Q}$ under $H_M$ and $K_M$ must at least be different.

\underline{Theorem (7.1-2):}

for $H_M=\left\{\mathbf{s}\in\mathbb{C}^N\left|
\bigvee\limits_{\mathbf{b}\in\mathbb{C}^M}\bigvee\limits_{\boldsymbol{\omega}\in\Omega}:\mathbf{s}:=\mathbf{A}\mathbf{b}\right.\right\}$ and $K_M=\mathbb{C}^N\backslash H_M$, it is true that
\begin{equation*}
\min\limits_{\boldsymbol{\omega}}\E_n\left\{\bar{Q}\left|\mathbf{s}_1\ldots\mathbf{s}_K\in K_M\right.\right\}>\min\limits_{\boldsymbol{\omega}}\E_n\left\{\bar{Q}\left|\mathbf{s}_1\ldots\mathbf{s}_K\in H_M\right.\right\}.
\end{equation*}

Proof:
It is clear that under $H_M \bigwedge\limits_{i\in\{1\ldots k\}}:\mathbf{s}^*_i\hat{\boldsymbol{\Gamma}}\mathbf{s}_i=0$, but under $K_M \bigvee\limits_{i\in\{1\ldots k\}}:\mathbf{s}_i^*\hat{\boldsymbol{\Gamma}}\mathbf{s}_i>0$, because all $s_i\notin H_M$.

The statistic $\bar{Q}$ can be computed after estimating $\hat{\boldsymbol{\omega}}$ by means of a stochastic approximating averaging $Q(\hat{\boldsymbol{\omega}})$ over a few few (for example two to four) samples of $\mathbf{z}$. In order to set a bound for a test to level $\alpha$, the distribution of $\bar{Q}$ must be computed.

Writing
\begin{equation*}
K\bar{Q}=\sum_{k=1}^K\left\lVert\hat{\mathbf{\Gamma}}\mathbf{z}_k\right\rVert^2
\end{equation*}
with
\begin{align*}
\mathbf{z}_i=&\mathbf{A}\mathbf{b}_i+\mathbf{n}_i\\
\mathbf{n}_1\sim&\mathcal{N}(\mathbf{0},\sigma^2\mathbf{I}).
\end{align*}
One can now make the approximation that the parameter estimation is good; that is 
\setcounter{equation}{2}
\begin{equation}
\hat{\boldsymbol{\Gamma}}\mathbf{A}\mathbf{b}_i\approx\mathbf{0}\label{Eq7-1-3}
\end{equation}
(for the stochastic approximation, this is asymptotically satisfied. Thus, one gets
\begin{align*}
K\bar{Q}=&\sum_{k=1}^K\left\lVert\hat{\boldsymbol{\Gamma}}\mathbf{n}_k\right\rVert^2\\
=&\sum_{k=1}^K\left\lVert\mathbf{x}_k\right\rVert^2
\end{align*}
for $\mathbf{x}_k\sim\mathcal{N}_{\mathbb{C}^{N-M}}(\mathbf{0},\sigma^2\mathbf{I})$, because $\boldsymbol{\Gamma}$ is a projection. This means that $\frac{2}{\sigma^2}K\bar{Q}$ is roughly $\chi^2$-distributed with $K(2N-2M)$ degrees of freedom. With this approximation, for $P\left\{\bar{Q}\leq\eta\right\}=1-\alpha$, one can choose
\begin{equation}
\eta=\frac{\sigma^2}{2K}\chi^2_{K(2N-2M);\alpha}\label{Eq7-1-4}
\end{equation}
for when $H_M$ is valid ($\chi^2_{r;a}$ is defined to be the $\alpha$-fractile of the $\chi^2$ distribution with $r$ degrees of freedom). In the case where $N\gg M$ (for example, it suffices when $N=21$ with $M=2$) or a large $K$, the $\chi^2$ distribution can be approximated using a normal distribution. In such an instance,
\begin{equation*}
\frac{\frac{2}{\sigma^2}K\bar{Q}-\E\left\{\frac{2}{\sigma^2}K\bar{Q}\right\}}{\sqrt{\var\left\{\frac{2}{\sigma^2}K\bar{Q}\right\}}}\sim\mathcal{N}_{\mathbb{R}}(0,1).
\end{equation*}
For a central $\chi^2$-distributed random variable $x$, $\E\{x\}=r$ and $\var\{x\}=2r$ so that
\begin{equation*}
\frac{\frac{2}{\sigma^2}K\bar{Q}-K(2N-2M)}{\sqrt{2K(2N-2M)}}\sim\mathcal{N}(0,1)
\end{equation*}
meaning that
\begin{equation}
\frac{\frac{1}{K}\sum_{k=1}^K\left\lVert\hat{\boldsymbol{\Gamma}}\mathbf{z}_k\right\rVert^2-\sigma^2(N-M)}{\sigma^2\sqrt{(N-M)/K}}\sim\mathcal{N}(0,1).\label{Eq7-1-5}
\end{equation}
For $P\{\bar{Q}\leq\eta\}=1-\alpha$, one should then use
\begin{equation}
\eta=\sigma^2(\sqrt{(N-M)/K}U_\alpha+N-M)\label{Eq7-1-6}
\end{equation}
($U_\alpha$ is the $\alpha$-fractile of $\mathcal{N}(0,1)$).

The test in (7.1-4), or respectively (7.1-6) asymptotically, that is for a large number of iterations by the stochastic approximation and for high SNR, provides a test at a level of $\alpha$.

Given Signal Model $1$ with fixed amplitudes $\mathbf{b}_k=\mathbf{b}$ for all $k$, then, using the theorems regarding the asymptotic distribution of the likelihood ratio, the likelihood ratio test can be viewed as an ``optimal test'' for comparison. In this instance $\mathbf{z}\sim\mathcal{N}_{\mathbb{C}^N}(\mathbf{s},\sigma^2\mathbf{I})$ with
\begin{itemize}
\item $\mathbf{s}=\mathbf{A}\mathbf{b}$ if $H$ is true
\item $\mathbf{c}\in\mathbb{C}^N\backslash H$ if $K$ is true.
\end{itemize}
For 
\begin{equation*}
p(\mathbf{z}_1\ldots\mathbf{z}_k;\mathbf{s})=\frac{1}{\pi^{NK}\sigma^{NK}}e^{-\frac{1}{\sigma^2}\sum_{i=1}^K(\mathbf{z}_1-\mathbf{s})^*(\mathbf{z}_i-\mathbf{s})}
\end{equation*}
it is true that 
\begin{itemize}
\item $\max\limits_{\mathbf{s}\in\mathbb{C}^N}p(\mathbf{z}_1\ldots\mathbf{z}_K;\mathbf{s})$ is assumed for $\hat{\mathbf{s}}=\frac{1}{K}\sum_{i=1}^K\mathbf{z}_i:=\bar{\mathbf{z}}$.
\item $\max\limits_{\mathbf{s}\in\mathbf{H}}p(\mathbf{z}_1\ldots\mathbf{z}_K;\mathbf{s})$ is assumed for $\hat{\mathbf{b}}=\left.\mathbf{A}\left(\mathbf{A}^*\mathbf{A}\right)^{-1}\mathbf{A}^*\right|_{\hat{\boldsymbol{\omega}}}\bar{\mathbf{z}}$
with $\hat{\boldsymbol{\omega}}$ so that $\sum_{i=1}^K\left\lVert\left.\mathbf{z}_i-\mathbf{A}\left(\mathbf{A}^*\mathbf{A}\right)^{-1}\mathbf{A}^*\right|_{\hat{\boldsymbol{\omega}}}\bar{\mathbf{z}}\right\rVert^2=\min !$.
\end{itemize}
Thus, the logarithm of the likelihood ratio according to (3.1-8) is

\begin{align}
\ln T=&\frac{1}{\sigma^2}\left(\sum_{i=1}^K\left\lVert\mathbf{z}_i-\mathbf{A}\left(\mathbf{A}^*\mathbf{A}\right)^{-1}\mathbf{A}^*\bar{\mathbf{z}}\right\rVert^2-\sum_{i=1}^K\lVert\mathbf{z}_i-\bar{\mathbf{z}}\rVert^2\right)\notag\\
=&\frac{K}{\sigma^2}\lVert\hat{\boldsymbol{\Gamma}}\bar{\mathbf{z}}\rVert^2\label{Eq7-1-7}
\end{align}
after multiplying out and combining terms. The quantity $2\ln T$ is asymptotically (meaning for $K\rightarrow\infty$) $\chi^2_r$-distributed (See, for example, \cite[Theorem 2.44, pg. 95]{ref3}) with
\begin{align*}
r=&\dim_{\mathbb{R}}\mathbb{C}^N-\dim_{\mathbb{R}}H\\
=&\left\{\begin{IEEEeqnarraybox}[\relax][c]{c's}
2N-3M&with linear arrays (hypotheses parameterized via $\mathbf{u},\mathbf{b})$\\
2N-4M&with planar arrays (hypotheses parameterized via $\mathbf{u},\mathbf{v},\mathbf{b})$
\end{IEEEeqnarraybox}\right.
\end{align*}
so that under $H_M$ for $P\{2\ln T\leq \eta\}=1-\alpha$,
\begin{equation}
\eta=\left\{\begin{IEEEeqnarraybox}[\relax][c]{c's}
\chi^2_{2N-3M; \alpha}&for a linear array\\
\chi^2_{2N-4M;\alpha}&for a planar array
\end{IEEEeqnarraybox}\right.\label{Eq7-1-8}
\end{equation}
should be chosen.

Through the likelihood ratio test, the number of degrees of freedom of the remaining noise after the signal extraction (projection) is further reduced by the minimization with respect to $\boldsymbol{\omega}$ that is simultaneously applied to all $\mathbf{n}_k$. This introduces a small variance in the statistic. Approximating $2\ln T$ as in (7.1-5) with a normal distribution, then one gets, for example with a linear array:
\begin{equation}
\frac{2\ln T-(2N-3M)}{\sqrt{2(2N-3M)}}=\frac{\left\lVert\hat{\boldsymbol{\Gamma}}\bar{\boldsymbol{z}}\right\rVert^2-\frac{\sigma^2}{K}\left(N-\frac{3}{2}M\right)}{\frac{\sigma^2}{K}\sqrt{N-\frac{3}{2}M}}\sim\mathcal{N}(0,1).\label{Eq7-1-9}
\end{equation}

The large variance compared with (7.1-5) is the price that must be paid for separating the test statistic from the minimization process and for the independence of the signal model. The smaller variance in (7.1-9) is an indication that the likelihood ratio test has a higher precision.

The knowledge of the thresholds $\eta_{\alpha;M}$ in (7.1-6) is more important in a multiple-hypothesis test than when simply detecting a target, because closely-spaced targets are generally seldom-occurring phenomena and thus one cannot experimentally set the detection thresholds $\eta_{\alpha;M}$ during radar applications.

\subsection{Calculation of the Detection Probability of Stochastic Signals}\label{Sec7-2}

If the number of targets $M$ is given, then the associated measurement vectors $\mathbf{z}$ are elements of the alternative $K_i$ of the tests $H_i\longleftrightarrow K_i$ for $i<M$. If Signal Model $2,$ $3$, or $4$ is given, then one can compute the distribution of $\bar{Q}$ for such measurement vectors from the distribution of the alternative. Thus, one gets the detection probability because according to (3.3-2) and (3.3-3)
\begin{equation*}
P_E=(1-\alpha)P\{\bar{Q}_{M-1}>\eta\}\ldots P\{\bar{Q}_{1}>\eta\}\quad\text{under $H_M$}.
\end{equation*}

To calculate the detection probability, assume $M^{\text{ex}}$ targets are given with $M<M^{\text{ex}}$, $\mathbf{z}\notin H_M$, $\mathbf{z}=\mathbf{s}+\mathbf{n}$ with $\mathbf{s}=\mathbf{A}\mathbf{b}$ and $\text{rank }\mathbf{A}=M^{\text{ex}}$, $\mathbf{n}\sim\mathcal{N}_{\mathbb{C}^N}(\mathbf{0},\sigma^2\mathbf{I})$. Then, $\E\{\mathbf{s}^*\boldsymbol{\Gamma}\mathbf{s}\}>0$ according to (7.1-2), and $\frac{2}{\sigma^2}Q$ is $\chi^2_{2N-2M}(\delta^2)$-distributed when conditioned on $\mathbf{s}$ with non-centrality parameter $\delta^2=\frac{2}{\sigma^2}\mathbf{s}^*\boldsymbol{\Gamma}\mathbf{s}$.

This is because $\frac{2}{\sigma^2}Q$ can be written as a sum of normally distributed values with equal variance as follows:
\begin{align*}
Q=&\left\lVert\boldsymbol{\Gamma}\mathbf{s}+\boldsymbol{\Gamma}\mathbf{n}\right\rVert^2\\
=&\lVert\mathbf{t}+\tilde{\mathbf{n}}\rVert^2\quad\text{ with $\mathbf{t},\tilde{\mathbf{n}}\in\mathbb{R}^{2N-2M},\tilde{n}_i\sim\mathcal{N}_{\mathbb{R}}\left(0,\frac{\sigma^2}{2}\right)$}\\
=&\sum_{i=1}^{2N-2M}\tilde{\tilde{n}}_i^2\quad\text{with $\tilde{\tilde{n}}_i\sim\mathcal{N}_{\mathbb{R}}\left(t_i,\frac{\sigma^2}{2}\right)$.}
\end{align*}
Thus,
\begin{equation*}
\frac{2}{\sigma^2}Q=\sum_{i=1}^{2N-2M}x_i^2\quad\text{with $x_i\sim\mathcal{N}_{\mathbb{R}}\left(\frac{t_i\sqrt{2}}{\sigma},1\right)$}
\end{equation*}
and
\begin{equation*}
\delta=\sum_{i=1}^{2N-2M}\left(\frac{t_i\sqrt{2}}{\sigma}\right)^2.
\end{equation*}

For large $N$, the distribution of $Q$ conditioned on $\delta^2$ can be approximated by a normal distribution and thus $\bar{Q}$ can be approximated with a normal distribution. The distribution of $\bar{Q}$ is thus
\begin{align*}
P\{\bar{Q}\leq\eta\}=&\E_{\delta^2_1\ldots\delta^2_K}\left\{P\left\{\bar{Q}\leq\eta\left|\delta_1^2\ldots\delta_k^2\right.\right\}\right\}\\
=&\int\phi\left(\frac{\eta-\E_n\{\bar{Q}\}}{\sqrt{\var_n\{\bar{Q}\}}}\right)dP_{\delta_1^2}\ldots dP_{\delta_K^2}
\end{align*}
where $P_{\delta^2_i}$ is the distribution of $\delta_i^2$ from $K_M$, $\phi$ the cumulative distribution function of the normal distribution, and $\E_n$, $\var_n$ represent the expected value and variance only with respect to $n$.

According to Appendix 4, (A.4-1), (A.4-2), one has
\begin{align*}
\E_n\{\bar{Q}\}=&\sigma^2(N-M)+\frac{\sum_{i=1}^K\mathbf{s}_i^*\boldsymbol{\Gamma}\mathbf{s}_i}{K}\\
=&:\delta^2(N-M)+q
\end{align*}
and
\begin{equation*}
\var_n\{\bar{Q}\}=\frac{\sigma^4}{K}(N-M).
\end{equation*}

The problem to be solved with $\eta=\sigma^2\left(\sqrt{\frac{N-M}{K}}U_\alpha+N-M\right)$ according to (7.1-6) is the integral
\setcounter{equation}{0}
\begin{equation}
\int_0^\infty\phi\left(\frac{\sigma^2\left(\sqrt{\frac{N_M}{K}}U_{\alpha}+N-M\right)-\left(\sigma^2(N-M)+q\right)}{\sigma^2\sqrt{\frac{N-M}{K}}}\right)dP_q=
\int_0^\infty\phi\left(U_\alpha-\frac{q}{\sigma^2\sqrt{\frac{N-M}{K}}}\right)dP_q.\label{Eq7-2-1}
\end{equation} 

For Signal Models $2$ and $3$, the integral (7.2-1) can probably only be solved numerically (as an integral over the phase $\phi_i$). For Signal Model $4$, however, one can compute the distribution $p_{\delta^2}(x)$ and solve the integral analytically. Consequently, for this unfavorable situation, the detection probability of the entire test can be computed. It is
\begin{align*}
Kq=&\sum^K\left\lVert\hat{\boldsymbol{\Gamma}}\mathbf{s}_k\right\rVert^2\\
=&\sum^K\mathbf{b}^*_k\mathbf{A}^*\hat{\boldsymbol{\Gamma}}\mathbf{A}\mathbf{b}_k\\
=&\sum^K\mathbf{b}_k^*\mathbf{G}\mathbf{b}_k\quad\mathbf{b}_k\sim\mathcal{N}(\mathbf{0},\mathbf{B}).
\end{align*}
The matrix $\mathbf{G}=\mathbf{A}^*\hat{\boldsymbol{\Gamma}}\mathbf{A}$ is generally a positive-definite Hermitian matrix of dimensions $(M+I)\times(M+I)$ for the case where $M^\text{ex}=M+I$ targets are present and one is solving for $M$ targets. $\mathbf{G}$ is then positive definite when the vectors $\hat{\boldsymbol{\Gamma}}\mathbf{a}_1,\ldots,\hat{\boldsymbol{\Gamma}}\mathbf{a}_{M+I}$ are linearly independent. This is almost always the case for large $N$. Otherwise, instead of $\mathbf{b}$, $\mathbf{x}=\mathbf{L}^{-1}\mathbf{b}$, if $\mathbf{B}=\mathbf{L}\mathbf{L}^*$, then $\mathbf{x}\sim\mathcal{N}(\mathbf{0},\mathbf{I})$ and if
\begin{equation*}
\mathbf{G}=\mathbf{U}\left(\begin{IEEEeqnarraybox}[\relax][c]{c'c'c'c'c}
\IEEEstrut
\lambda_1&		&			&	&\\
		&\ddots	&			&	&\\
		&		&\lambda_r	&	&\\
		&		&			&0	&\\
		&		&			&	&0
\IEEEstrut
\end{IEEEeqnarraybox}\right)\mathbf{U}^*
\end{equation*}
with $\mathbf{U}$ unitary, then $\mathbf{y}=\mathbf{U}^*\mathbf{x}\sim\mathcal{N}(\mathbf{0},\mathbf{I})$ so that instead of considering $\mathbf{b}^*\mathbf{G}\mathbf{b}$, one can consider\\ $\left(\begin{IEEEeqnarraybox}[\relax][c]{c'c'c}
\mathbf{y}_1^*&\ldots&\mathbf{y}_r^*
\end{IEEEeqnarraybox}\right)\left(\begin{IEEEeqnarraybox}[\relax][c]{c'c'c}
\IEEEstrut
\lambda_1&		&\\
		&\ddots	&\\
		&		&\lambda_r
\IEEEstrut
\end{IEEEeqnarraybox}\right)\left(\begin{IEEEeqnarraybox}[\relax][c]{c}
\IEEEstrut
\mathbf{y}_1\\
\vdots\\
\mathbf{y}_r
\IEEEstrut
\end{IEEEeqnarraybox}\right)$ and once again a positive-definite, Hermitian form is given.

The probability density of such averaged, positive-definite forms can now be computed:\\
\underline{Theorem (7.2-2):} (Probability density of averaged Hermitian forms)
\begin{itemize}
\item \textbf{Preconditions}

Let $\mathbf{G},\mathbf{B}$ be positive definite Herminitan $I\times I$ matrices, $\mathbf{b}\sim\mathcal{N}(\mathbf{0},\mathbf{B})$,$\lambda_1\ldots\lambda_I$, the eigenvalues of $\mathbf{G}\mathbf{B}$, $\mu_i=\frac{1}{\lambda_i}$ $(i=1\ldots I)$, and
\begin{equation*}
q=\sum_{i=1}^K\mathbf{b}_i^*\mathbf{G}\mathbf{b}_i.
\end{equation*}
\item \textbf{Assertions}

\begin{itemize}
\item (i):
\begin{equation*}
p_{I,K}(q;\mu_1\ldots\mu_I)=(-1)^{IK+1}\frac{\prod_{i=1}^I\mu_i^K}{(K-1)!}\left.\sum_{l=1}^I\frac{\partial^{K-1}}{\partial z^{K-1}}\frac{e^{-qz}}{\prod_{t\neq l}(z-\mu_t)^K}\right|_{z=\mu_l}
\end{equation*}

\item (ii): Special for $I=1$:

\begin{equation*}
p_{1,K}(q;\mu)=\mu^K\frac{q^{K-1}}{(K-1)!}e^{-q\mu}\quad(\text{$\chi^2_{2K}$-distribution})
\end{equation*}

\item (iii): Special for $K=1$:

\begin{equation*}
p_{I,1}(q;\mu_1\ldots\mu_I)=-\prod_{i=1}^I\mu_i\sum_{t=1}^I\frac{e^{-q\mu_t}}{
\prod_{l\neq t}(\mu_t-\mu_l)}
\end{equation*}

\item (iv) Special for $I=2$:

\begin{align*}
p_{21}(q;\mu_1,\mu_2)=&l_{12}p_{11}(q;\mu_2)+l_{21}p_{11}(q;\mu_1)\\
p_{22}(q;\mu_1,\mu_2)=&l_{12}^2p_{12}(q\mu_2)+l_{21}^2p_{12}(q;\mu_1)+2l_{12}l_{21}p_{21}(q;\mu_1,\mu_2)\\
p_{23}(q;\mu_1,\mu_2)=&l_{12}^3p_{13}(q;\mu_2)+l_{21}^3p_{13}(q;\mu_1)+3l_{12}l_{21}p_{22}(q;\mu_1,\mu_2)
\end{align*}
where $l_{ij}=\frac{\mu_i}{\mu_i-\mu_j}$
\end{itemize}

The distribution (iii) agrees with that in \cite{ref29}, which is derived in a different manner.

\item \textbf{Proof}

See Appendix 5 for the proof.

\end{itemize}

Thus, the integral in (7.2-1) becomes
\setcounter{equation}{2}
\begin{align}
P_{M,K}\left\{\bar{Q}\leq\eta\right\}=&\int_0^\infty\phi\left(U_\alpha-\frac{q}{\text{\ae}}\right)Kp_{M,K}(Kq;\mu_1\ldots\mu_M)dq\notag\\
=&K\text{\ae}\int_{-\infty}^{U_\alpha}\phi(x)p_{M,K}(K\text{\ae}(U_\alpha-x);\mu_1\ldots\mu_M)dx\label{Eq7-2-3}
\end{align}
with
\begin{equation*}
\text{\ae}:=\sigma^2\sqrt{\frac{N-\hat{M}}{K}}
\end{equation*}
for $M$ targets present and $H_{\hat{M}}$ being tested, $\hat{M}<M$.

The probability of a Type-$2$ error can be computed with the above expression.

\underline{Corollary 7.2.4:}

\begin{itemize}
\item \textbf{Preconditions}

Let $\beta_{M,K}:=P_{M,K}\{\bar{Q}\leq\eta\}$ for $\eta=\sigma^2\left(U_\alpha\sqrt{\frac{N-\hat{M}}{K}}+N-\hat{M}\right)$ according to \eqref{Eq7-2-3} and
\begin{align*}
l_{ij}:=&\frac{\mu_i}{\mu_i-\mu_j}\\
\text{\ae}:=&\sigma^2\sqrt{\frac{N-\hat{M}}{K}}
\end{align*}
and let $\frac{1}{\mu_1}\ldots\frac{1}{\mu_M}$ be the eigenvalues of $\mathbf{A}^*\hat{\boldsymbol{\Gamma}}\mathbf{A}\mathbf{B}$.

\item \textbf{Assertions}

\begin{itemize}
\item (i)

\begin{align*}
\beta_{1,1}(\mu\text{\ae})=&1-\alpha-\psi f(\Delta(\mu\text{\ae}))\\
\beta_{1,2}(\mu2\text{\ae})=&\beta_{1,1}(\mu 2\text{\ae})-\mu2\text{ae}\psi\left(\left.\Delta f(\Delta)\right|_{\mu2\text{\ae}}+\frac{1}{\sqrt{2\pi}}\right)\\
\beta_{1,3}(\mu3\text{\ae})=&\beta_{1,2}(\mu3\text{\ae})-\frac{1}{2}(\mu3\text{\ae})^2\left.\psi\left((1+\Delta^2)f(\Delta)+\frac{\Delta}{\sqrt{2\pi}}\right)\right|_{\mu3\text{\ae}}
\end{align*}

\item (ii)

\begin{align*}
\beta_{2,1}(\mu_1\text{\ae},\mu_2\text{\ae})=&l_{12}\beta_{1,1}(\mu_2\text{\ae})+l_{21}\beta_{1,1}(\mu_1\text{\ae})\\
\beta_{2,2}(\mu_12\text{\ae},\mu_22\text{\ae})=&l_{12}^2\beta_{1,2}(\mu_22\text{\ae})+l_{21}^2\beta_{1,2}(\mu_12\text{\ae})+2l_{12}l_{21}\beta_{2,1}(\mu_12\text{\ae},\mu_22\text{\ae})\\
\beta_{2,3}(\mu_13\text{\ae},\mu_23\text{\ae})=&l_{12}^3\beta_{1,3}(\mu_23\text{\ae})+l_{21}^3\beta_{1,3}(\mu_13\text{\ae})+3l_{12}l_{21}\beta_{2,2}(\mu_13\text{\ae},\mu_23\text{\ae})
\end{align*}
with
\begin{align*}
\psi:=&e^{-\frac{U_\alpha^2}{2}}\\
f(x):=&e^{\frac{x^2}{2}}\phi(x)\\
\Delta(\mu K\text{\ae}):=&U_\alpha-\mu K\text{\ae}
\end{align*}
\end{itemize}

\item \textbf{Proof}

See Appendix 6 for the proof.

\end{itemize}

Because the probabilities only depend on the eigenvalues of $\mathbf{G}\mathbf{B}$ in (7.2-2), these formulae also allow the detection probabilities in the presence of subarrays and coupling (if one knows the coupling-scattering matrix), because only the modified eigenvalues need be used.

With (7.2-4), one can investigate particular two-target configurations with equal amplitudes. The estimate for $\hat{M}=1$ would lie directly between both targets so that the matrix $\mathbf{G}$ in (7.2-2) is known. The eigenvalues of such a $2\times2$ matrix are known so that the probability of exceeding the threshold at level $\hat{M}=1$ can be easily computed from (7.2-4). The detection probability is thus
\begin{align*}
P_E=&P\left\{\bar{Q}_{\hat{M}=2}\leq\eta\right\}P\left\{\bar{Q}_{\hat{M}=1}>\eta\right\}\\
=&(1-\alpha)(1-\beta_{2,K})\quad \text{if the test meets level $\alpha$}
\end{align*}
In order to compute the detection probability for more than two targets or for two targets with different magnitudes, the estimated direction $\hat{\boldsymbol{\omega}}$ with $\E\{Q(\hat{\boldsymbol{\omega}})\}=\min!$ for $M<M^{\text{ex}}$ must first be numerically determined in order to compute the eigenvalues of $\bar{G}$.

The test (7.1-6) and the detection probability (7.2-4) are also valid with general normally distributed noise as long as the parameter estimation is still good in the sense of (7.1-3). $\bar{Q}$ is still asymptotically normally distributed as long as the elements of the covariance matrix of the noise are constrained. In particular, if one had unequal receiver noise in the $N$ channels with $\frac{\sum_{i=1}^N\sigma_i^2}{N}=\sigma^2,$ $0<c_1\leq\sigma_i^2\leq c_2$, then the approximation of (7.1-5) or, respectively, (7.2-1) can be used, though with the quantities
\begin{align*}
\E_n\{\bar{Q}\}=&\sum_{i=1}^{N-M}\tilde{\sigma}_i^2+q\rightarrow(N-M)\sigma^2+q\text{ for $N\rightarrow\infty$}\\
\var_n\{\bar{Q}\}=&\frac{1}{K}\sum_{i=1}^{N-M}\tilde{\sigma}_i^4\quad\text{Does not converge to $\frac{\sigma^4}{K}(N-M)$}
\end{align*}
whereby $\tilde{\sigma}_i^2$ are the eigenvalues of $\boldsymbol{\Gamma}\mathbf{D}$ and $\mathbf{D}=\diag(\sigma_i^2)$. For example, for $\tilde{\sigma}_i^2\sim\mathcal{R}(\sigma^2(1-c),\sigma^2(1+c))$, then $\var_n\{\bar{Q}\}=\frac{\sigma^4}{K}\left(1+\frac{c^2}{3}\right)(N-M)$.

\subsection{The Behavior of the Multihypothesis Test}
\begin{figure}
\centering
\includegraphics[width=0.6\textwidth]{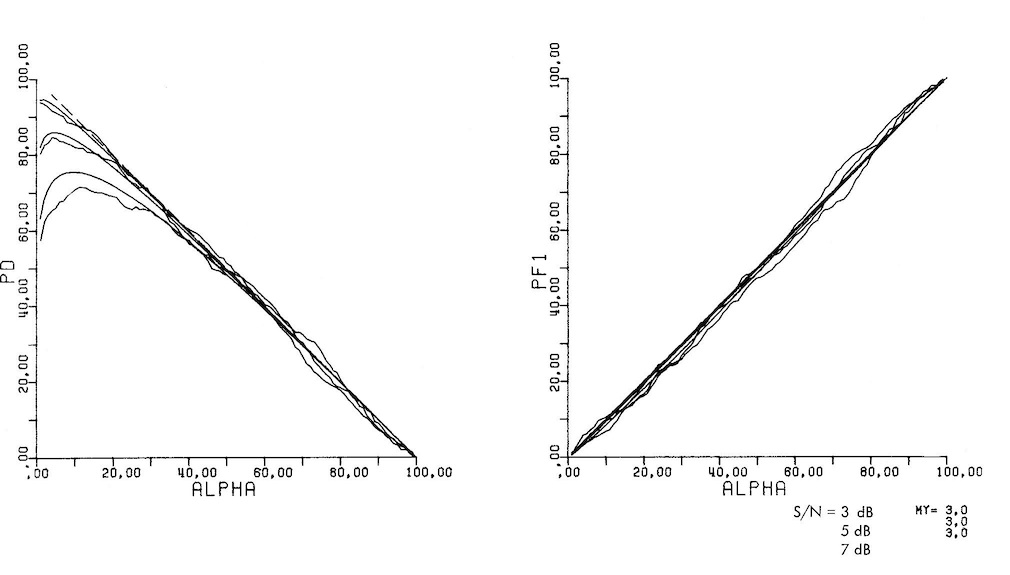}
\caption{Detection characteristics for two targets using the ELAN 21L array with $du=0.55BW$.
\label{Fig7-1}}
\end{figure}

\begin{figure}
\centering
\includegraphics[width=0.6\textwidth]{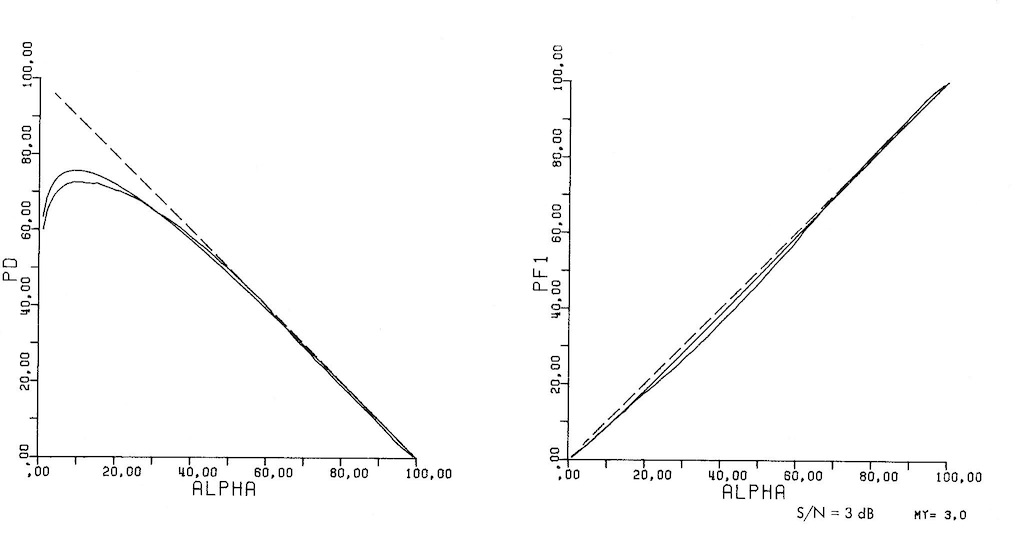}
\caption{Detection characteristics for two targets using the ELAN 21L array with $du=0.55BW$.
\label{Fig7-2}}
\end{figure}

Simulations of the tests developed here with the threshold $\eta_{\alpha;M}$ as in (7.1-6) are rather complex, because many simulations are needed in order to make a statement regarding $\alpha$, particularly when the detection probability is high. Thus, for each specified target scenario, $500$ minimizations of $\bar{Q}$ for $M=1,\ldots, M_{\text{ex}}$ are performed. For each, $17$ iterations of the original algorithm, with bounding, are used. The accuracy of these simulations is shown in Fig. \ref{Fig7-1}. The plots show the complete test $PD=P\{M=2\}$ for two targets with an SNR of $3,5,7\,\text{dB}$ and $PF1=P\{M>2\}$, which is set by $\alpha$ (The targets lie $0.55 BW$ apart, ELAN 21 L, Signal Model 4). Thus, given an estimated target number of $(M=1$, the estimated direction is assumed to lie in the middle between both targets, with $M=2$, the exact directions are assumed to be found; that is, precisely the conditions of Chapter 7.2. As helping lines, the curves of the optimal Neyman-Pearson test are shown for $PF2=0$ so that $PF1=\alpha$ and $PD=1-\alpha$. That the computed detection probability is not exactly reached by the simulations, is due to the small number of simulations $(500)$ and also the approximation of the $\chi^2$ distribution by the normal distribution. This is shown in Fig. \ref{Fig7-2} for the curves with $SNR=3\,\text{dB}$. These curves were made with $5000$ trials and still show the influence of the approximation via the normal distribution. This seems to already be sufficiently accurate for $N=21$.

\begin{figure}
\centering
\includegraphics[width=0.6\textwidth]{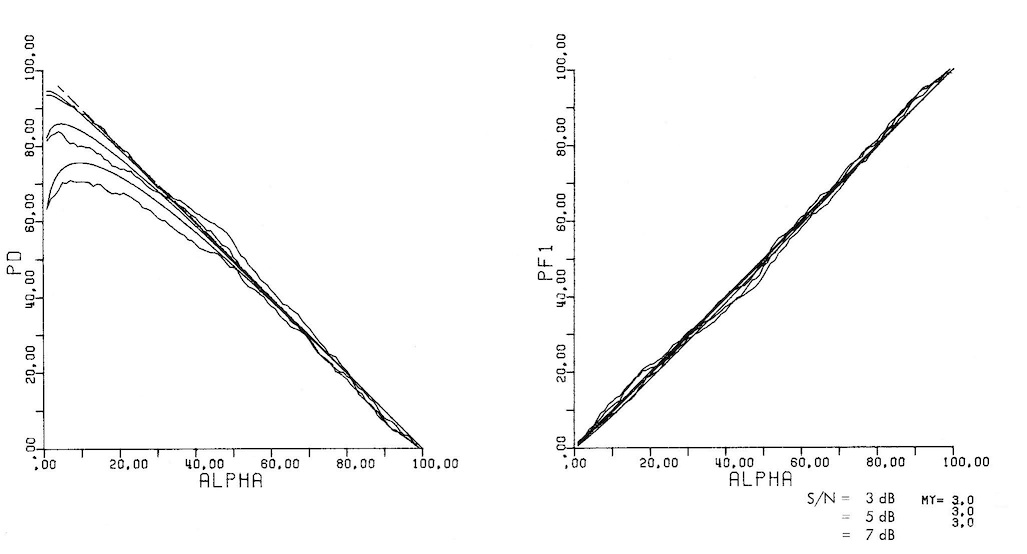}
\caption{Detection characteristics for two targets using the ELAN 21L array with $du=0.55BW$.
\label{Fig7-3}}
\end{figure}

For the same situation as Fig. \ref{Fig7-1}, Fig. \ref{Fig7-3} shows the actual detection probability with estimated parameters. The  test agrees quite well within the bounds of the simulation accuracy with the exact curves in Fig. \ref{Fig7-1}. Figure \ref{Fig7-3}  also shows that by demanding too small a value of $\alpha$, the detection probability can be dramatically reduced. A value of $\alpha$ between $5\%$ and $10\%$ seems to be a good choice. The standard deviation for the direction estimation is with estimated target number $M=2$, $0.04BW, 0.05BW, 0.06BW$ for the respective SNR values $7$, $5$, and $3\,\text{dB}$. This is within the bounds of the predicted dispersion according to Fig. \ref{Fig4-8} (b), Curve 5, because the CRLB for $\Delta\psi=90^\circ$ corresponds to Signal Model $4$ and the variance is well described by the stochastic approximation of Fig. \ref{Fig5-3}. Applying Curve 5 for the SNR of $3\,\text{dB}$ means that the stochastic approximation converging with 17 iterations corresponds to ML-estimation with averaging over $9$ samples.

\begin{figure}
\centering
\includegraphics[width=0.6\textwidth]{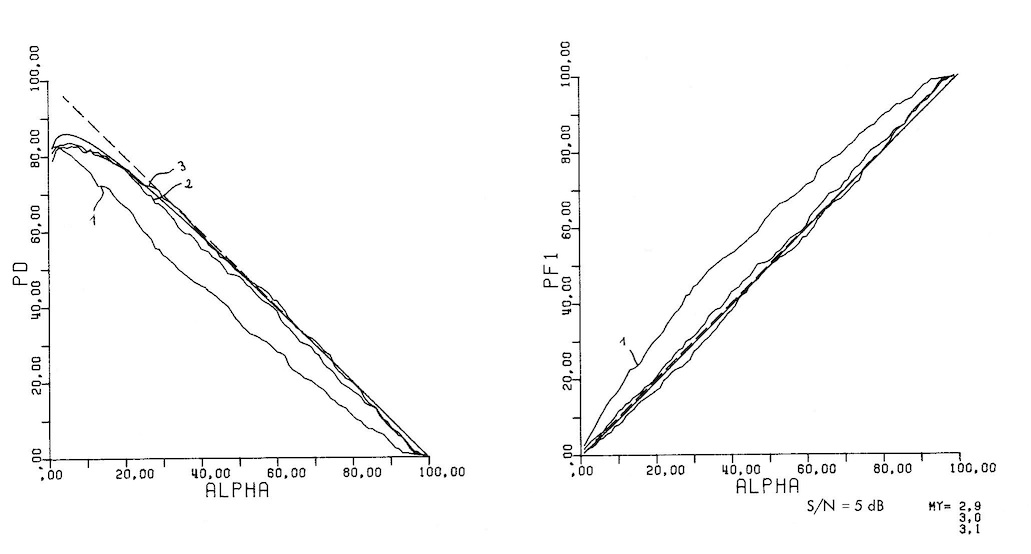}
\caption{Detection characteristics for two targets using the ELAN 21L array with $du=0.55BW$.
\label{Fig7-4}}
\end{figure}

The influence of different noise powers on the receiver (the same target situation as in Fig. \ref{Fig7-1}, SNR=$5\,\text{dB}$) is shown in Fig. \ref{Fig7-4}. For the noise power, fluctuations of $1\,\text{dB}$ were assumed and $\sigma_i^2\sim\mathcal{R}\left(\sigma^2\left(1-\frac{1}{4}\right),\sigma^2\left(1+\frac{1}{4}\right)\right)$ was randomly drawn. Curve 1 shows the results for the test where $\sigma^2$ is used without knowledge of the fluctuations. In comparison to the theoretical curves, the detection probability drops and the test no longer attains a level $\alpha$. Therefore, the estimates $\hat{\boldsymbol{\omega}}$ are assumed to be exact. Curve 2 shows the test, also for exact $\hat{\boldsymbol{\omega}}$, but $\E\{\bar{Q}\}$ and $\var\{\bar{Q}\}$ corresponding to (7.2-5) for the threshold $\eta$ are used. Curve 3 shows the case where the parameter $\hat{\boldsymbol{\omega}}$ is estimated. The estimation error due to the fluctuations  in the noise power is negligible. The state of the amplifiers should be taken into account for the test if possible. For large numbers of elements, the difference should be even smaller.

\begin{figure}
\centering
\includegraphics[width=0.6\textwidth]{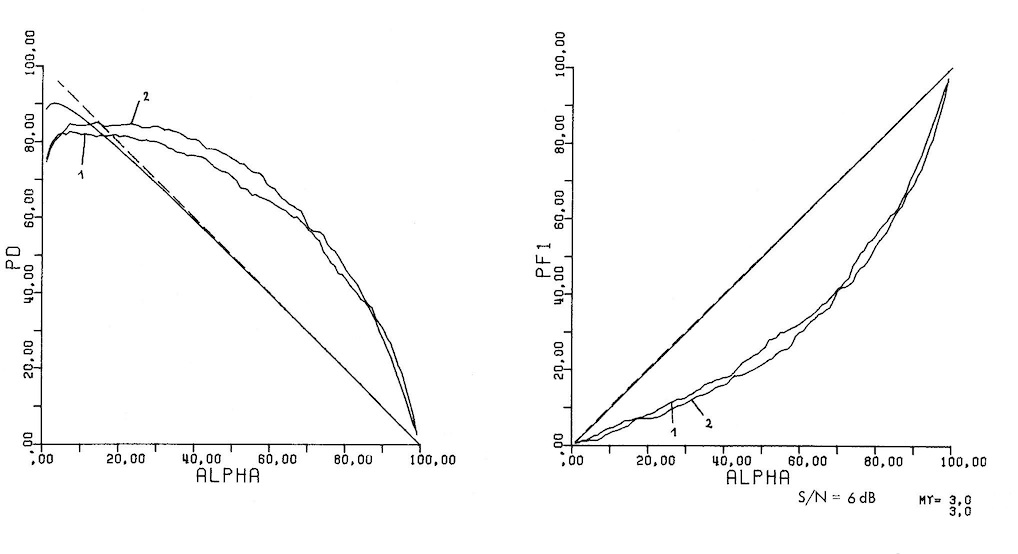}
\caption{Detection characteristics for two targets using the ELAN 21L array with $du=0.55BW$.
\label{Fig7-5}}
\end{figure}

That the test is sensitive to deviations from the assumption that the noise in all of the channels is the same is shown by the behavior in the presence of a jammer. The test with the same target position as in Fig. \ref{Fig7-1} is shown in Fig. \ref{Fig7-5} for the case where there is a jammer in direction $u=-0.5$ as in (2.2-2) with finite width of $s=0.1$ with the effective value $\wp=\sigma/2$. The effective value of the targets is $\sqrt{\E\left\{|b_i|^2\right\}}=1.6\sigma$ so that the signal-to-internal noise ratio is $7\,\text{dB}$ and the SNR is $6\,\text{dB}$.

White noise having power $\sigma^2+\sigma^2/4$ is assumed for the test. Curve 1 shows the  test for the exact parameter values, Curve 2  for the estimated value $\hat{\boldsymbol{\omega}}$.The influence of the error in the estimator is negligible. The deviation of the given noise background, in which the signal is to be detected, is the primary source of the increased estimation error. If one were to only consider the internal noise $\sigma^2$ when computing the threshold $\eta$ as in (7.1-6), the detection probability would sink by $50\%$.

The worsening of the detection probability in the presence of jamming can only be solved if one computes $\E\{\bar{Q}\}, \var\{\bar{Q}\}$ for the given jammer angle and estimation target direction; that is, by estimating the direction and power of the jammer and assuming noise according to Model $4$ or by estimation of the jamming covariance matrix $\mathbf{R}$ and computation of $\E\{\bar{Q}\}=\trace\boldsymbol{\Gamma}\mathbf{R},\var\{\bar{Q}\}=\frac{1}{K}\trace\boldsymbol{\Gamma}\mathbf{R},\boldsymbol{\Gamma}\mathbf{R}$. Another possibility is that one takes all jammers (at a particular range) into account in the $Q$ function and estimates them as targets.

\begin{figure}
\centering
\includegraphics[width=0.6\textwidth]{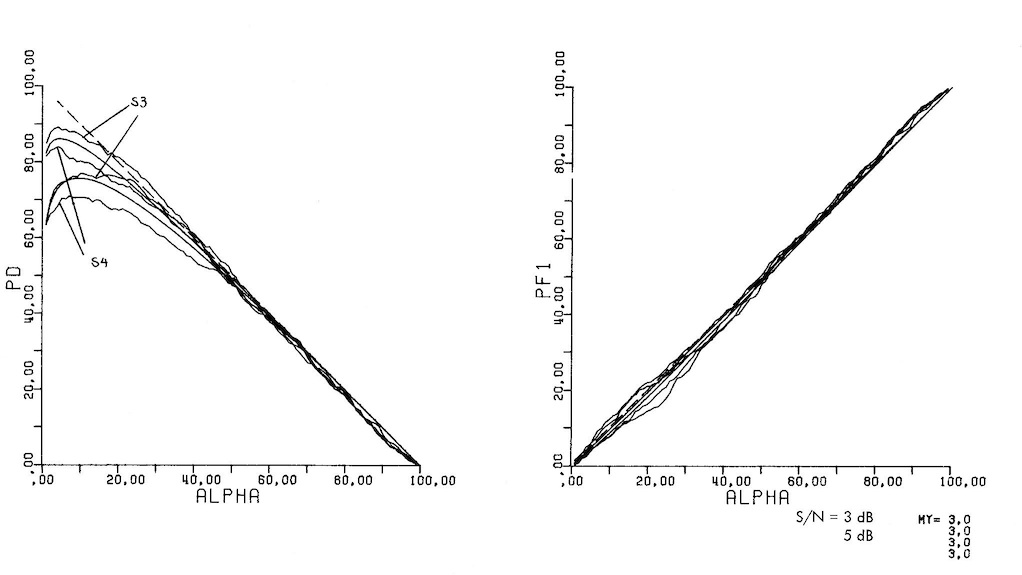}
\caption{Detection characteristics for two targets using the ELAN 21L array with $du=0.55BW$.
\label{Fig7-6}}
\end{figure}

If instead of using Signal Model $4$, Signal Model $3$ were used, the test (and the estimation according to Chapter 5.3) improves. This is shown in Figure \ref{Fig7-6} for the case of exact estimation (target situation and estimate values as in Fig. \ref{Fig7-1} for SNR=$3,5\,\text{dB}$). The curves for Signal Model $3$ (S3) are above the theoretical detection probability for Signal Model $4$ (S4). The level $\alpha$, however, is still held. The maintenance of level $\alpha$ is thus (with sufficiently accurate estimates) really independent of the signal model. The higher detection probability comes from the higher probability of $\bar{Q}>\eta$ at the level $M=1$.

One can thus view Signal Model $4$ as being unfavorable in the instances where one excludes the phase positions of $\Delta\varphi=0^\circ,180^\circ$ under Models $1$ and $2$. The curves for the theoretical detection probabilities can be viewed as lower limits on the detection probabilities under these other models. One can thus skip timely simulations and study the influence of other parameters.

\begin{figure}
\centering
\includegraphics[width=0.6\textwidth]{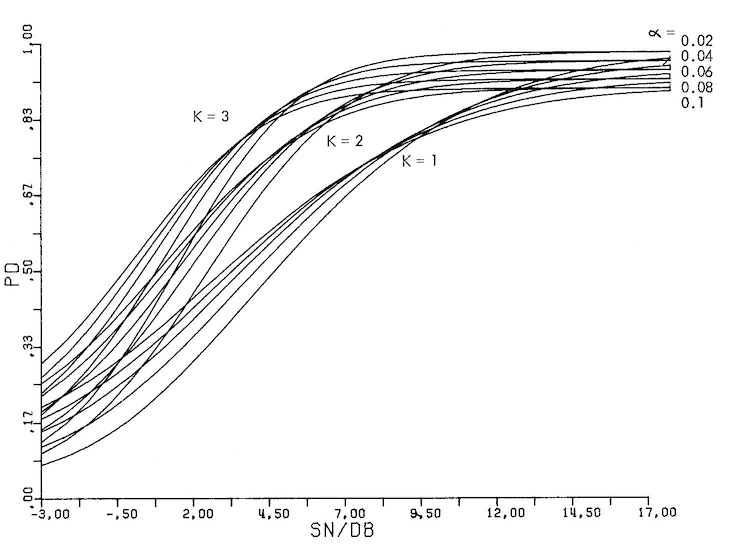}
\caption{The resolution of two targets using the ELAN 21L array with $du=0.5BW$.
\label{Fig7-7}}
\end{figure}

\begin{figure}
\centering
\includegraphics[width=0.6\textwidth]{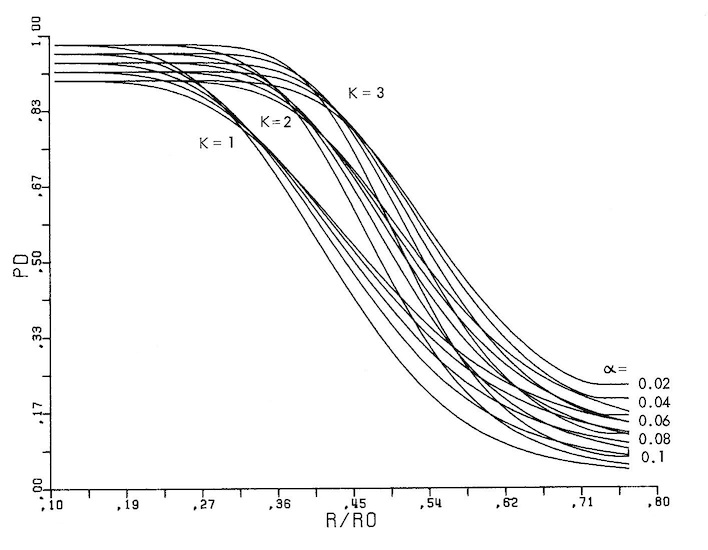}
\caption{The resolution of two targets with $du=0.5BW$.
\label{Fig7-8}}
\end{figure}

For ELAN 21 L, Fig. \ref{Fig7-7} shows the theoretical detection probability of two targets (separation of the targets is $0.5BW$) as a function of SNR for $\alpha=2,4,6,8,10\,\%$ and the number of samples of $\bar{Q}$ averaged $K=1,2,3$. The same detection probabilities as a function of the distance between targets are shown in Fig. \ref{Fig7-8}, whereby the signal power $\E\{|S|^2\}$ is proportional to $\frac{1}{R^4}$ and $R0$ is the $3$-$\text{dB}$ distance of the whole SNR, where
\begin{equation}
\frac{\E\{2N|b|^2\}}{\E\{|n|^2\}}=2.
\end{equation}
Even a small number of samples of $Q$ result in large improvements.

\begin{figure}
\centering
\includegraphics[width=0.6\textwidth]{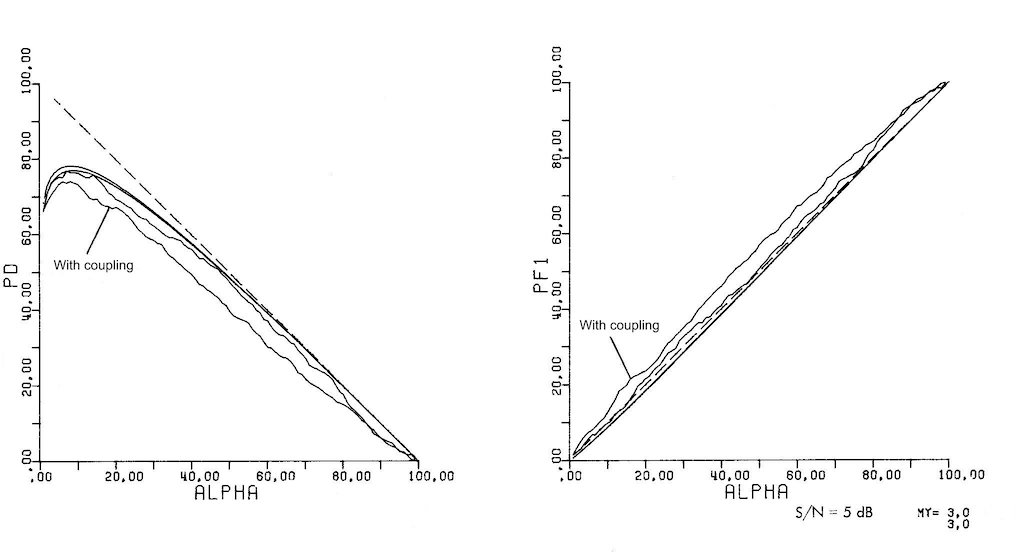}
\caption{Detection characteristics for two targets using the ELAN 25 array with $du=0.5BW$.
\label{Fig7-9}}
\end{figure}

\begin{figure}
\centering
\includegraphics[width=0.6\textwidth]{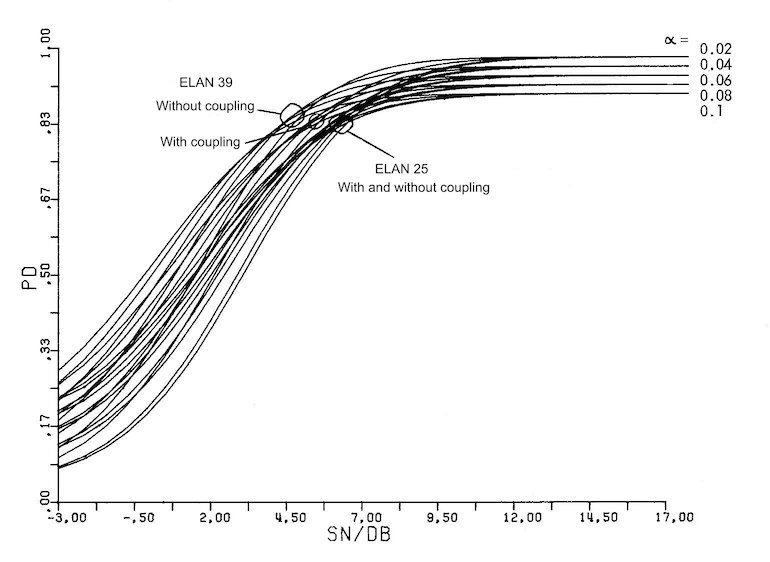}
\caption{The resolution of two targets with $du=0.5BW$.
\label{Fig7-10}}
\end{figure}

The coupling only has a small influence. This is shown in Fig. \ref{Fig7-9} for ELAN 25 for two targets with a separation of $BW/2$ and SNR$=5\,\text{dB}$. However, the level $\alpha$ is not quite held anymore, which can be explained by the fact that at the level $M=2$, $\boldsymbol{\Gamma}\mathbf{s}\neq\mathbf{0}$ even when using exact direction estimates. The standard deviation of the directional estimates at the level $M=2$ is, with and without coupling, only about $0.08BW$. For the same situation, Fig. \ref{Fig7-10} shows the SNR-loss due to coupling for ELAN 25 and 29. For ELAN 39, with the highest element density, the estimation is better, but the loss due to stronger coupling is larger at about $1\,\text{dB}$.

In all, the theoretical detection probabilities provide the ability to investigate the behavior of the test, because the parameter estimation is generally sufficient and the error level $\alpha$ is independent of the signal model and is thus attained. However, the resolution of target in arbitrary Gaussian noise requires an exact knowledge of the first two moments of this noise.

\section{Summary}\label{Sec8}

\begin{enumerate}
\item The resolution problem was formulated for spatially and temporally sampled, overlapping, monochromatic waves with temporally fluctuating amplitudes and phases in Gaussian noise. A solution is theoretically possible using a sequence of estimates from performing likelihood maximization in likelihood ratio tests.
\item To assure the existence and uniqueness of a solution, an appropriate restriction of the angle of incidence of the wave is necessary. When only sampling spatially,  
a strong $M$-regular antenna layout is necessary to resolve $M$ targets. When also sampling temporally, only a weak $M$-regular layout is needed. These conditions are particularly important when one wishes to resolve targets using as few elements as possible.
\item The condition of the estimate can be analyzed by studying the curvature of the $Q$-function and $\E\{Q\}$ at the minimum point. This essentially corresponds to an application of the CRLB.
\item Resolution using only a single impulse, which means only with spatial sampling, demands a very high SNR (over $25\,\text{dB}$). Additionally, there are target constellations (depending primarily upon the phase difference of the targets) that even with high SNRs can only be poorly resolved. Phase fluctuations by multiple temporal samples significantly improve the direction estimation. Thus, an increase in the angular resolution necessitates a worsening of the Doppler estimate, corresponding to a direction-frequency ambiguity relation.
\item A simple, sequential estimation algorithm for direction estimation using the stochastic approximation is given. This procedure can still be refined. The procedure uses only the current sum-and-difference beam outputs in each iteration as its inputs. The asymptotic behavior of the algorithm is well approximated by the CRLB.
\item The stochastic approximation algorithm is relatively robust to deviations from the assumed signal and noise model. Interference powers on the order of the target amplitudes have little influence on the estimates as long as they are located far from the beam.
\item A sequential multihypothesis tests was developed to determine the number of targets present. The test can be used with the stochastic approximation and is computationally efficient. Additionally, the test maintains a firm probability that the number of targets present is overestimated.
\item Detection and false-alarm probabilities for the resolution of noise sources (Signal Model $4$) can be calculated for this test. Because this model can be viewed as being pessimistic, such probabilities have meanings for the other models as well.
\item The test is not very sensitive against signal distortions, as may result from mutual coupling for thinned arrays. It is however sensitive against deviations from the assumed noise model, because this is the basis of the detection criterion for the different signals. This sensitivity can be mitigated by interference pre-filtering, or by including jammers as additional targets to be estimated, or by estimating the first two moments of the interference. Another promising way could be the development of a non-parametric test procedure. 
\item The attainable resolution can be shown through two examples:
\begin{itemize}
\item For the planar array ELAN 25, two targets located $0.5$ beamwidths apart were correctly detected $76\%$ of the time. In $5\%$ of cases, the number of targets was overestimated (a hard level of $\alpha=0.05$). The target directions were estimated with a standard deviation of 0.08 beamwidths in 17 iterations with an average SNR of $5\,\text{dB}$ per element.
\item For the linear antenna ELAN 21 L, two targets located $0.5$ beamwidths apart were correctly detected $95\%$ of the time with $\alpha=0.02$. The standard deviation of the angular estimates was $0.04$ beamwidths after $17$ iterations.

\end{itemize}
\end{enumerate}

\renewcommand\appendixname{Appendices}
\appendices
\section{}\label{Appen1}
\subsection{Proofs for Chapter 4}
\underline{Proof of (4.1-6)}\hfill(A.1-1)\\
\setcounter{equation}{1}
Without loss of generality, let $\mathbf{A}=(\mathbf{a}(\omega_1),\ldots,\mathbf{a}(\omega_M),\mathbf{a}(\omega_M),\ldots,\mathbf{a}(\omega_M))$ with $\omega_i\neq\omega_k$ for all $i,k\in\{1,\ldots, M\}$, $i\neq k$. $\mathbf{A}$ is an $N\times(M+k)$ matrix.

For such singular $\mathbf{A}$ matrices, one can solve $\min\limits_{\mathbf{b}}\lVert\mathbf{a}-\mathbf{A}\mathbf{b}\rVert^2$ using a Householder transformation \cite{ref19}. That is, through the use of a unitary matrix $\mathbf{U}$ that brings $\mathbf{A}$ into a triangular form. Then $\lVert\mathbf{z}-\mathbf{A}\mathbf{b}\rVert^2=\lVert\mathbf{U}\mathbf{z}-(\mathbf{U}\mathbf{A})\mathbf{b}\rVert^2$.

Let $\mathbf{A}$ be partitioned in the form $\mathbf{A}=(\mathbf{A}_1\vline\mathbf{A}_2)$, where $\mathbf{A}_1$ is a regular $N\times M$ matrix, and $\mathbf{A}_2$ is an $N\times k$ matrix with rank $\mathbf{A}_2=1$. The corresponding partitioning is applied to the other terms:

\begin{align*}
\mathbf{b}=&\left(\begin{IEEEeqnarraybox}[\relax][c]{c}
\IEEEstrut
\mathbf{b}_1\\
\hline\\
\mathbf{b}_2
\IEEEstrut
\end{IEEEeqnarraybox}\right)&\mathbf{y}:=&\mathbf{U}\mathbf{z}&\mathbf{y}=&\left(\begin{IEEEeqnarraybox}[\relax][c]{c}
\IEEEstrut
\mathbf{y}_1\\
\hline\\
\mathbf{y}_2
\IEEEstrut
\end{IEEEeqnarraybox}\right)
\end{align*}
where $\mathbf{b}_1\in\mathbb{C}^M$, $\mathbf{b}_2\in\mathbb{C}^k$, $\mathbf{y}_1\in\mathbb{C}^M$, and $\mathbf{y}_2\in\mathbb{C}^{N-M}$. Thus, $\mathbf{U}\mathbf{A}$ has the form
\begin{equation*}
\mathbf{U}\mathbf{A}=\left(\begin{array}{cccc|cccc}
*		&* 		&\ldots	&*		&*		&* 		&\ldots	&*\\
0		&*		&\ldots	&*		&*		&* 		&\ldots	&*\\	
\vdots	&\vdots	&\ddots	&\vdots	&\vdots	&\vdots	&\ddots	&\vdots\\
0		&0		&\ldots  	&*		&*		&*		&\ldots  	&*\\
0		&0		&\ldots  	&0		&0		&0		&\ldots  	&0\\	
\vdots	&\vdots	&\ddots	&\vdots	&\vdots	&\vdots	&\ddots	&\vdots\\
0		&0		&\ldots	&0		&0		&0		&\ldots	&0
\end{array}\right)=\left(\begin{array}{c|c}
\mathbf{B}_1&\mathbf{B}_2\\
\mathbf{0}&\mathbf{0}
\end{array}\right)
\end{equation*}
where the asterisks denotes nonzero values. Consequently,
\begin{equation*}
\lVert\mathbf{U}\mathbf{z}-\mathbf{u}\mathbf{A}\mathbf{b}\rVert^2=\left\lVert\left(\begin{IEEEeqnarraybox}[\relax][c]{c}
\IEEEstrut
\mathbf{y}_1-\mathbf{B}_1\mathbf{b}_1-\mathbf{B}_2\mathbf{b}_2\\
\mathbf{y}_2
\IEEEstrut
\end{IEEEeqnarraybox}\right)\right\rVert^2.
\end{equation*}
This expression is minimized for $\mathbf{y}_1=\mathbf{B}_1\mathbf{b}_1+\mathbf{B}_2\mathbf{b}_2$ (For example, $\mathbf{b}_2=\mathbf{0}$, $\mathbf{b}_1=\mathbf{B}_1^{-1}\mathbf{y}_1$) and has the minimal value $\lVert\mathbf{y}_2\rVert^2$, which is determined by the Householder transformation $\mathbf{U}$. However, $\mathbf{U}$ is only determined by $\mathbf{A}_1$.

\underline{Proof of (4.1-7)}\hfill

Since $\boldsymbol{\Gamma}=\mathbf{I}-\mathbf{A}\left(\mathbf{A}^*\mathbf{A}\right)^{-1}\mathbf{A}^*$ and $\mathbf{A}$ is infinitely differentiable, it remains to show that $(\mathbf{A}^*\mathbf{A})^{-1}$ is infinitely differentiable.

Define
\begin{equation*}
\mathbf{S}\left(\boldsymbol{\omega}\right):=\mathbf{A}^*(\boldsymbol{\omega})\mathbf{A}(\boldsymbol{\omega})
\end{equation*}
and consider
\begin{equation*}
\mathbf{F}:{\left(V^M \backslash \gamma\right) \times \mathbb{C}^{M\times M}\rightarrow \mathbb{C}^{M\times M}\atop (\boldsymbol{\omega},\mathbf{T})\mapsto \mathbf{S}(\boldsymbol{\omega})\mathbf{T}-\mathbf{I}}.
\end{equation*}
Then $\mathbf{F}$ implicitly defines the function $\mathbf{S}^{-1}(\boldsymbol{\omega})$ because
\begin{equation*}
\mathbf{F}(\boldsymbol{\omega},\mathbf{T})=\mathbf{0}\Longleftrightarrow \mathbf{T}=\mathbf{S}^{-1}(\boldsymbol{\omega}).
\end{equation*}

For real differentiability, let $\mathbf{F}$ be written as a real function:
\begin{equation*}
\mathbf{F}(\boldsymbol{\omega},\mathbf{T})=\left(\begin{IEEEeqnarraybox}[\relax][c]{c'c}
\IEEEstrut
\text{Re } \mathbf{S}&-\text{Im } \mathbf{S}\\
\text{Im } \mathbf{S}&\text{Re }  \mathbf{S}
\IEEEstrut
\end{IEEEeqnarraybox}\right)\left(\begin{IEEEeqnarraybox}[\relax][c]{c'c}
\IEEEstrut
\text{Re }  \mathbf{T}&-\text{Im } \mathbf{T}\\
\text{Im } \mathbf{T}&\text{Re }  \mathbf{T}
\IEEEstrut
\end{IEEEeqnarraybox}\right)-\left(\begin{IEEEeqnarraybox}[\relax][c]{c'c}
\IEEEstrut
\mathbf{I}&\mathbf{0}\\
\mathbf{0}&\mathbf{I}
\IEEEstrut
\end{IEEEeqnarraybox}\right).
\end{equation*}
Thus, $\mathbf{F}$ is infinitely differentiable with regard to $(\boldsymbol{\omega},\text{Re }\mathbf{T},\text{Im }\mathbf{T})$ and the Jacobian determinants are
\begin{equation*}
\det\left(\frac{\partial F_{ij}}{\partial t_{kl}}\right)_{i,j=1\ldots 2M\atop k,l=1\ldots 2M}=\det\left(
\begin{IEEEeqnarraybox}[\relax][c]{c'c'c'c}
\IEEEstrut
\mathbf{S}&\mathbf{0}&\ldots&\mathbf{0}\\
\mathbf{0}&\mathbf{S}&\ldots&\mathbf{0}\\
\vdots	&\vdots	&\ddots&\vdots\\
\mathbf{0}&\mathbf{0}&\ldots&\mathbf{S}
\IEEEstrut
\end{IEEEeqnarraybox}\right)
\end{equation*}
because $\mathbf{S}$ is regular on $V^M\backslash \gamma$.

Thus, the implicit function theorem holds that $\mathbf{S}^{-1}(\boldsymbol{\omega})$ is likewise infinitely differentiable, though the derivatives from $\mathbf{S}^{-1}$ and $\mathbf{S}$ can indeed be singular matrices (For example, $\mathbf{S}_{\omega_i\omega_k}$ ($i\neq k$) is singular).

\underline{Computation of the First and Second Derivatives of $Q$}

\begin{itemize}
\item \underline{Lemma 1}

\begin{equation*}
\frac{\partial}{\partial u_i}(\mathbf{A}^*)=\left(\frac{\partial \mathbf{A}}{\partial u_i}\right)^*\quad \text{for $i=1,\ldots, M$}
\end{equation*}
An analogous statement can be made for $v_i$. The proof is through substitution.

\item \underline{Lemma 2}

For all regular differentiable matrices $\mathbf{M}(\mathbf{u})$ and $ \mathbf{M}^{-1}(\mathbf{u})$, the following holds
\begin{equation*}
\frac{\partial}{\partial u_i}(\mathbf{M}^{-1})=-\mathbf{M}^{-1}\frac{\partial \mathbf{M}}{\partial u_i}\mathbf{M}^{-1}.
\end{equation*}

Proof:
\begin{equation*}
\mathbf{M}\mathbf{M}^{-1}=\mathbf{I}\Longrightarrow \frac{\partial\mathbf{M}}{\partial u_i}\mathbf{M}^{-1}+\mathbf{M}\frac{\partial\mathbf{M}^{-1}}{\partial u_i}=\mathbf{0}.
\end{equation*}
\end{itemize}

In the following the abbreviation $f_{u_i}:=\frac{\partial f}{\partial u_i}$. Then one can write
\begin{equation*}
Q_{u_i}=\mathbf{z}^*\boldsymbol{\Gamma}_{u_i}\mathbf{z}
\end{equation*}
and
\setcounter{equation}{1}
\begin{align}
\boldsymbol{\Gamma}_{u_i}=&-\frac{\partial}{\partial u_i}\mathbf{A}\left(\mathbf{A}^*\mathbf{A}\right)^{-1}\mathbf{A}^*\notag\\
=&-\left(\mathbf{A}_{u_i}\left(\mathbf{A}^*\mathbf{A}\right)^{-1}\mathbf{A}^*-\mathbf{A}\left(\mathbf{A}^*\mathbf{A}\right)^{-1}\left(\mathbf{A}_{u_i}^*\mathbf{A}+\mathbf{A}^*\mathbf{A}_{u_i}\right)\left(\mathbf{A}^*\mathbf{A}\right)^{-1}\mathbf{A}^*+\mathbf{A}\left(\mathbf{A}^*\mathbf{A}\right)^{-1}\mathbf{A}_{u_i}^*\right)\notag\\
=&-\left(\boldsymbol{\Gamma}\mathbf{A}_{u_i}\left(\mathbf{A}^*\mathbf{A}\right)^{-1}\mathbf{A}^*+\mathbf{A}\left(\mathbf{A}^*\mathbf{A}\right)^{-1}\mathbf{A}^*_{u_i}\boldsymbol{\Gamma}\right).\label{eqA-1-2}
\end{align}
In particular, because $\boldsymbol{\Gamma}\mathbf{A}=\mathbf{0}$,
\begin{align*}
\mathbf{A}^*\boldsymbol{\Gamma}_{u_i}\mathbf{A}=\mathbf{0}\\
\trace\boldsymbol{\Gamma}_{u_i}=0
\end{align*}
and
\begin{align}
\E\left\{Q_{u_i}\right\}=&\trace\boldsymbol{\Gamma}_{u_i}\E\left\{\mathbf{z}\mathbf{z}^*\right\}\notag\\
=&\trace\boldsymbol{\Gamma}_{u_i}\left(\mathbf{I}+\mathbf{A}\mathbf{B}\mathbf{A}^*\right)\label{eqA-1-3}
\end{align}
so that
\begin{equation*}
\left.\E\left\{Q_{ui}\right\}\right|_{\bar{\boldsymbol{\omega}}}=\mathbf{0}
\end{equation*}

\begin{align*}
Q_{u_i}=&-2\text{Re }\mathbf{z}^*\mathbf{A}\left(\mathbf{A}^*\mathbf{A}\right)^{-1}\mathbf{A}_{u_i}^*\boldsymbol{\Gamma}\mathbf{z}\\
=&-2\text{Re }\hat{\mathbf{b}}^*\mathbf{A}^*_{u_i}\left(\mathbf{z}-\mathbf{A}\hat{\mathbf{b}}\right)
\end{align*}
with 
\begin{align*}
\hat{\mathbf{b}}:=&\left(\mathbf{A}^*\mathbf{A}\right)^{-1}\mathbf{A}^*\mathbf{z}\\
\mathbf{A}_{u_i}=&\left(\begin{IEEEeqnarraybox}[\relax][c]{c'c'c'c'c}
\IEEEstrut
0		&\ldots	&-jx_ia_{11}	&\ldots	&0\\
\vdots	&\ddots	&\vdots		&\ddots	&0\\
0		&\ldots	&-jx_Na_{Ni}	&\ldots	&0
\IEEEstrut
\end{IEEEeqnarraybox}\right)
\end{align*}
where only the $i$th column of $\mathbf{A}_{u_i}$ is nonzero. Thus,
\begin{align*}
\hat{\mathbf{b}}^*\mathbf{A}^*_{u_i}=j\hat{b}_i^*\mathbf{a}_i^*\mathbf{D}_x\\
\mathbf{D}_x=\diag\left(x_k \right)
\end{align*}
and then
\begin{align}
Q_{u_i}=&-2\text{Re }j\hat{b}^*_i\left(\mathbf{a}_i^*\mathbf{D}_x\mathbf{z}-\mathbf{a}_i^*\mathbf{D}_x\mathbf{A}\hat{\mathbf{b}}\right)\notag\\
=&2\text{Im }\hat{b}_i^*\left(\mathbf{a}_i^*\mathbf{D}_x\mathbf{z}-\mathbf{a}_i^*\mathbf{D}_x\mathbf{A}\hat{\mathbf{b}}\right)\notag\\
Q_{v_i}=&2\text{Im }\hat{b}_i^*\left(\mathbf{a}_i^*\mathbf{D}_y\mathbf{z}-\mathbf{a}_i^*\mathbf{D}_y\mathbf{A}\hat{\mathbf{b}}\right)\label{eqA-1-4}\\
Q_{\mathbf{u}}=&2\text{Im }\diag(\hat{b}_i^*)\left(\mathbf{A}^*\mathbf{D}_x\mathbf{z}-\mathbf{A}^*\mathbf{D}_x\mathbf{A}\hat{\mathbf{b}}\right)\notag
\end{align}
and $Q_{\mathbf{v}}$ is the same as $Q_{\mathbf{u}}$ with the appropriate substitutions.

For the second derivative, the following holds:
\begin{equation*}
Q_{\omega_i\omega_k}=\mathbf{z}^*\boldsymbol{\Gamma}_{\omega_i\omega_k}\mathbf{z}\quad(\omega_i=u_i\text{ or }v_i)
\end{equation*}
and
\begin{align}
\boldsymbol{\Gamma}_{\omega_i\omega_k}=&-\boldsymbol{\Gamma}\left(\mathbf{A}_{\omega_i}\mathbf{A}^{-1}\mathbf{A}_{\omega_k}^*+\mathbf{A}_{\omega_k}\mathbf{S}^{-1}\mathbf{A}_{\omega_i}^*\right)\boldsymbol{\Gamma}\notag\\
&+\mathbf{A}\mathbf{S}^{-1}\left(\mathbf{A}^*_{\omega_k}\boldsymbol{\Gamma}\mathbf{A}_{\omega_i}+\mathbf{A}^*_{\omega_i}\boldsymbol{\Gamma}\mathbf{A}_{\omega_k}\right)\mathbf{S}^{-1}\mathbf{A}^*\notag\\
&-\boldsymbol{\Gamma}\left(\mathbf{A}_{\omega_i\omega_k}-\mathbf{A}_{\omega_i}\mathbf{S}^{-1}\mathbf{A}^*\mathbf{A}_{\omega_k}\right)\mathbf{S}^{-1}\mathbf{A}^*\notag\\
&-\mathbf{A}\mathbf{S}^{-1}\left(\mathbf{A}^*_{\omega_i\omega_k}-\mathbf{A}^*_{\omega_k}\mathbf{A}\mathbf{S}^{-1}\mathbf{A}^*_{\omega_i}\right)\boldsymbol{\Gamma}\notag\\
&+\left(\boldsymbol{\Gamma}\mathbf{A}_{\omega_k}\mathbf{S}^{-1}\mathbf{A}^*\mathbf{A}_{\omega_i}\mathbf{S}^{-1}\mathbf{A}^*+\mathbf{A}\mathbf{S}^{-1}\mathbf{A}_{\omega_i}^*\mathbf{A}\mathbf{S}^{-1}\mathbf{A}^*_{\omega_k}\boldsymbol{\Gamma}\right)\label{eqA-1-5}
\end{align}
where
\begin{equation*}
\mathbf{S}:=\mathbf{A}^*\mathbf{A}.
\end{equation*}

In particular, because $\boldsymbol{\Gamma}\mathbf{A}=\mathbf{0}$,
\begin{align}
\left.\mathbf{s}^*\boldsymbol{\Gamma}_{\omega_i\omega_k}\mathbf{s}\right|_{\bar{\boldsymbol{\omega}}}=&\mathbf{b}^*\mathbf{A}^*\boldsymbol{\Gamma}_{\omega_i\omega_k}\mathbf{A}\mathbf{b}\notag\\
=&\mathbf{b}^*\left(\mathbf{A}^*_{\omega_k}\boldsymbol{\Gamma}\mathbf{A}_{\omega_i}+\mathbf{A}^*_{\omega_i}\boldsymbol{\Gamma}\mathbf{A}_{\omega_k}\right)\mathbf{b}\notag\\
=&2\text{Re }\left(b_k^*\mathbf{a}^*_{k,\omega_k}\boldsymbol{\Gamma}\mathbf{a}_{i,\omega_i}b_i\right)\notag\\
=&2\text{Re }\left(b_k^*\mathbf{a}^*_k\mathbf{D}\boldsymbol{\Gamma}\mathbf{D}\mathbf{a}_ib_i\right)\quad (\text{$\mathbf{D}=\mathbf{D}_x$ or $\mathbf{D}_y$}).\label{eqA-1-6}
\end{align}

Similarly,
\begin{align}
\mathbf{a}_i^*\boldsymbol{\Gamma}_{\omega_i\omega_k}\mathbf{a}_k=&\delta_{ik}2\text{Re }\mathbf{a}^*_{i,\omega_i}\boldsymbol{\Gamma}\mathbf{a}_{i,\omega_i}\notag\\
 =&\delta_{ik}2\text{Re }\mathbf{a}_i^*\mathbf{D}\boldsymbol{\Gamma}\mathbf{D}\mathbf{a}_i\label{eqA-1-7}
\end{align}
\begin{equation}
\trace\boldsymbol{\Gamma}_{\omega_i,\omega_k}=0\label{eqA-1-8}
\end{equation}
because then the first two terms in (A.1-5) vanish.

\subsection{Proof of Theorem (4.3-2):}\label{Appen2}
\setcounter{equation}{0}

Given for $\mathbf{z}=\mathbf{s}+\mathbf{n}$, $\mathbf{z}\sim\mathcal{N}(\mathbf{s},\mathbf{I})$:
\begin{align*}
\E\left\{Q\right\}=&\E\left\{\mathbf{s}^*\boldsymbol{\Gamma}\mathbf{s}+\mathbf{n}^*\boldsymbol{\Gamma}\mathbf{s}+\mathbf{s}^*\boldsymbol{\Gamma}\mathbf{n}+\mathbf{n}^*\boldsymbol{\Gamma}\mathbf{n}\right\}\\
=&\mathbf{s}^*\boldsymbol{\Gamma}\mathbf{s}+\trace\boldsymbol{\Gamma}\E\left\{\mathbf{n}\mathbf{n}^*\right\}\\
=&\mathbf{s}^*\boldsymbol{\Gamma}\mathbf{s}+N-M
\end{align*}
because
\begin{equation*}
\trace\boldsymbol{\Gamma}=\trace\mathbf{I}-\trace\left(\mathbf{A}^*\mathbf{A}\right)^{-1}\mathbf{A}^*\mathbf{A}.
\end{equation*}

At the point $\boldsymbol{\omega}_0$, where $Q_{\boldsymbol{\omega}}(\boldsymbol{\omega}_0)=\mathbf{0}$, the matrix of the second F-tensor is
\begin{equation*}
\mathbf{S}(\boldsymbol{\omega}_0)=\mathbf{s}^*\boldsymbol{\Gamma}_{\boldsymbol{\omega}\boldsymbol{\omega}}(\boldsymbol{\omega}_0)\mathbf{s}.
\end{equation*}

Furthermore using (3.1-10) for $\boldsymbol{\vartheta}=(\mathbf{u},\mathbf{v},\text{Re }b_1,\text{Im }b_2,\ldots. \text{Re }g_m,\text{Im }b_M)$,
\begin{align}
F_{ik}=&\int\frac{\partial \ln p}{\partial \vartheta_i}(\mathbf{z};\mathbf{s}(\boldsymbol{\vartheta}))\frac{\partial \ln p}{\partial \vartheta_k}(\mathbf{z};\mathbf{s}(\boldsymbol{\vartheta}))p(\mathbf{z};\mathbf{s}(\boldsymbol{\vartheta}))d\mathbf{z}\notag\\
=&-\int\frac{\partial^2 \ln p}{\partial \vartheta_i\partial \vartheta_k}(\mathbf{z};\mathbf{s}(\boldsymbol{\vartheta}))p(\mathbf{z};\mathbf{s}(\boldsymbol{\vartheta}))d\mathbf{z}\label{eqA-2-1}
\end{align}
because
\begin{equation*}
\frac{\partial^2 \ln p}{\partial \vartheta_i \partial \vartheta_k}=-\frac{1}{p^2}\frac{\partial p}{\partial\vartheta_i}\frac{\partial p}{\partial\vartheta_k}+\frac{1}{p}\frac{\partial^2 p}{\partial \vartheta_i\partial \vartheta_k}
\end{equation*}
and
\begin{equation*}
\int\frac{\partial^2 p}{\partial \vartheta_i \partial \vartheta_k}d\mathbf{z}=0
\end{equation*}
assuming that of the integrals exist.

Because
\begin{equation*}
p(\mathbf{z};\mathbf{s}(\boldsymbol{\vartheta}))=\frac{1}{\pi^N}e^{-\left(\mathbf{z}-\mathbf{s}\left(\boldsymbol{\vartheta}\right)\right)^*\left(\mathbf{z}-\mathbf{s}\left(\boldsymbol{\vartheta}\right)\right)}
\end{equation*}
one must find the derivatives of
\begin{equation*}
\tilde{Q}(\boldsymbol{\vartheta})=\left(\mathbf{z}-\mathbf{s}\left(\boldsymbol{\vartheta}\right)\right)^*\left(\mathbf{z}-\mathbf{s}\left(\boldsymbol{\vartheta}\right)\right)
\end{equation*}
for
\begin{equation*}
\mathbf{s}\left(\boldsymbol{\vartheta}\right)=\mathbf{A}\mathbf{b}
\end{equation*}

\begin{align*}
\tilde{Q}_{\vartheta_i}=&-2\text{Re }\mathbf{s}^*_{\vartheta_i}(\mathbf{z}-\mathbf{s})\\
\tilde{Q}_{\vartheta_i\vartheta_k}=&-2\text{Re}\left[\mathbf{s}_{\vartheta_i\vartheta_k}(\mathbf{z}-\mathbf{s})-\mathbf{s}^*_{\vartheta_i}\mathbf{s}_{\vartheta_k}\right]
\end{align*}
and also for $\mathbf{z}\sim\mathcal{N}(\mathbf{s},\mathbf{I})$
\begin{equation*}
\E\left\{\tilde{Q}_{\vartheta_i\vartheta_k}\right\}=F_{ik}=2\text{Re }\mathbf{s}^*_{\vartheta_i}\mathbf{s}_{\vartheta_k}
\end{equation*}
or
\begin{equation*}
\mathbf{F}=2\text{Re }\mathbf{s}^*_{\boldsymbol{\vartheta}}\mathbf{s}_{\boldsymbol{\vartheta}}
\end{equation*}
when
\begin{equation*}
\left(\mathbf{s}_{\boldsymbol{\vartheta}}\right)_{i,k}=\frac{\partial s_i}{\partial \vartheta_k}.
\end{equation*}

Now,
\begin{equation*}
\mathbf{s}_{\omega_i}=\mathbf{a}_{i,\omega_i}b_i
\end{equation*}
and
\begin{equation*}
\mathbf{s}_{\alpha_i}=\mathbf{a}_i
\end{equation*}
for $\alpha_i=\text{Re }b_i$ and
\begin{equation*}
\mathbf{s}_{\beta_i}=i\mathbf{a}_i
\end{equation*}
for $\beta_i=\text{Im }b_i$. With that, $\mathbf{F}$ has the form
\begin{equation*}
\mathbf{F}=2\text{Re }\left(\begin{array}{c|cc}
\mathbf{B}^*\mathbf{A}^*_{\boldsymbol{\omega}}\mathbf{A}_{\boldsymbol{\omega}}\mathbf{B}&\mathbf{B}^*\mathbf{A}^*_{\boldsymbol{\omega}}\mathbf{A}&j\mathbf{B}^*\mathbf{A}^*_{\boldsymbol{\omega}}\mathbf{A}\\
\hline
\mathbf{A}^*\mathbf{A}_{\boldsymbol{\omega}}\mathbf{B}&\mathbf{A}^*\mathbf{A}&j\mathbf{A}^*\mathbf{A}\\
-j\mathbf{A}^*\mathbf{A}_{\boldsymbol{\omega}}\mathbf{B}&-j\mathbf{A}^*\mathbf{A}&\mathbf{A}^*\mathbf{A}
\end{array}\right)
\end{equation*}
for
\begin{align*}
\mathbf{B}:=&\diag\limits_{k=1}^M(b_k)&\left(\mathbf{A}_{\boldsymbol{\omega}}\right)_{ik}=&\frac{\partial a_{ik}}{\partial \omega_k}.
\end{align*}
The matrix $\mathbf{F}$ can be simplified to
\begin{equation*}
\mathbf{F}=\left(\begin{array}{c|cc}
\text{Re }\mathbf{M}&\text{Re }\mathbf{R}&-\text{Im }\mathbf{R}\\
\hline
\text{Re }\mathbf{R}^T&\text{Re }\mathbf{G}&-\text{Im} \mathbf{G}\\
\text{Im }\mathbf{R}^T&\text{Im}\mathbf{G}&\text{Re }\mathbf{G}
\end{array}\right)
\end{equation*}
with
\begin{align*}
\mathbf{M}:=&2\mathbf{B}^*\mathbf{A}^*_{\boldsymbol{\omega}}\mathbf{A}_{\boldsymbol{\omega}}\mathbf{B}&\mathbf{R}:=&2\mathbf{B}^*\mathbf{A}^*_{\boldsymbol{\omega}}\mathbf{A}_{\boldsymbol{\omega}}&\mathbf{G}:=&2\mathbf{A}^*\mathbf{A}.
\end{align*}

It is to be shown that the block corresponding to $\text{Re }\mathbf{M}$ in $\mathbf{F}^{-1}$ equals $\mathbf{S}^{-1}\left(\boldsymbol{\omega}_0\right)$. For every regular matrix, one can write
\begin{equation}
\left(\begin{IEEEeqnarraybox}[\relax][c]{c'c}
\IEEEstrut
\mathbf{M}&\mathbf{R}\\
\mathbf{R}'&\mathbf{G}
\IEEEstrut
\end{IEEEeqnarraybox}\right)^{-1}=\left(\begin{IEEEeqnarraybox}[\relax][c]{c'c}
\IEEEstrut
\mathbf{E}&\mathbf{F}\\
\mathbf{F}'&\mathbf{H}
\IEEEstrut
\end{IEEEeqnarraybox}\right)\rightarrow\mathbf{E}^{-1}=\mathbf{M}-\mathbf{R}\mathbf{G}^{-1}\mathbf{R}^T.\label{eqA2-2}
\end{equation}
Thus, it can be shown that
\begin{equation*}
\mathbf{s}^*\boldsymbol{\Gamma}_{\boldsymbol{\omega}\boldsymbol{\omega}}\mathbf{s}=\text{Re }\mathbf{M}-\left(\text{Re }\mathbf{R},-\text{Im }\mathbf{R}\right)\left(\begin{IEEEeqnarraybox}[\relax][c]{c'c}
\IEEEstrut
\text{Re }\mathbf{G}&-\text{Im } \mathbf{G}\\
\text{Im }\mathbf{G}&\text{Re }\mathbf{G}
\IEEEstrut
\end{IEEEeqnarraybox}\right)^{-1}\left(\begin{IEEEeqnarraybox}[\relax][c]{c}
\IEEEstrut
\text{Re }\mathbf{R}^T\\
\text{Im }\mathbf{R}^T
\IEEEstrut
\end{IEEEeqnarraybox}\right).
\end{equation*}
Now there is
\begin{equation*}
\left(\begin{IEEEeqnarraybox}[\relax][c]{c'c}
\IEEEstrut
\text{Re }\mathbf{G}&-\text{Im } \mathbf{G}\\
\text{Im }\mathbf{G}&\text{Re }\mathbf{G}
\IEEEstrut
\end{IEEEeqnarraybox}\right)^{-1}=\left(\begin{IEEEeqnarraybox}[\relax][c]{c'c}
\IEEEstrut
\text{Re }(\mathbf{G}^{-1})&-\text{Im }(\mathbf{G}^{-1})\\
\text{Im }(\mathbf{G}^{-1})&\text{Re }(\mathbf{G}^{-1})
\IEEEstrut
\end{IEEEeqnarraybox}\right)
\end{equation*}
due to the real representation of complex numbers $\alpha+j\beta$ in $2\times 2$ matrices $\left(\begin{IEEEeqnarraybox}[\relax][c]{c'c}
\IEEEstrut
\alpha&-\beta\\
\beta&\alpha
\IEEEstrut
\end{IEEEeqnarraybox}\right)$.

Thus, it can be shown that
\begin{equation*}
\left(\begin{IEEEeqnarraybox}[\relax][c]{c'c}
\IEEEstrut
\text{Re }\mathbf{R}&-\text{Im }\mathbf{R}
\IEEEstrut
\end{IEEEeqnarraybox}\right)\left(\begin{IEEEeqnarraybox}[\relax][c]{c'c}
\IEEEstrut
\text{Re }\mathbf{G}&-\text{Im } \mathbf{G}\\
\text{Im }\mathbf{G}&\text{Re }\mathbf{G}
\IEEEstrut
\end{IEEEeqnarraybox}\right)^{-1}\left(\begin{IEEEeqnarraybox}[\relax][c]{c}
\IEEEstrut
\text{Re }\mathbf{R}^T\\
\text{Im }\mathbf{R}^T
\IEEEstrut
\end{IEEEeqnarraybox}\right)=\text{Re }\mathbf{R}\mathbf{G}^{-1}\mathbf{R}^*
\end{equation*}
but
\begin{align}
\text{Re }\left(\mathbf{M}-\mathbf{R}\mathbf{G}^{-1}\mathbf{R}^*\right)=&2\text{Re }\left(\mathbf{B}^*\mathbf{A}^*_{\boldsymbol{\omega}}\mathbf{A}_{\boldsymbol{\omega}}\mathbf{B}-\mathbf{B}^*\mathbf{A}^*_{\boldsymbol{\omega}}\mathbf{A}\left(\mathbf{A}^*\mathbf{A}\right)^{-1}\mathbf{A}^*\mathbf{A}_{\boldsymbol{\omega}}\mathbf{B}\right)\notag\\
=&2\text{Re }\mathbf{B}^*\mathbf{A}^*_{\boldsymbol{\omega}}\boldsymbol{\Gamma}\mathbf{A}_{\boldsymbol{\omega}}\mathbf{B}\notag\\
=&\mathbf{S}(\boldsymbol{\omega}_0)\label{Eq-A-2-3}
\end{align}
where the last step uses (A.1-6).

\subsection{Proof of Theorem (5.1-3):}
\setcounter{equation}{0}

\underline{Relating to (i)}

It is sufficient, as in Chapter \ref{Sec4-2} to show the linear independence of three components of three vectors $\mathbf{a}(u_1)$, $\mathbf{a}(u_2)$, $\mathbf{a}(u_3)$. Letting the element positions be $x_1=0$, $x_2=d$, $x_3=rd$ with $r>1$, then it is to be shown that
\begin{equation*}
\left|\begin{IEEEeqnarraybox}[\relax][c]{c'c'c}
\IEEEstrut
1&1&1\\
e^{-jdu_1}&e^{-jdu_2}&e^{-jdu_3}\\
e^{-jdru_1}&e^{-jdru_2}&e^{-jdru_3}
\IEEEstrut
\end{IEEEeqnarraybox}\right|\neq 0.
\end{equation*}
Without loss of generality, one can choose $u_1=0$ (eliminating $e^{-jdu_1}$ and $e^{-jdru_1}$), so that with $\gamma_i:=e^{-jdu_i}$, one gets
\begin{equation*}
\left|\begin{IEEEeqnarraybox}[\relax][c]{c'c'c}
\IEEEstrut
1&1&1\\
1&\gamma_2&\gamma_3\\
1&\gamma_2^r&\gamma_3^r
\IEEEstrut
\end{IEEEeqnarraybox}\right|=0\Longleftrightarrow \frac{\gamma_2^r-1}{\gamma_2-1}=\frac{\gamma_3^r-1}{\gamma_3-1}
\end{equation*}
Additionally, considering the function
\begin{equation*}
f(u)=e^{-j\frac{r-1}{2}du}\frac{\sin r\frac{d}{2}u}{\sin \frac{d}{2}u}
\end{equation*}
one can then write
\begin{equation*}
f(u_i)=\frac{\gamma^r_i-1}{\gamma_i-1},\quad i=2,3.
\end{equation*}
But
\begin{equation*}
f(u_2)=f(u_3)\Longleftrightarrow\left(\frac{r-1}{2}du_2=\frac{r-1}{2}du_2+2\pi k,k\in\mathbb{Z}\right)
\end{equation*}
\underline{and} $(u_2=-u_3)$ if $u$ falls within the beamwidth of the $3$-element pattern, $|u|\leq \frac{\lambda}{rd}$.

For $u=u_2=-u_3$ and $d$ in centimeters $\frac{2\pi}{\lambda}d$ instead of $d$, compare to (2.1-2) then $\frac{2\pi}{\lambda}(r-1)du=2\pi k$, meaning that $u=\frac{\lambda}{(r-1)d}k$. Because $|u|\leq\frac{\lambda}{rd}$, this condition is not fulfilled, and so the determinant is never $0$.

\underline{Relating to (ii)}

\underline{For the Forward Equivalence $\Longrightarrow$}:

Given three directions $\omega_0,\omega_1,\omega_2\in \Omega$ with the same elevation $v$, meaning that $\omega_i=(u_i,v)$, then
\begin{align*}
\mathbf{a}(\omega_i)=\mathbf{D}\left(\begin{IEEEeqnarraybox}[\relax][c]{c}
\IEEEstrut
e^{-jx_1u_i}\\
\vdots\\
e^{-jx_Nu_i}
\IEEEstrut
\end{IEEEeqnarraybox}\right)\\
=\mathbf{D}\mathbf{a}_x(u_i)
\end{align*}
where $\mathbf{D}=\diag\left(e^{-jy_kv}\right)$ so that
\begin{equation*}
\sum_{i=0}^2b_i\mathbf{a}(\omega_i)=\mathbf{0}\Longleftrightarrow\sum_{i=0}^2b_i\mathbf{D}\mathbf{a}_x(u_i)=\mathbf{0}.
\end{equation*}

\underline{For the Reverse Equivalence $\Longleftarrow$}:
\begin{figure}
\centering
\includegraphics[width=0.3\textwidth]{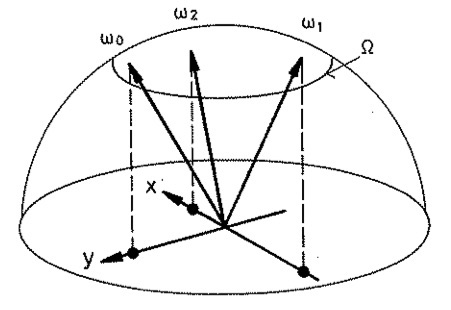}
\caption{\label{FigAppen31}}
\end{figure}

\begin{figure}
\centering
\includegraphics[width=0.5\textwidth]{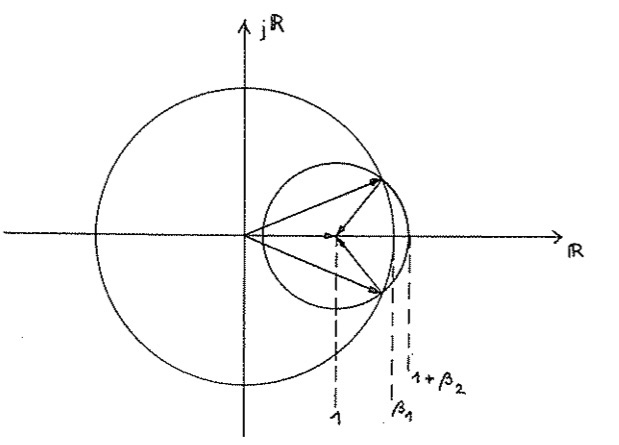}
\caption{\label{FigAppen32}}
\end{figure}

Assuming that $\mathbf{x},\mathbf{y}$ are not weakly $2$-regular on $\Omega$, there exists $\omega_0,\omega_1,\omega_2\in\Omega$ so that $\mathbf{a}(\omega_0)=b_1\mathbf{a}(\omega_1)+b_2\mathbf{a}(\omega_2)$. Without loss of generality, it can be assumed that $\omega_0=(0,v)$, $\omega_1=(u_1,0)$, $\omega_2=(u_2,0)$, and that the coordinate system for $\mathbf{x}, \mathbf{y}$ are appropriately chosen as in Fig. \ref{FigAppen31}. Then, for all $k=1\ldots N$:
\begin{equation*}
e^{jy_kv}=\beta_1e^{j(x_ku_1+\psi_1)}+\beta_2e^{j(x_ku_2+\psi_2)}
\end{equation*}
if $b_i=\beta_ie^{j\psi_i}$, and respectively
\begin{equation*}
1=\beta_1e^{j\xi_1}+\beta_2e^{j\xi_2}\tag{*}
\end{equation*}
with $\xi_1=x_ku_i-y_kv+\psi_i$. For a given $\beta_1$, $\beta_2$, these equations have only two solutions $(\mu_1,\mu_2)$ and $(-\mu_1,-\mu_2)$ as in Fig. \ref{FigAppen32}.

Thus,
\begin{equation*}
\left\{\begin{IEEEeqnarraybox}[\relax][c]{c'l}
\IEEEstrut
x_ku_1-y_kv+\psi_1=\mu_1+2\pi l&\\
x_ku_2-y_kv+\psi_2=\mu_2+2\pi l&(l=\ldots-2,-1,0,1\ldots)
\IEEEstrut
\end{IEEEeqnarraybox}\right.
\end{equation*}
or
\begin{equation*}
\left\{\begin{IEEEeqnarraybox}[\relax][c]{c}
\IEEEstrut
x_ku_1-y_kv+\psi_1=-\mu_1+2\pi l\\
x_ku_2-y_kv+\psi_2=-\mu_2+2\pi l
\IEEEstrut
\end{IEEEeqnarraybox}\right.
\end{equation*}
meaning that
\begin{equation*}
x_k=\frac{\pm(\mu_1-\mu_2)-(\psi_1-\psi_2)}{u_1-u_2}+\frac{2\pi l}{u_1-u_2}
\end{equation*}
independently of $v$. Thus, within a multiple of $2\pi$, such a non-regular positioning only has two different $x$ (and $y$) coordinates, (because $\left(\begin{IEEEeqnarraybox}[\relax][c]{c'c}
\IEEEstrut
u_1&v\\
u_2&v
\IEEEstrut
\end{IEEEeqnarraybox}\right)$ is regular if $u_1\neq u_2$).
These $x$ (or respectively $y$) coordinates are, however, not $2$-regular, because for the direction $u=0$, $u_1,u_2$, the associated direction vectors $\mathbf{a}(u)$ are also linearly independent due to (*) (given the same $\beta_1,\beta_2,\psi_1\psi_2$).

For $|u_1-u_2|\leq BW\leq\frac{\lambda}{D}$ ($D$ being the aperture diameter in centimeters) there is no $2\pi$ repetition, because then with
\begin{equation*}
\frac{2\pi}{\lambda}|x_k-x_l|=\frac{2\pi r}{|u_1-u_2|}
\end{equation*}
with $k,l,r\in\mathbb{Z}$, and $x_k$ in centimeters, it must be that
\begin{equation*}
|x_k-x_l|\geq rD.
\end{equation*}

\underline{Relating to (iii)}

It is clear that $\mathbf{a}(\omega_0+\omega)=\mathbf{D}_0\mathbf{a}(\omega)$ with $\mathbf{D}_0=\diag e^{-j(x_ku_0+y_kv_0)}$.

\subsection{The First and Second Moments of $Q$}\label{Appen4}
\setcounter{equation}{0}

\begin{align*}
\mathbf{Q}=&\mathbf{z}^*\boldsymbol{\Gamma}\mathbf{z}\\
=&\trace \boldsymbol{\Gamma}\mathbf{z}\mathbf{z}^*.
\end{align*}
Therefore,
\begin{align}
\E\left\{Q\right\}=&\E\{\trace \boldsymbol{\Gamma}\mathbf{z}\mathbf{z}^*\}\label{EqA4-1}\\
=&\trace\boldsymbol{\Gamma}\E\{\mathbf{z}\mathbf{z}^*\}\notag
\end{align}
because the integral can be pulled into the sum.

\underline{Lemma (A.4-2):}

Let $\mathbf{A},\mathbf{B}\in\mathbb{R}^{N\times N}, \mathbf{C}, \mathbf{D}\in\mathbb{C}^{N\times N},\mathbf{u}\in\mathbb{R}^N,\mathbf{w}\in\mathbb{C}^N$.
\begin{enumerate}
\item $\mathbf{n}\sim\mathcal{N}_{\mathbb{R}^N}(\mathbf{0},\mathbf{R})\Longrightarrow\E\left\{\mathbf{n}^T\mathbf{A}\mathbf{n}\mathbf{n}^T\mathbf{u}\right\}=\mathbf{0}$ and $\E\left\{\mathbf{n}^T\mathbf{A}\mathbf{n}\mathbf{n}^T\mathbf{B}\mathbf{n}\right\}=\trace(\mathbf{A}\mathbf{R})\trace(\mathbf{B}\mathbf{R})+\trace(\mathbf{A}\mathbf{R}\mathbf{B}^T\mathbf{R})+\text{sp}(\mathbf{A}\mathbf{R}\mathbf{B}\mathbf{R})$.
\item $\mathbf{n}\sim\mathcal{N}_{\mathbb{C}^N}(\mathbf{0},\mathbf{R})\Longrightarrow\E\left\{\mathbf{n}^*\mathbf{C}\mathbf{n}\mathbf{n}^*\mathbf{w}\right\}=0$ and $\E\left\{\mathbf{n}^*\mathbf{C}\mathbf{n}\mathbf{n}^*\mathbf{D}\mathbf{n}\right\}=\trace(\mathbf{C}\mathbf{R})\trace(\mathbf{D}\mathbf{R})+\trace(\mathbf{C}\mathbf{R}\mathbf{D}\mathbf{R})$.
\end{enumerate}

Proof: It suffices to prove the solution for $\mathbf{n}\sim\mathcal{N}_{\mathbb{C}^N}(\mathbf{0},\mathbf{I})$, because then for $\tilde{\mathbf{n}}=\mathbf{L}\mathbf{n}$ with $\mathbf{L}\mathbf{L}^*=\mathbf{R}$ holds $\tilde{\mathbf{n}}\sim\mathcal{N}(\mathbf{0},\mathbf{R})$, so that the statements for $\mathbf{L}^*\mathbf{A}\mathbf{L}$, $\mathbf{L}\mathbf{u}$ can be applied.

Initially, let $\mathbf{n}\sim\mathcal{N}_{\mathbb{R}^N}\left(\mathbf{0},\mathbf{I}\right)$ then
\begin{equation*}
\mathbf{n}^T\mathbf{D}\mathbf{n}\mathbf{n}^T\mathbf{u}=\sum_{\nu,\mu,\lambda}n_\nu D_{\nu\mu}n_{\mu}n_{\lambda}u_{\lambda}
\end{equation*}
where if two indices are the same, the expected value of the remaining factors is zero and if three indices are the same, then the third moment is zero. Thus, for real as well as for complex $\mathbf{n}$, the third moment is always zero.
\begin{align*}
\mathbf{n}^T\mathbf{A}\mathbf{n}\mathbf{n}^T\mathbf{B}\mathbf{n}=&\sum_{\nu,\mu}n_\nu A_{\nu\mu}n_{\mu}\sum_{\lambda,\tau}n_{\lambda}B_{\lambda\tau}n_{\tau}\\
=&\sum_{\nu}n^4\nu A_{\nu\nu}B_{\nu\nu}+\sum_{\nu\neq\lambda}n_{\nu}^2A_{\nu\nu}n_{\lambda}^2B_{\lambda\lambda}+\underbrace{\sum_{\nu,\lambda\neq\tau}n_{\nu}^2A_{\nu\nu}n_{\lambda}B_{\lambda\tau}n_{\tau}}_{\text{expected value $=0$}}+\underbrace{\sum_{\lambda,\nu\neq\mu}n_\nu A_{\nu\mu}n_{\mu}n_{\lambda}^2B_{\lambda\lambda}}_{\text{expected value $=0$}}\notag\\
&+\underbrace{\sum_{\nu\neq\mu,\lambda\neq\tau}n_\nu A_{v\mu}n_{\mu}n_{\lambda}B_{\lambda\tau}n_{\tau}}_{\text{Expected value $\neq0$ only if $\nu=\lambda\wedge\mu=\tau$ or $\nu=\tau\wedge\mu=\lambda$}}.
\end{align*}
Thus,
\begin{align*}
\E\left\{\mathbf{n}^T\mathbf{A}\mathbf{n}\mathbf{n}^T\mathbf{B}\mathbf{n}\right\}=&\sum_{\nu}3A_{\nu\nu}B_{\nu\nu}+\sum_{\nu\neq\lambda}A_{\nu\nu}B_{\lambda\lambda}+\sum_{\nu\neq\mu}A_{\nu\mu}B_{\nu\mu}+\sum_{\nu\neq\mu}A_{\nu\mu}B_{\mu\nu}\\
=&\sum_{\nu,\mu}\left(A_{\nu\nu}B_{\mu\mu}+A_{\nu\mu}B_{\nu\mu}+A_{\nu\mu}B_{\mu\nu}\right)\\
=&\trace(\mathbf{A})\trace(\mathbf{B})+\trace(\mathbf{A}\mathbf{B}^T)+\trace(\mathbf{A}\mathbf{B})
\end{align*}
so Condition 1 of the lemma holds.

Now let $\mathbf{n}=\mathbf{a}+j\mathbf{b}$ with $\mathbf{a},\mathbf{b}\sim\mathcal{N}_{\mathbb{R}^N}(\mathbf{0},\mathbf{I})$ Then
\begin{align*}
(\mathbf{a}-j\mathbf{b})^T\mathbf{C}(\mathbf{a}+j\mathbf{b})(\mathbf{a}-j\mathbf{b})\mathbf{D}(\mathbf{a}+j\mathbf{b})=&
\mathbf{a}^T\mathbf{C}\mathbf{a}\mathbf{a}^T\mathbf{D}\mathbf{a}+\mathbf{a}^T\mathbf{C}\mathbf{a}\mathbf{b}^T\mathbf{D}\mathbf{b}-j\underbrace{\mathbf{a}^T\mathbf{C}\mathbf{a}(\mathbf{b}^T\mathbf{D}\mathbf{a}-\mathbf{a}^T\mathbf{D}\mathbf{b})}\notag\\
&+\mathbf{b}^T\mathbf{C}\mathbf{b}\mathbf{a}^T\mathbf{D}\mathbf{a}+\mathbf{b}^T\mathbf{C}\mathbf{b}\mathbf{b}^T\mathbf{D}\mathbf{b}-j\underbrace{\mathbf{b}^T\mathbf{C}\mathbf{b}(\mathbf{b}^T\mathbf{D}\mathbf{a}-\mathbf{a}^T\mathbf{D}\mathbf{b})}\notag\\
&-.j\underbrace{\mathbf{a}^T\mathbf{D}\mathbf{a}(\mathbf{b}^T\mathbf{C}\mathbf{a}-\mathbf{a}^T\mathbf{C}\mathbf{b})}-j\underbrace{\mathbf{b}^T\mathbf{D}\mathbf{b}(\mathbf{b}^T\mathbf{C}\mathbf{a}-\mathbf{a}^T\mathbf{C}\mathbf{b})}\notag\\
&-(\mathbf{b}^T\mathbf{C}\mathbf{a}-\mathbf{a}^T\mathbf{C}\mathbf{b})(\mathbf{b}^T\mathbf{D}\mathbf{a}-\mathbf{a}^T\mathbf{D}\mathbf{b})
\end{align*}
where the braced terms (all of the complex terms) have an expected value of zero.

Using Part 1 of the lemma, one gets
\begin{align*}
&\E\left\{(\mathbf{a}-j\mathbf{b})^T\mathbf{C}(\mathbf{a}+j\mathbf{b})(\mathbf{a}-j\mathbf{b})^T\mathbf{D}(\mathbf{a}+j\mathbf{b})\right\}=2(\trace\mathbf{C}\trace\mathbf{D}+\trace\mathbf{C}\mathbf{D}^T+\trace\mathbf{C}\mathbf{D})+2\trace\mathbf{C}\trace\mathbf{D}\notag\\
&-\E\{\underbrace{\mathbf{b}^T\mathbf{C}\mathbf{a}\mathbf{a}^T\mathbf{D}^T\mathbf{b}}_{=\trace\mathbf{C}\mathbf{D}^T}-\underbrace{\mathbf{b}^T\mathbf{C}\mathbf{a}\mathbf{a}^T\mathbf{D}\mathbf{b}}_{\trace\mathbf{C}\mathbf{D}}-\underbrace{\mathbf{a}^T\mathbf{C}\mathbf{b}\mathbf{b}^T\mathbf{D}\mathbf{a}}_{\trace\mathbf{C}\mathbf{D}}+\underbrace{\mathbf{a}^T\mathbf{C}\mathbf{b}\mathbf{b}^T\mathbf{D}^T\mathbf{a}}_{\trace\mathbf{C}\mathbf{D}^T}\}\\
=&4\left(\trace(\mathbf{C})\trace(\mathbf{D})+\trace\mathbf{C}\mathbf{D}\right)
\end{align*}
for $\mathbf{a}\mathbf{b}\sim\mathcal{N}_{\mathbb{R}^N}(\mathbf{0},\frac{1}{2}\mathbf{I})$. Consequently, $\mathbf{a}+j\mathbf{b}\sim\mathcal{N}_{\mathbb{C}^N}(\mathbf{0},\mathbf{I})$. One thus has
\begin{equation*}
\E\left\{\mathbf{n}^*\mathbf{C}\mathbf{n}\mathbf{n}^*\mathbf{D}\mathbf{n}\right\}=\trace(\mathbf{C})\trace(\mathbf{D})+\trace\mathbf{C}\mathbf{D}
\end{equation*}
which proves Part 2 of the lemma. Thus, one can write
\begin{align*}
\E\left\{Q^2\right\}=&\E\left\{\mathbf{z}^*\boldsymbol{\Gamma}\mathbf{z}\mathbf{z}^*\boldsymbol{\Gamma}\mathbf{z}\right\}\quad\text{for $\mathbf{z}\sim\mathcal{N}_{\mathbb{C}^N}(\mathbf{0},\mathbf{R})$}\\
=&\left(\trace(\boldsymbol{\Gamma}\mathbf{R})\right)^2+\trace\boldsymbol{\Gamma}\mathbf{R}\boldsymbol{\Gamma}\mathbf{R}
\end{align*}
and
\begin{align}
\var\left\{\mathbf{Q}\right\}=&\E\left\{Q^2\right\}-\E\left\{Q\right\}^2\notag\\
=&\trace\boldsymbol{\Gamma}\mathbf{R}\boldsymbol{\Gamma}\mathbf{R}.\label{eqA4-3}
\end{align}

\subsection{The Probability Distribution of Averaged Hermite Forms of Complex Normally Distributed Random Variables}
\setcounter{equation}{0}

Let
\begin{equation*}
q=\sum_{i=1}^K\mathbf{b}_i^*\mathbf{G}\mathbf{b}_i
\end{equation*}
where $\mathbf{b}_i\sim\mathcal{N}_{\mathbb{C}^N}(\mathbf{0},\mathbf{B})$, $\mathbf{G},\mathbf{B}$ are positive-definite $N\times N$ Hermitian matrices.

\begin{figure}[h]
\centering
\includegraphics[width=0.5\textwidth]{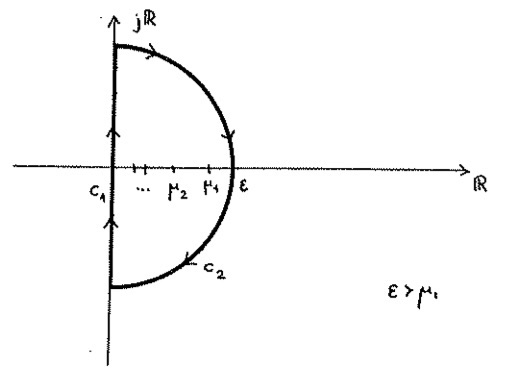}
\caption{\label{Appendix5Fig1}}
\end{figure}

The characteristic function of $q$ is according to Goodman \cite{ref30}
\begin{align*}
\Psi_q(\theta)=&\frac{1}{|\mathbf{I}-j\theta\mathbf{B}\mathbf{G}|^K}\\
=&\frac{1}{\prod_{i=1}^N(1-j\theta\lambda_i)^K}
\end{align*}
when $\lambda_1,\ldots,\lambda_N$ are the eigenvalues from $\mathbf{B}\mathbf{G}$ (or $\mathbf{G}\mathbf{B}$ or $\mathbf{L}^*\mathbf{G}\mathbf{L}$ for $\mathbf{B}=\mathbf{L}\mathbf{L}^*$). Continuing the simplification,
\begin{align*}
\Psi_q(\theta)=\frac{\prod_{i=1}^N\mu_i^K}{\prod_{i=1}^N(\mu_i-j\theta)^K}
\end{align*}
with $\mu_i=\frac{1}{\lambda_i}$. Thus,
\begin{align*}
p_q(x)=&\frac{1}{2\pi}\int_{-\infty}^\infty e^{-j\theta x}\Psi_q(\theta)d\theta\\
=&\frac{\prod\mu_i^K}{2\pi j}\int_{-\infty}^\infty\frac{e^{-j\theta x}}{\prod_{i=1}^K(\mu_i-j\theta)^K}jd\theta\\
=&\frac{\prod\mu_i^K}{2\pi j}\lim\limits_{\epsilon\rightarrow\infty}\int\limits_{c_\epsilon}\frac{e^{-xz}}{\prod (\mu_i-z)^K}dz
\end{align*}
where $c_\epsilon$ is the way shown in Fig. \ref{Appendix5Fig1}
where $c_\epsilon=c_1+c_2$ with $c_1=jt,t\in(-\epsilon,\epsilon)$ and $c_2=\epsilon e^{-jt},t\in\left[-\frac{\pi}{2},\frac{\pi}{2}\right]$, and $\int\limits_{c_2}$ goes to zero for $\epsilon\rightarrow\infty$. Thus,
\begin{equation*}
p_q(x)=\prod_{i=1}^N\mu_i^K\sum_{k=1}^N\text{Residual }(f,\mu_k)\cdot\text{number of rotations }(c,\mu_k)
\end{equation*}
for
\begin{equation*}
f(z)=\frac{e^{-xz}}{\prod_{i=1}^N(\mu_i-z)^K}
\end{equation*}
with rotation number $(c,\mu_k)=-1$ for all $k=1,\ldots, N$ (negative rotation direction) and residual $\left(\frac{g(z)}{(z-a)^K},a\right)=\frac{g^{(k-1)}(a)}{(k-1)!}$, the function $p_q$ can finally be computed as:
\begin{equation*}
p_q(x)=\left.\prod_{i=1}\mu_i^K\frac{(-1)^{NK+1}}{(K-1)!}\sum_{l=1}^N\frac{\partial^{K-1}}{\partial z^{K-1}}\frac{e^{-xz}}{\prod_{i\neq l}(z-\mu_l)^K}\right|_{z=\mu_l}.
\end{equation*}

\underline{Thus, for $N=1$}:
\begin{equation*}
\left.\frac{\partial^{K-1}}{\partial z^{K-1}}e^{-xz}\right|_{\mu}=(-x)^{K-1}e^{-x\mu}
\end{equation*}
so
\begin{equation*}
p_{1,K}=\mu^K\frac{x^{K-1}}{(K-1)!}e^{-x\mu}.
\end{equation*}

\underline{$K=1$; $N$ Arbitrary}

\begin{equation*}
\text{Res }\left(\frac{e^{-xz}}{\prod_l(z-\mu_l)},\mu_k\right)=\frac{e^{-x\mu_k}}{\prod_{l\neq k}(\mu_k-\mu_l)}.
\end{equation*}
Thus,
\begin{equation*}
p_{N,1}(x)=-\prod_{i=1}^N\mu_i\sum_{k=1}^N\frac{e^{-x\mu_k}}{\prod_{l\neq k}(\mu_k-\mu_l)}
\end{equation*}
and for $N=2$
\begin{equation*}
p_{2,1}(x)=\frac{\mu_1\mu_2}{\mu_1-\mu_2}\left(e^{-x\mu_2}-e^{-x\mu_1}\right).
\end{equation*}

\underline{$N=2,K=2$}

\begin{align*}
\text{Res }\left(\frac{e^{-xz}}{(z-\mu_2)^2},\mu_1\right)=-\frac{e^{-x\mu_1}}{(\mu_1-\mu_2)^3}(2+(\mu_1-\mu_2)x)\\
\text{Res }\left(\frac{e^{-xz}}{(z-\mu_1)^2},\mu_2\right)=-\frac{e^{-x\mu_2}}{(\mu_2-\mu_1)^3}(2-(\mu_1-\mu_2)x)
\end{align*}
so
\begin{equation*}
p_{2,2}(x)=\frac{\mu_1^2\mu_2^2}{(\mu_1-\mu_2)^2}x\left(e^{-x\mu_2}+e^{-x\mu_1}\right)-\frac{2\mu_1^2\mu_2^2}{(\mu_1-\mu_2)^3}\left(e^{-x\mu_2}-e^{-x\mu_1}\right).
\end{equation*}

\underline{$N=2,K=3$}
\begin{align*}
\text{Res }\left(\frac{e^{-xz}}{(z-\mu_2)^3},\mu_1\right)=&\left.\frac{1}{2}\left(\frac{e^{-xz}}{(z-\mu_2)^3}\right)''\right|_{z=\mu_1}\\
=&\frac{1}{2}\frac{e^{-x\mu_1}}{(\mu_1-\mu_2)^3}\left(\frac{12}{(\mu_1-\mu_2)^2}+\frac{6x}{\mu_1-\mu_2}+x^2\right).
\end{align*}
Correspondingly,
\begin{equation*}
\text{Res }\left(\frac{e^{-xz}}{(z-\mu_1)^3},\mu_2\right)=-\frac{1}{2}\frac{e^{-x\mu_2}}{(\mu_1-\mu_2)^3}\left(\frac{12}{(\mu_1-\mu_2)^2}-\frac{6x}{\mu_1-\mu_2}+x^2\right).
\end{equation*}
Thus,
\begin{align*}
p_{2,3}(x)=&\frac{1}{2}\frac{\mu_1^3\mu_2^3}{(\mu_1-\mu_2)^3}\left[\left(\frac{12}{(\mu_1-\mu_2)^2}+x^2\right)\left(e^{-x\mu_2}-e^{-x\mu_1}\right)-\frac{6x}{\mu_1-\mu_2}\left(e^{-x\mu_2}+e^{-x\mu_1}\right)\right]\\
=&\left(\frac{\mu_1}{\mu_1-\mu_2}\right)^3\frac{\mu_2^3}{2}x^2e^{-x\mu_2}-\left(\frac{\mu_2}{\mu_1-\mu_2}\right)^3\frac{\mu_1^3}{2}x^2e^{-x\mu_1}\notag\\
&-3\frac{\mu_1\mu_2}{(\mu_1-\mu_2)^2}\left[\frac{\mu_1^2\mu_2^2}{(\mu_1-\mu_2)^2}x\left(
e^{-x\mu_2}+e^{-x\mu_1}\right)-2\frac{\mu_1^2\mu_2^2}{(\mu_1-\mu_2)^3}\left(e^{-x\mu_2}-e^{-x\mu_1}\right)\right].
\end{align*}

\subsection{The Computation of the Integral in (7.2-3)}
\setcounter{equation}{0}

One only has to compute the integrals in (7.2-4, i) because part (ii) follows from (7.2-2, iv).

For $M=1$, according to (7.2.2, ii)
\begin{align*}
\beta_{1,K}=&K\text{\ae}\int_{-\infty}^{U_\alpha}\phi(x)\frac{\mu^K}{(K-1)!}(K\text{\ae}(U_\alpha-x))^{K-1}e^{-K\text{\ae}(U_\alpha-x)\mu}dx\\
=&\frac{K\text{\ae}\mu^K}{(K-1)!}(-1)^{K-1}\frac{\partial^{K-1}}{\partial \mu^{K-1}}\underbrace{\int_{-\infty}^{U_{\alpha}}\phi(x)e^{-\mu K\text{\ae}(U_{\alpha}-x)}dx}_{=:I}
\end{align*}
where
\begin{equation*}
\text{\ae}=\sigma^2\sqrt{\frac{N-\hat{M}}{K}}.
\end{equation*}

Regarding the integral $I$,
\begin{align*}
I=&e^{-\tau U_{\alpha}}\int_{-\infty}^{U_{\alpha}}\phi(x)e^{\tau x} dx,\quad(\tau=\mu K \text{\ae})\\
=&e^{-\tau U_{\alpha}}\left[\left.\frac{1}{\tau}e^{\tau x}\phi(x)\right|_{-\infty}^{U_{\alpha}}-\frac{1}{\tau}\int_{-\infty}^{u_\alpha}e^{\tau x}\phi'(x)dx\right]\quad\text{integration by parts}\\
=&\frac{1-\alpha}{\tau}-\frac{e^{-\tau U_{\alpha}}}{\tau\sqrt{2\pi}}\underbrace{\int_{-\infty}^{U_{\alpha}}e^{-\frac{x^2}{2}+\tau x-\frac{\tau^2}{2}}dx}_{=\int_{-\infty}^{U_{\alpha}-\tau}e^{-\frac{x^2}{2}}dx}e^{\frac{\tau^2}{2}}\\
=&\frac{1}{\tau}\left(1-\alpha-e^{\frac{\tau^2}{2}-\tau U_{\alpha}}\phi(U_{\alpha}-\tau)\right)\\
=&\frac{1}{\tau}\left(1-\alpha-e^{-\frac{U_{\alpha}^2}{2}}e^{\frac{(U_{\alpha}-\tau)^2}{2}}\phi(U_{\alpha}-\tau)\right).\tag{*}
\end{align*}
Thus,
\begin{equation*}
\beta_{11}=1-\alpha-e^{-\frac{U_{\alpha}^2}{2}}e^{\frac{(U_{\alpha}-\mu\text{\ae})^2}{2}}\phi(U_{\alpha}-\mu\text{\ae}).
\end{equation*}

\underline{For $K=2$:}

\begin{align*}
\frac{\partial I}{\partial \mu}=&K\text{\ae}\frac{\partial I}{\partial \tau}\\
=&K\text{\ae}\left[\frac{e^{-\frac{U_{\alpha}^2}{2}}}{\tau}\left(\left(U_{\alpha}-\tau\right)e^{\frac{(U_{\alpha}-\tau)^2}{2}}\phi(U_{\alpha}-\tau)+e^{\frac{(U_{\alpha}-\tau)^2}{2}}\frac{e^{-\frac{(U_{\alpha}-\tau)^2}{2}}}{\sqrt{2\pi}}\right)-\frac{I}{\tau}\right]\\
=&\frac{e^{-\frac{U_{\alpha}^2}{2}}}{\mu}\left((U_{\alpha}-\tau)e^{\frac{(U_{\alpha}-\tau)^2}{2}}\phi(U_{\alpha}-\tau)+\frac{1}{\sqrt{2\pi}}\right)-\frac{I}{\mu}.
\end{align*}
Thus,
\begin{equation*}
\beta_{12}=2\text{\ae}\mu I-2\text{\ae}\mu e^{-\frac{U_{\alpha}^2}{2}}\left[(U_{\alpha}-\tau)e^{\frac{(U_{\alpha}-\tau)^2}{2}}\phi(U_{\alpha}-\tau)+\frac{1}{\sqrt{2\pi}}\right].
\end{equation*}

\underline{For $K=3$:}
\begin{align*}
\frac{\partial^2I}{\partial \mu^2}=&(K\text{\ae})^2\frac{\partial^2 I}{\partial \tau^2}\quad\text{define $\Delta=U_{\alpha}-\tau$}\\
=&(K\text{\ae})^2\left[\frac{\frac{\partial}{\partial \tau}e^{-\frac{U_{\alpha}^2}{2}}\left(\Delta e^{\frac{\Delta^2}{2}}\phi(\Delta)+\frac{1}{\sqrt{2\pi}}\right)-\frac{\partial I}{\partial \tau}}{\tau}-\frac{e^{-\frac{U_{\alpha}^2}{2}}\left(\Delta e^{\frac{\Delta^2}{2}}\phi(\Delta)+\frac{1}{\sqrt{2\pi}}\right)-I-I}{\tau^2}\right]\\
=&(K\text{\ae})^2\left[\frac{2 I}{\tau^2}-\frac{2e^{-\frac{U_{\alpha}^2}{2}}\left(\Delta e^{\frac{\Delta^2}{2}}\phi(\Delta)+\frac{1}{\sqrt{2\pi}}\right)}{\tau^2}+\frac{e^{-\frac{U_{\alpha}^2}{2}}}{\tau}\left(\Delta_{\tau}e^{\frac{\Delta^2}{2}}\phi(\Delta)+\Delta^2\Delta_\tau e^{\frac{\Delta^2}{2}}\phi(\Delta)+\Delta e^{\frac{\Delta^2}{2}}e^{-\frac{\Delta^2}{2}}\frac{\Delta_\tau}{\sqrt{2\pi}}\right)\right]\\
=&(K\text{\ae})^2\left[\frac{2I}{\tau}-\frac{e^{-\frac{U_{\alpha}^2}{2}}}{\tau}\left(e^{\frac{\Delta^2}{2}}\phi(\Delta)(1+\Delta^2)+\frac{\Delta}{\sqrt{2\pi}}\right)\right].
\end{align*}
Thus,
\begin{align*}
\beta_{13}=&K\text{\ae}\mu\left[I-\frac{K\text{\ae}\mu}{2}e^{-\frac{U_{\alpha}}{2}}\left(e^{\frac{\Delta^2}{2}}\phi(\Delta)(1+\Delta^2)+\frac{\Delta}{\sqrt{2\pi}}\right)\right]\\
=&1-\alpha-\psi f(\Delta)-\frac{\tau^2}{2}\psi\left(f(\Delta)(1+\Delta^2)+\frac{\Delta}{\sqrt{2\pi}}\right)
\end{align*}
with
\begin{align*}
f(x)=&e^{\frac{x^2}{2}}\phi(x)\\
\psi=&e^{-\frac{U_{\alpha}^2}{2}}.
\end{align*}

\newpage
\section*{Résumé}

\begin{tabular}{ll}
14 March 1948				&Born in Burg on the island of Fehmarn\\
1954--1958				&Attended primary school (Volkschule) in Hanau am Main\\
1958--1966				&Attended linguistic secondary school (neusprachliches Gymnasium) in Neuß am Rhein\\
5 Nov. 1966				&Passed the Abitur on the linguistic secondary school in Neuß am Rhein\\
1 Jan. 1967 -- 1 Jan. 1969		&Service with the German military\\
April 1969					&Beginning of a degree program in Mathematics with a minor in Physics at the University\\
						&of Cologne\\
16 July 1975				&Passed the primary test for the degree of Diplom in Mathematics at the University of\\
						&Cologne\\
Since 1 Sep. 1975			&Researcher at the Research Institute for Radio and Mathematics (Forschungsinstitut für\\
						&Funk and Mathematik) in Wachtberg-Werthhoven
\end{tabular}

\bibliographystyle{IEEEtran}
\bibliography{ThesisTrans}

\begin{thebibliography}{10}
\providecommand{\url}[1]{#1}
\csname url@samestyle\endcsname
\providecommand{\newblock}{\relax}
\providecommand{\bibinfo}[2]{#2}
\providecommand{\BIBentrySTDinterwordspacing}{\spaceskip=0pt\relax}
\providecommand{\BIBentryALTinterwordstretchfactor}{4}
\providecommand{\BIBentryALTinterwordspacing}{\spaceskip=\fontdimen2\font plus
\BIBentryALTinterwordstretchfactor\fontdimen3\font minus
  \fontdimen4\font\relax}
\providecommand{\BIBforeignlanguage}[2]{{%
\expandafter\ifx\csname l@#1\endcsname\relax
\typeout{** WARNING: IEEEtran.bst: No hyphenation pattern has been}%
\typeout{** loaded for the language `#1'. Using the pattern for}%
\typeout{** the default language instead.}%
\else
\language=\csname l@#1\endcsname
\fi
#2}}
\providecommand{\BIBdecl}{\relax}
\BIBdecl

\bibitem{ref1}
M.~I. Skolnik, \emph{Radar Handbook}.\hskip 1em plus 0.5em minus 0.4em\relax
  McGraw Hill, 1970.

\bibitem{ref2}
K.~{von Schlachta}, ``Über eine {K}lassifizierungsmöglichkeit von
  {R}adarzielen mittels kohärent gemessener {E}chosignale,'' Ph.D.
  dissertation, TU Berlin, 1977.

\bibitem{ref18}
J.~A. Stuller, ``Generalized likelihood signal resolution,'' \emph{IEEE Trans.
  IT}, vol.~21, no.~3, May 1975.

\bibitem{ref3}
H.~Witting and G.~Nölle, \emph{Angewandte mathematische {S}tatistik}.\hskip
  1em plus 0.5em minus 0.4em\relax Stuttgart: Teubner, 1970.

\bibitem{ref30}
N.~R. Goodman, ``Statistical analysis based on a certain multivariate complex
  gaussian distribution (an introduction),'' \emph{Ann. of Math. Stat.},
  vol.~34, 1963.

\bibitem{ref23}
L.~Ljung, ``Analysis of recursive stochastic algorithms,'' \emph{IEEE
  Transactions on Automatic Control}, vol.~22, no.~4, Aug. 1977.

\bibitem{ref4}
R.~T. Lacoss, ``Data adaptive spectral analysis,'' \emph{Geophysics}, vol.~36,
  pp. 661--675, Aug. 1971.

\bibitem{ref5}
T.~J. Ulrych and T.~N. Bishop, ``Maximum entropy spectral analysis and
  autoregressive decomposition,'' \emph{Rev. Geophysics and Space Phys.},
  vol.~13, pp. 183--200, Feb. 1975.

\bibitem{ref6}
G.~O. Young, ``Optimum space-time signal processing and parameter estimation,''
  \emph{IEEE Transactions on Aerospace and Electronic Systems}, vol.~4, no.~3,
  May 1968.

\bibitem{ref7}
G.~O. Young and J.~E. Howard, ``Applications of space-time decision and
  estimation theory to antenna processing system-design,'' \emph{Proceedings of
  the IEEE}, vol.~38, no.~5, May 1970.

\bibitem{ref8}
A.~A. Ksienki and R.~B. McGhee, ``A decision theoretic approach to the angular
  resolution and parameter estimation of multiple targets,'' \emph{IEEE
  Transactions on Aerospace and Electronic Systems}, vol.~4, no.~3, May 1968.

\bibitem{ref9}
W.~D. Wirth, ``Entdeckung und {P}arameterschätzung bei {Z}ielen mit geringem
  {W}inkelabstand,'' FFM, Tech. Rep. 192, Oct. 1972.

\bibitem{ref10}
W.~D. White, ``Low-angle radar tracking in the presence of multipath,''
  \emph{IEEE Transactions on Aerospace and Electronic Systems}, vol.~10, no.~6,
  Nov. 1974.

\bibitem{ref11}
G.~E. Pollon and G.~W. Lank, ``Angular tracking of two closely spaced radar
  targets,'' \emph{IEEE Transactions on Aerospace and Electronic Systems},
  vol.~4, no.~4, Jul. 1968.

\bibitem{ref12}
F.~G. Willwerth and I.~Kupiec, ``Array aperture sampling technique for
  multipath compensation,'' \emph{Microwave Journal}, Jun. 1976.

\bibitem{ref13}
K.~Bauer, ``Direkt-lösende {A}lgorithmen für die {I}nterferenzanalyse {T}eil
  {I} und {II},'' \emph{Frequenz}, vol.~30, no. 4 and 5, 1976.

\bibitem{ref14}
D.~C. Rife and R.~R. Boorstyn, ``Multiple tone parameter estimation from
  discrete-time observations,'' \emph{Bell Syst. Techn. J.}, vol.~55, no.~9,
  Nov. 1976.

\bibitem{ref15}
J.~E. Ehrenberg, T.~E. Ewart, and R.~D. Morris, ``Signal processing techniques
  for resolving individual pulses in a multipath signal,'' \emph{J. Acoust.
  Soc. Am.}, vol.~63, no.~6, Jun. 1978.

\bibitem{ref16}
I.~N. {El Behery} and R.~H. McPhie, ``Maximum likelihood estimation of the
  number, directions, and strengths of point radio sources from variable
  baseline interferometer data,'' \emph{IEEE Trans. AP}, vol.~26, no.~2, Mar.
  1978.

\bibitem{ref17}
R.~A. Birgenheier, ``Parameter estimation of multiple signals,'' Ph.D.
  dissertation, University of California, Los Angeles, 1972.

\bibitem{ref20}
T.~W. Anderson, \emph{An introduction to Multivariate Statistical
  Analysis}.\hskip 1em plus 0.5em minus 0.4em\relax J. Wiley, 1958.

\bibitem{ref21}
W.~Klingenberg, \emph{Eine {V}orlesung über {D}ifferentialgeometrie}.\hskip
  1em plus 0.5em minus 0.4em\relax Heidelberger Taschenbücher, Springer, 1973.

\bibitem{ref22}
D.~Gromoll, W.~Klingenberg, and W.~Meyer, ``Riemannsche {G}eometrie im
  {G}roßen,'' \emph{Lecture Notes in mathematics}, vol.~55, 1975.

\bibitem{ref24}
M.~T. Wasan, \emph{Stochastic Approximation}.\hskip 1em plus 0.5em minus
  0.4em\relax Cambridge University Press, 1969.

\bibitem{ref26}
I.~Kanter, ``Multiple {G}aussian targets: the track-on-jam problem,''
  \emph{IEEE Transactions on Aerospace and Electronic Systems}, vol.~13, no.~6,
  Nov. 1977.

\bibitem{ref25}
------, ``The ratio of functions of random variables,'' \emph{IEEE Transactions
  on Aerospace and Electronic Systems}, vol.~13, no.~6, Nov. 1977.

\bibitem{ref27}
R.~D. Martin and C.~J. Masreliez, ``Robust estimation via stochastic
  approximation,'' \emph{IEEE Transactions on Information Theory}, vol.~21,
  no.~3, May 1975.

\bibitem{ref28}
G.~Hüschelrath, ``Ein rechnergesteuertes {N}ahfeldvermessungsverfahren für
  elektronisch steuerbare {G}ruppenantennen,'' Ph.D. dissertation, TH Aachen,
  Feb. 1978.

\bibitem{ref29}
J.~C. Demaret and A.~Garcet, ``Sum of exponential random variables,''
  \emph{AEÜ}, vol.~31, no.~11, 1977.

\bibitem{ref19}
J.~Stoer, \emph{Einführung in die {N}umerische {M}athematik {I}}.\hskip 1em
  plus 0.5em minus 0.4em\relax Springer, 1973.

\end{thebibliography}

\end{document}